\documentclass[a4paper,12pt]{article}
\usepackage{graphicx}
\usepackage[latin1]{inputenc}
\usepackage{pstricks,pst-node,pst-text,pst-3d}
\usepackage{colordvi}
\usepackage{axodraw}
\usepackage{epsfig}
\usepackage{moreverb}

\begin{document}
\bibliographystyle{unsrt}


\newcommand{\be}{\begin{equation}}
\newcommand{\beq}{\begin{equation}}
\newcommand{\eeq}{\end{equation}}
\newcommand{\ee}{\end{equation}}

\newcommand{\refeq}[1]{Eq.\ref{eq:#1}}
\newcommand{\refig}[1]{Fig.\ref{fig:#1}}
\newcommand{\refsec}[1]{Sec.\ref{sec:#1}}

\newcommand{\beqn}{\begin{eqnarray}}
\newcommand{\eeqn}{\end{eqnarray}}
\newcommand{\bea}{\begin{eqnarray}}
\newcommand{\ena}{\end{eqnarray}}
\newcommand{\ra}{\rightarrow}
\newcommand{\susy}{{{\cal SUSY}$\;$}}
\newcommand{\su}{$ SU(2) \times U(1)\,$}

\newcommand{\gag}{$\gamma \gamma$ }
\newcommand{\gagt}{\gamma \gamma }
\newcommand{\gam}{\gamma \gamma }
\def\W{{\mbox{\boldmath $W$}}}
\def\B{{\mbox{\boldmath $B$}}}
\def\V{{\mbox{\boldmath $V$}}}
\newcommand{\np}{{\em Nucl.\,Phys.\,}}
\newcommand{\pl}{{\em Phys.\,Lett.\,}}
\newcommand{\pr}{{\em Phys.\,Rev.\,}}
\newcommand{\prl}{{\em Phys.\,Rev.\,Lett.\,}}
\newcommand{\prep}{{\em Phys.\,Rep.\,}}
\newcommand{\zp}{{\em Z.\,Phys.\,}}
\newcommand{\sovjnp}{{\em Sov.\, J.\ Nucl.\, Phys.\, }}
\newcommand{\nuclinst}{{\em Nucl.\, Instrum.\, Meth.\, }}
\newcommand{\annp}{{\em Ann.\, Phys.\, }}
\newcommand{\intjmp}{{\em Int.\, J.\, of Mod.\,  Phys.\, }}

\newcommand{\eps}{\epsilon}
\newcommand{\mw}{M_{W}}
\newcommand{\mww}{M_{W}^{2}}
\newcommand{\mwmw}{M_{W}^{2}}
\newcommand{\mhmh}{M_{H}^2}
\newcommand{\mz}{M_{Z}}
\newcommand{\mzz}{M_{Z}^{2}}

\newcommand{\cw}{c_W}
\newcommand{\sw}{s_W}
\newcommand{\tw}{\tan\theta_W}
\def\tww{\tan^2\theta_W}
\def\stw{s_{2w}}

\newcommand{\smw}{s_M^2}
\newcommand{\cmw}{c_M^2}
\newcommand{\seff}{s_{{\rm eff}}^2}
\newcommand{\ceff}{c_{{\rm eff}}^2}
\newcommand{\seffl}{s_{{\rm eff\;,l}}^{2}}
\newcommand{\sww}{s_W^2}
\newcommand{\cww}{c_W^2}
\newcommand{\swo}{s_W}
\newcommand{\cwo}{c_W}

\newcommand{\epm}{$e^{+} e^{-}\;$}
\newcommand{\epemt}{$e^{+} e^{-}\;$}
\newcommand{\epem}{e^{+} e^{-}\;}
\newcommand{\ememt}{$e^{-} e^{-}\;$}
\newcommand{\emem}{e^{-} e^{-}\;}

\newcommand{\lra}{\leftrightarrow}
\newcommand{\tr}{{\rm Tr}}
\def\ls1{{\not l}_1}
\newcommand{\cms}{centre of mass\hspace*{.1cm}}


\newcommand{\dkg}{\Delta \kappa_{\gamma}}
\newcommand{\dkz}{\Delta \kappa_{Z}}
\newcommand{\dz}{\delta_{Z}}
\newcommand{\dgz}{\Delta g^{1}_{Z}}
\newcommand{\dgzt}{$\Delta g^{1}_{Z}\;$}
\newcommand{\la}{\lambda}
\newcommand{\lag}{\lambda_{\gamma}}
\newcommand{\lambdae}{\lambda_{e}}
\newcommand{\laz}{\lambda_{Z}}
\newcommand{\lnl}{L_{9L}}
\newcommand{\lnr}{L_{9R}}
\newcommand{\lt}{L_{10}}
\newcommand{\lu}{L_{1}}
\newcommand{\ld}{L_{2}}
\newcommand{\eeww}{e^{+} e^{-} \ra W^+ W^- \;}
\newcommand{\eewwt}{$e^{+} e^{-} \ra W^+ W^- \;$}
\newcommand{\epemww}{e^{+} e^{-} \ra W^+ W^- }
\newcommand{\epemwwt}{$e^{+} e^{-} \ra W^+ W^- \;$}
\newcommand{\eennhht}{$e^{+} e^{-} \ra \nu_e \bar \nu_e HH\;$}
\newcommand{\eennhh}{e^{+} e^{-} \ra \nu_e \bar \nu_e HH\;}
\newcommand{\eennht}{$e^{+} e^{-} \ra \nu_e \bar \nu_e H\;$}
\newcommand{\eennh}{e^{+} e^{-} \ra \nu_e \bar \nu_e H\;}
\newcommand{\eettht}{$e^{+} e^{-} \ra t \bar t H\;$}
\newcommand{\eetth}{e^{+} e^{-} \ra t \bar t H\;}
\newcommand{\eezhht}{$e^{+} e^{-} \ra Z H H\;$}
\newcommand{\eezhh}{e^{+} e^{-} \ra Z H H\;}
\newcommand{\eeeeht}{$\epem \ra e^+ e^- H \;$}
\newcommand{\eeeeh}{$\epem \ra e^+ e^-  H$}
\newcommand{\eenngt}{$\epem \ra e^+ e^- \gamma \;$}
\newcommand{\eenng}{$\epem \ra e^+ e^-  \gamma$}

\newcommand{\ppwg}{p p \ra W \gamma}
\newcommand{\wwhh}{W^+ W^- \ra HH\;}
\newcommand{\wwhht}{$W^+ W^- \ra HH\;$}
\newcommand{\ppwz}{pp \ra W Z}
\newcommand{\ppwgt}{$p p \ra W \gamma \;$}
\newcommand{\ppwzt}{$pp \ra W Z \;$}
\newcommand{\gamgamt}{$\gamma \gamma \;$}
\newcommand{\gamgam}{\gamma \gamma \;}
\newcommand{\egamt}{$e \gamma \;$}
\newcommand{\egam}{e \gamma \;}
\newcommand{\gamgamwwt}{$\gamma \gamma \ra W^+ W^- \;$}
\newcommand{\gamgamwwht}{$\gamma \gamma \ra W^+ W^- H \;$}
\newcommand{\gamgamwwh}{\gamma \gamma \ra W^+ W^- H \;}
\newcommand{\gamgamwwhht}{$\gamma \gamma \ra W^+ W^- H H\;$}
\newcommand{\gamgamwwhh}{\gamma \gamma \ra W^+ W^- H H\;}
\newcommand{\ggww}{\gamma \gamma \ra W^+ W^-}
\newcommand{\ggwwt}{$\gamma \gamma \ra W^+ W^- \;$}
\newcommand{\ggwwht}{$\gamma \gamma \ra W^+ W^- H \;$}
\newcommand{\ggwwh}{\gamma \gamma \ra W^+ W^- H \;}
\newcommand{\ggwwhht}{$\gamma \gamma \ra W^+ W^- H H\;$}
\newcommand{\ggwwhh}{\gamma \gamma \ra W^+ W^- H H\;}
\newcommand{\ggwwz}{\gamma \gamma \ra W^+ W^- Z\;}
\newcommand{\ggwwzt}{$\gamma \gamma \ra W^+ W^- Z\;$}

\newcommand{\veps}{\varepsilon}

\newcommand{\ptu}{p_{1\bot}}
\newcommand{\vecptu}{\vec{p}_{1\bot}}
\newcommand{\ptd}{p_{2\bot}}
\newcommand{\vecptd}{\vec{p}_{2\bot}}
\newcommand{\ie}{{\em i.e.}}
\newcommand{\cm}{{{\cal M}}}
\newcommand{\cl}{{{\cal L}}}
\newcommand{\cd}{{{\cal D}}}
\newcommand{\cv}{{{\cal V}}}
\def\slashc{c\kern -.400em {/}}
\def\slashp{p\kern -.400em {/}}
\def\slashq{q\kern -.450em {/}}
\def\slashL{L\kern -.450em {/}}
\def\slashcl{\cl\kern -.600em {/}}
\def\slashr{r\kern -.450em {/}}
\def\slashk{k\kern -.500em {/}}
\def\Ww{{\mbox{\boldmath $W$}}}
\def\B{{\mbox{\boldmath $B$}}}
\def\noi{\noindent}
\def\nn{\noindent}
\def\sm{${\cal{S}} {\cal{M}}\;$}
\def\smn{${\cal{S}} {\cal{M}}$}
\def\smp{${\cal{S}} {\cal{M}}$}
\def\mssm{${\cal{M}} {\cal{S}} {\cal{S}} {\cal{M}}\;$}
\def\mssmp{${\cal{M}} {\cal{S}} {\cal{S}} {\cal{M}}$}
\def\nph{${\cal{N}} {\cal{P}}\;$}
\def\sb{$ {\cal{S}}  {\cal{B}}\;$}
\def\ssb{${\cal{S}} {\cal{S}}  {\cal{B}}\;$}
\def\ssbe{{\cal{S}} {\cal{S}}  {\cal{B}}}
\def\cviol{${\cal{C}}\;$}
\def\pviol{${\cal{P}}\;$}
\def\cpviol{${\cal{C}} {\cal{P}}\;$}

\newcommand{\lgg}{\lambda_1\lambda_2}
\newcommand{\lww}{\lambda_3\lambda_4}
\newcommand{\ppin}{ P^+_{12}}
\newcommand{\pmin}{ P^-_{12}}
\newcommand{\ppout}{ P^+_{34}}
\newcommand{\pmout}{ P^-_{34}}
\newcommand{\sinsq}{\sin^2\theta}
\newcommand{\cossq}{\cos^2\theta}
\newcommand{\yt}{y_\theta}
\newcommand{\hppll}{++;00}
\newcommand{\hpmll}{+-;00}
\newcommand{\hpplt}{++;\lambda_30}
\newcommand{\hpmlt}{+-;\lambda_30}
\newcommand{\hpptt}{++;\lambda_3\lambda_4}
\newcommand{\hpmtt}{+-;\lambda_3\lambda_4}
\newcommand{\dk}{\Delta\kappa}
\newcommand{\klam}{\Delta\kappa \lambda_\gamma }
\newcommand{\kac}{\Delta\kappa^2 }
\newcommand{\lac}{\lambda_\gamma^2 }
\def\gamgamtzz{$\gamma \gamma \ra ZZ \;$}
\def\gamgamtww{$\gamma \gamma \ra W^+ W^-\;$}
\def\gamgamtwwe{\gamma \gamma \ra W^+ W^-}

\def\intfd{ \int \frac{d^4 r}{(2\pi)^4} }
\def\intnd{ \int \frac{d^n r}{(2\pi)^n} }
\def\intnmu{ \mu^{4-n} \int \frac{d^n r}{(2\pi)^n} }
\newcommand{\Dkm}{[(r+k)^2-m_2^2]}
\newcommand{\Dkom}{[(r+k_1)^2-m_2^2]}
\newcommand{\Dkotm}{[(r+k_1+k_2)^2-m_3^2]}
\def\piggt{$\Pi_{\gamma \gamma}\;$}
\def\pigg{\Pi_{\gamma \gamma}}
\newcommand{\mn}{{\mu \nu}}
\newcommand{\mzb}{M_{Z,0}}
\newcommand{\mzbs}{M_{Z,0}^2}
\newcommand{\mwb}{M_{W,0}}
\newcommand{\mwbs}{M_{W,0}^2}
\newcommand{\dgg}{\frac{\delta g^2}{g^2}}
\newcommand{\dee}{\frac{\delta e^2}{e^2}}
\newcommand{\dss}{\frac{\delta s^2}{s^2}}
\newcommand{\dmw}{\frac{\delta \mww}{\mww}}
\newcommand{\dmz}{\frac{\delta \mzz}{\mzz}}
\def\pigz{\Pi_{\gamma Z}}
\def\pizz{\Pi_{Z Z}}
\def\piww{\Pi_{WW}}
\def\pioo{\Pi_{11}}
\def\pitt{\Pi_{33}}
\def\pitq{\Pi_{3Q}}
\def\piqq{\Pi_{QQ}}
\def\delr{\Delta r}
\def\calm{{\cal {M}}}
\def\gww{G_{WW}}
\def\gzz{G_{ZZ}}
\def\goo{G_{11}}
\def\gtt{G_{33}}
\def\szz{s_Z^2}
\def\estk{e_\star^2(k^2)}
\def\sstk{s_\star^2(k^2)}
\def\cstk{c_\star^2(k^2)}
\def\sstz{s_\star^2(\mzz)}
\def\mzst{{M_Z^{\star}}(k^2)^2}
\def\mwst{{M_W^{\star}}(k^2)^2}
\def\epo{\varepsilon_1}
\def\epd{\varepsilon_2}
\def\ept{\varepsilon_3}
\def\dro{\Delta \rho}
\def\gmu{G_\mu}
\def\alpz{\alpha_Z}
\def\danpmz{\Delta\alpha_{{\rm NP}}(\mzz)}
\def\danpk{\Delta\alpha_{{\rm NP}}(k^2)}
\def\calt{{\cal {T}}}
\def\piggh{\pigg^h(s)}
\def\cuv{C_{UV}}
\def\pilr{G_{LR}}
\def\pill{G_{LL}}
\def\dak{\Delta \alpha(k^2)}
\def\damz{\Delta \alpha(\mzz)}
\def\dahmz{\Delta \alpha^{(5)}_{{\rm had}}(\mzz)}
\def\sth{s_{\theta}^2}
\def\cth{c_{\theta}^2}
\newcommand{\siki}[1]{Eq.\ref{eq:#1}}
\newcommand{\zu}[1]{Fig.\ref{fig:#1}}
\newcommand{\setu}[1]{Sec.\ref{sec:#1}}
\newcommand{\anlg}{\tilde\alpha}
\newcommand{\bnlg}{\tilde\beta}
\newcommand{\dnlg}{\tilde\delta}
\newcommand{\enlg}{\tilde\varepsilon}
\newcommand{\knlg}{\tilde\kappa}
\newcommand{\xiw}{\xi_W}
\newcommand{\xiz}{\xi_Z}
\newcommand{\dbr}{\delta_B}
\newcommand{\bothd}{{ \leftrightarrow \atop{\partial^{\mu}} } }

\newcommand{\BARE}[1]{\underline{#1}}
\newcommand{\ZF}[1]{\sqrt{Z}_{#1}}
\newcommand{\ZFT}[1]{\tilde{Z}_{#1}}
\newcommand{\ZH}[1]{\delta Z_{#1}^{1/2}}
\newcommand{\ZHb}[1]{\delta Z_{#1}^{1/2\,*}}
\newcommand{\DM}[1]{\delta M^2_{#1}}
\newcommand{\DMS}[1]{\delta M_{#1}}
\newcommand{\Dm}[1]{\delta m_{#1}}
\newcommand{\tree}[1]{\langle {#1}\rangle}

\newcommand{\Cuv}{C_{UV}}
\newcommand{\logw}{\log M_W^2}
\newcommand{\logz}{\log M_Z^2}
\newcommand{\logh}{\log M_H^2}
\newcommand{\swt}{s_W^2}
\newcommand{\cwt}{c_W^2}
\newcommand{\swf}{s_W^4}
\newcommand{\cwf}{c_W^4}
\newcommand{\MWt}{M_W^2}
\newcommand{\MZt}{M_Z^2}
\newcommand{\MHt}{M_H^2}

\newcommand{\VECsl}[1]{\not{#1}}

\newcommand{\Bphi}{\mbox{\boldmath$\phi$}}

\newcommand{\alphat}{\tilde\alpha}
\newcommand{\betat}{\tilde\beta}
\newcommand{\kappat}{\tilde\kappa}
\newcommand{\deltat}{\tilde\delta}
\newcommand{\epsilont}{\tilde\epsilon}
\newcommand{\mhh}{M_H^2}

\newcommand{\cha}{{\tt CHANEL}$\;$}

\def\al{\alpha}
\def\bt{\beta}
\def\gm{\gamma}
\def\Gm{\Gamma}
\def\et{\eta}
\def\del{\delta}
\def\Del{\Delta}
\def\kp{\kappa}
\def\lm{\lambda}
\def\Lm{\Lambda}
\def\th{\theta}
\def\zt{\zeta}
\def\ro{\rho}
\def\sig{\sigma}
\def\Sig{\Sigma}
\def\eps{\epsilon}
\def\vare{\varepsilon}
\def\vphi{\varphi}
\def\om{\omega}
\def\Om{\Omega}
\def\bar{\overline}
\def\d{{\rm d}}
\def\pdf{\partial}
\def\Int{\int\nolimits}
\def\det{{\rm det}}
\def\non{\nonumber}
\def\eqn{\begin{equation}}
\def\eqne{\end{equation}}
\def\eqa{\begin{eqnarray}}
\def\eqae{\end{eqnarray}}
\def\ary{\begin{array}}
\def\arye{\end{array}}
\def\dsc{\begin{description}}
\def\dsce{\end{description}}
\def\itm{\begin{itemize}}
\def\itme{\end{itemize}}
\def\enu{\begin{enumerate}}
\def\enue{\end{enumerate}}
\def\ct{\begin{center}}
\def\cte{\end{center}}
\def\D{{\cal D}}
\def\bfD{\D}

\def\brique{}
\def\Black{}
\def\noir{}
\def\Blue{}
\def\Green{}
\def\rouge{}
\def\bleu{}
\def\vert{}
\def\vertb{}
\def\vertvert{}

\def\bleuc{}
\def\gris{}
\def\grisn{}
\def\jaune{}
\def\orange{}
\def\violet{}
\def\bleucc{}
\def\bleuv{}
\def\bleuvv{}

\pagestyle{empty}
\begin{titlepage}

\begin{center}

\vspace*{5cm}

{\Large {\bf Automatic Calculations in High Energy Physics
 \\
and }\\ {\tt GRACE} {\bf at one-loop }}

\vspace{8mm}

{\large G. B\'elanger${}^{1)}$, F. Boudjema${}^{1)}$, J.
Fujimoto${}^{2)}$, T. Ishikawa${}^{2)}$, \\ T. Kaneko${}^{2)}$, K.
Kato${}^{3)}$,  Y.
Shimizu${}^{2,4)}$ }\\

\vspace{4mm}

{\it 1) LAPTH$^\dagger$, B.P.110, Annecy-le-Vieux F-74941, France}
\\ {\it
2) KEK, Oho 1-1, Tsukuba, Ibaraki 305--0801, Japan} \\
{\it 3) Kogakuin University, Nishi-Shinjuku 1-24, Shinjuku, Tokyo
163--8677, Japan} \\
{\it 4) Graduate University for Advanced Studies, Hayama,
Miura-gun, Kanagawa, 240-0193, Japan}\\

\vspace{10mm}

\end{center}

\vspace*{\fill}


\vspace*{1cm}

$^\dagger${\small UMR 5108 du CNRS, associ\'ee  \`a l'Universit\'e
de Savoie.} \normalsize

\end{titlepage}

\vspace*{2cm}

\centerline{ {\bf Abstract} } \baselineskip=14pt \noindent

{\small  We describe the main building blocks of a generic
automated package for the calculation of Feynman diagrams. These
blocks include  the generation and creation of a model file, the
graph generation, the symbolic calculation at an intermediate
level of the Dirac and tensor algebra, implementation of the loop
integrals, the generation of the matrix elements or helicity
amplitudes, methods for the phase space integrations and
eventually the event generation. The report focuses on the fully
automated systems for the calculation of physical processes based
on the experience in developing  {\tt GRACE-loop} which is a
general purpose code applicable to one-loop corrections in the
Standard Model. As such, a detailed description of the
renormalisation procedure in the Standard Model is given
emphasizing the central role played by the non-linear gauge fixing
conditions for the construction of such automated codes. These new
gauge-fixing conditions are used as a very efficient means to
check the results of large scale automated computations in the
Standard Model. Their need is better appreciated when it comes to
devising efficient and powerful algorithms for the reduction of
the tensorial structures of the loop integrals and the reduction
of the $N>4$ point-function to lower rank integrals. A new
technique for these reduction algorithms is described. Explicit
formulae for all two-point functions in a generalised non-linear
gauge are given, together with the complete set of counterterms.
We also show how infrared divergences are dealt with in the
system.  We give a comprehensive presentation of some systematic
test-runs which have been performed at the one-loop level for a
wide variety of two-to-two processes to show the validity of the
gauge check. These cover fermion-fermion scattering, gauge boson
scattering into fermions, gauge bosons and Higgs bosons scattering
processes. Comparisons with existing results on some one-loop
computation in the Standard Model show excellent agreement. These
include $e^+ e^- \ra t \bar t, W^+ W^-, ZH$; $\gamma \gamma \ra t
\bar t, W^+ W^-$; $e \gamma \ra e Z, \nu W$ and $W^+ W^- \ra W^+
W^-$. We also briefly recount some recent development concerning
the calculation of one-loop corrections to $3$ body final states
cross sections in \epemt with the help of an automated system.
\newpage
\pagestyle{plain}
\pagenumbering{roman}
\tableofcontents

\newpage


\renewcommand{\theequation}{\thesection.\arabic{equation}}

\setcounter{equation}{0}

\setcounter{page}{1}
\pagenumbering{arabic}

\setcounter{equation}{0}


\renewcommand{\theequation}{\thesection.\arabic{equation}}

\setcounter{equation}{0}

\setcounter{equation}{0}
\section{Introduction}
\subsection{The need for automation  in the
Standard Model}

\def\psl{p\kern -.500em {/}}
\def\ksl{k\kern -.500em {/}}

\newcommand{\grc}{{\tt GRACE}$\;$}
\newcommand{\grcp}{{\tt GRACE}}
\newcommand{\grcl}{{\tt GRACE-loop$\;$}}
Much of the success of the Standard Model, \sm, of the electroweak
interaction rests on the results of the various precision
measurements, notably those of LEP and SLC. These precision
measurements required the knowledge of higher order quantum
corrections. Although the latter are rather involved, calculations
are still under control since the bulk of the observables pertain
to two-body final states. In fact due to the present available
energy, the most precise predictions relate to fermion pair
production, a calculation which is far easier to handle than that
for $W$ pair production even if one leaves out the fact  that for
the latter one needs a full 4-fermion final state calculation.
Next generation machines will involve much higher energies and
luminosities opening up the thresholds for multiparticle
production and/or the need to go beyond one and two-loop radiative
corrections. On the other hand even when one {\em only} considers
three particles in the final state, the complexity increases
tremendously especially within the electroweak framework. So much
so that even a process like $\epem \ra \nu_e \bar{\nu}_e H$ which
would be the main production mechanism for the Higgs at the next
linear collider and where the tree-level calculation receives a
contribution from only a  single (non-resonant) diagram,  a full
one-loop  calculation has only very recently been
completed\cite{eennhradcor2002,eennhletter,Dennereennh1}. For such
processes, hand calculations become quickly intractable and very
much prone to error. Moreover, a complete hand calculation for
such processes is not possible, even for the tree-level cross
sections, as one has to resort to numerical methods for the phase
space integration. Especially for QCD processes, to alleviate some
of the major hurdles in the calculation of matrix elements for
physical observables beyond leading and next-to-leading order, one
has devised some powerful alternatives to the standard
diagrammatic Feynman approach\cite{Bernreview}, with most recently
the development of the
twistor-space\cite{Witten-twistors,Branfhuber-mhv,Khoze-twistors,Bern-twistors}.
However most of them, if not all,  involve at most one massive
particle and a single parameter, the QCD coupling constant.
Moreover the techniques work because of the exact gauge symmetry
of QCD and thus, apart from a handful processes, these methods can
not be carried over to the electroweak theory where the
computations involve a variety of masses and scales. Faced with
these difficulties the need for computers is even more evident for
the calculation of electroweak processes.

Ideally one would like to automatise the complete process of
calculating radiative corrections and multi-particle production
starting from the Lagrangian or the Feynman rules to the cross
section. Automation is, in principle, feasible since most of the
ingredients of perturbation theory are based  on well established
algorithms. With the increase in computer power and storage,
together with possible parallelization, one could deal, in a
relatively short time, with more and more complex projects thus
bypassing the problem of huge output files being produced, at
least in the intermediate stages. The idea of automation of the
calculations in high-energy physics is not new. It dates back to
the 1960's when packages such as
\texttt{SCHOONSCHIP}\cite{schoonschip} and then
\texttt{REDUCE}\cite{Hearn,reduce} had been developed. These are
symbolic manipulation codes that automatise the algebraic parts of
a matrix element evaluation, like traces on Dirac matrices and
contraction of Lorentz indices. Such codes have evolved quite a
lot with applications not only optimised for high-energy physics
like \texttt{FORM}\cite{form} but also more general purpose
programs like \texttt{Mathematica}\cite{mathematica} and
\texttt{Maple}\cite{maple}. Generation of QED Feynman graphs at
any order in the coupling constant was automatised in the late
70's\cite{Sasaki}. One of the first major application of these
early developments in this field was the calculation of the
anomalous magnetic moments of the electron and the
muon\cite{Kinoshita}. The first automatic system incorporating all
the steps for the calculation of a cross section, from Feynman
graph generation, amplitude generation through a \texttt{REDUCE}
source code that produces a FORTRAN code, phase space integration
and event generation with \texttt{BASES/SPRING}\cite{bases} is
\texttt{GRAND}\cite{grand}. It was limited to tree-level processes
in QED. In the early nineties, a few groups  started to develop
packages aiming at the automatisation in the \sm \cite{ailyon}.

\subsection{Different approaches and levels of automation}
A hand calculation of a process at a certain order in perturbation
can follow different methods, approaches, tricks and sometimes
relies on approximations. It is no wonder then that these same
issues and variations have translated into the automation of these
calculations and have led to the implementation of a few softwares
with varying degrees of automation, different domain of
application while exploiting different programming languages, see
Ref.~\cite{autocalreview} for a survey of some of these systems.
Some of the systems are collections of software tools used to help
only certain aspects of the hand calculation. Example are codes
that only generate the Feynman diagrams, like {\tt
QGRAF}\cite{qgraf} or codes for the algebraic and analytic
manipulations on loop diagrams but on a diagram by diagram basis
like {\tt XLOOPS}\cite{xloops}. Others are designed for specific
applications\cite{matad,mincer}, like QCD corrections to some
electroweak processes for example. The report will concentrate
only on the  fully automatised systems that are able to output a
source code for the numerical calculation of cross sections
without any intervention by the user, apart of course from
providing the input which consists in specifying the process. In
reviewing the characteristics of these codes and the different
steps that go into building these tools, we will see that some of
the specialised codes we have just mentioned could be considered
as a module in the long chain that goes from the Lagrangian to the
cross section. To go into the details of how the various steps are
implemented we will have to be more specific, since there is
hardly any standardisation of either the methods, the algorithms
or the computer language. Therefore we will most of the time refer
to the experience we gained in developing \grcp. We therefore
present the case of automation in the \sm at the tree-level and
concentrate more on the one-loop level. A fully automatic system
beyond one-loop has not been constructed yet.

\subsubsection{Automation at tree-level}
In the usual diagrammatic Feynman approach followed by most of the
automated systems, the cross sections can be obtained by computing
directly the unpolarised squared matrix elements or in terms of
the  (helicity) amplitudes using spinor techniques. Although the
computer algorithms for these two techniques can be quite
different, in both cases one needs the Feynman rules and Feynman
graphs to be generated. The  automatic systems, \grcp\cite{grace},
{\tt CompHEP}\cite{comphep}, the {\tt FeynArts-FeynCalc-FormCalc}
package\cite{feynart,FeynArts,feyncalc,formcalc1,formcalc2}, {\tt
Madgraph}\cite{madgraph} and {\tt fdc}\cite{fdc}, follow the
diagrammatic approach with applications to both the \sm and its
supersymmetric version. A detailed description of the Minimal
Supersymmetric Standard Model, \mssmp, of \grc is found in
\cite{chanel-susy}.\\
\noi  It is also possible to make do without  Feynman diagrams and
arrive even more directly at the cross sections. This can be
achieved through an iterative solution of the equations of motion
or by solving the Dyson-Schwinger
equation\cite{code-alpha,helac,code-omega}. This approach leads to
faster codes. However, it has not been extended beyond tree-level.
Two\cite{grace,feyncalc} of the diagrammatic approach codes have
on the other hand been extended to one-loop.\\
\noi Depending on the method in the diagrammatic approach, Dirac
matrices and spinors can be treated as symbolic objects before
being converted to numerical quantities with the help of a
symbolic manipulation system as will be explained in
section~\ref{tree-sqme}. One can also take a more ``numerical"
approach where these objects are combined into numerical entities.
This is treated in section~\ref{tree-sqme}. \\
\noi Even at tree-level, one problem is the size of the output
file when one is dealing with  multi-leg processes. Automatic
systems produce, for a complicated process, a huge source
code\footnote{The size of the problem grows rapidly as the number
of external  particles increases. For example, \(2 \rightarrow N\)
tree process in \(\phi^3\) model has \((2 N - 1)!!\) Feynman
graphs.} which in turn requires a large amount of CPU time for the
numerical evaluation. This necessitates a large disk space, a
human  control over  a large number of batch jobs or could even
necessitate to split the source code into small pieces in order to
make compilers work. In addition to the problem of size, it may be
necessary to write specific kinematics routines when the amplitude
has a complicated structure of singularities, see section
\ref{sec:kinem}. Of course, the non-diagrammatic approach also
requires a proper phase space integration routine.

\subsubsection{Automation at one-loop level}
The problem with the size of the output files and the integration
over phase space are exacerbated for one-loop  processes. These
are however not the major hurdles for extending a tree-level code
to one-loop. One first needs to master all the theoretical
background related to the renormalisation of a model or a theory.
A consistent renormalisation procedure that gives all possible
counterterms, which would have to be implemented in the automatic
code to tackle any one-loop process, needs to be clearly defined.
A symbolic treatment of space-time dimension is inevitable for the
regularisation of ultraviolet divergences. Infrared divergences
will have to be regularised either by a small fictitious mass or
through  dimensional regularisation, DR, \cite{DimRegthV,DimReg2}.
A major investment has to do with a fast and efficient algorithm
for the loop integrations, especially the algorithm for the
reduction of the tensorial structures to the scalar $N$-point
functions and the reduction of the $N>4$ scalar functions to lower
$N$-point functions for codes that allow multi-leg one-loop
integrals. Here also, almost each code reverts to a
specific technique and algorithm. \\
\noi For one-loop calculation in the electroweak theory,  2
\(\rightarrow\) 2 processes are now easily and fully automatised
as will be made evident in the report. Although there is a large
number of Feynman graphs for 2 \(\rightarrow\) 3 processes,
automatic systems such as the package {\tt
FeynArts-FeynCalc-FormCalc} with the extension\cite{newNtofour} of
the one-loop library {\tt LoopTools}\cite{looptools} or \grcl have
shown the feasibility of an automatic one-loop calculation for $2
\ra 3$ processes in the \smp, where human intervention is kept to
management of the large number of files and batch jobs. Most
important processes for Higgs production at the linear collider,
\eennht\cite{eennhradcor2002, eennhletter,Dennereennh1},
\eeeeht\cite{eeeehgrace}, \eezhht\cite{eezhhgrace,eezhhchinese},
\eettht\cite{eetthgrace,eetthdenner,eetthchinese}, $\gamma \gamma
\ra t \bar t H$\cite{ggtthchinese} as well as
\eenngt\cite{eennggrace} have been computed thanks to the
automatic systems.  This is also the case of the most recent
calculations of the complete one-loop electroweak corrections to
$2 \ra 4$ processes, namely $\epem \ra \nu \bar{\nu} H
H$\cite{eennhhgrace} and some specific 4-fermion channels in
$\epem \ra 4f$\cite{eeto4fdenner}\footnote{A first investigation
of these processes using an automatic code was done
in\cite{eemnudgrace}.}.

\subsubsection{Checking the correctness of the results}
An automatic system produces some numbers as the results of a
black-box calculation which may not necessarily be correct. A user
may feed some input data which the authors of the system have not
thought of. There may be bugs in the program which have not been
detected with the tests made by the developers. Compilers may have
problems,  especially with  highly optimizing options. Even if the
generated program is logically correct, numerical cancellations
may produce wrong results. The Monte-Carlo integration package may
give some  false value just because of insufficient sampling
points. Thus systematic procedures of checking the results are
indispensable for automatic systems. These procedures will be
classified into three categories:
\begin{enumerate}
\item[i)] Checks by the computer system\\
      If the generated code is set-up so that it can be run with
      different accuracies (double, quadruple precision), it would
      be easy to detect problems related to numerical cancellations.
      Good FORTRAN compilers supply options for changing precisions
      without modification of the source code.
      It is also a good idea to run the program on other machines with
      different compilers or architecture.
      For the Monte-Carlo integration, one can increase the number of sampling points in
      order to test the stability of the results.

\item[ii)] Self-consistency checks within the automatic system\\
      If the theoretical model has a free parameter, related to its
      symmetry,
      which does
      not alter the physical results, it will be used to check the
      results.
      Physical quantities in gauge theories are independent under the
      change of gauge parameters.
      When an automatic system includes gauge parameters as variables
      in the generated code of the numerical calculation, one can
      explicitly check the gauge invariance of the obtained results.
      When the system keeps the regularisation parameters of the ultraviolet or
      infrared divergences, one can confirm the cancellation of these
      divergences explicitly.
      The incorporation of these self-checking procedures is one of the most important feature for
      the reliability of the automatic calculation.

\item[iii)] Comparison with other calculations\\
      This is a standard procedure provided another independent calculation exists or can be performed using
      a different automatic code.

\end{enumerate}


Although the ultraviolet and infrared tests are rather
straightforward to implement, the gauge parameter check requires a
very careful and judicious choice of the gauge-fixing function.
This is  especially true in automatic codes for one-loop
amplitudes and cross sections. A few tree-level codes have the
gauge parameter check incorporated through the usual linear
't~Hooft-Feynman gauges or give the possibility to switch to the
unitary gauge. For example, in \grc and  for tree-level processes,
the gauge-parameter independence check has been applied
successfully by comparing the results in the unitary gauge to
those of the 't~Hooft-Feynman gauge.  An agreement up to $\sim
15$($\sim 30$) digits  in double (quadruple) precision has been
reached  for a few selected points in phase space therefore
  confirming, at this stage, that the system works very well
for tree-level processes. However, as we will see, these types of
gauges are not suited at all for one-loop calculations in the
electroweak sector and explain, in part, why there are at the
moment only two general purpose codes for one-loop calculations.
None of them exploits or is defined for a general {\em linear}
$R_\xi$ gauge. The latter tends to considerably increase an
already very large size of the file corresponding to each of the
numerous one-loop diagrams. Not only the expressions get large
compared to the usual $\xi=1$ 't~Hooft-Feynman gauge but also call
for extending the algorithms for the reduction of some new tensor
and $N$-point function integrals. These are at the heart of a
one-loop calculation especially that their evaluation is very much
time consuming. Generalised non-linear
gauges\cite{NonLinear,nlg-generalised}, still defined with the
gauge parameter $\xi=1$, are on the other hand very well suited
for an automatic code of one-loop amplitudes. This will become
clear when we will go through the different stages of a one-loop
calculation and the different modules that are required for the
construction, or the extension to, one-loop amplitudes in the
electroweak theory. The implementation of the generalised
non-linear gauge is therefore crucial in \grc at one-loop. This is
also the reason it has a quite central place in this report and
deserves that we summarise, already at this stage, some of its
salient features and the simplifications it brings when
implemented in an automatic code for one-loop processes. Note that
the package {\tt FeynArts-FeynCalc} has the \sm defined in the
background-field gauge\cite{backgroundfield} beside the usual
linear $\xi=1$ Feynman gauge.

\subsection{Importance of judicious gauge-fixing for automated
one-loop calculations}

A computation in a general $R_\xi$ gauge or unitary gauge brings
about unnecessary complications and sometimes troublesome
numerical unstabilities especially when one deals with several
gauge bosons.  Take for instance the propagator of the $W$ gauge
boson, of momentum $k$ and mass $M_W$. In a general $R_\xi$ gauge,
with the gauge fixing parameter $\xiw$, it writes
\beqn
\label{propagatorexplain}
\displaystyle{ \frac{1}{k^2-M_W^2}
 \left( g_{\mu\nu}-(1-\xiw)\frac{k_{\mu}k_{\nu}}{k^2-\xiw M_W^2}\right)
 }.
\eeqn
In the 't~Hooft-Feynman gauge, $\xiw=1$,  only the ``transverse"
part consisting of the metric tensor $g_{\mu \nu}$ contributes and
leads to a straightforward contraction of neighbouring vertices.
Numerical instabilities are due to the contribution of the
``longitudinal" $k_\mu k_\nu$ part of the gauge propagators.
Moreover, the longitudinal tensor structure considerably inflates
the size of each intermediate result, for example with $n$
intermediate heavy gauge bosons instead of performing $n$
operations one performs $2^n$ operations. Since the longitudinal
expressions involve momenta, they can contribute terms that
increase with energy and which require a subtle cancellation among
various diagrams. A situation which is most acute in the unitary
gauge,  obtained by formally taking $\xiw \ra \infty$ in
Eq.~\ref{propagatorexplain}.
 These problems are of course exacerbated in loop
calculations and, as is known, calculations and renormalisability
itself are arduous if not problematic in the unitary
gauge\cite{Passarinobook}. Within {\tt GRACE} one of the problems
in these gauges (unitary, or general linear type gauges) is that
the library containing the various loop integrals is designed
assuming that the numerator for the propagator of the vector
particles is $g^{\mu\nu}$. For instance, the library for the
three-point vertex functions is implemented with only up-to the
third-rank tensor
and therefore the library applies equally well with fermion loops,
gauge loops or a mixture of these if the calculation is performed
in the 't~Hooft-Feynman gauge. In any other gauge one would have,
for the vertex functions alone, had to deal with a $9$th rank
tensor! Again this not only creates very large expressions but
also introduces terms with large superficial divergences that
eventually need to be canceled extremely precisely between many
separate contributions. Fortunately one can also work  with a
class of gauge-fixing conditions that maintain all the advantages
of the usual 't~Hooft-Feynman gauge with exactly the same simple
structure for the gauge propagators. The new gauge parameters
modify some vertices involving the gauge, scalar and ghost sector
and at the same time introduce new vertices. In fact by
judiciously choosing some of these parameters the structure of the
vertices can get even simpler than with the usual linear
gauge-fixing conditions. The class of gauges we are referring to
exploit non-linear gauge fixing
conditions\cite{NonLinear,nlg-generalised}. Apart from the
possible simplifications that these gauges bring, we have
considered a generalised class of non-linear gauges so as to
perform in an efficient way the gauge-parameter independence
checks within the \grc system. Actually the generalised gauge we
choose depends on $5$ parameters\cite{nlg-generalised}. Therefore
not only this allows for a wide range of checks but since the
different parameters affect different parts of the bosonic sector
one can check different classes of contributions to a single
process and thus more easily track down any bug. There are other
welcome features of these checks. They serve as powerful tools on
every step of the automated computation, from the correct
implementation of the model file which in fact can be checked
mostly at tree-level to  the correct implementation of the tensor
integrals. The reduction of the latter to scalar integrals is most
prone to error.  We will show how the tensor reduction is carried
out in \grcp. The gauge check allows therefore to test that the
reduction of these integrals into the scalar integrals is
implemented properly\cite{eennhletter}. Additional tests like
those of infrared finiteness further verify the scalar integrals.
Another advantage of the non-linear gauge checks over those that
may be attempted within a linear $R_\xi$ gauge is that on a
diagram by diagram basis, the gauge-parameter dependence in our
checks are polynomials in the non-linear gauge parameters whereas
in the linear  $R_\xi$ gauge the dependence is buried within
logarithms and rational functions. We will show how one can
exploit this fact for a very powerful gauge check.

\subsection{Plan and outline of the review}
The aim of this paper is to describe in some detail the workings
of a code for the automatic calculation of cross sections at
one-loop in the \sm based on the Feynman diagrammatic approach.
Though a few codes will be reviewed, the details and algorithms
that enter the construction of such codes are based primarily on
the experience we gained while developing \grcp. The next section,
section~\ref{sec:overview}, will first give a general overview of
the main building blocks of a generic automated package for the
calculation of Feynman diagrams starting from a Lagrangian down to
an integrated cross section. A short review of some specialised
purpose software is given in section~\ref{sec:special-codes}. We
then discuss in some detail how the model file is implemented in
\ref{sec:mdlfile} and how Feynman graph generation is achieved
automatically in \ref{sec:graph-gen-new}. Amplitude generation at
tree-level, both through the spinor technique and directly through
the squared matrix, is described in
section~\ref{sec:automatic-tree}. This section will also very
briefly present alternative automatic codes that make do without
Feynman graphs at tree-level. It is followed by the implementation
of phase space integration and event generation. The extension to
a one-loop automatic system is described in
section~\ref{sec:automaticloop}. The issue of the large size of
the generated code especially at one-loop and file management is
discussed in this section. Section~\ref{sec:automaticloop}  will
also emphasise the importance of internal self-checks on the
correctness of the results of an automatic code, in particular the
importance of the non-linear gauge. This naturally leads the way
to two chapters serving as the theoretical background. First, in
section~\ref{sec:sm-nlg} quantisation in a non-linear gauge is
briefly outlined while a detailed description of the
renormalisation procedure in the \sm within a non-linear gauge is
exposed in section~\ref{sec:renormalisationx}. Explicit formulae
for all two-point functions in a generalised non-linear gauge are
given, together with the complete set of counterterms. We will
also point out some of the issues of renormalisation and gauge
dependence when dealing with unstable particles and the problems
with the implementation of the width of these particles in
section~\ref{unstableparticles}. All of this is needed for the
implementation of the model file in the \smp. The need for a
non-linear gauge is better appreciated when it comes to devising
efficient and powerful algorithms for the reduction of the
tensorial structures of the loop integrals and the reduction of
the $N>4$ point-function to lower rank integrals. A new technique
for these reduction algorithms is described in
section~\ref{num-para-intg}. Section~\ref{sec:looptests} is
devoted to how the ultraviolet and infrared finiteness checks are
dealt with in the system. In this same section we also give a
comprehensive presentation of some systematic test-runs which have
been performed at the one-loop level for a wide variety of
two-to-two processes to show the validity of the gauge check.
These cover fermion-fermion scattering, gauge boson scattering
into fermions, gauge bosons and Higgs bosons scattering processes.
Section~\ref{testxs} is a convincing testimony of the power of a
fully automatic system at one-loop since comparisons with existing
results on some one-loop computations in the \sm  show excellent
agreement. These include $e^+ e^- \ra t \bar t, W^+ W^-, ZH$;
$\gamma \gamma \ra t \bar t, W^+ W^-$; $e \gamma \ra e Z, \nu W$
and $W^+ W^- \ra W^+ W^-$. The final section,
section~\ref{sec:conclude}, contains our conclusions. We also
provide some detailed appendices. In particular we provide the
full set of Feynman rules within the generalised non-linear gauge
as well as the library for the counterterms. Full results for all
the self-energy diagrams of all the particles in the model
including the Goldstone sector is also relegated to an appendix.

\setcounter{equation}{0}
\section{Overview of an automatic system:\\ GRACE as an example}
\label{sec:overview}
\begin{figure}[htbp]
\caption{\label{chartflow} {\em {\tt GRACE} System Flow.}}
\begin{center}
    \includegraphics[width=16cm,height=0.9\textheight]{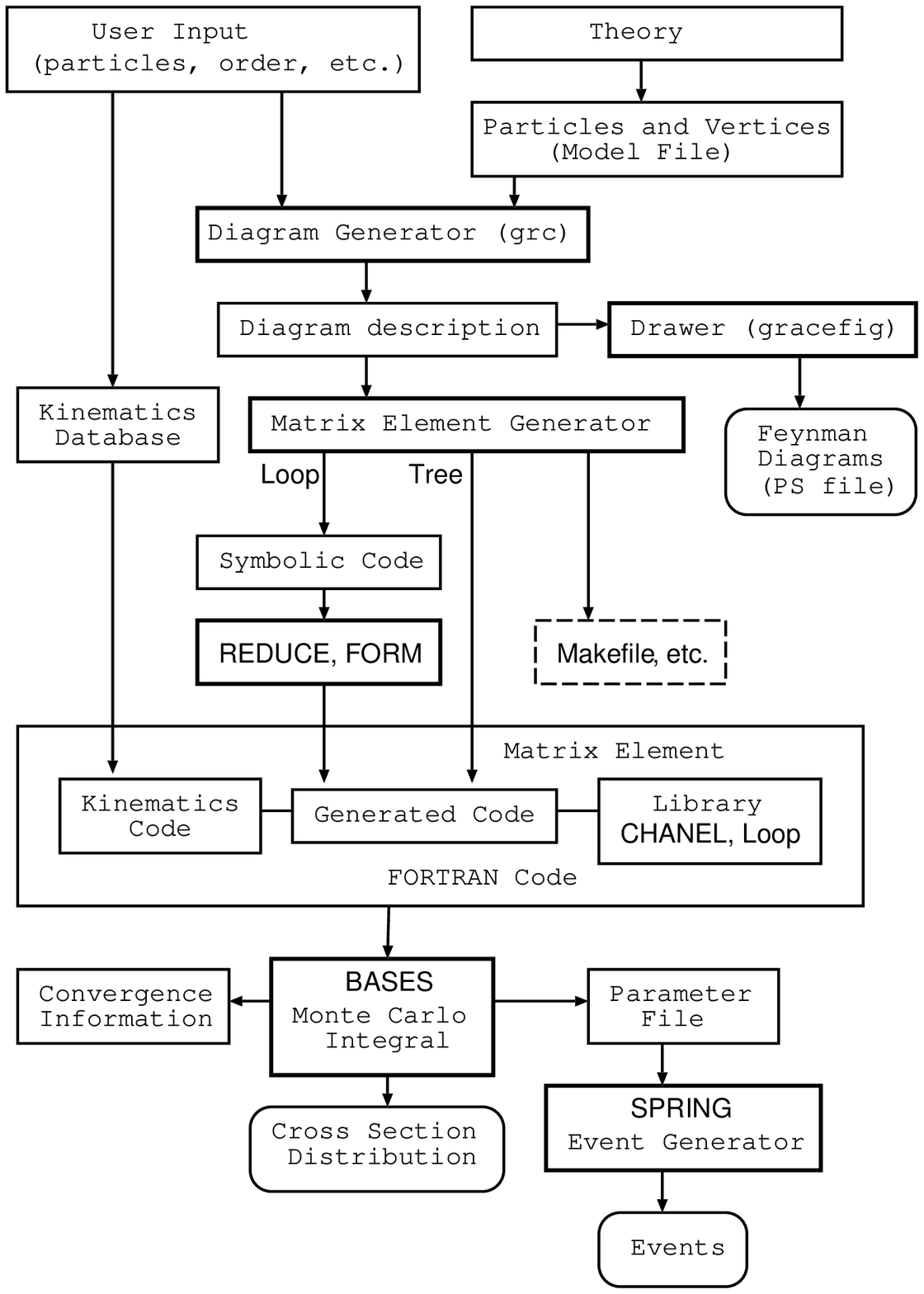}
    \end{center}
\end{figure}


The different components and steps that go into the calculation of
a cross section, or even the corresponding Monte-Carlo event
generator, in a code for the automatic evaluation of Feynman
diagrams are depicted in Fig.~\ref{chartflow}. Automatic systems
not based on the Feynman diagrammatic approach have a different
architecture, moreover the technique they are based on have not
been extended beyond the tree-level. We take  \grcl as an example.
Although not all the modules are present in all such codes,
especially as concerns the one-loop part, this should give an
overall view of how such systems work and what inputs are required
to make them function. Details of the different parts and
components of the packages will be reviewed in the next sections
together with the theoretical background.  In the following, the
most important modules
will be enclosed in boxes for easy  reference to the flow chart of Fig.~\ref{chartflow}.\\

\noi \fbox{Theory} \fbox{Particles and Vertices} \\
 The system first requires the implementation and definition of a
 model through a model file that gives the particle content, the
 parameters as well as the vertices as derived from the
 Lagrangian. The implementation of the model file
 is described in more details in section~\ref{sec:mdlfile}.\\

\noi  \fbox{User Input} \\
 The user, on the other hand, sets
 as input the incoming particles and the final particles and
 specifies the order, in perturbation theory, at which the cross
 section is to be calculated.\\

\noi \fbox{Diagram Generator} \fbox{Diagram Drawer} \\
\grc will first generate, through its own routine {{\tt grc}} the
full set of diagrams (tree and loop as well as counterterms for
the latter) with the possibility of a {\tt Postscript} output of
the Feynman diagrams with the help of the utility diagram-drawer
{{\tt gracefig}}. Most of the postscript files for the Feynman
graphs in this report have been produced by {{\tt gracefig}}. The
{\tt FeynArts}\cite{FeynArts} package based on {\tt Mathematica}
performs similar steps. Section~\ref{sec:graph-gen-new} will
review some of the issues and part of the algorithms for the
automatic generation of the Feynman diagrams while
appendix~\ref{sec:graph-gen-app} will give some
more technical details about graph generation.\\

\noi \fbox{Matrix Element Generator}  \\
The matrix element generator encodes all the information on the
diagrams. For the tree-level process the system generates a {\tt
FORTRAN} code which describes the helicity amplitudes using the
\cha library and routines\cite{chanel}. More details about the
algorithm together with a worked-out example will be presented in
section~\ref{sec:automatic-tree}. We will only briefly describe
the approach that avoids Feynman diagrams altogether in
section~\ref{sec:alphacode}.
\\
For the computations at one-loop, one first generates a symbolic
manipulation source code, based on {{\tt REDUCE}\cite{reduce} or
{{\tt FORM}}\cite{form} that writes, for each set of tree and loop
diagrams, the interference term ${{\cal T}}^{\rm loop}_i {{\cal
T}}^{{\rm tree}\dagger}_j$. A helicity formalism option is also
possible here. Only then the \fbox{{\tt FORTRAN} source code} is
generated and the cross section computed with the help of the
\fbox{loop library} and the counterterm library that performs the
integration over the Feynman parameters and takes into account the
counterterm constants. The symbolic manipulation for the loop
calculation performs a number of important tasks, such as Dirac
algebra (taking traces for fermion loops) and tensor manipulations
 in $n=4-2\epsilon$ dimension if DR\cite{DimRegthV,DimReg2} is used\footnote{ Some
issues related to regularisation and the treatment of $\gamma_5$
are discussed in section~\ref{regg5}.},  introducing Feynman
parametric integrals, shifting loop momenta appropriately etc..
{\tt FeynCalc/FormCalc}\cite{feyncalc,formcalc1,formcalc2} carries
out a similar function with the help, for the one-loop integrals,
of the {\tt
LoopTools} library\cite{looptools}.\\
The implementation of all these intermediate steps that are
necessary at one-loop is discussed in more detail in
section~\ref{sec:automaticloop}. To fully appreciate the issues at
stake, it is important also to refer to the theory sections on
renormalisation, section~\ref{sec:renormalisationx}, and also on
the algorithm for the loop integrals in
section~\ref{num-para-intg}. As already mentioned the latter
module is a critical part of an automatic code at one-loop. This
also explains why almost every code comes with its own solution to
the problem. The implementation of this part has a bearing on the
size and hence performance of the system, see
section~\ref{num-para-intg} and section~\ref{sec:size-code} . This
calls for parallelisation and vectorisation of the codes as
discussed in section~\ref{num-para-intg}.\\

\noi \fbox{Phase space integration}\\
 The integration over phase space
is carried {\it via} {\tt BASES}\cite{bases},  a Monte-Carlo
integration package. One can test the convergence of the
integration routine and get cross sections and distributions. The
simulation and \fbox{event generation} is done through the package
{\tt SPRING}\cite{bases}. \grc includes a number of kinematics
routines, through a \fbox{kinematics database},  for processes
with up to $6$ particles in the final state. The user can select
the appropriate kinematics routine from the available library
depending on the singular behaviour of the process.  This singular
behaviour can be due to a $t$-channel photon exchange for example
or some other peaking behaviour like the crossing of a resonance.
Some of the issues that need to be addressed in the code for the
automatic calculations of cross sections as regards integration
over phase space will be described in section~\ref{sec:kinem}.

\def\aitalc{{\sc \textit{a}{\r{\i}}\raisebox{-0.14em}{T}alc}}
\subsection{Specialised codes and building blocks}
\label{sec:special-codes}
 As mentioned earlier, some specialised
codes exist that only tackle one of the steps above. {\tt
QGRAF}\cite{qgraf} is a very powerful Feynman diagram generator.
Most of the codes however are matrix element generators that work
once a {\em specific} diagram is supplied to the code. Examples
include {\tt MINCER}\cite{mincer}, {\tt MATAD}\cite{matad} and
{\tt SHELL2}\cite{shell2}, all using {\tt FORM} as a symbolic
language. Though being devised for up to three-loop diagrams,
their domain of application is limited to graphs with a restricted
hierarchy of masses and momenta. {\tt Xloops}\cite{xloops} treats
more general graphs, up to two-loop, but again on a
diagram-by-diagram basis. {\tt GEFICOM}\cite{autocalreview} can be
considered as a master program that interfaces {\tt QGRAF} with
{\tt MINCER}\cite{mincer} and {\tt MATAD}\cite{matad}. {\tt
DIANA}\cite{diana}  is another Feynman diagram analyser with
graphics facilities that exploits the power of {\tt FORM} and is
based on {\tt QGRAF} for the generation of the Feynman diagrams.
However it does not include, for applications to loop
calculations, all the ingredients we listed in the diagram
generators of the  complete automatic system since it lacks the
module for loop integration. For a full review of such packages, see~\cite{autocalreview}.\\
The new code \aitalc\cite{aitalc} on the other hand, could be
considered as a fully automated code for the calculation of cross
sections. Built on {\tt DIANA}, it goes a step further by adding
the {\tt LOOPTOOLS}\cite{looptools} library. The code is
restricted to  the evaluation of $2 \ra 2$ processes at one-loop
with only external
fermions. It does not include hard bremmstrahlung.\\

In a different context, {\tt MicrOMEGAs}\cite{micromegas-all} is a
code written in C for the automatic evaluation of the relic
density of dark matter in supersymmetry. By default the lightest
neutralino is assumed to be the lightest supersymmetric particle,
but the user can set any supersymmetric particle (not including
Higgses) to be the LSP. The code is {\em built on} {\tt
CalcHEP}\cite{calchep} which generates, while running, the
subprocesses needed for a given set of  \mssm
parameters\footnote{{\tt CalcHEP}\cite{calchep} is an outgrowth of
{\tt CompHEP}\cite{comphep}.}.

\subsection{Implementation of the model file}
\label{sec:mdlfile} The model file contains all the information
about the theoretical model where the calculation is to be
performed. In particular  the whole set of particles of the model
and their interactions must be transcribed in a machine readable
format. To perform the calculation one then needs to define the
particles and write down the Feynman rules in terms of all
possible vertices needed to build up the Feynman graphs. For the
electroweak theory this means, among other things, writing all the
set of Feynman rules as listed in Appendix B. At the loop order,
the definition requires that one specifies counterterms after
having set the renormalisation procedure. This means for instance
that the loop order of a vertex be also specified. The set of
one-loop vertices that need to be generated at one-loop in the
electroweak theory is listed in Appendix~\ref{sec:vtxcnt}.
\begin{figure}[htb]
\caption{{\em An example of a model file in {\tt GRACE}.}}
    \label{fig:mdl}
\begin{verse}
{\footnotesize
\begin{verbatim}
%=======================================
 Order={ELWK, QCD[qcd]};
 Version={2,2,0};
 PPhase=2;
%=======================================
% gauge bosons
%---------------------------------------
 Particle=W-plus["W+"]; Antiparticle=W-minus["W-"];
     Gname={"W", "W^+", "W^-"};
     PType=Vector; Charge=1; Color=1; Mass=amw; Width=agw;
     PCode=2; KFCode=24; Gauge="wb";
 Pend;
%
 Particle=Z["Z0"]; Antiparticle=Particle;
     Gname={"Z^0"};
     PType=Vector; Charge=0; Color=1; Mass=amz; Width=agz;
     PCode=4; KFCode=23; Gauge="zb";
 Pend;
    ...
 Particle=Higgs["H"]; Antiparticle=Particle;
     Gname={"H"};
     PType=Scalar; Charge=0; Color=1; Mass=amh; Width=agh;
     PCode=31; KFCode=25; PSelect="higgs";
 Pend;
    ...
 Vertex={Higgs,     W-plus, W-minus}; ELWK=1; FName=chww; Vend;
 Vertex={Higgs,     Z,      Z      }; ELWK=1; FName=chzz; Vend;
 Vertex={chi-minus, W-plus, Z      }; ELWK=1; FName=cwzm; Vend;
    ...
\end{verbatim}
}
\end{verse}
\end{figure}

Most automatic systems  read model files in which the information
on the particles and the vertices, through the Feynman rules, are
coded manually. Figure \ref{fig:mdl} shows an example of a model
file for the \texttt{GRACE} system. At first some options are
specified. In this example, the name of the coupling constants are
defined by \texttt{Order=...}. Each particle is described through
a set of properties such as the name of the particle, spin,
electric charge, representation of internal symmetries, whether it
is massive or massless and so on. The interactions of the
particles are defined by a set of vertices. This example shows the
definition of some scalar-vector-vector, {\tt SVV}, vertices which
consist of the list of interacting particles, the order of the
coupling constants and the name  of the coupling constants used in
the generated code. For a model such as the \sm or any
renormalisable model, all types of {\tt SVV} vertices have a
common Lorentz structure. As with other types of vertices {\tt
FFV},{\tt VVV} {\it etc..}, see Appendix B, the Lorentz structure
is exploited in building up the Feynman amplitude, as  will be
shown below.

 Although it is easy to code a model file by hand for a simple
model such as QED, it is not always an easy task for a more
complicated model. For example, the \mssm consists of more than
$80$ particles resulting, at tree-level alone, in more than
$3,000$ vertices. This is the reason why it is more desirable to
construct model files automatically from a Lagrangian, with a
minimum of human intervention. In this case the set of Feynman
rules and vertices is generated automatically. Using dedicated
programming languages, software packages have been developed for
the automatic generation of Feynman rules. One can cite beside
\texttt{LanHEP}\cite{lanhep} originally designed to work with {\tt
CompHEP}, the codes included in \texttt{FDC}\cite{fdc} or
\texttt{gss}\cite{gss}. In this case, a Lagrangian is usually
given in a quite compact and symmetric form. Gauge fixing terms
and ghost terms are added to it. The latter can even be
implemented automatically by first defining the symmetries of the
theory. For instance, in \texttt{LanHEP}, using the BRST
transformation~\cite{BRS,Tyutin} (see Appendix A) as done in
section~\ref{quantisation} one can make the system automatically
derive the ghost Lagrangian and the corresponding Feynman rules.
In the following we sketch some of the steps in automatically
deriving the Feynman rules from an algebraic implementation of the
Lagrangian and by applying some simple set of rules. We take {\tt
LanHEP} as an example without going into the details of the
procedure, the interested reader should consult
the manual of {\tt LanHEP}\cite{lanhep}.\\
\noi One first needs to define the particles of the model as shown
in Fig.~\ref{fig:lanhep1} for the bosons of the electroweak model.
\begin{figure}[htb]
\caption{{\em Particle description in {\tt LanHEP}}}
    \label{fig:lanhep1}
\begin{verse}
{\footnotesize
\begin{verbatim}

 vector
    A/A: (photon, gauge),
    Z/Z:('Z boson', mass MZ = 91.1875, gauge),
    'W+'/'W-': ('W boson', mass MW = MZ*CW, gauge).
scalar  H/H:(Higgs, mass MH = 115).
\end{verbatim}
}
\end{verse}
\end{figure}

\noi Figure~\ref{fig:lanhep2} shows how parts of the Higgs
Lagrangian are entered once the Higgs doublet {\tt pp} ({\tt .f}
refers to the Goldstones), and the covariant derivative {\tt
Dpp\^{}mu\^{}a}
 are defined. The command {\tt lterm} specifies a term in the
Lagrangian.
\begin{figure}[htb]
\caption{{\em Entering the Higgs interaction in {\tt LanHEP}}}
    \label{fig:lanhep2}
\begin{verse}
{\footnotesize
\begin{verbatim}
let pp = { -i*'W+.f',  (vev(2*MW/EE*SW)+H+i*'Z.f')/Sqrt2 },
PP=anti(pp).
lterm -2*lambda*(pp*anti(pp)-v**2/2)**2
lambda=(EE*MH/MW/SW)**2/16, v=2*MW*SW/EE .

let Dpp^mu^a = (deriv^mu+i*g1/2*B0^mu)*pp^a +
     i*g/2*taupm^a^b^c*WW^mu^c*pp^b.
let DPP^mu^a = (deriv^mu-i*g1/2*B0^mu)*PP^a
    -i*g/2*taupm^a^b^c*{'W-'^mu,W3^mu,'W+'^mu}^c*PP^b.
lterm DPP*Dpp.
\end{verbatim}
}
\end{verse}
\end{figure}
\noi Shifts to introduce the wave function renormalisation are
performed through the command {\tt transform}, see
Fig.~\ref{fig:lanhep3}.
\begin{figure}[htb]
\caption{{\em Introducing wave function counterterms in {\tt
LanHEP}}}
    \label{fig:lanhep3}
\begin{verse}
{\footnotesize
\begin{verbatim}
transform A->A*(1+dZAA/2)+dZAZ*Z/2, Z->Z*(1+dZZZ/2)+dZZA*A/2,
    'W+'->'W+'*(1+dZW/2),'W-'->'W-'*(1+dZW/2).
transform H->H*(1+dZH/2), 'Z.f'->'Z.f'*(1+dZZf/2),
    'W+.f'->'W+.f'*(1+dZWf/2),'W-.f'->'W-.f'*(1+dZWf/2).
\end{verbatim}
}
\end{verse}
\end{figure}
\noi It is also possible in {\tt LanHEP} to introduce a command
{\tt brst} for the BRST transformations~\cite{BRS,Tyutin}. In the
example of Fig.~\ref{fig:lanhep4}, by acting on the gauge-fixing
function, this generates the ghost Lagrangian.

\begin{figure}[h!]
\caption{{\em Introducing gauge-fixing and ghosts in {\tt
LanHEP}}}
    \label{fig:lanhep4}
\begin{verse}
{\footnotesize
\begin{verbatim}
let G_Z = deriv*Z+(MW/CW+EE/SW/CW/2*nle*H)*'Z.f'.
lterm- G_Z**2/2.
lterm -'Z.C'*brst(G_Z).
\end{verbatim}
}
\end{verse}
\end{figure}

\subsection{Feynman Diagram Generation}
\label{sec:graph-gen-new}
 The automatic generation of Feynman diagrams
necessary for the computation of a process within a model proceeds
after the user has defined the process through an input file.
Fig.\ref{fig:prc} shows an example of an input file specifying a
process. In this example it is $\epem \ra W^+ W^- \gamma$ at
tree-level . The input data specifies a theoretical model (here
{\tt sm.mdl}), the order of perturbation by counting the power in
the coupling constants (here {\tt ELWK=3}) and by choosing the
initial ($e^+,e^-$) and final particles ($\gamma, W^+, W^-$). In
the \grc input file, {\tt Kinem="2302"} identifies a choice of
kinematics in the kinematics library to be used for the integrated
cross section\footnote{For details about the format of the input
file, see Ref~\cite{chanel-susy}.}.

\begin{figure}[htb]
\caption{{\em An example of the input file in the electroweak \sm
for the
    scattering process $\epem \ra W^+ W^-
\gamma$ at tree-level.}}
    \label{fig:prc}
\begin{verse}
{\footnotesize
\begin{verbatim}

%%%%%%%%%%%%%%%%%%%%%%%%%%%%%%%%%%%%%%%%
Model="sm.mdl";
%%%%%%%%%%%%%%%%%%%%%%%%%%%%%%%%%%%%%%%%
Process;
  ELWK=3;
  Initial={electron, positron};
  Final  ={photon, W-plus, W-minus};
  Kinem="2302";
Pend;
\end{verbatim}
}
\end{verse}
\end{figure}

A typical algorithm for the generation of the Feynman diagrams
will be the following:
\begin{enumerate}
\item Generate the number of vertices.\\
      The number of vertices is restricted by the order of the coupling
      constants for the physical process and is given by the input file, see Fig.\ref{fig:prc}.
      Each vertex has a fixed number of  propagators
      and external particles to be connected.
\item Connect vertices with propagators or external particles.\\
      There are multiple ways to connect vertices.
      All possible configurations are to be generated.
\item Particle assignment.\\
      Particles are assigned to propagators confirming that the connected
      vertex is defined in the model.
      As there will be many ways to assign particles to propagators,
      all possible configurations are to be generated. In generating the Feynman diagrams,
conservation laws such as electric charge and fermion number
conservation will be employed in order to avoid fruitless trials.
\end{enumerate}

The generation of Feynman diagrams borrows heavily from graph
theory. In the following let us call a vertex or an external
particle a \textsl{node}. Similarly, let an \textsl{edge} be a
connection between two nodes, which may be a propagator or a
connection between a vertex and an external particle. Thus an edge
is expressed by a pair of two nodes (which are connected by the
edge). The diagram or graph generation process is to construct
edges in all possible ways.

\begin{figure}[hbt]
 \caption{{\em An example of information about a generated diagram in
    \texttt{GRACE}. The first part below defines the external particles. It is followed by defining the connections
    between the nodes and the vertices in the form ``node number=\{edge numbers \}".}}
    \label{fig:grf}
\begin{verse}
{\footnotesize
\begin{verbatim}
Process=1; External=5;
   0= initial electron;
   1= initial positron;
   2= final photon;
   3= final w-plus;
   4= final w-minus;
Eend; elwk=3;Loop=0;

Graph=1; Gtype=1; Sfactor=-1; Vertex=3;
   0={    1[positron]};
   1={    2[electron]};
   2={    3[photon]};
   3={    4[w-plus]};
   4={    5[w-minus]};
   5[order={1,0}]={    1[electron],    2[positron],    6[photon]};
   6[order={1,0}]={    4[w-minus],    6[photon],    7[w-plus]};
   7[order={1,0}]={    3[photon],    5[w-plus],    7[w-minus]};
Vend; Gend;

Graph=2;
    ...
\end{verbatim}
}
\end{verse}
\end{figure}

%
%

We show an example of an output file generated by the
\texttt{GRACE} system in Fig.~\ref{fig:grf} for the process we
defined in the input file of Fig.~\ref{fig:prc}, namely $\epem \ra
W^+ W^- \gamma$. The first part of this file describes the
information about the physical process.
The first generated diagram, {\tt Graph=1}, consists of the $5$
external particles and the $3$ vertices. To this corresponds $7$
edges that connect these {\em nodes}; for example, the final
photon (labelled as node $2$ and 3rd external particle) is
connected to the
\(\gamma W^+ W^-\) vertex (node $7$) by edge $3$. Particles are
defined as incoming to the node. This information is used by
another program such as the code for drawing  diagrams. This
reconstructs the structure of diagrams, places nodes on a graphic
device and connects them by edges as shown in
Fig.\ref{fig:grcfig}.

\begin{figure}[htb] \caption{{\em An example of a diagram drawn by
\texttt{GRACE} based on the file in Fig.~\ref{fig:grf}. Here each
dot is a node. In this figure we have only labelled  those nodes
that correspond to the external particles according to the listing
in Fig.~\ref{fig:grf}. The numbers in parentheses correspond to
the edges as defined in Fig.~\ref{fig:grf} also.}}
    \label{fig:grcfig}
    \begin{center}
    \includegraphics[width=8cm,clip=true]{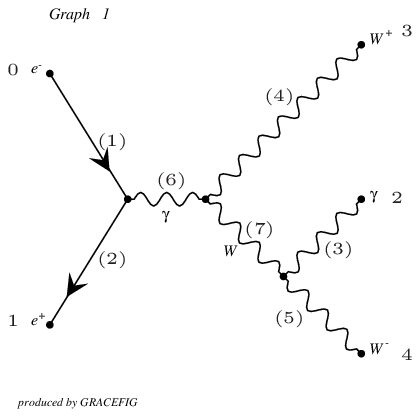}
    \end{center}
\end{figure}

It is to be noted that although vertices of the same kind are not
distinguished from each other, for instance in our example the
vertex $W^+ W^- \gamma$ appears twice in {\tt Graph 1} (see
Fig.~\ref{fig:grcfig}), they will be distinguished in a program,
usually through a sequence of numbered labels, in our case node
$6$ and $7$ (see Fig.~\ref{fig:grf}). On the other hand, since a
Feynman graph is a topological object, it is independent of the
way one assigns the sequence of numbers to nodes. Such simple
algorithms will produce diagrams, which are topologically the
same, many times. The problem of diagram generation is not so much
to generate diagrams but to avoid such duplications. Although it
is not so difficult to avoid generation of duplicated diagrams
when limited to special cases such as QED\cite{grand} or tree
processes, the problem of redundancy gets exasperated when loop
diagrams are generated.

A general method of graph generation avoiding duplicated graphs,
applicable to any process with any order in the coupling
constants, has been developed by graph theorists \cite{orderly}.
With this method, one can calculate the symmetry factors of the
Feynman diagrams at the same time. Unfortunately,  such a method
is not efficient enough for a large number of external particles
or loops
\cite{isograph}. With some optimization, this method was first
applied to the generation of Feynman diagrams in the code
\texttt{QGRAF}\cite{qgraf}.  Another technique of optimisation and acceleration was
proposed in \cite{grc}. The
\texttt{GRACE} system exploits the latter algorithm and acceleration for the generation of Feynman diagrams.
Appendix~\ref{sec:graph-gen-app} presents in some detail the issue
of graph duplication and calculation of the symmetry factors and
the need for optimising and accelerating the generation of Feynman
diagrams. Examples beyond one-loop and outside the electroweak
model are given to illustrate these issues.

\section{Automatic systems at tree-level}
\label{sec:automatic-tree} We first start by giving the main lines
of how the automatic code builds up and calculates Feynman
amplitudes from the  Feynman graphs at tree-level. It is very much
a transcription of a calculation by hand.

\subsection{Squared matrix elements}
\label{tree-sqme}
One method is to calculate squared amplitudes.
Here projection operators, Dirac's gamma matrices and the sum over
the polarizations of vector particles can be handled symbolically.
This symbolic treatment of the mathematical expression is
performed with a symbolic manipulating system such as
\texttt{REDUCE} or \texttt{FORM}\cite{form} in the case of \grc or
a specially developed package in the case of
\texttt{CompHEP}\cite{comphep}. Symbolic calculation in
\texttt{FeynCalc/FormCalc}\cite{feyncalc} uses
\texttt{MATHEMATICA} and \texttt{FORM}. An automatic system
generates a source code for a symbolic manipulating system in
accordance with the Feynman rules of the model. As the number of
particles increases, the number of Feynman graphs grows very
rapidly. This means that this method is not suitable in these
cases as it requires computing the square of the number of Feynman
diagrams. Moreover this method is clearly unsuitable if one
requires information on the polarisations.

\subsection{Spinor technique: a worked out example in GRACE}
\label{tree-spinor} Another method employs the spinor technique.
Spinors and gamma matrices are dealt with in a numerical way. A
library of all possible types of vertices and propagators are
defined as subroutines to be called for a numerical calculation.
The \texttt{CHANEL} \cite{chanel} library for \texttt{GRACE}
\cite{grace} and \texttt{HELAS} \cite{helas} library for
\texttt{MadGraph} \cite{madgraph} are examples of such
subroutines. The automatic system calls these libraries according
to the structure of the Feynman graphs. Output of the system is a
complete code for the numerical calculation of differential cross
sections without calling any other package. Since this method
calculates helicity amplitudes directly, it is natural to
calculate polarized cross sections. The necessary CPU time is
proportional to the number of Feynman graphs in this method.

We will here show in some detail how the helicity amplitude method
can be implemented in an automatic code such as {\tt GRACE}. The
method is purely numerical. The amplitude corresponding to each
Feynman graph is first decomposed into vertex sub-amplitudes. Each
of these sub-amplitudes is read from a pre-defined model file
library so that one only has to call the library, in this case
{\tt CHANEL}, where these sub-amplitudes are defined. In order to
achieve this, the propagators that appear
as internal lines are expressed as a product of wave functions.\\

\begin{figure}[htb]
\caption{\label{hel-eetowwg}{\em A Feynman graph contributing to $\epem \ra W^+
W^-\gamma$.}}
\begin{center}
\includegraphics[width=10cm,height=5cm]{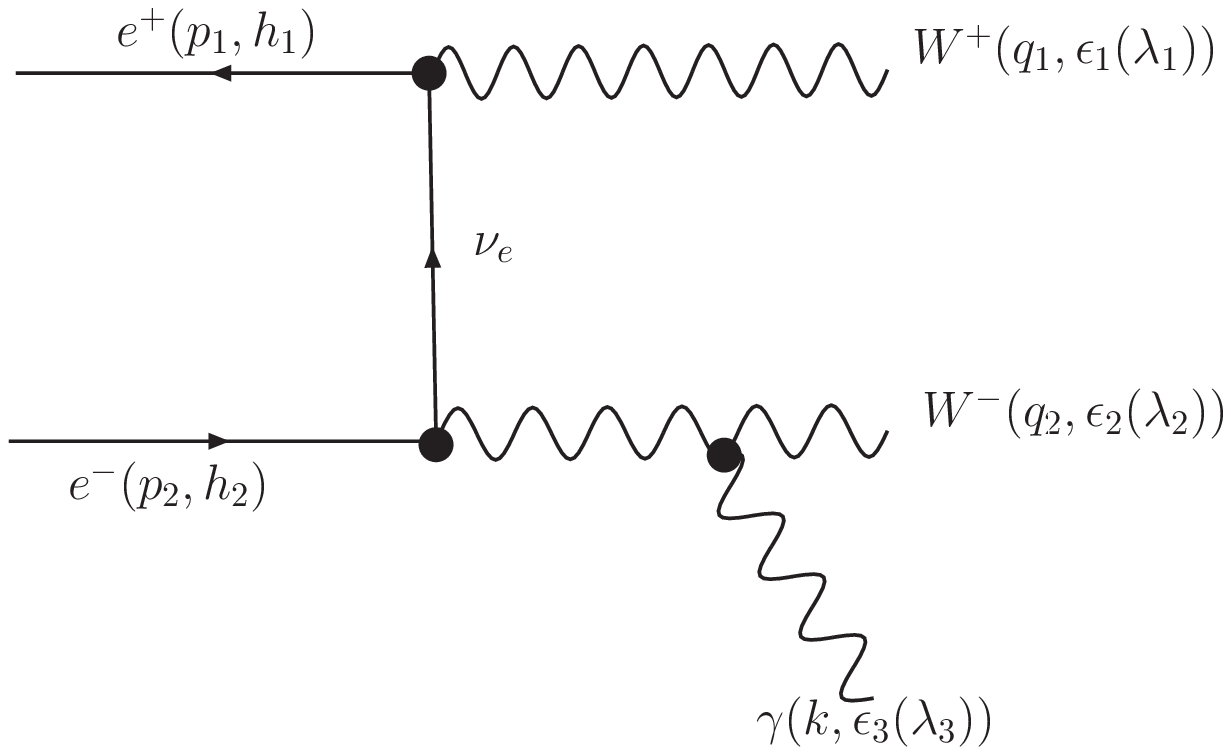}
\end{center}
\end{figure}
\noi Consider the scattering amplitude corresponding to the
Feynman graph shown in Fig.~\ref{hel-eetowwg} as an example.
\(p_1\), \(p_2\), \(q_1\) \(q_2\) and \(k\) are momenta of
\(e^+\), \(e^-\), \(W^+\), \(W^-\) and \(\gamma\), and \(h_1\) and
\(h_2\) are helicities of \(e^+\) and \(e^-\), and
\(\epsilon_1(q_1,\lambda_1)\), \(\epsilon_2(q_2,\lambda_2)\) and
\(\epsilon_3(k,\lambda_3)\) are polarization vectors of \(W^+\),
\(W^-\) and \(\gamma\), respectively with the corresponding
helicity $\lambda_{1,2,3}$.

The scattering amplitude for this contribution is given by
\begin{eqnarray}
  T_{fi} & = &
  \overline{v}(p_1,h_1) \, c_{eW}^\eta \, \epsilon_{1\eta}(q_1) \,
  S_F(-p_1+q_1)\, c_{eW}^\mu \, u(p_2, h_2) \nonumber\\
  & & \times D_{F\,\mu\nu}(q_2+k) \,
  c_{WW\gamma}^{\nu\rho\sigma}(q_2+k,-q_2,-k) \,
  \epsilon_{2\rho}(q_2) \, \epsilon_{3\sigma}(k),
\end{eqnarray}
where \(c_{eW}^\eta\) and \(c_{WW\gamma}^{\nu\rho\sigma}\) express
electron-\(W\) and photon-\(W\) couplings, respectively, and they
are given by:
\begin{equation}
c_{eW}^\eta = \frac{e M_Z}{\sqrt{2(M_Z^2 - M_W^2)}} \,
               \gamma^\eta \, \frac{1 - \gamma_5}{2}
\end{equation}
and
\begin{equation}
c_{WW\gamma}^{\nu\rho\sigma}(p,q,r) = e [(p-q)^\sigma g^{\nu \rho}
+ (q-r)^\nu g^{\rho \sigma}  +(r-p)^\rho g^{\sigma \nu}] .
\end{equation}

In  {\tt CHANEL}, propagators are expressed as a  bi-linear form
of wave functions:
\begin{eqnarray}
 S_F(p) = \frac{\sum_{\alpha i} w_{\alpha,i} \, U^{\alpha}(h^{(i)}, p^{(i)}) \,
                \overline{U}^{\alpha}(h^{(i)}, p^{(i)})}
               {p^2 - m^2}
\end{eqnarray}
and
\begin{eqnarray}
 D_{F\,\mu\nu}(p) =
          \frac{\sum_{i} \, w_i \,
          \epsilon^{(i)}_\mu(p) \, \epsilon^{(i)}_\nu(p)}
               {p^2 - m^2},
\end{eqnarray}
where \(w_{i,j}\) and \(w_i\) are c-numbers and  weight factors
for the decomposition of propagator. \(U^{\alpha}\) represents
either a spinor \(u\) or \(v\) depending on the value of index
\(\alpha\). Momenta \(p^{(i)}\) are calculated from the
(off-shell) fermion momentum \(p\).

 The amplitude then writes as a product of vertex sub-amplitudes
\begin{eqnarray}
  T_{fi} & = &
  {\displaystyle D(-p_1 + q_1, 0) \; D(q_2 + k, m_W)\;
  \sum_{\alpha,i} w_{\alpha,i} \, \sum_l w_l } \nonumber\\
  & & \times V_{eW^+}^{(\alpha, i)} \, V_{eW^-}^{(\alpha, i, l)} \,
       V_{WW\gamma}^{(l)}, \nonumber
\end{eqnarray}
where
\begin{eqnarray}
  D(p, m) & = & \frac{1}{p^2 - m^2},  \nonumber\\
  V_{eW^+}^{(\alpha, i)} & = &
  \overline{v}(p_1,h_1) \, c_{eW}^\eta \, \epsilon_{1\eta}(q_1)\,
  U^{\alpha}({(-p_1+q_1)}^{(i)}, h^{(i)}), \\
  V_{eW^-}^{(\alpha, i, l)}  & =  &
  \overline{U}^{\alpha}(p^{(i)}, h^{(i)}) \,
  c_{eW}^\mu \, \epsilon^{(l)}_\mu(q_2+k) \, u(p_2, h_2), \nonumber
\end{eqnarray}
and
\begin{eqnarray}
V_{WW\gamma}^{(l)}  & = &
  c_{WW\gamma}^{\nu\rho\sigma}(q_2+k,-q_2,-k) \, \epsilon^{(l)}_\nu(q_2+k) \,
  \epsilon_{2\rho}(q_2)  \, \epsilon_{3\sigma}(k) .
\end{eqnarray}

Further details of how the spinors and polarisation vectors are
represented, together with the weight factors can be found in
\cite{grace}. The fermion-fermion-vector (FFV) of the \sm vertex
parts \( V_{eW^+}^{(\alpha, i)} \) and \(V_{eW^-}^{(\alpha, i)}\)
are calculated with the help of the {\tt CHANEL} subroutine {\tt
SMFFV} while the 3-vector (VVV) vertex part \( V_{WW\gamma}^{(l)}
\) by the subroutine {\tt SMVVV}. These subroutines calculate all
combinations of helicity states and of spinors \(u\) and \(v\).
Therefore given a set of momenta and helicities, these vertex
parts return a number automatically.
\\
\noi Summation over indices pertaining to the fermion propagators is
made through a call to the  subroutine {\tt SMCONF}:
\[  V_{eeW^+W^-}^{(l)} = \sum_{\alpha,i} w_{\alpha,i}
    V_{eW^+}^{(\alpha, i)} \, V_{eW^-}^{(\alpha, i, l)}
\]
The last free index \(l\), which corresponds to the propagator of
a vector particle is through a call to the subroutine {\tt
SMCONV}:
\[
    \sum_l w_l V_{eeW^+W^-}^{(l)}
    V_{WW\gamma}^{(l)}.
\]
\par
\smallskip
\noindent
The program generates a sequence of subroutine calls in the
following order:
\begin{itemize}
\item[1)] For a given process, subroutines {\tt SMEXTF} and {\tt SMEXTV} are called
first, encoding information on the external fermions ({\tt
SMEXTF}) and external bosons ({\tt SMEXTV}) in a form suitable for
the calculation of the vertices. Since this part is common to all
graphs, it is generated once before calculating an amplitude of
the first graph.

\item[2)] Momenta of propagators are defined by taking linear combinations of
    external momenta based on momentum conservation.
    Then the denominators of the propagators are calculated by the subroutine
    {\tt SNPRPD}.

\item[3)] The decomposition of the numerator of the propagators as
a bi-linear product of wave functions is carried through {\tt
SMINTF} for fermions and {\tt SMINTV} for vectors.

\item[4)]  The different subroutines
    such as {\tt SMFFV} or {\tt SMVVV} are called for calculating
    vertices sub-amplitudes.

\item[5)]  These vertices are connected in correspondence with the
    propagators of the graph, which is realized by appropriate
    subroutines
    such as {\tt SMCONF} and {\tt SMCONV}.
\item[6)] Before the amplitudes are summed over all graphs a
reordering of the particles, so that they appear in the same
sequence for all the graphs, is carried through a special routine.

\item[7)] After summing over all diagrams, the helicity amplitudes
are squared. Summation over spin states can also be performed
automatically by a call to a dedicated routine.
\end{itemize}

The generated code for this example is shown in
Fig.~\ref{fig:amps}. A more detailed description of the  {\tt
CHANEL} library is given in \cite{grace}.

\begin{figure}[htbp]
   \caption{ {\em The code for the helicity amplitude corresponding to the graph in Fig.~\ref{hel-eetowwg}.}}
    \label{fig:amps}
\begin{verse}
{\scriptsize
\begin{verbatim}

************************************************************************
 *             Graph No. 25 - 1
 *         Generated No. 25
************************************************************************
       subroutine ag25
       implicit real*8(a-h,o-z)

       include 'incl1.h'
     ...
       complex*16 atmp
 *-----------------------------------------------------------------------
 * Denominators of propagators
       aprop = 1.0d0
       call snprpd(pphase,aprop,vn21,amnu(1)**2,0.0d0)
       call snprpd(pphase,aprop,vn23,amw**2,amw*agw)
 * Internal momenta
       call smintf(amnu(1),pf21,vn21,ex21i,pt21i,cf21i)
       call smintv(lepinv,amw,pf23,eq23b,ev23b,vn23,igauwb)
 * Vertices (8)
       call smffv(lextrn,lintrn,lepinv,ex2l,ex21i,amlp(1),amnu(1),
      &           cwnl(1,1),cf2l,cf21i,pt2l,pt21i,eq23b,lt5,av5)
       call smffv(lintrn,lextrn,lepexv,ex21i,ex4l,amnu(1),amlp(1),
      &           cwln(1,1),cf21i,cf4l,pt21i,pt4l,eq17b,lt6,av6)
       call smvvv(lepexa,lepinv,lepexv,-1,-1,-1,caww,pf9,pf23,pf30,eq9d,
      &           eq23b,eq30b,lt7,av7)
       call smconf(lt5,lt6,2,1,ex21i,av5,av6,lt8,av8)
       call smconv(lt7,lt8,2,2,ev23b,av7,av8,lt,av)
       sym = - 1.0d0
       cf  = + 1.0d0
       aprop         = cf*sym/aprop
       indexg(1) = 3
          ...
       indexg(5) = 4
       if(jcpol(3).ne.0) call smcpol(1, lt, av)
       call ampord(lt, av, indexg, agcwrk)
       ancp(jgraph) = 0.0d0
       do 500 ih = 0 , lag-1
          atmp    = agcwrk(ih)*aprop
          agc(ih,0) = agc(ih,0) + atmp
          ancp(jgraph) = ancp(jgraph) + atmp*conjg(atmp)
   500 continue
       return
       end
\end{verbatim}
}
\end{verse}
\hrule
\end{figure}

\subsection{Dealing with QCD and supersymmetry}
Extension of both the squared matrix elements and spinor
techniques to supersymmetric models requires a proper treatment of
Majorana particles. Calculation with Majorana particles involves
the charge conjugation operator. When a symbolic manipulation
system is employed, the symbolic treatment of this operator must
be implemented. A simple algorithm\cite{feynMajorana} has been
constructed where this operator is confined to the conjugated
vertices , so in effect we only deal with the usual Dirac
propagator while the vertices do not explicitly invoke charge
conjugation. In the case of the spinor technique, we only need to
add new appropriate subroutines to a library, such as {\tt
CHANEL}.

For QCD, the automatic calculation  refers to the calculation of
the partonic hard scattering part of the matrix elements. The
colour factor of a graph is separated out. However, Feynman rules
for the four point vertex of the gluon is expressed in a mixed
form of colour matrices, through the $SU(3)$ structure constants,
$f^{abc}$ and Lorentz parts:
\begin{eqnarray*}
&    g_s^2 \left[ \left(f^{ace} f^{bde} - f^{ade} f^{cbe} \right)
                        g_{\alpha\beta} g_{\gamma\delta}
                + \left(f^{abe} f^{cde} - f^{ade} f^{bce} \right)
                        g_{\alpha\gamma} g_{\beta\delta}
      \right. & \\  & \left.
                + \left(f^{ace} f^{bde} - f^{abe} f^{cde} \right)
                        g_{\alpha\delta} g_{\gamma\beta} \right]
\end{eqnarray*}
There are two ways for the separation of these factors. One is
that the Feynman graph is generated with the original quartic
vertex but the  amplitude generator expands this vertex into a sum
of three terms where the colour index is factored out. The other
method, also used in {\tt CompHEP} for example, introduces an
auxiliary field which interacts through a vertex with two gluons.
The Feynman graph generator constructs the quartic gluon vertex as
the sum of graphs consisting of  \(s\), \(t\) and \(u\) type
exchanges of the auxiliary field. The auxiliary field method is
also best suited in the case of the \mssm where additional
coloured particles with 4-point vertices exhibiting  a variety of
colour structures are needed.

\subsection{Checking the result at tree-level} The \sm
and the \mssm are gauge theories. One can exploit the gauge
freedom to check the result of the calculation of a cross section
or helicity amplitude. For the calculation based on the spinor
technique, it is easy to keep the gauge parameters as variables in
the generated code without increasing the CPU time. For  the
squared matrix technique that relies on symbolic manipulation,
checks that are based on varying the usual 't Hooft-Feynman
parameter lead to very complicated and very large expressions.
Checks are much more manageable with generalised non-linear gauges
that will be discussed in detail later in
Section~\ref{sec:sm-nlg}.

%

\subsection{Automatic tree calculations without Feynman graphs}
\label{sec:alphacode} Let us briefly mention that there exists
codes for the automatic calculation of  amplitudes that do not
require the standard textbook approach based on Feynman graphs.
\texttt{ALPHA}\cite{code-alpha} uses an algorithm based on an
iterative solution of the equation of motion whereas
\texttt{HELAC} \cite{helac} is based on solving the
Schwinger-Dyson equation. An acceleration method is developed in
\cite{code-omega} and is applicable also to the methods using
Feynman amplitudes. All these methods exploits a binary
representation of the momenta involved in the tree-level amplitude
calculation.

 Let \(p_i\), \(i = 1,
..., N\) be the momenta of the external particles for a certain
amplitude. Momentum conservation imposes the restriction \(\sum_i
p_i = 0\). A momentum \(q\) of a propagator in a tree graph is
expressed as \(q = \sum_i a_i p_i\), where \(a_i\) is either 0 or
1. Thus the possible number of momenta appearing in the
calculation is bounded by \(2^{N-1}-1\) and only a finite limited
number of vertices appear in the calculation. This helps to
construct tree \(N\)-point Green's function combining these
vertices in a kind of matrix operation.


\subsection{Kinematics and event generation}
\label{sec:kinem}

Having obtained the amplitudes or the squared matrix elements, the
last step is the integration over phase space to obtain the cross
section. One may also want to generate events for the analysis of
experimental data. For multi-particle final states, phase space
integration means integrating over many variables. Since the
amplitudes themselves, the integrand, are most often lengthy and
have a complicated structure in these variables, Monte Carlo
integration packages such as \texttt{VEGAS}\cite{vegas} and
\texttt{BASES}\cite{bases} are used.  Even if the integrand were a
smooth function over the whole phase space without narrow peaks,
naive use of a numerical integration package may cause problems as
it would be difficult to reach the required accuracy especially
for a large multi-dimensional space or when one deals with
complicated boundary configurations.  Narrow peaks can occur for
example where the momentum of an internal particle is such that it
approaches its  mass-shell. The amplitude could blow up. As these
peaks give an important contribution to the results, it is
necessary to catch the detailed structure of these singularities
and to integrate around these peaks within good accuracy. In this
case the Monte Carlo integration package needs to accumulate
enough sampling point around these peaks. When singularities run
parallel to integration axes, \texttt{VEGAS} or \texttt{BASES}
optimizes the distribution of sampling points looking at the
projected values on the axes.

It is fair to say that it is this stage of the automatic
calculation of cross sections in high energy physics which
requires most intervention by the user. This is due to the fact
that there is at the moment no general purpose integrator that can
automatically catch the different types of singularities and
suitably adapt itself to properly treat all the singularities
present in a process with enough accuracy. Several integration
packages have been developed \cite{dice,parint,whizard} for
handling many singular situations but none is general enough that
it can be trusted to run independently and give an accurate result
in any situation. The user should select good integration
variables so that the singularities are arranged along the
integration variables. This selection of the integration variables
depends, of course, on the process. Some systems, such as \grc,
include  a library of kinematics suitable to handle a number of
topologies for the singularities for a given number of external
legs.

In the case of QCD, the partonic cross sections and amplitudes
need, for example, to be convoluted with parton distribution
functions. The code can also provide packages for parton shower
and hadronisation.

Once a numerical code of a differential cross section is obtained,
one can generate unweighted events by a program package such as
\texttt{SPRING} \cite{bases} using the result of the integration
calculated by \texttt{BASES}. Unweighted simulated events,
generated in accordance with the theoretical predictions are very
useful tools to analyse the experimental data and take into
account the complex structure of the detectors.

\section{Automatic systems for one-loop processes}
\label{sec:automaticloop} Extension of an automatic system from
tree-level to one-loop processes is, in principle, straightforward
because all the ingredients needed for one-loop  calculations in a
renormalisable theory are known. In practice, however, quite a few
of the new features not met when handling  processes at tree-level
pose a real challenge and are fraught with technical difficulties.
This explains why although there has been a proliferation of
automatic systems for tree processes since the beginning of the
1990s, only a couple can tackle one-loop processes. Currently two
systems are available, {\tt GRACE-loop} and {\tt
FeynArt/FeynCalc/FormCalc}\cite{feynart, feyncalc,
formcalc1,formcalc2} which apply both to the electroweak sector of
the \sm and the \mssmp. Application to QCD at one-loop is at the
stage of development by some groups but a general purpose code is
still not completed yet. In this section we describe what kind of
problems one encounters in the construction of an automatic system
at one-loop based on the experience we gained while developing
{\tt GRACE-loop} for electroweak processes in $e^+e^-$ colliders.

An automatic system for one-loop should be composed of the
following ingredients:
\begin{enumerate}
\item Proper graph generator. This should provide all one-loop diagrams
for a given process. Most graph generators that work for
tree-level handle at least one-loop diagrams. We have already
described the general algorithm of graph generation in
Section~\ref{sec:graph-gen-new}.
\item Amplitude generator. This very much depends on how to calculate one-loop
amplitudes. Both the helicity amplitude formulation and/or the
squared matrix elements technique are possible as explained later.
\item The model file should include a library for counterterms.
This should contain all terms necessary to compensates ultraviolet
divergences. For a given theoretical model, one can have different
libraries reflecting the fact that one can choose between
different renormalisation and regularisation schemes. In {\tt
GRACE},  ultraviolet(UV) divergences are regularised through
dimensional regularisation and the electroweak sector is
renormalised on-shell, see section~\ref{sec:rconst}. Note in
passing that there should also be a prescription for handling
infrared(IR) divergences. In {\tt GRACE} this is done by giving
the photon an infinitesimal mass, see section~\ref{irtest}. These
libraries have a structure similar to those defined for the
tree-level model files.
\item A library for one-loop integrals.  This should include analytic
formulas for 2-, 3- and 4-point scalar integrals, see
section~\ref{scalarintl4}. 5- and 6-point functions can be
expressed by a sum of 4-point functions, section~\ref{fivesixint}.
 As we will see in section~\ref{scalarintl4}, a scalar
integral is an integral over the loop momentum where the integrand
is a product of the  denominators of the various propagators
inside the loop, so that no loop momentum appears in the numerator
of the integrand. When products of loop momenta are involved one
speaks of higher rank tensors.

\item Beside the aforementioned reduction of 5- or 6-point diagrams
to 4-points, one needs another library which decomposes higher
rank tensors of a box diagram to relevant scalar integrals and
surface integrals, namely 3-point functions. The reduction
formulas for higher rank tensors of vertex diagrams is needed as
well. There are a few algorithms dealing with the last two points.
One needs very efficient algorithms and numerical routines here,
since this steps can take up a large fraction of the CPU time.
This is developed further in this article in
section~\ref{reducboxtensor}.
\item Ways to check the results. This is essential and must be
performed. We consider it as the most important part of the system
because the calculation is much more complicated than the tree
processes. Lacking these tests one cannot be convinced that the
obtained results are correct. Usually, one  checks UV and IR
finiteness. Gauge invariance is also a very powerful tool, however
the familiar 't Hooft gauges are not suitable for multi-leg
one-loop processes. \grc incorporates a series of parameters
derived from a generalised non-linear gauge.
\end{enumerate}

The theoretical background and the algorithms necessary for the
construction of the libraries related to the points 3,4,5 above
will be developed in detail in sections~\ref{cuvtest},
\ref{irtest}, \ref{scalarintl4}, \ref{fivesixint} and
\ref{reducboxtensor} respectively. The gauge-parameter
independence check in the last point above is crucial and could be
considered as pivotal. Beside being a most powerful check on the
calculation it also has a bearing on how  the libraries for the
reduction of the  one-loop integrals are constructed. The choice
of special gauges here is almost mandatory. This is the reason we
pay so much attention to the implementation of the non-linear
gauge in this article. Although we will, in the next sessions, get
back to the details of all the points listed above and before we
point to some features related to the technical implementation of
the one-loop radiative corrections in the electroweak part of the
\sm in the automatic code \grcp, it is worth to briefly expose the
gauge-parameter check.

\subsection{Gauge parameter independence}
The most stringent test for the consistency of the calculation
will be provided by the gauge-parameter independence of the cross
section. In tree-level calculations with an automatic code any
gauge fixing will do. This is because the inclusion of any
modifications brought by the gauge fixing Lagrangian and
associated ghost Lagrangian is always easy since no loop
integration is needed. Typically, the standard $R_\xi$ gauge
fixing is used where the gauge parameter hides in the propagators
of the gauge particles $\gm,Z,W^\pm$, and unphysical scalars of
the theory. For example in a general $R_\xi$ gauge,  the $Z$
propagator writes

\eqn D_Z^{\mu\nu}(q^2)={1\over q^2-M_Z^2+i\eps}\left[g^{\mu\nu}
-(1-\xi_Z){q^\mu q^\nu\over q^2-\xi_Z M_Z^2+i\eps}\right]. \eqne

This formulation of gauge fixing, however, is not well suited when
loop calculations are involved. The ``longitudinal" part, in the
second term of  the propagator,  constitutes an  obstacle in
carrying out the loop integration, due to the appearance of not
only unphysical thresholds, $\xi_Z M_Z^2$, functions but  also the
introduction of  higher rank tensors, $q^\mu q^\nu$, and higher
order $N$-point functions since this part involves ``two"
denominators. Since as we will see, the reduction of tensor loop
integrals and higher $N$-point functions is very much time
consuming it is  desirable to find another formalism for gauge
fixing which allows the propagators to take as simple form as
possible, $\xi_Z=1$. This can in fact be realized in the
non-linear-gauge(NLG) fixing formulation\cite{nlg-generalised}.
With this kind of gauge fixing, the propagators can be taken to
correspond to $\xi_i=1$ for all gauge particles but additional
gauge parameters, through derivatives in the gauge-fixing, modify
a large number of vertices where the gauge and unphysical
(Goldstones and ghosts) are involved. We will take a generalised
gauge with   five gauge parameters in total,
$\tilde\al,\tilde\bt,\tilde\del, \tilde\vare,\tilde\kp$. Although
this means some more work in redefining the model file as compared
to the linear $R_\xi$ gauge which only modifies a few propagators
and also because renormalisation is slightly more involved than in
the by-now textbook linear $R_\xi=1$ gauge, implementation of the
non-linear gauge  is worth the investment if one is interested in
a general purpose automatic loop calculator. One then has the
possibility of performing gauge-parameter independence checks,
involving the $5$ parameters, on a large scale calculation. We
will see in section~\ref{sec:nlg-tests} that, though each diagram
out of a set of some hundreds diagrams shows a dependence in a
gauge parameter, summing over all diagrams gives a gauge
independent result. This is a very powerful check on the
calculation and on the automatic system.

\subsection{Generation of matrix elements}
As is in the case of tree processes, in general  the matrix
elements can be obtained  either through the helicity formalism or
one  can obtain directly the squared matrix elements with the help
of some symbolic manipulation programs. {\tt GRACE-loop} has both
versions.

The procedure to get the helicity amplitudes is as follows. Once a
diagram is generated a {\tt REDUCE} code is produced which
contains an expression for the corresponding  amplitude. This is
just a ``reading of the amplitude" in {\tt REDUCE}.  We do not
take any trace of $\gm$ matrices unless a fermion loop is formed
inside the diagram. The role of {\tt REDUCE} is only to rearrange
the generated terms. First, the Feynman parameters for the loop
integration are introduced, see Section~\ref{num-para-intg} and
Eq.~\ref{tensorintl4}. The loop momentum $l$, see
Eq.~\ref{int-M-N} is shifted so that one has denominators
containing $l$ only through $l^2$. The amplitude now contains not
only the loop, external momenta and Feynman
parametric variables but also strings of $\gm$ matrices.\\
\noi Operations are done  in $n$-dimension. All the Lorentz
contractions are taken in $n$-dimension if the pair of indices
does not bridge two fermion lines. Then, without taking the trace,
every product of $\gm$ matrices along the fermion lines is
replaced by a corresponding symbol which works as a function
 when the code is converted to a {\tt FORTRAN} source code.
The remaining contractions and the calculation of the products of
$\gm$ matrices are left to be done numerically in particular by
the {\tt CHANEL} library that we already described in the case of
tree-level amplitudes. As stated, {\tt REDUCE} is used, in this
approach, to get the final form of the amplitude, but only for the
rearrangement of terms and the functions such as {\tt spur}
(Trace) and {\tt index} (contraction) are applied only to quite a
limited part of the manipulation. \\
Now the obtained amplitude can be regarded as a polynomial with
respect to the loop integration parameters. Each monomial is
replaced by a symbol which represents a one-loop integral with the
numerator corresponding to that monomial. Thus the amplitude can
be calculated when a library of loop integrals is supplied, see
Section~\ref{num-para-intg}. Since we do substantially nothing
with the amplitudes, it is quite easy to get them.

On the other hand, in the second approach traces of all $\gm$
matrices and contractions of indices are applied systematically on
the product of a  tree amplitude with a loop amplitude. Again
dimensional regularisation is assumed in the case of UV divergent
diagrams. The coefficient of a monomial of the parameters for the
loop integration is merely a polynomial composed of external
momenta. That monomial is also replaced by a symbol which is a
substitute for the corresponding loop integral. This approach
generates much larger expressions than the helicity approach, see
Table~\ref{tab:size}. It may happen that  {\tt REDUCE} is not able
to complete all the manipulations because too huge intermediate
expressions are produced.

The helicity amplitude formalism is certainly desirable because it
can provide the spin information. However, the sums on spin states
required for external and internal particles produce a certain
number of arithmetic operations. This would not cause any problem
for $2\to2$ processes where the total cross section is obtained by
integrating the matrix element with respect to only one variable,
the scattering angle. When the final state contains more than
2-body, however, one needs to perform a multi-dimensional
integration over phase space with, for example, 4 variables for
$2\to3$ and 7 for $2\to4$ processes. If one relies on integration
packages that are based on Monte-Carlo algorithms, one has to
compute the loop matrix element many times, typically in the order
of 50K $-$1M times in total. To get the ${\cal {O}}(\al)$
corrections the contribution of the one-loop diagrams must be
combined with the cross section for real photon emission. Usually
the latter almost compensate the virtual corrections. Thus the
number of points also should be properly chosen to achieve good
accuracy.  Hence, at least at present, it is not always realistic
especially for complicated processes such as $2 \ra 3$ and $2\ra
4$ processes to use the method based on the helicity amplitudes
for the computation of the total cross section as this will
require a lot of CPU time. The helicity amplitude approach is,
however,  helpful for comparing and checking, at several points in
phase space,  the squared matrix elements obtained the symbolic
way, through {\tt REDUCE} for example.  Another limitation of the
symbolic way, however,  is that it cannot be applied to a process
that has no tree-level contribution such as $\gm\gm\to \gm \gm$
and $\gm \gm \ra \gm Z$. For such cases one has to use the
helicity amplitude approach.

\subsection{Regularisation scheme and the issue of $\gamma_5$}
\label{regg5}  An important step in the proof of renormalisability
of the \sm is that the symmetries that are present at tree-level
are still preserved at the loop level. With a  theory like the \sm
that involves  both vector and axial currents, it  has been known
that all gauge symmetries may not be preserved at the loop level
due to the Adler-Bell-Jackiw anomaly\cite{Anomly1}. In the case of
the \smp, the charge assignments of all particles within a family
is such that the anomaly is not present\cite{AnomalySMfree}. It is
important to stress that the existence of a {\em genuine} anomaly
has nothing to do with regularisation in the sense that there
could be no regularisation that can make the anomaly disappear
while maintaining all other quantum symmetries.  Nonetheless an
unfortunate choice of regularisation can induce an {\em apparent}
anomaly that would violate some Ward identities that could stand
in the way of a proper renormalisation programme. The use, even in
QED,  of a naive cut-off  is such an example. These {\em apparent}
anomalies can be removed by the introduction of extra counterterms
usually not obtained through multiplicative renormalisation. This,
of course, unduly complicates  the renormalisation procedure and
can make the implementation at the automation level more
problematic. Pauli-Villars regularisation\cite{Pauli-Villars} that
works so well in QED fails in the case of non-Abelian gauge
theories because a mass term is introduced in a naive way.
Dimensional Regularisation\cite{DimRegthV,DimReg2} is a very
powerful and extremely {\em practical} scheme for regularising
divergences in non-Abelian theories\footnote{For a step-by-step
presentation of the method, see for example \cite{Ryder} or
\cite{PeskinBook}. The chiral anomaly is also nicely exposed in
these textbooks.}. In DR, the loop integrals as well as the Dirac
and tensor algebra are calculated in a space-time with arbitrary
$n$ dimensions. It is relatively easy to implement in an automatic
code for one-loop calculations. The divergences are isolated as
poles in $n-4$ through, for example,  the variable $C_{UV}$ as is
done in {\tt GRACE}, with
\beqn
\label{cuvdef} C_{UV}=\frac{1}{\varepsilon} -\gamma_E+\log 4\pi,
\quad n=4-2 \varepsilon.
\eeqn
\noi For the electroweak \smp, the problem with DR is the
treatment of $\gamma_5$ which is a $4$-dimensional object and
hence can not be continued to an arbitrary $n$-dimension. With
calculations involving $\gamma_5$ one therefore needs some
additional scheme or prescription. In the original 't~
Hooft-Veltman prescription\cite{DimRegthV,gamma5maison},
$\gamma_5$ anticommutes with the $\gamma$ matrices that carry  one
of the $n=4$ dimensions but commutes with the rest. This split is
not satisfactory as it breaks Lorentz invariance in the full
$n$-dimensional space and does not manifestly respect the
conservation of the gauge current\cite{Bonneaug5,PeskinBook}. Many
variations on the original scheme, that may be considered as not
being fully mathematically consistent, have been applied in
different calculations, see\cite{gamma5jegerlehner} for a review.
For applications to the anomaly-free \smp, the most efficient and
practical scheme, especially from the point of view of
implementation in a computer code, is the so-called Naive
Dimensional Regularisation whereby $\gamma_5$ is taken as fully
anticommuting with all the $n$-dimensional $\gamma$ matrices. In
\grcl this is what is implemented. Let us also point out that the
cyclicity of the trace is not used. For loop calculations in
supersymmetry DR is not quite appropriate since it breaks
supersymmetry by splitting the number of fermionic and bosonic
degrees of freedom. An alternative is Dimensional
Reduction\cite{DRED1,DRED-inconsistency}. A very nice review is
given in\cite{RegSUSY-review,DREDpedagogical}. At the one-loop
level an equivalent prescription is Constrained Differential
Renormalisation\cite{CDR}, CDR.
Reference\cite{formcalc1,formcalc2} shows how both DR and CDR can
be implemented in the automatic code, {\tt FormCalc} for the
calculation of one-loop diagrams.

\subsection{Size of the generated programs}
\label{sec:size-code} One of the difficulties with running codes
of one-loop processes produced by an automatic system has to do
with the huge increase in the number of Feynman diagrams, as
compared to tree-level processes, combined with the equally large
size of the generated computer file for the   matrix elements of
{\em each} diagram. We could then easily end up with a total size
of a program for a given process which is so huge that it is not
always easy to compute the cross section in a realistic time
scale. To give a feel for the size of such programs, we show in
Table~\ref{tab:size} examples for some processes that have been
calculated so far with up to 6-leg final states ($2 \ra 4$
processes). Here the total number of diagrams  in a general
non-linear gauge is shown. Once the gauge parameter check is
successfully passed, for the cases of $2\to3$ and $2\to4$
processes the amplitudes are  generated again in the linear gauge,
which corresponds to a particular choice of the non-linear gauge
as will be explained and by switching off the electron Yukawa
coupling. This determines the set of amplitudes for the
``production job" to be supplied to a phase space integrator.
 We note in passing that switching off the electron Yukawa
coupling does not mean that all occurrences of the electron mass
are set to zero. In particular the electron mass is kept in order
to regularise collinear divergences, see section~\ref{seckctest},
or for the renormalisation of the fine structure constant in the
charge counterterm, see Appendix~\ref{sec:deltaY}.  \newline \noi
The size of the generated codes refers only to the amplitude and
does not take into account common necessary libraries. In the case
of 4-fermion production, to save CPU time the masses of the light
external fermions are neglected, as long as they appear in the
numerator of the matrix element. At some points in phase space the
full set and the production set are compared to confirm that the
latter is correctly generated. The actual CPU time for $e^+e^-\to
e^+e^-H$, for example, amounts to 150K hours equivalent to a
single IBM RS/6000(375MHz, Power 3) CPU in quadruple precision.

\vspace{0.5cm}
\begin{table}[htb]
\caption{{\em Size of the generated code for some cross sections
measured by the number of lines. We also show the total number of
tree-level and one-loop diagrams for each process in some general
gauge. The numbers in parenthesis have the small electron Yukawa
couplings switched off. ``helicity" refers to the method of
generating the helicity amplitude. {\tt REDUCE} refers to the
direct computation of the full matrix elements squared.}}
\label{tab:size}
\ct
\begin{tabular}{|c|c|c|c|c|c|}
\hline &&\multicolumn{2}{c|}{number of graphs}&\multicolumn{2}{c|}
{program size}\\
\cline{3-6}
$2\to n$-body&process&tree&one-loop&helicity&{\tt REDUCE} \\
&&&&(lines)&(lines)\\
\hline
$2\to2$&$e^+e^-\to t\bar t$&4(2)&150(54)&48K&54K\\
&$e^+e^-\to W^+W^-$&4(3)&334(153)&151K&282K\\
&$W^+W^-\to W^+W^-$&7(7)&925(921)&432K&1.28M\\
\hline
$2\to3$&$e^+e^-\to\nu_e\bar\nu_e H$\cite{eennhletter}&12(2)&1350(249)&92K&696K\\
&$e^+e^-\to e^+e^-H$\cite{eeeehgrace}&42(2)&4470(510)&------&154K\\
&$e^+e^-\to t\bar tH$\cite{eetthgrace}&21(6)&2327(771)&581K&1.50M\\
&$e^+e^-\to ZHH$\cite{eezhhgrace}&27(6)&5417(1597)&------&6.55M\\
&$e^+e^-\to \nu_e\bar\nu_e\gm$\cite{eennggrace}&10(5)&1099(331)&339K&1.06M\\
\hline
$2\to4$&$e^+e^-\to\nu_e\bar\nu_e HH$&81(12)&19638(3416)&------&50.5M\\
&$e^+e^-\to\mu^-\bar\nu_\mu u\bar
d$\cite{eemnudgrace}&44(10)&6094(668)
&1.22M&75.8M\\
\hline
\end{tabular}
\cte
\end{table}

\setcounter{equation}{0}
\section{The Standard Model in a general non-linear gauge}
\label{sec:sm-nlg}
\subsection{The classical Lagrangian}
To help define our conventions and notations we first introduce
the classical Lagrangian of the \sm which is fully gauge invariant
under $SU(2)\times U(1)$.

We denote the gauge fields of the theory of the $SU(2)\times U(1)$
group as $W^a_{\mu} (a=0, 1, 2, 3)$. The weak isospin triplet
refers to $1,2,3$ and the hypercharge singlet to the $0$
component. The  corresponding gauge couplings are $g^a$. The gauge
invariant field strength $F^a_{\mu\nu}$  writes in a compact form
as

\begin{equation}
F^a_{\mu\nu}=\partial_{\mu}W^a_{\nu} - \partial_{\nu}W^a_{\mu}
+g^a f^{abc}\ W^b_{\mu}W^c_{\nu}\;\;,\;\; g^a=\left\{
\begin{array}{ll}
g & (a=1, 2, 3) \\
g'& (a=0)
\end{array}
\right. , \;f^{abc}=\left\{
\begin{array}{cc}
\varepsilon^{abc} & (a, b, c \ne 0) \\
0                 & (\mathrm{otherwise})
\end{array}
\right. .
\end{equation}

\noi This leads to the pure gauge contribution ${{\cal L}}_G$
\begin{equation}
{{\cal L}}_G=-\frac{1}{4} F^a_{\mu\nu} F^{a\;\mu\nu}  \;.
\end{equation}

\noi The gauge interaction of the matter fields is completely
specified by their isospin and hypercharge ($Y$) quantum numbers,
such that the electromagnetic charge is $Q=T^3+Y$, and the
covariant derivative, $D_\mu$

\beqn D_\mu  = \partial_{\mu}- i g\sum_{a=1}^3 W^a_{\mu} T^a  -i
g^\prime Y W^0_{\mu} \;. \eeqn

\noi with $T^a=\sigma^a/2$, where $\sigma^a$ are the usual Pauli
matrices. Left-handed fermions $L$ of each generation belong to a
doublet while right-handed fermions $R$ are in a $SU(2)$ singlet.
The fermionic gauge Lagrangian is just

\begin{equation}
{{\cal L}}_F=i \sum \bar{L} \gamma^{\mu}D_{\mu} L
          +i \sum \bar{R} \gamma^{\mu}D_{\mu} R \; ,
\label{eq:fermion}
\end{equation}

\noi where the sum is assumed over all doublets and singlets of
the three generations.\\

\noi Mass terms for both the gauge bosons and fermions are
generated in a gauge invariant way through the Higgs mechanism. To
that effect one introduces a scalar doublet with hypercharge
$Y=1/2$ that spontaneously  breaks the symmetry of the vacuum
through a non-zero vacuum expectation value $v$

\begin{equation}
\phi=\frac{1}{\sqrt{2}} \left(\begin{array}{c} i\chi_1+\chi_2 \\
v+H-i\chi_3
\end{array}\right)=
\left(\begin{array}{c} i\chi^+ \\ (v+H-i\chi_3)/\sqrt{2}
\end{array}\right), \;\;\; \langle  0|\phi|0 \rangle= \left(\begin{array}{c} 0 \\ v/\sqrt{2} \end{array}\right) \;.
\label{eq:sccomp}
\end{equation}

\noi The scalar Lagrangian writes

\begin{equation}
\label{eq:hpot} {{\cal L}}_S=(D_\mu \phi)^{\dagger}(D^\mu
\phi)+{{\cal L}}_{\rm pot}\;\;,\;\;{{\cal L}}_{\rm
pot}=-V(\phi)=\mu^2\phi^{\dagger}\phi-\lambda(\phi^{\dagger}\phi)^2
\equiv - \lambda \left( \phi^{\dagger}\phi
-\frac{v^2}{2}\right)^2+\frac{\mu^4}{4\lambda} \;.
\end{equation}

\noi The Nambu-Goldstone bosons $\chi^\pm,\chi_3$ in ${{\cal
L}}_S$ get absorbed by the $Z$ and $W^\pm$ to give the latter
masses ($M_{Z,W}$), while the photon $A$ remains massless. The
physical fields $W^\pm, A,Z$ relate to the original $W$ quartet as

\begin{equation}
\left\{
\begin{array}{l}
\displaystyle{A_{\mu}=\frac{g'W^3_{\mu}+gW^0_{\mu}}{\sqrt{g^2+g'^2}}
= s_WW^3_{\mu}+c_WW^0_{\mu} } \\
{ } \\
\displaystyle{Z_{\mu}=\frac{gW^3_{\mu}-g'W^0_{\mu}}{\sqrt{g^2+g'^2}}
= c_WW^3_{\mu}-s_WW^0_{\mu} } \\
{ } \\
\displaystyle{W_{\mu}^{\pm}=\frac{W^1_{\mu}\mp i W^2_{\mu}}{\sqrt{2}} } \; ,\\
\end{array}
\right.
\end{equation}
with
\begin{equation}
c_W=\frac{g}{\sqrt{g^2+g'^2}},\quad s_W=\frac{g'}{\sqrt{g^2+g'^2}}
\; , \label{eq:wangle1}
\end{equation}
the electromagnetic coupling $e$
\begin{equation}
e=\frac{gg'}{\sqrt{g^2+g'^2}}, \quad g=\frac{e}{s_W}, \quad
g'=\frac{e}{c_W}, \label{eq:wangle2}
\end{equation}
and the masses
\begin{equation}
M_W=\frac{ev}{2s_W},\qquad M_Z=\frac{ev}{2s_Wc_W}.
\label{eq:gmass}
\end{equation}

\noi ${{\cal L}}_S$ also defines the mass of the Higgs
\begin{equation}
M_H^2=2\mu^2. \label{eq:hmass}
\end{equation}

\noi $\mu,\lambda$ and $v$ are not all independent parameters. $v$
is defined to be the minimum of the scalar potential. This is
equivalent to requiring no tadpole in ${{\cal L}}_S$. In other
words we require the coefficient, $T$,
\begin{equation}
T=v(\mu^2-\lambda v^2), \label{eq:tadp}
\end{equation}
\noi of the term linear in $H$ in Eq.~\ref{eq:hpot} to be zero,
$T=0$. We will impose this requirement to all orders.

\noi Fermion masses require the introduction of a corresponding
Yukawa coupling, $f_U$ ($f_D$) for an up-type fermion $f$ (for a
down-type fermion)
\begin{equation}
{{\cal L}}_M= -\sum_{{\rm up}} f_U \bar{L}_U\tilde{\phi}R_U
-\sum_{{\rm down}} f_D \bar{L}_D\phi R_D + (h.c.),
\;\;\;\tilde{\phi}=i \sigma ^2 \phi^{*} \;\;,\;\;
m_{U,D}=\frac{f_{U,D}\; v}{\sqrt{2}} \;. \label{eq:scferm}
\end{equation}

Instead of the original set of independent parameters
$\{g,g^\prime,\lambda,\mu^2, f_{U,D}\}$, it is much more
advantageous to revert to an equivalent set of physical parameters
that is directly related to physical observables, namely $\{e,
M_W, M_Z,M_H, m_U, m_D\}$. In this respect note that the weak
mixing angle is just a book-keeping quantity that will be defined,
at all orders of perturbation, in terms of the masses of the
vector bosons:
\beqn
\label{defswmwmz}
\cw=\frac{M_W}{M_Z}.
\eeqn
\noi If one allows $v$ to be an independent parameter we will
trade it for the tadpole, $T$, which we will add to the list of
independent parameters that specify the theory.

\def\db{\delta_{\rm BRS}}
\subsection{Quantisation: Gauge-fixing and Ghost Lagrangian}
\label{quantisation}
 \noi As known because of the gauge freedom in the
classical Lagrangian ${{\cal L}}_C$, ${{\cal L}}_C={{\cal
L}}_G+{{\cal L}}_F+{{\cal L}}_S+{{\cal L}}_M$, a Lorentz invariant
quantisation requires a gauge fixing. We generalise the usual
't~Hooft linear gauge condition to a more general non-linear gauge
that involves five extra parameters\cite{nlg-generalised},
$\zeta=({\tilde{\alpha},\tilde{\beta},\tilde{\delta},\tilde{\kappa},\tilde{\epsilon}})$.

\beqn
\label{fullnonlineargauge} {{\cal L}}_{GF}&=&-\frac{1}{\xi_W}
\overbrace{|(\partial_\mu\;-\;i e \tilde{\alpha} A_\mu\;-\;ig c_W
\tilde{\beta} Z_\mu) W^{\mu +} + \xi_W^\prime \frac{g}{2}(v
+\tilde{\delta} H +i \tilde{\kappa} \chi_3)\chi^{+}}^{G^+}|^{2} \nonumber \\
& &\;-\frac{1}{2 \xi_Z} (\overbrace{\partial.Z + \xi_Z^\prime
\frac{g}{ 2 c_W}
(v+\tilde\varepsilon H) \chi_3}^{G^Z})^2 \;-\frac{1}{2 \xi_A} (\overbrace{\partial.A }^{G_A})^2 \nonumber \\
&\equiv &-\frac{1}{\xi_W} G^+ G^- -\frac{1}{2 \xi_Z} (G^Z)^2
-\frac{1}{2 \xi_A} (G^A)^2
\eeqn

Note that it is not essential for the Feynman parameters
$\xi_{W,Z}^\prime$ that appear within the functions $G^{\pm,Z}$ to
be equal to those that appear as factors of $G^+ G^-$ $(\xi_{W})$
and $G_Z^2$$(\xi_{Z})$. However in this case ${{\cal L}}_{GF}$
does not cancel, at tree-level, the mixing terms $\chi\mbox{-}W,
\chi_3\mbox{-}Z$, see for
instance\cite{PeskinBook,Aitchison-littlebook}.  To avoid this
unnecessary complications we will stick to
$\xi_{W,Z}^\prime=\xi_{W,Z}$.

To construct the ghost Lagrangian ${{\cal L}}_{Gh}$, we will
require that the full effective Lagrangian, or rather the full
action, be invariant under the BRST transformation (the measure
being invariant). The required set of transformations needed to
construct the ghost Lagrangian together with the definition of the
ghost fields can be found in Appendix \ref{app-brstrans}. This is
a much more appropriate procedure than the usual Fadeev-Popov
approach especially when dealing with the quantum symmetries of
the generalised non-linear gauges we are studying. This implies
that the full quantum Lagrangian

\beqn
{\cal L}_Q={\cal L}_{C} + {\cal L}_{GF} + {\cal L}_{Gh}\; ,
\eeqn

\noi be such that $\db {\cal L}_Q=0$ and therefore $\db {\cal
L}_{GF}= - \db {\cal L}_{Gh}$.

\noi Moreover  we appeal to the auxiliary $B$-field formulation of
the gauge-fixing Lagrangian ${\cal L}_{GF}$. We will see later
that this formulation is also very useful to extract some
Ward-Takahashi identities. Within this approach

\beqn
\label{lgfB} {\cal L}_{GF}=\xi_W B^+B^- + \frac{\xi_Z}{2}|B^Z|^2 +
 \frac{\xi_A} {2}|B^A|^2 + B^-G^+ + B^+G^- + B^Z G^Z + B^A
 G^A.
\eeqn
From the equations of motion for the $B$-fields one recovers the
usual ${\cal L}_{GF}$ together with the condition
$B^i=-\frac{G^i}{\xi_i}$ ($\xi=\{\xi_W,\xi_Z,\xi_A\}$).

Defining the anti-ghost, $\bar c^i$, from the gauge fixing
functions, we write
\beqn
\label{antigtransf} \db \bar c^i=  B^i \, .
\eeqn

\noi Then by identification
\beqn {\cal L}_{Gh}&=&-\left(\bar
c^+\db G^+ + \bar c^-\db G^- + \bar
c^Z\db G^Z + \bar c^A \db G^A \right) + \db \tilde{{\cal L}}_{Gh} \nonumber \\
&\equiv & {\cal L}_{FP}+ \db \tilde{{\cal L}}_{Gh}\; .
\eeqn

\noi That is, one recovers the Fadeev-Popov prescription, ${\cal
L}_{FP}$, but only up to an overall function, $\db \tilde{{\cal
L}}_{Gh}$, which is BRST invariant. The complete Feynman rules we
list assume $\tilde{{\cal L}}_{Gh}=0$ which is sufficient for
one-loop calculations. For higher orders a counterterm not of the
Fadeev-Popov type, but which is BRST invariant on its own, may be
required to renormalise a quartic ghost vertex, in this case one
can take $\tilde{{\cal L}}_{Gh}=\lambda \epsilon_{ijk} \db (\bar
c^i \bar c^j c^k)$. The full set of Feynman rules derived from
${\cal L}_{Q}$ is relegated to Appendix \ref{sec:frule}. These
Feynman rules are derived with an arbitrary set
$\zeta=(\alphat,\betat,\deltat,\kappat,\epsilont),
\xi_i=(\xi_W,\xi_Z,\xi_A$) although for one-loop applications and
for all our tests we stick with the 't~Hooft-Feynman gauge
$\xi_W=\xi_Z=\xi_A=1$ where the gauge boson propagators take a
very simple form. Only their ``transverse" part $g_{\mu \nu}$
contributes, see Eq.~\ref{propagatorexplain}. This also greatly
simplifies the calculations not only because the expressions get
more compact but also because of the fact that the longitudinal
parts introduce a high degree of (superficial) ultraviolet
divergences. Although these greatly simplify when adding, all
diagrams the cancellations are very subtle and may be not
efficiently handled when implemented numerically. In practical
calculations one can also tune $\alphat, \betat,..$ so that one
reduces the number of diagrams and simplify some of the vertices.
For instance for photonic vertices $\alphat=1$ is to be preferred
since there is no $W^\pm \chi^\mp A$ vertex and also because the
$WW\gamma$ simplifies considerably. One can also choose $\betat$
so that $W^\pm \chi^\mp Z$ vanishes.

\setcounter{equation}{0}
\section{Renormalisation and counterterms}
\label{sec:renormalisationx}
\subsection{Renormalisation constants}
\label{sec:rconst}

The renormalisation procedure follows very closely the on-shell
renormalisation scheme, carried in\cite{kyotorc} in the case of
the usual linear gauge. The set of physical input parameters
includes all the masses of the model together with the value of
the electromagnetic coupling as defined in the Thomson limit. As
explained above we   also add the tadpole, $T$, to this list.
Renormalisation of these parameters would then lead to finite
S-matrix elements. For the mass eigenstates and thus a proper
identification of the physical particles that appear as external
legs in our processes, field renormalisation is needed. S-matrix
elements obtained from these rescaled Green's functions will lead
to external legs with unit residue. Therefore one also needs wave
function renormalisation of the fields. In the linear gauge with
all $\xi=1$ this also renders Green's functions finite. Especially
for the unphysical sector of the theory, the precise choice of the
fields redefinition is not essential if one is only interested in
S-matrix elements of physical processes. We will therefore
concentrate essentially on the renormalisation of the physical
parameters and physical fields, although we also introduce field
renormalisation for the Goldstone bosons.

All fields and parameters introduced so far in section~3 are
considered as bare parameters with the exception of the gauge
fixing Lagrangian which we choose to write in terms of {\em
renormalised fields}.  Care should then be exercised when we split
the tree-level contributions and the counterterms. In Appendix
\ref{app-ren-gf} we also present the alternative approach where
the gauge-fixing term is also written in terms of bare parameters.
Differences between the two approaches, of course, only affect the
unphysical scalar sector. In Appendix \ref{ward-id-goldstones} we
derive some useful Ward identities that constrain the two-point
functions in this sector.

For the renormalised quantity $X$, the corresponding  bare value
will  be defined by an underlined {\LARGE $_{-}$} symbol,
$\BARE{X}$, and its  counterterm by $\delta X$ .

\noi For the physical parameters, and the tadpole, we define

\beqn
\label{ctparameters}
\BARE{M}^2_{W}&=&{M}^2_{W}+\DM{W},   \nonumber \\
\BARE{M}^2_{Z}&=&{M}^2_{Z}+\DM{Z}, \nonumber \\
\BARE{m}_f&=&m_f+\Dm{f}, \nonumber \\
\BARE{M}^2_H &= &M^2_H + \DM{H}, \nonumber \\
\BARE{e}&=&Y e = (1 + \delta Y) e, \nonumber \\
\BARE{T}&=& T + \delta T.
\eeqn

\noi We now turn to the fields and the wave function
renormalisation constants.

\begin{enumerate}
\item Gauge fields
\begin{equation}
\label{wfrct}
\begin{array}{l}
\displaystyle{\BARE{W}^{\pm}_{\mu}=\ZF{W} \;{W}^{\pm}_{\mu} } \;\;,\;\; \ZF{W}=1+\ZH{W}, \nonumber \\
{ } \nonumber \\
\displaystyle{ \left( \begin{array}{c} \BARE{Z}_{\mu} \nonumber \\
\BARE{A}_{\mu} \end{array} \right) =
\left( \begin{array}{cc} \ZF{ZZ} & \ZF{ZA} \nonumber \\
                         \ZF{AZ} & \ZF{AA} \end{array} \right)
\left( \begin{array}{c} {Z}_{\mu} \nonumber \\ {A}_{\mu}
\end{array} \right) } \;\;,\;\;
\nonumber \\
\ZF{AA,ZZ}=1+\ZH{AA,ZZ}\;\;,\;\;
\ZF{AZ,ZA}=\ZH{AZ,ZA} \; .\nonumber \\
{ }
\end{array}
\end{equation}

\item Fermions

For simplicity we will assume no fermion mixing and therefore no
\cpviol violation. In this case the wave functions renormalisation
constants $\ZH{fL,fR}$ can be taken real as we will see later.
\begin{equation}
\begin{array}{l}
\BARE{\stackrel{(-)}{f}}_{L,R}= \ZF{f_{L,R}}\; \stackrel{(-)}{f}_{L,R} \;\;,\;\; \ZF{f_{L,R}}=1+\ZH{f_{L,R}} \; .\nonumber \\
\end{array}
\end{equation}

\item Scalars
\begin{equation}
\begin{array}{l}
\BARE{S}= \ZF{S}\;S  \;\;,\;\; \ZF{S}=1+\ZH{S} \;\;,\;\; S=H,\chi^{\pm},\chi_3\nonumber \; .\\
\end{array}
\end{equation}
\end{enumerate}

Because we are only presenting an application to processes at
one-loop, there is no need to be specific about the
renormalisation of the ghost sector. This is sketched in Appendix
\ref{app:rcountergh}. Suffice to say that to generate the ghost
Lagrangian including counterterms one needs to re-express ${{\cal
L}}_{GF}$ in terms of bare fields to first generate, through BRST
transformations, ${{\cal L}}_{Gh}$ with bare fields. This is
because ${\cal L}_{GF}$ is written in terms of renormalised fields
and as such does not induce any counterterm. However the BRST
transformations are defined for bare fields.

The generated counterterm library for all 3 and 4-point vertices
is listed in Appendix~\ref{sec:vtxcnt}. The counterterms are fixed
through renormalisation conditions that are set, with the
exception of the $e^+ e^- A$ vertex\footnote{In fact, as we will
see, due to a Ward identity, the counterterm for the charge can be
expressed in terms of two-point functions also, see
Eq.~\ref{deltayct}.}, from the two-point functions to which we now
turn.

\subsection{Two-point functions at one-loop including counterterms}
\label{sec:ct2pt}

We work in the on-shell scheme closely following \cite{kyotorc}
for the determination of the renormalisation constants. The
renormalisation conditions on the parameters are essentially
derived by properly defining the masses of  all the physical
particles $A,Z,W^\pm,f$ and the electromagnetic constant. They are
all set from the propagator and the $e^+e^-A$ vertex. Let us first
turn to the  propagators of  the fields of the theory.

The counterterm contribution will be denoted by a caret while the
full contribution (counterterm and one-loop diagrams contribution)
is denoted by a tilde, so that for the vector boson we may write

\beqn
\label{defpi} \tilde{\Pi}=\Pi+\hat{\Pi} \;.
\eeqn

\noi Moreover it is necessary to decompose these contributions
according to their Lorentz structure. For our purpose we will only
consider the case of no mixing (and hence no \cpviol violation) in
the fermionic sector. The decomposition of two-point functions is
as follows.

\begin{center}
\begin{tabular}{ll}
\hline
type & formula \\
\hline
\rule[-5mm]{0mm}{12mm} vector-vector & $
\displaystyle{\Pi_{\mu\nu}(q^2)=
  \left(g_{\mu\nu}-\frac{q_{\mu}q_{\nu}}{q^2} \right) \Pi_T(q^2)
 + \frac{q_{\mu}q_{\nu}}{q^2}  \Pi_L(q^2) } $ \\
\hline
\rule[-5mm]{0mm}{12mm} scalar-scalar & $ \Pi(q^2) $ \\
\hline
\rule[-5mm]{0mm}{12mm} vector-scalar & $iq_{\mu} \Pi(q^2) $ ($q$ is the momentum of the incoming scalar ) \\
\hline \rule[-1mm]{0mm}{6mm} fermion-fermion & $\Sigma(q^2)= K_1 I
+ K_5 \gamma_5
 + K_{\gamma} \slashq
 + K_{5\gamma} \slashq \gamma_5 $ \\
\hline
\end{tabular}
\end{center}

\vspace{3mm}

Complete one-loop results for all two-point functions in the
generalised non-linear gauge are collected in
Appendix~\ref{sec:prop-corr}.

\noi The contribution of the counterterms to the two-point
functions writes

\begin{enumerate}
\item Vector-Vector

\begin{center}
\begin{tabular}{ll}
\hline
$WW$ & $\hat{\Pi}_T^W = \DM{W} + 2(M_W^2-q^2)\ZH{W} $ \\
     & $\hat{\Pi}_L^W = \DM{W} + 2M_W^2\ZH{W} $ \\
\hline
$ZZ$ & $\hat{\Pi}_T^{ZZ} = \DM{Z} + 2(M_Z^2-q^2)\ZH{ZZ} $ \\
     & $\hat{\Pi}_L^{ZZ} = \DM{Z} + 2M_Z^2\ZH{ZZ} $ \\
\hline
$ZA$ & $\hat{\Pi}_T^{ZA} = (M_Z^2-q^2)\ZH{ZA}-q^2\ZH{AZ} $ \\
     & $\hat{\Pi}_L^{ZA} = M_Z^2\ZH{ZA} $ \\
\hline
$AA$ & $\hat{\Pi}_T^{AA} = -2q^2\ZH{AA} $ \\
     & $\hat{\Pi}_L^{AA} = 0 $ \\
\hline
\end{tabular}
\end{center}

\item Scalar-Scalar

\begin{center}
\begin{tabular}{ll}
\hline \rule[-2mm]{0mm}{8mm} $HH$ & $\hat{\Pi}^{H}
 = 2(q^2-M_H^2)\ZH{H} - \DM{H} + \frac{3\delta T}{v}$ \\
\hline \rule[-2mm]{0mm}{8mm} $\chi_3\chi_3$ & $\hat{\Pi}^{\chi_3}
 = 2q^2\ZH{\chi_3} + \frac{\delta T}{v}$ \\
\hline \rule[-2mm]{0mm}{8mm} $\chi\chi$ & $\hat{\Pi}^{\chi}
 = 2q^2\ZH{\chi} + \frac{\delta T}{v}$ \\
\hline
\end{tabular}
\end{center}

\item Vector-Scalar

\begin{center}
\begin{tabular}{ll}
\hline $W\chi$ & $\hat{\Pi}^{W\chi}
 = M_W (\delta M_W/M_W + \ZH{W}+\ZH{\chi}) $ \\
\hline $Z\chi_3$ & $\hat{\Pi}^{Z\chi_3}
 = M_Z (\delta M_Z/M_Z + \ZH{ZZ}+\ZH{\chi_3}) $ \\
\hline $A\chi_3$ & $\hat{\Pi}^{A\chi_3}
 = M_Z \ZH{ZA} $ \\
\hline
\end{tabular}
\end{center}


\def\ree{\Re e}
\def\im{\Im m}

\item Fermion-Fermion

At one-loop, this sector is unaffected by the parameters of the
non-linear gauge and thus all functions are as in the usual linear
gauge case. As mentioned earlier, all wave functions constants are
real since we do not consider mixing in the fermionic sector.

\beqn
\label{zctfermions}
\hat{K}_1  &=&-m_f \left(\ZH{fL} + \ZH{fR}
  \right)-\Dm{f},  \nonumber \\
\hat{K}_5&=&0, \nonumber \\
\hat{K}_{\gamma} &=&
\left(\ZH{fL} + \ZH{fR} \right), \nonumber \\
\hat{K}_{5\gamma} &=&
=-\left(\ZH{fL} - \ZH{fR}
  \right) \; .
\eeqn
\end{enumerate}

\subsection{Renormalisation Conditions}
\label{sec:ren-cdts}
Leaving aside the renormalisation of the
electromagnetic charge, these two-point functions give all other
renormalisation constants. Before deriving these let us first turn
to the tadpole.

\begin{enumerate}
\item Tadpole \par
The counterterm for the tadpole contribution, $\delta T$, is
defined such that the tadpole loop contribution $T^{loop}$ and the
counterterm $\delta T$ combine such that the tadpole
$\tilde{T}=T^{loop}+\delta T$=0. Then

\begin{equation}
\delta T= - T^{loop}. \label{eq:crnrmtp}
\end{equation}

 The tadpole contribution in the electroweak \sm is
sometimes necessary. An example is the loop two-point functions of
the massive vector bosons and the Higgs in order to check the
BRST\cite{BRS,Tyutin} or the Slavnov-Taylor\cite{SlavnovTaylor}
identities, see for example Appendix~\ref{ward-id-goldstones}.

\item Charged vector \par
The conditions specify that the pole-position of the propagator is
$M_W^2$, and that the residue of the propagator at the pole is 1.
\begin{equation}
\left. \Re e \tilde{\Pi}^W_T(M_W^2)=0,\quad \frac{d}{d q^2}\Re e
\tilde{\Pi}^W_T(q^2)\right|_{q^2=M_W^2}=0 \; .\label{eq:rnrmw}
\end{equation}
This gives the following relations.
\begin{equation}
\left. \DM{W}=-\Re e \Pi^W_T(M_W^2),\quad \ZH{W}=\frac{1}{2}
\frac{d}{d q^2}\Re e \Pi^W_T(q^2)\right|_{q^2=M_W^2}\; .
\label{eq:crnrmw}
\end{equation}

\item Neutral vector \par
The conditions to be imposed on  the photon-photon and $Z-Z$
self-energies are the same as for the $W\mbox{-}W$ transition. In
addition we require that there should be no mixing between $Z$ and
the photon at the poles $q^2=0, M_Z^2$.
\begin{equation}
\left. \Re e \tilde{\Pi}^{ZZ}_T(M_Z^2)=0,\quad \frac{d}{d q^2} \Re
e \tilde{\Pi}^{ZZ}_T(q^2)\right|_{q^2=M_W^2}=0 \; ,
\label{eq:rnrmz}
\end{equation}
\begin{equation}
\left. \tilde{\Pi}^{AA}_T(0)=0,\quad \frac{d}{d
q^2}\tilde{\Pi}^{AA}_T(q^2)\right|_{q^2=0}=0 \; ,\label{eq:rnrma}
\end{equation}
\begin{equation}
\tilde{\Pi}^{ZA}_T(0)=0,\quad \Re e \tilde{\Pi}^{ZA}_T(M_Z^2)=0\;
. \label{eq:rnrmza}
\end{equation}
Among these 6 conditions, $\tilde{\Pi}^{AA}_T(0)=0$ produces
nothing, except that it ensures that the loop calculation does
indeed give $\Pi^{AA}_T(0)=0$. One then derives,
\begin{equation}
\left. \DM{Z}=-\Re e\Pi^{ZZ}_T(M_Z^2),\quad \ZH{ZZ}=\frac{1}{2}
\Re e \frac{d}{d q^2}\Pi^{ZZ}_T(q^2)\right|_{q^2=M_Z^2}\; ,
\label{eq:crnrmz}
\end{equation}
\begin{equation}
\ZH{AA}=\frac{1}{2}\frac{d}{d q^2}\Pi^{AA}_T(0) \; ,
\label{eq:crnrma}
\end{equation}
\begin{equation}
\ZH{ZA}=-\Pi^{ZA}_T(0)/M_Z^2,\quad \ZH{AZ}=\Re e
\Pi^{ZA}_T(M_Z^2)/M_Z^2 \; .\label{eq:crnrmza}
\end{equation}

\item Higgs \par
The conditions specify that the pole-position of the propagator is
$M_H^2$, and that the residue of the propagator at the pole is 1,
\begin{equation}
\left. \Re e \tilde{\Pi}^H(M_H^2)=0,\quad \frac{d}{d q^2} \Re e
\tilde{\Pi}^H(q^2)\right|_{q^2=M_H^2}=0 \; .\label{eq:rnrmh}
\end{equation}
This gives the following relations.
\begin{equation}
\left. \DM{H}=\Re e\Pi^H(M_H^2)+\frac{3 \delta T}{v},\quad
\ZH{H}=-\frac{1}{2}\frac{d}{d q^2}\Re e
\Pi^H(q^2)\right|_{q^2=M_H^2} \; .\label{eq:crnrmh}
\end{equation}

\item Fermion \par
The conditions for pole-positions and residues are the same as for
the other physical particles. Also the vanishing of $\gamma_5$ and
$\gamma^{\mu}\gamma_5$ terms at the pole is required. These
conditions read
\begin{equation}
m_f \Re e \tilde{K}_{\gamma}(m_f^2)+ \Re e \tilde{K}_1(m_f^2)=0,
\quad \left.\frac{d}{d \slashq} \Re e \left( \slashq
\tilde{K}_{\gamma}(q^2)+\tilde{K}_1(q^2)\right)
\right|_{\slashq=m_f}=0, \quad \label{eq:rnrmf}
\end{equation}
\begin{equation}
 \Re e \tilde{K}_{5}(m_f^2)=0, \quad
\Re e  \tilde{K}_{5\gamma}(m_f^2)=0. \label{eq:rnrmf2}
\end{equation}

\cpviol invariance leads to $K_5=0$.
In this case, one can take both  $\ZH{fL}$ and $\ZH{fR}$ to be
real using the invariance under a phase rotation. We obtain the
following relations.
\begin{equation}
\begin{array}{l}
\Dm{f}=  \Re e \left( m_f {K}_{\gamma}(m_f^2) + {K}_1(m_f^2) \right)  \; ,\\
\ZH{fL}= \frac{1}{2} \Re e( {K}_{5\gamma}(m_f^2) -
{K}_{\gamma}(m_f^2) ) \left.
       - m_f \frac{d}{d q^2} \left(m_f \Re e K_{\gamma} (q^2)
       +\Re e K_{1} (q^2)\right)\right|_{q^2=m_f^2} \; ,\\
\ZH{fR}=-\frac{1}{2} \Re e( {K}_{5\gamma}(m_f^2)+
{K}_{\gamma}(m_f^2) ) \left.
       - m_f \frac{d}{d q^2} \left(m_f \Re e K_{\gamma} (q^2)
       + \Re e K_{1} (q^2)\right)\right|_{q^2=m_f^2} \; .\\
\end{array}
\label{eq:crnrmf}
\end{equation}

\item Charge \par
While there are many vertices in the theory, if the charge $e$ is
properly renormalised, we do not need any further renormalisation
conditions. The condition can be imposed on any vertex. The most
natural one is to fix the $e^+e^-A$ vertex as is usually done in
QED by relating it to the Thomson limit. The condition requests
that the coupling is $-e$ when $q$, the momentum of the photon, is
0, while the $e^\pm$ with momentum $p_\pm$ are one shell,
\begin{equation}
\left. (e^+e^-A\ \mathrm{one\ loop\ term} + e^+e^-A\
\mathrm{counter\ term}) \right|_{q=0,p_\pm^2=m_e^2} = 0 \; .
\label{eq:chargerc}
\end{equation}
The counterterm is  defined in Appendix~\ref{sec:vtxcnt}. From
this, we obtain $\delta Y$. In fact we will see that due to a Ward
identity, see for example \cite{Bohmrc}, $\delta Y$ writes as a
combination of $\ZH{AA}$ and $\ZH{ZA}$ which is valid in all
gauges.

\item The unphysical sector \par
Because we are interested in applications to physical processes,
the renormalisation of this sector is not adamant. Nonetheless one
may choose to work, as far as  possible, with finite Green's
functions involving the Goldstones and the longitudinal modes of
the vector bosons. With a linear gauge-fixing condition in the
't~Hooft-Feynman gauge, and in the approach we are taking where
the gauge-fixing Lagrangian is written in terms of renormalised
fields from the outset, all divergences in this sector are taken
care of by properly choosing $Z_{\chi_3,\chi^\pm}$. Therefore,
following\cite{kyotorc}, we define the wave-function
renormalisation for $\chi=\chi_3,\chi^\pm$,

\beqn
\left. \ZH{\chi}=-\frac{1}{2}\frac{d}{d q^2}
\left(\Pi^\chi(q^2)\right)\right|_{\cuv-{\rm part}}\; .
\eeqn

where $\left.\Pi^\chi(q^2)\right|_{\cuv-{\rm part}}$ is the
divergent part of the Goldstone two-point functions. We extend the
same definition in the case of the non-linear gauge, see for
example Eq.~\ref{explicitzchi}. In our approach, where the gauge
fixing term is expressed in terms of renormalised quantities from
the onset, this is not sufficient to make all the unphysical
scalar two-point functions and mixing finite in the non-linear
gauge. In fact in our approach and with the non-linear gauge, even
$\tilde{\Pi}_L^{W^\pm}$, which does not involve $\ZH{\chi^\pm}$,
is not finite. However as shown in
Appendix~\ref{ward-id-goldstones}, there is a strong constraint on
the two-point functions of the unphysical scalars. For our purpose
of using these kinds of gauge fixing to check the gauge-parameter
independence of the results, this also is a non trivial test on
the finiteness and gauge independence of the results. In Appendix
\ref{app-ren-gf} we show explicitly how one may choose to have
finite two-point functions in the Goldstone sector at the expense
of renormalising the gauge parameters. This method could be
followed but it introduces a few extra renormalisation constants,
which may slow the code for the cross section evaluation. In any
case, although we can calculate the cross sections in any gauge,
the gauge-parameter independence check is systematically applied
on some random points in phase space. When this is passed we
generally calculate the cross section in the linear gauge, with
all 't~Hooft-Feynman parameters being equal to one. In this case,
linear gauge condition with $\xi=1$, both approaches are
equivalent.

\end{enumerate}

\subsection{Some remarks on the explicit form of the renormalisation constants}
\label{remarks-wfr-one}
The renormalisation procedure outlined above together with the
exact and complete computation of all two-point functions permits
to derive in a straightforward manner the explicit expressions for
all parameters and wave function renormalisation constants. The
complete expressions for all two-point functions are defined in
Appendix \ref{sec:prop-corr}. From the conditions imposed in
section \ref{sec:ren-cdts} one immediately extracts all the
necessary counterterms. Since the general expressions for these
are lengthy and can be read off from Appendix \ref{sec:prop-corr},
we do not list all of them here, but just comment on some
important general features.

\subsubsection{Mass shifts and charge renormalisation}
We first verify that all counterterms to the input parameter of
the physical particles, namely the masses of all particles,
$\delta M_{W,Z,H,f}$, are gauge-parameter independent. This also
applies to the tadpole counterterm. This constitutes a strong
check on our results.

The same holds for the charge renormalisation constant, $\delta
Y$. Although this may be derived from the knowledge of $\ZH{AA}$
and $\ZH{ZA}$ through a Ward identity, it is easy to compute it
directly. This is done explicitly in Appendix \ref{sec:deltaY}. We
find the gauge-parameter independent result
\begin{equation}
\label{deltayct}
\delta Y = - \ZH{AA} + \frac{s_W}{c_W}\ZH{ZA}\; .
\end{equation}
While both $\ZH{AA}$ and  $\ZH{ZA}$ are gauge-parameter dependent,
see Eqs.~\ref{zaa-explicit}-\ref{zza-explicit} below, the above
combination is universal.
\begin{equation}
\delta Y= \frac{\alpha}{4\pi} \left\{
-\frac{7}{2}(\Cuv-\logw)-\frac{1}{3} +\frac{2}{3}\sum_f Q_f^2
(\Cuv-\log m_f^2) \right\}.
\end{equation}
\noi $\Cuv$ is defined in Eq.~\ref{cuvdef}.
\subsubsection{Wave function renormalisation constants}
Since $\ZH{AA}$ and $\ZH{ZA}$ are crucial for charge
renormalisation and since their expressions are rather simple we
give them explicitly.\\

\noindent We have
\beqn
\label{zaa-explicit} \ZH{AA}&=& \frac{\alpha}{4\pi} \left[
\left(\frac{3}{2}+2\anlg\right)\left(\Cuv-\logw\right)
+\frac{1}{3} -\frac{2}{3}
\sum_f Q_f^2 \left(\Cuv-\log m_f^2\right) \right] \nonumber \\
&\equiv& \frac{\alpha}{4\pi} \left[ \left(-2
(1-\alphat)+\frac{7}{2} \right) \left(\Cuv-\logw\right)
+\frac{1}{3} -\frac{2}{3} \sum_f Q_f^2 \left(\Cuv-\log
m_f^2\right) \right],\nonumber \\
\eeqn

\noi where a summation on all fermions of charge $Q_f$ is
performed.

\begin{equation}
\label{zza-explicit} \ZH{ZA}= -\frac{\alpha}{2 \pi}
\frac{c_W}{s_W} (1-\alphat) \; \left(\Cuv-\logw\right) \;
.\label{eq:zfacza}
\end{equation}
%

\noi This shows that although both $\ZH{AA}$ and $\ZH{ZA}$ are
gauge-parameter dependent, the combination that appears in the
charge renormalisation is not. Moreover, observe that the choice
$\alphat=1$ gives a vanishing $Z$-$A$ transition at one-loop. This
is due to the residual $U(1)$ gauge symmetry which remains after
gauge fixing the charged sector, with this particular choice of
the gauge parameter.

The remaining wave functions constants are not very illuminating
and involve lengthy expressions that we can extract from Appendix
\ref{sec:prop-corr}. Here we just list the gauge-parameter
dependence of those of the physical particles which can be
expressed in a rather compact form as

\beqn
\label{eq: zhnlgdependence}
\ZH{AZ}&=& \widetilde{\ZH{AZ}} + \frac{\alpha}{2\pi}\frac{
c_W}{s_W} \betat
\left(\Cuv-\Re e F_0(W,W,Z) \right), \nonumber \\
\ZH{ZZ}&=&  \widetilde{\ZH{ZZ}} + \frac{\alpha}{2\pi}\frac{
\cww}{\sww} \betat
\left(\Cuv-\Re e F_0(W,W,Z) \right), \nonumber \\
\ZH{W}&=&  \widetilde{\ZH{W}} -\frac{\alpha}{4 \pi}\frac{1}{\sww}
\left\{ \sww \alphat \left(\Cuv-\Re e F_0(A,W,W) \right) +\cww
\betat
\left(\Cuv-\Re e F_0(Z,W,W) \right) \right\}, \nonumber \\
\ZH{H}&=&  \widetilde{\ZH{H}} +\frac{ \alpha }{8\pi}\frac{1}{\sww}
\left\{ \deltat
 \left(\Cuv-\Re e F_0(W,W,H) \right) +\frac{\epsilont}{2\cww}
\left(\Cuv-\Re e F_0(Z,Z,H) \right) \right\}\; .\nonumber \\
\eeqn

\noi where the quantities with $\widetilde{\;\;\;\;\;\;\;}$
correspond to the linear gauge result with all Feynman parameters
set to 1. The function $F_0$ is defined in
Appendix~\ref{app-onetwoptfct}. As
known\cite{eezhsmrcDenner,eezhsmrcKniehl,eezhsmrcJeger} the
requirement of having the residues of the renormalised propagators
of all physical particles to be unity leads to a (very sharp)
threshold singularity in the wave function of the Higgs at the
thresholds corresponding to $M_H=2 M_W,2 M_Z$. This singularity is
all contained in the explicit derivative term in
$\widetilde{\ZH{H}}$ and is therefore gauge-parameter independent.
Solutions to smooth this behaviour\cite{threshold-sing}, like the
inclusion of the finite width of the $W$ and $Z$, do exist but we
have not implemented them yet in the present version of \grcp.
Therefore when scanning over $M_H$ it is sufficient to avoid these
regions within $1$GeV around the thresholds.

\newcommand{\wfr}{wave function renormalisation$\;$}
\newcommand{\wfre}{wave function renormalisation}

\subsection{Issues of renormalisation for unstable particles}
\label{unstableparticles}
\subsubsection{Wave function renormalisation for unstable particles: absorptive part and gauge dependence}
\label{subsec:imwfr} As can be explicitly seen in the previous
paragraph \wfr constants, contrary to the counterterms for the
physical parameters such as the masses and couplings, are
gauge-parameter dependent. This is a reflection that fields are
not physical observables. In fact, since at the end we are only
interested in S-matrix observables, we could have defined an
approach where we could have done without the introduction of \wfr
but at the expense of not having finite Green's
functions\footnote{ In fact, as stated in
section~\ref{sec:ren-cdts}, not all two-point functions and mixing
involving the unphysical scalars are finite in the general
non-linear gauge in our approach starting with a renormalised
gauge-fixing term.}.  We discussed this aspect when we defined the
wave function renormalisation of the unphysical Goldstone
particles in section~\ref{sec:ren-cdts} and we argued that the
\wfr of the Goldstones, which only appear as internal particles,
cancels out exactly. The argument applies to all particles that
only appear as internal particles. Indeed, the \wfr constant would
appear both in the correction to the propagator and to the two
vertices to which this propagator attaches. It is easy to see that
the effect from the propagator cancels that from the vertices. The
\wfr therefore only applies to the external particles. However
there is a problem when one is dealing with the \wfr of unstable
particles. One aspect of this problem was just pointed out in the
previous paragraph, section~\ref{remarks-wfr-one}, related to the
singularity brought about by the \wfr of the Higgs near the
threshold for $WW$ and $ZZ$ production. The other problem has to
do with the fact that the \wfr has been defined to be real so that
the Lagrangian be Hermitian. But obviously one is, for unstable
particles, applying Hermitian quantum field theory to
non-Hermitian problems\cite{Kleefeld:2002au}.

In the standard on-shell approach, because of \wfre,  loop and
counterterm insertions on external legs are simply and
conveniently not taken into account since they are thought to
cancel each other. The fact is, if one insists on only using the
real part of the \wfr then one does not completely cancel the
self-energy and counterterm insertions on the external legs of an
unstable particle. There remains in particular a contribution from
the absorptive part of the self-energy. These absorptive parts
occur for unstable massive particles when thresholds are crossed,
 they could also correspond to unphysical thresholds that occur in
gauges where the gauge parameter is $\xi \neq 1$.

\begin{figure}[!h]
\caption{{\em Selected $\tilde\del$ dependent diagrams for
$\chi^\pm$
             loop(left) and the vertex counterterm(right).}}
    \label{fig:hww}
    \begin{center}
    \includegraphics[width=14cm,height=7cm]{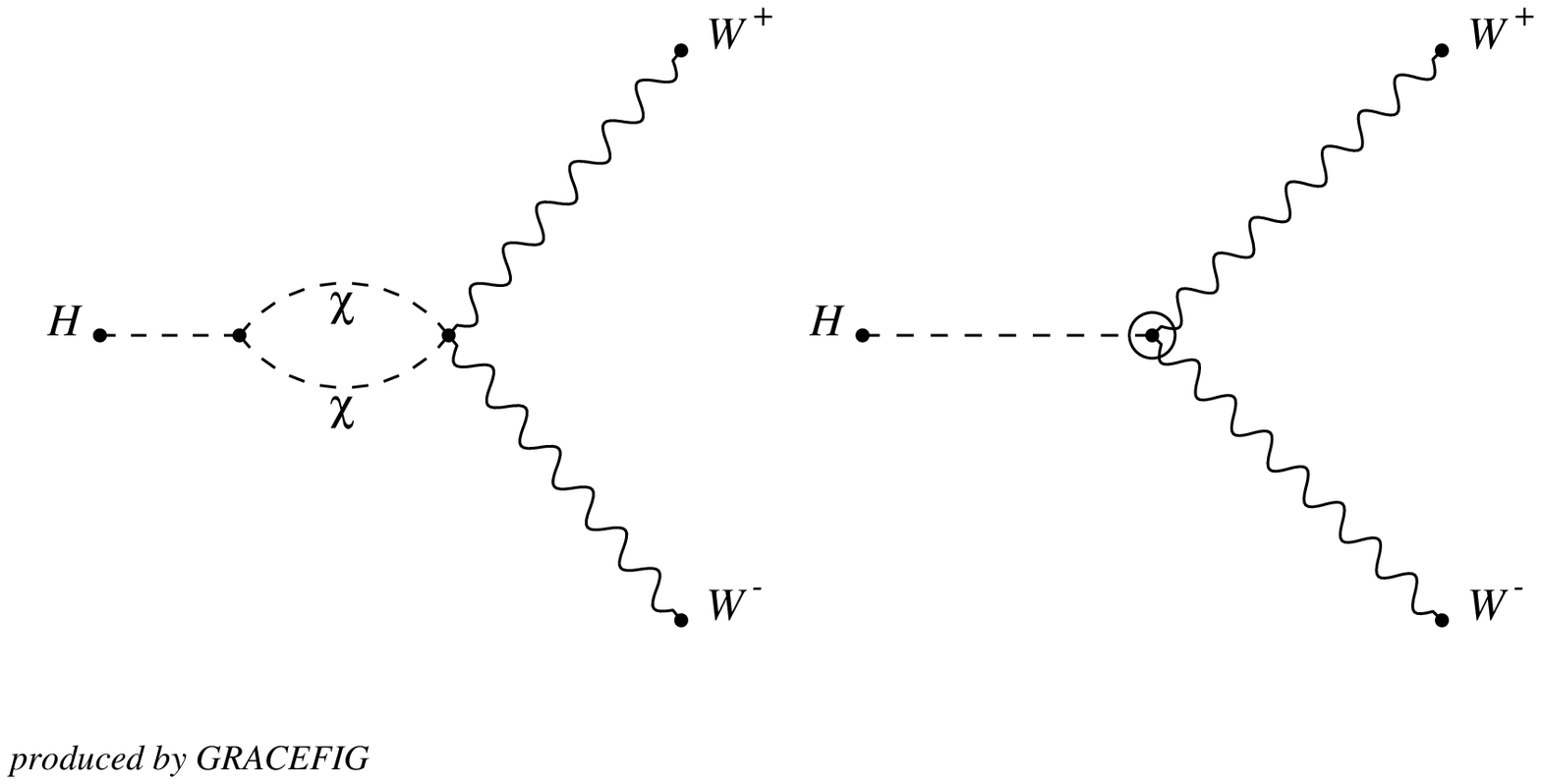}
    \end{center}
\end{figure}
Let us  show an example where a gauge dependent threshold appears
explicitly even for physical thresholds in the non-linear gauge.
We take again a Higgs heavy enough to decay into a pair of $W$ and
consider the one-loop {\em amplitude}. In particular we
concentrate here only on the eventual $\tilde{\delta}$ dependence.
This is contained in the two selected graphs in Fig.~\ref{fig:hww}
consisting of the fish-type one-loop diagram and the $HWW$
counterterm.

The fish-type diagram develops  a $\tilde{\delta}$ dependent
absorptive part which, at the amplitude level, does not cancel
against the counterterm contribution if one insists on real \wfr
for the Higgs as given by Eq.~\ref{eq: zhnlgdependence}. The
absorptive $\tilde{\delta}$ dependent contribution from the fish
would cancel on the other hand if $\delta Z_H$ is defined to
contain both the real and imaginary part. The inclusion of the
imaginary part of the \wfr is in fact just a convenient short-cut.
More correctly had we taken into account the Higgs self-energy
insertion (together with the counterterm) on the external leg of
the Higgs \footnote{Some care should be exercised for external leg
insertions, such as factors of $1/2$ and apparent $0/0$ divisions.
For detailed worked out examples see \cite{bailin-book}.}, see
Fig.~\ref{fig:hww-self-ext}, the $\tilde{\delta}$ dependence would
drop completely. In the non-linear gauge we found a few examples
for other non-linear gauge-fixing parameters and for other
particles besides the Higgs, like the top in $t \ra b W^+$.

\begin{figure}[!h]
\caption{{\em Self-energy insertion on the external leg of the
Higgs (right) with its counterterm (left) for $H \ra W^+ W^-$}.}
    \label{fig:hww-self-ext}
    \begin{center}
\mbox{
\includegraphics[width=8cm,height=7cm]{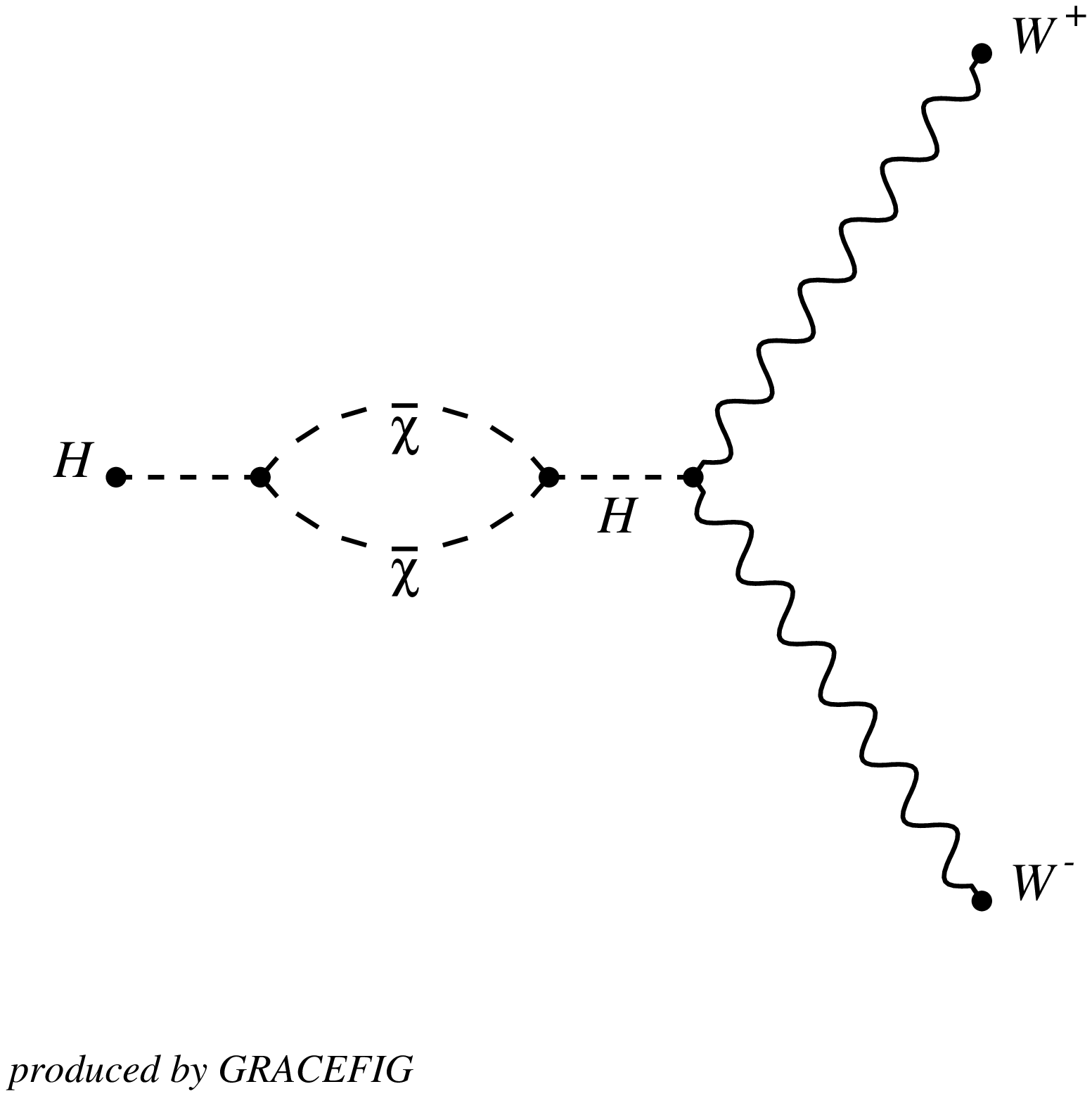}
\includegraphics[width=8cm,height=7cm]{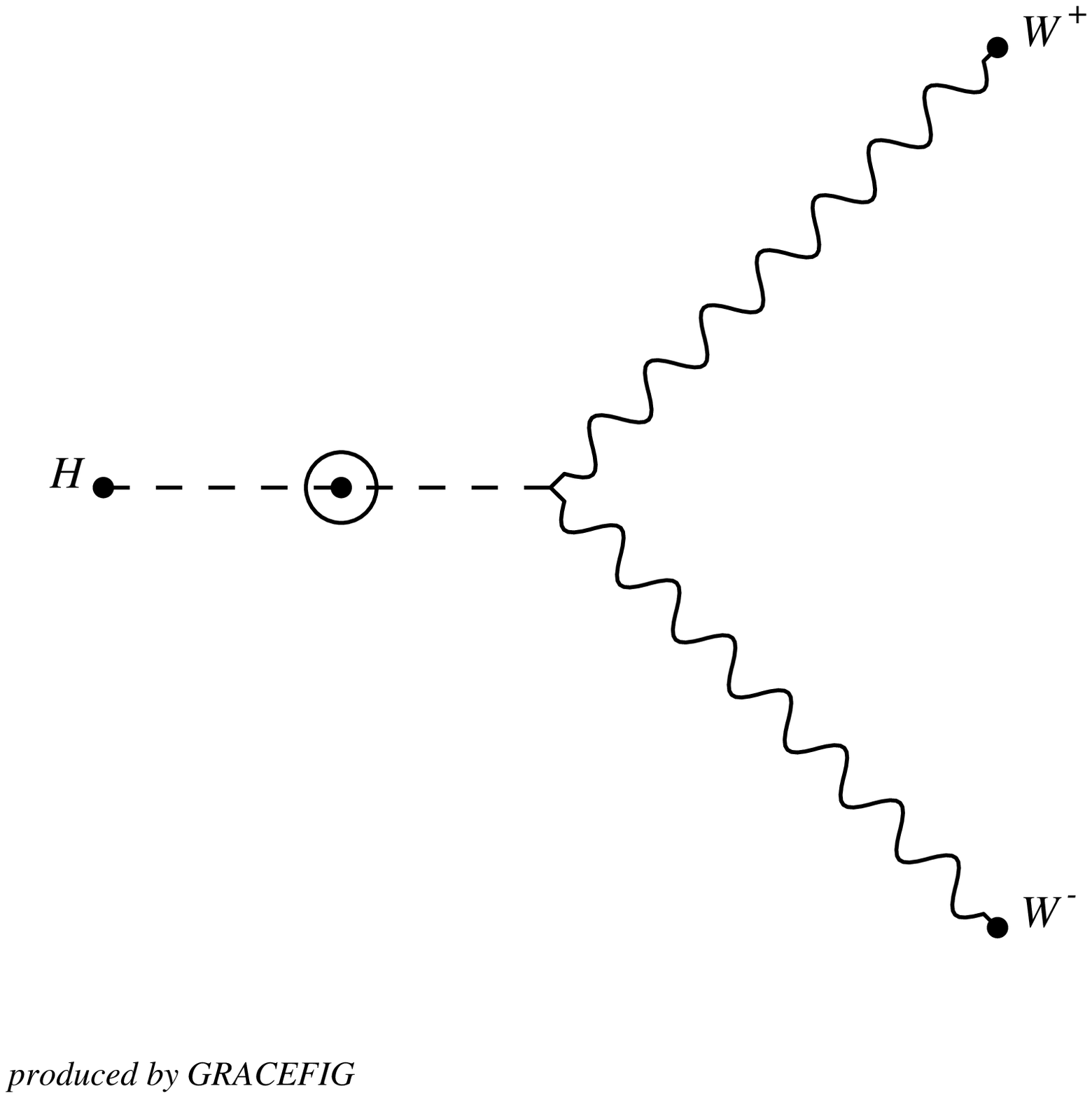}}
\end{center}
\end{figure}

In principle, if one is only interested in corrections to the
total cross section, at the one-loop level it is irrelevant
whether the imaginary part of the \wfr is included or not. This is
because the effect of the \wfr is an overall multiplicative factor
to the amplitude. For the total cross section one only needs $\Re
e (T_{\rm tree} \times T_{1-loop}^\dagger)$, where $T_{\rm tree}$
and $T_{1-loop}$   refer to the tree and one-loop amplitude
respectively. Since the \wfr contribution, $Z_p$ for a particle
$p$, to the one-loop amplitude is $Z_p \; T_{\rm tree}$, only $\Re
e Z_p$ would be picked up. But in applications  where CP violation
is an issue \cite{wfr-espriu} one might need the correct gauge
invariant one-loop {\rm amplitude}. This effectively requires that
the  imaginary part of the \wfr be implemented if one is to avoid
external leg insertions. This confirms that it is  essential to
require gauge independence at the level of the amplitude rather
than in the cross section. The gauge dependence argument makes the
inclusion of the absorptive part of the self-energy a necessity,
although the gauge dependent part must cancel against other
contributions, in all generality this should not be the case for
the gauge independent part which should then be observable. For
yet another aspect of the \wfre, see\cite{Kniehl-wfr}.

\subsubsection{Width implementation for resonant diagrams and gauge
invariance}

Since it is difficult to reconcile the concept of an asymptotic
state with an unstable particle, it has been
argued\cite{VeltmanWidthThesis} not to consider S-matrix elements
for external unstable particles. This provides an easy way out to
the problem we have just discussed concerning the wave function
renormalisation, especially that such an unstable particle decays
into stable particles. Nonetheless the treatment of unstable
particles even as internal particles poses problems. Even when
calculating tree-level processes one has to regularise the
propagator of an unstable particle if one is close to the
resonance region. This regularisation brings in elements which are
outside the order at which the perturbative calculation is being
carried out. This problem is  exacerbated when performing loop
calculations with unstable particles.  Take a tree-level matrix
element where part of the contribution is due to the exchange of a
massive particle which, in the following, we will take to be a
gauge boson,

\beqn
\label{wgi0} {\cal M}^0=\frac{R(s)}{s-M^2}+ T(s)
\eeqn
$s$ is some invariant mass and $M$ is the mass of the particle.
$R(s)$ represents the ``resonant" contribution and the remainder
$T$ some non resonant contribution. Although there are a few
instances where $R(s)$ and $T(s)$ are separately gauge
independent, in the most general case this need not be the case.
${\cal M}^0$ in Eq.~\ref{wgi0} is gauge independent. The residue
of the pole $R(M^2)$ and the non pole remnant $\tilde{T}(s)$, see
below, are however independently gauge
invariant\cite{StuartWidthPole,Sirlin-polemass,HVeltman-mass,AeppliWidth2}
\beqn
\label{wgipole0} {\cal M}^0=\frac{R(M^2)}{s-M^2}+
\overbrace{\left( \frac{R(s)-R(M^2)}{s-M^2} + T(s)
\right)}^{\tilde{T}(s)}.
\eeqn
Therefore the naive way of implementing a constant width $\Gamma$,
\beqn
\label{wgd} {\cal M}^0_\Gamma=\frac{R(s)}{s-M^2+i M \Gamma}+ T(s),
\eeqn
although numerically regulating the resonant behaviour, breaks, in
general,  gauge invariance and can lead to disastrous results, see
Ref.~\cite{widthminami} for such an example. Often the use of a
running width is made $\Gamma \ra \Gamma(s)$. The latter is
justified on the basis that it emerges from Dyson summation.
However this summation, that moreover mixes the orders in
perturbation theory, is only made on the self-energy two-point
function. It therefore breaks gauge invariance since only one part
of the total contribution to the amplitude is corrected. It has
been found\cite{width-schemes-review}, for tree-level process,
that in fact in most cases the running width does so much worse
than the constant width. One proposal to remedy this situation
while insisting on using the running width was to generalise the
Dyson summation to include accompanying corrections to vertices.
This scheme, the fermion pole
scheme\cite{Baurwidth,width-schemes-review,fermionscheme,fermionschemeImprove},
is in fact only part of the full one-loop calculation which is
quite unpractical especially from the point of view of an
automatic code. Moreover the scheme only takes into account
fermion loops for the gauge boson propagators and is only meant as
an  effective means to correct tree-level predictions. There is
also an effective Lagrangian approach to implement this scheme
with some considerations to the unstable top quark and
Higgs\cite{BeenakkerWidthEffective2,BeenakkerWidthEffective}.

Another scheme that is easily adapted to an automatic
implementation and that can be carried beyond tree-level is the
factorisation scheme. Starting with the gauge independent matrix
element of Eq.~\ref{wgi0} one endows the total contribution with
the {\em overall} factor \newline $(s-M^2)/(s-M^2+i M \Gamma)$,
such that
\beqn
{\cal M}^0 \ra \frac{s-M^2}{s-M^2+i M \Gamma} {\cal M}^0 .
\eeqn
Although gauge invariant, this  is unsatisfactory as it puts all
non-resonant contributions to zero close to the resonance. The
pole scheme
\cite{StuartWidthPole,Sirlin-polemass,HVeltman-mass,AeppliWidth2}
based on an expansion around the pole offers some insight. At
tree-level only the pole term is regulated by the introduction of
a width, while the non-resonant gauge invariant remnants are not
put to zero at the resonance. This amounts to consider
\beqn
\label{pole-tree}
{\cal M}^0_p=\frac{R(M^2)}{s-M^2+i M \Gamma}+
\left( \frac{R(s)-R(M^2)}{s-M^2} + T(s) \right)
\eeqn
In practice however, especially when one deals with many gauge
bosons and for multi-leg processes, this procedure becomes
untractable taking into account  that there might be a clash in
reconstructing the invariant $s$ from other momenta especially if
kinematical cuts are to be applied\cite{AeppliWidth2}. A variant
of the pole scheme\cite{BenekeWidth} in an effective Lagrangian
approach has also been advocated but its effectiveness has not
been fully demonstrated for processes of practical interest.

The majority of the schemes we have reviewed so far have mainly
been applied to tree-level processes although some implementations
like the fermion loop scheme are attempts at including parts of
the higher order corrections as are some of the effective
Lagrangian approaches. It is fair to say that there is as yet no
fully satisfactory solution for one-loop process. Nonetheless,
especially for neutral current processes that require the
introduction of a width, an automatic implementation can be
performed.
\\

\newcommand{\sff}{s_{f\bar f}}
\newcommand{\calmz}{{\cal {M}}^{(0)}}
\newcommand{\calmo}{{\cal {M}}^{(1)}}
\newcommand{\calnz}{{\cal {N}}^{(0)}}
\newcommand{\gz}{\Gamma_Z}
\newcommand{\propz}{\sff-\mzz+i\gz \mz}
\newcommand{\propfz}{\sff-\mzz+i\gzz \mz}
\newcommand{\propng}{\sff-\mzz}
\newcommand{\sts}{\tilde{\Pi}_T^{ZZ}(\sff)}
\newcommand{\stz}{\tilde{\Pi}_T^{ZZ}(M_Z^2)}
\newcommand{\azg}{A_{Z\gamma}\frac{\tilde{\Pi}_T^{Z\gamma}(\sff)}{\sff}}
\newcommand{\eezht}{$\epem \ra Z H \;$}

\underline{Treatment of the $Z$ width in {\tt GRACE-loop}: $\epem
\ra \epem H$ as an example}
\\

\begin{figure*}[hbtp]
\caption{{\em Contributing diagrams at tree-level in terms of the
s-channel type, left panel, obtained from $\epem \ra ZH$, and the
$t$-channel type from $ZZ$ fusion.}} \label{tree-level.fig}
\begin{center}
\includegraphics[width=10cm,height=3cm]{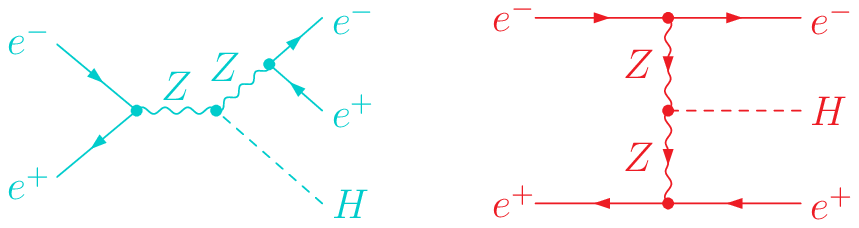}
\end{center}
\end{figure*}
Let us take as an example the case of \eeeeht at
one-loop\cite{eeeehgrace}. At tree-level, in the unitary gauge,
the \eeeeht process is built up from an $s$-channel diagram
originating from \eezht and a $t$-channel diagram which is a
fusion type, see Fig.~\ref{tree-level.fig}. Each type constitutes,
on its own, a gauge independent process. In fact the former
(neglecting lepton masses) can be defined as $\epem \ra \mu^+
\mu^- H$. This case therefore falls in the category where $R(s)$
($s$-channel here) and $T(s)$ ($t$-channel here) of Eq.~\ref{wgi0}
are separately gauge invariant. In principle it is only the $Z$
coupling to the outgoing lepton, in this $s$-channel contribution,
which can be resonating and thus requires a finite width.
Nonetheless in our code we dress both $Z$ in the $s$-channel type
diagrams with a constant $Z$ width. We apply no width to the $Z$
taking part in the $ZZ$-fusion diagrams.
\\
To help understand our implementation at one-loop it is
instructive to display a selection of some of the contributing
diagrams at one-loop.
\begin{figure*}[htbp]
\caption{\label{one-loop-diagrams} {\em A small selection of
different classes of loop diagrams contributing to \eeeeht. We
keep the same graph numbering as that produced by the system. {\tt
Graph 4311} belongs to the corrections from self-energies, here
both the virtual and counterterm contributions are generated and
counted as one diagram. {\tt Graph 349} shows a vertex correction.
Both graphs are considered as $s$-channel resonant Higgs-strahlung
contributions. {\tt Graph 762} represents a box correction, it is
a non resonant contribution, which  can not be deduced from
\eezht, but applies also to the correction to the $s$-channel
$\epem \ra \mu^+\mu^- H$. {\tt Graph 1481} is also a box
correction counted  as a correction to the $ZZ$ fusion. {\tt Graph
1575}, {\tt Graph 1741} and {\tt Graph 1757} are fusion type
corrections involving $\gamma \gamma$, $Z\gamma$ and $ZZ$ fusion.
{\tt Graph 2607} shows a pentagon correction which also counts as
an $s$-channel since it is induced for $\epem \ra \mu^+ \mu^- H$.
{\tt Graph 3157} on the other hand is a  pentagon correction that
only applies to \eeeeht \/.}}
\begin{center}
\includegraphics[width=16cm,height=14cm]{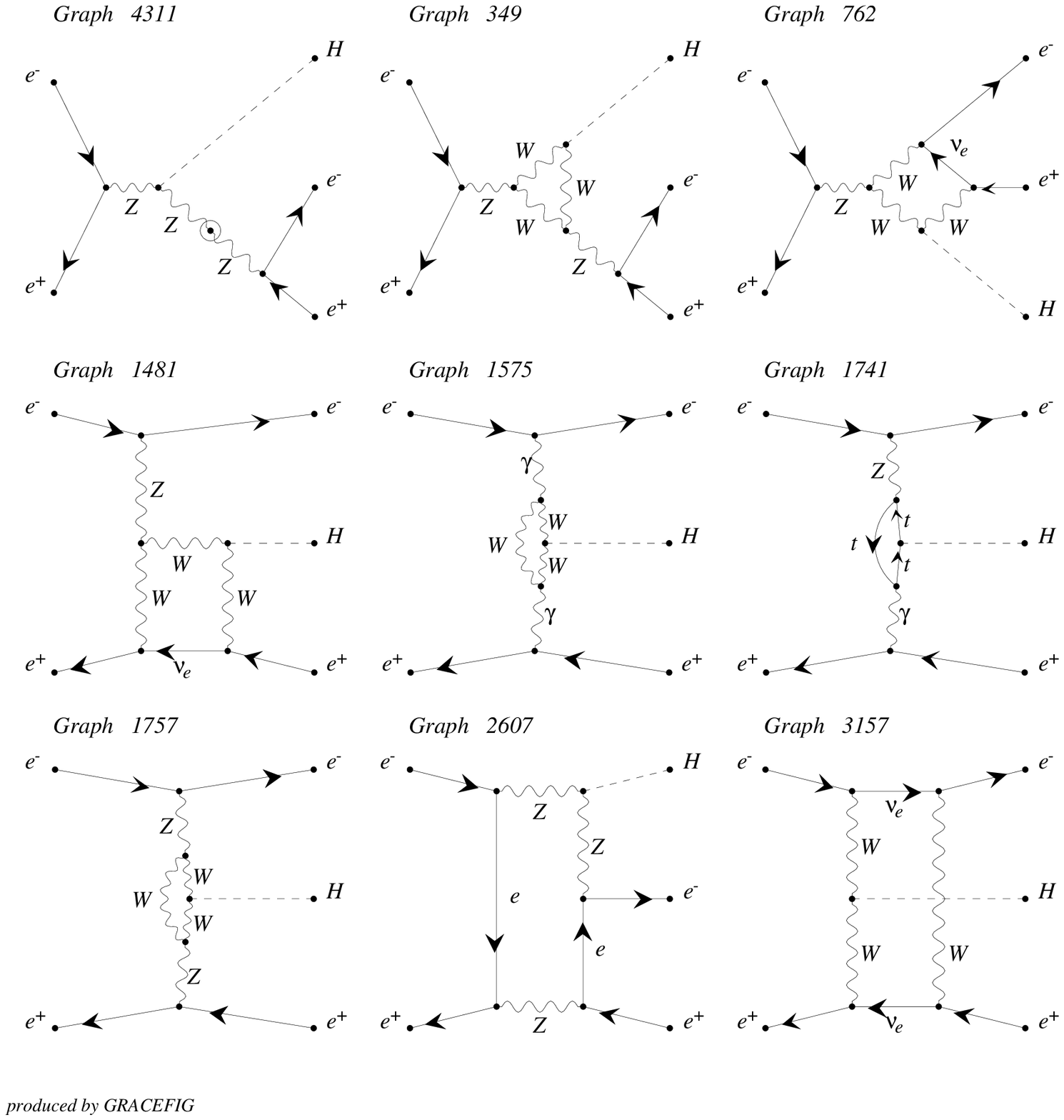}
\end{center}
\end{figure*}


\noi The introduction of a width is required only for the $Z$
coupling to the final electron pair. This $s$-channel contribution
is much smaller than the $t$-channel contributions for which we do
not endow the $Z$ propagators with a width. Building up on the
implementation of the width at tree-level, we include a constant
width to all $Z$ propagators {\em not circulating in a loop} for
the  $s$-channel type diagrams. For example we add a width to all
$Z$ in {\tt graphs 349,762,4311} of Fig.~\ref{one-loop-diagrams}.
For those one-loop diagrams with a self-energy correction to any
$Z$ propagator, represented by {\tt graph 4311} in
Fig.~\ref{one-loop-diagrams}, we follow a procedure along the
lines described in \cite{supplement100}. We will show how this is
done with a single $Z$ exchange coupling to a fermion pair of
invariant mass $\sff$.\\
\noi First, it is important to keep in mind that our tree-level
calculation of the $s$-channel is done by supplying the width in
the $Z$ propagator. Therefore it somehow also includes parts of
the higher order corrections to the $Z$ self-energy which should
be subtracted when performing a higher order calculation. The case
at hand is as simple as consisting, at tree-level, of one single
diagram. The simplest way to exhibit this subtraction is to
rewrite, the zero-th order amplitude, $\calmz$, before inclusion
of a width, in terms of what we call the tree-level (regularised)
amplitude, ${\widetilde {\cal M}}^{{\rm tree}}$
\beqn
\label{calmz}
\calmz=\frac{\calnz}{\propng}=\overbrace{\frac{\calnz}
{\propz}}^{{\widetilde{\cal M}}^{{\rm tree}}}
\left(1+\frac{i(\gz^0+\Delta\gz)\mz}{\sff-\mzz}\right), \quad
\Delta\gz=\gz-\gz^0.
\eeqn
The $\gz^0$ contribution will be combined with the one-loop
correction while $\Delta\gz$ will be counted as being beyond
one-loop.

 At the one-loop level, before the summation {\em \`a la} Dyson and
the inclusion of any ``hard" width, the amplitude is
gauge-invariant and can be decomposed as
\beqn
\label{1loopgi} \label{calmo} \calmo&=&\frac{\calnz}{\propng}
\frac{\sts}{\propng} +
\frac{\azg}{\propng}  + \frac{R_Z}{\propng} + C \nonumber \\
&=& \frac{1}{\propng}\left\{ \calnz \frac{\sts}{\propng} +
\left(\azg+R_Z\right) +(\sff-\mzz) C  \right\}.
\eeqn
The different contributions in $\calmo$ are the following. The
first term proportional to the tree-level contribution is due to
the renormalised transverse part of the $Z$ self-energy correction
$\tilde{\Pi}_T^{ZZ}$, including counterterms. Such a transition is
shown in {\tt Graph 4311} of Fig.~\ref{one-loop-diagrams}. The
term proportional to $\azg$ comes from the renormalised transverse
part of the $Z$-$\gamma$ self-energy, $\tilde{\Pi}_T^{Z\gamma}$,
with the photon attaching to the final fermion (this type is
absent for neutrinos in \eennht). The $R_Z$ terms combine one-loop
corrections  which nevertheless still exhibit a $Z$-exchange that
couples to the final fermions and hence these types of diagrams
can be resonant, an example is {\tt Graph 349} of
Fig.~\ref{one-loop-diagrams}. We can write $R_Z=Z_{ZH}+V_Z^f$,
where $Z_{ZH}$ corresponds to the part containing the correction
to $\epem \ra Z^\star H$, while $V_Z^f$ contains the corrections
to the final $Z_{f\bar f}$ vertex. $Z_{ZH}(s_{f\bar f}=M_Z^2)$
corresponds to $\epem \ra Z H$ and is gauge invariant at the pole.
The term $C$ contains all the rest which are apparently
non-resonant\footnote{Strictly speaking we, here, deal only with
the pure weak corrections. In the infrared limit some of the QED
diagrams can be resonant and require a $Z$ width even in a loop.
This is discussed in Ref.~\cite{eeeehgrace}.}, an example here is
{\tt Graph 762} of Fig.~\ref{one-loop-diagrams}. Both $\calmo$ and
$\calmz$ are gauge invariant.

Our procedure, in effects, amounts to first regularising the
overall propagator in Eq.~\ref{1loopgi} by the implementation of a
constant $Z$ width and then combining the renormalised $Z$
self-energy part in Eq.~\ref{1loopgi} with the $\gz^0$ part of
Eq.~\ref{calmz}. Since our on-shell renormalisation procedure is
such that $Re\tilde{\Pi}_T^{ZZ}(M_Z^2)=0$ and since
$\Gamma_Z^0=-Im \Pi_{T}^{ZZ}(M_Z^2)$, see \cite{nlgfatpaper}, our
prescription is to write
\beqn \label{width-rearrange}
\calmz+\calmo&\ra&\overbrace{\frac{\calnz}{\propz}}^{{\widetilde
{\cal M}}^{{\rm tree}}} + \overbrace{\frac{1}{\propz}
\widetilde{{\cal N}}^{1}}^{{\widetilde {\cal M}}^{{\rm 1-loop}}}
\nonumber \\
\widetilde{{\cal N}}^{1}&=&\calnz\frac{\left(\sts
-\stz\right)}{\propng} +
\left(\azg+R_Z\right) \nonumber \\
&+&(\sff-\mzz) C.\nonumber \\
\eeqn
The above prescription is nothing else but the factorisation
procedure avoiding double counting. It is gauge invariant but puts
the non-resonant terms to zero on resonance. In practice in the
automatic code, we supply a constant $Z$ width to all $Z$ not
circulating in a loop and by treating the one-loop $ZZ$
self-energy contribution as in Eq.~\ref{width-rearrange}. Up to
terms of order ${\cal{O}}(\Gamma_Z \alpha)$ this is equivalent to
Eq.~\ref{width-rearrange}. In particular the contribution of the
$C$ term  does not vanish on resonance, since its overall factor
is unity rather than the factor $(\sff-\mzz)/(\sff-\mzz+i\gz \mz)$
that would be present in the original factorisation prescription.
\\

\underline{The complex mass scheme}
\\
This scheme stems from the very simple observation that the
parameter $M^2$ in Eq.~\ref{wgi0} could be taken from the outset
as having an imaginary part. This imaginary part should be
included consistently even when it enters through couplings and
mixings as is the case for the electroweak mixing defined  in
Eq.~\ref{defswmwmz} through the ratio of the mass of the
(unstable) $W$ and $Z$. All the algebraic relations will therefore
be maintained in this analytic continuation, in particular gauge
invariance is not broken. Identifying the imaginary part with the
implementation of a width, at tree-level this scheme has been used
in \cite{complexmass-denner-tree}. This simple analytical
continuation is rather easily implemented in an automatic code for
the calculation of Feynman diagrams. The roots of this
idea\cite{StuartWidthComplex} pre-date its first practical usage
in a tree-level calculation and, in fact, emerged from
considerations at the loop level. The suggestion is that when
splitting the {\em real} bare parameters, as is done in
Eq.~\ref{ctparameters}, one may well take the renormalised and the
counterterm parameters to be both complex and similarly for the
fields and the wave function renormalisation as suggested already
by our discussion in section \ref{subsec:imwfr}. The only subtle
problem now is that the renormalised Lagrangian is not Hermitian.
This poses then the problem of perturbative unitarity and how one
defines the Cutksoky cutting rules\cite{CutkoskyRule}. Barring
this issue, a full implementation of this scheme has been worked
out at one-loop and first  applied to $\epem \ra
4f$\cite{eeto4fdenner}. Its implementation in an automatic code at
one-loop is rather straightforward since it amounts to define the
counterterms for the parameters of Eq.~\ref{ctparameters} and wave
function renormalisation of Eq.~\ref{wfrct} by including the
imaginary parts of the two-point functions, whereas the usual
on-shell scheme is based on taking only the real parts at the
appropriate renormalisation scale. The use of complex masses as we
have stressed needs to be carried out consistently by analytical
continuation. This means that even the corresponding masses that
enter the loop diagrams need to be complex. This calls for the
extension of the loop function, which we treat in the next chapter
(Chapter~\ref{num-para-intg}), to include complex mass arguments.
This automatically regularises some of the infrared resonant loop
integrals, a result that had been also arrived at by the direct
inclusion of the width in the loop so that factorisation of the
infrared factor be maintained\cite{supplement100}. For more
details about this scheme applied at one-loop see
Ref.~\cite{eeto4fdenner}. As it can be inferred by this
presentation this scheme lends itself to an easy implementation in
a code for the automatic calculations of Feynman diagrams.

\setcounter{equation}{0}
\setcounter{equation}{0}
\section{Evaluation of the loop integrals}
\label{num-para-intg}
\begin{figure}[htb]
\caption{\label{n-point-fig}{\em General structure of the N-point
loop integral. $l$ is the loop momentum, $M_i$ are the masses of
the particles circulating in the loop. $p_i$ are the external
momenta. $s_2$ is a combination of external momenta, see
Eq.~\ref{Di}. }}
\begin{center}
\includegraphics[width=8cm,height=8cm]{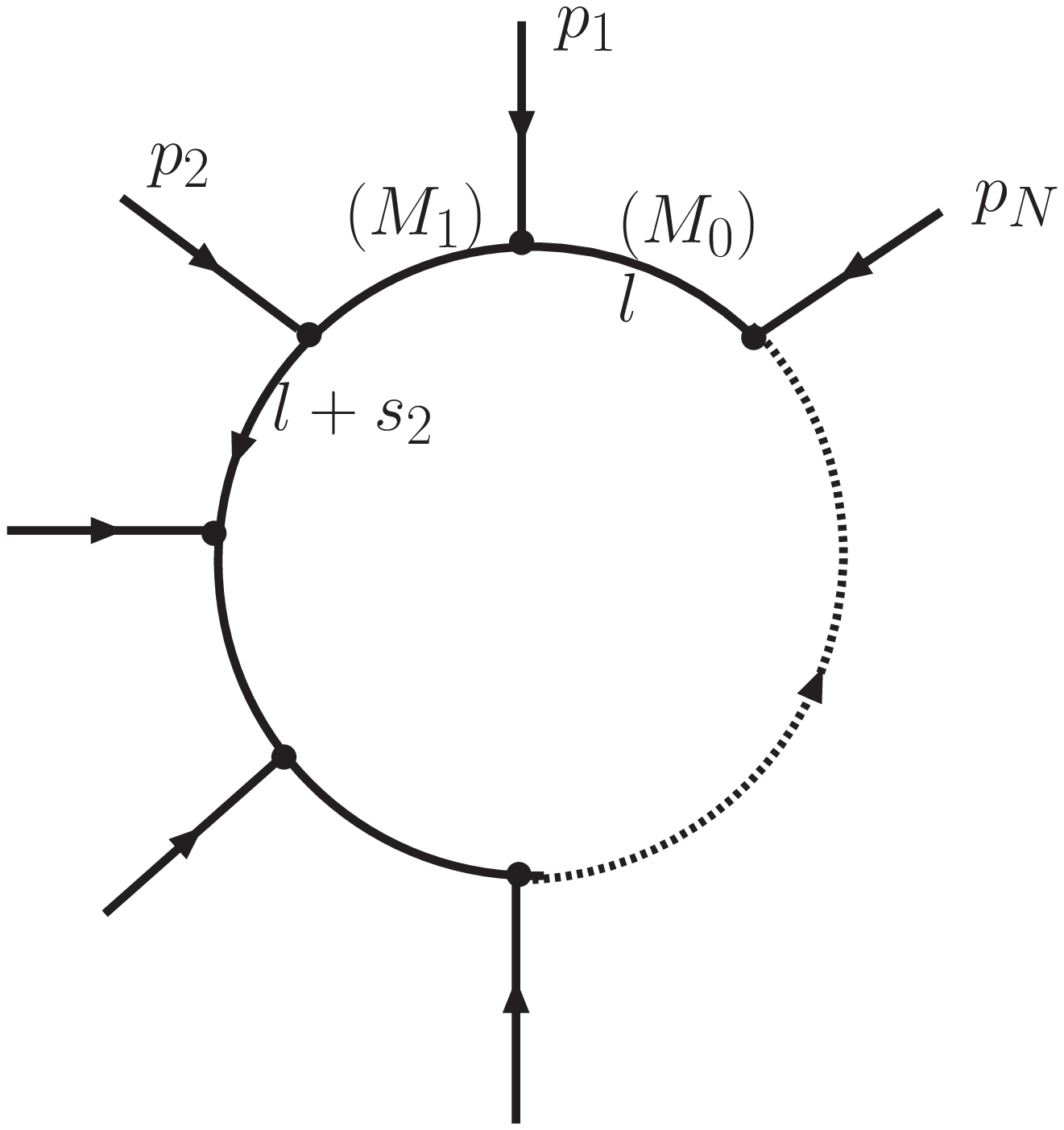}
\end{center}
\end{figure}

The evaluation of the loop integrals is one of the most important
ingredients of a loop calculation. This is also one of the most
time consuming especially as the number of external legs
increases. A generic loop integral involving $N$ external
particles is depicted in Fig.~\ref{n-point-fig}. The tensor
integral of rank $M$ corresponding to a $N$-point graph that we
encounter in the general non-linear gauge but with Feynman
parameters $\xi=1$ are such that $M \leq N$.  The object in
question writes in DR as

\beqn
\label{int-M-N}
T_{\underbrace{\mu \nu \cdots \rho}_{M}}^{(N)}=\int
\frac{d^n l}{(2\pi)^n} \; \frac{l_\mu l_\nu \cdots l_\rho}{D_0 D_1
\cdots D_{N-1}}, \quad M \leq N,
\eeqn
\noi where

\beqn
\label{Di}
D_i=(l+s_i)^2-M_i^2, \quad s_i=\sum_{j=1}^{i} p_j, \quad s_0=0.
\eeqn
\noi $M_i$ are the internal masses, $p_i$ the incoming momenta
and $l$ the loop momentum. \\

The $N$-point  scalar integrals correspond to $M=0$. All higher
rank tensors for a $N$-point function, $M\geq 1$, can be deduced
recursively from the knowledge of the $N$-point (and lower) scalar
integrals. In {\tt GRACE-loop} all tensor reductions of two, three
and four-point functions are performed by solving a system of
equations obtained by taking derivatives with respect to the
Feynman parameters. All higher orders parametric integrals
corresponding to the tensor integrals can then be recursively
derived from the scalar integral, as will be described below. It
is important to stress that this reduction is different to what is
usually done through the Passarino-Veltman \cite{PassarinoVeltman}
or the Brown-Feynman\cite{BrownFeynman} reductions. It is also
different from the approach of Bern, Dixon and
Kosower\cite{Bernparameters} who exploit differentiation of the
scalar integral with respect to a set of kinematical
variables.\\
\noi Although the present review mainly describes the methods of
one-loop calculations for up to $2 \ra 2$ processes, where only
$N\leq 4$ loop integrals are needed, we will describe briefly the
very recent development in the calculation of one-loop processes
with $5$ external legs and in one instance $6$ legs and in
particular how the $N=5,6$-point functions are treated in {\tt
GRACE-loop}. For $N\geq 5$ all integrals can be reduced to
$N=4$-point functions.

Since the computation of the scalar integrals, especially for
$N\leq 4$ is central let us first describe their implementation in
{\tt GRACE-loop}. Let us note that, in the intermediate stage of
the symbolic calculation dealing with loop integrals (in
$n$-dimension), we extract the regulator constant $C_{UV}$ defined
in Eq.~\ref{cuvdef}. We treat $C_{UV}$ as a parameter in the
subsequent (numerical) stages. We regularise any infrared
divergence by giving the photon a fictitious mass, $\lambda$. By
default we set this at $\lambda=10^{-15}$GeV.

\subsection{Scalar integrals for $N\leq 4$}
\label{scalarintl4} The two-point  integrals are implemented using
simple analytical formulae and evaluated numerically. This allows
to achieve a quite high precision. The scalar 3-point function and
all but the infrared divergent 4-point scalar functions are
evaluated through a call to the {\tt FF} package\cite{ff}.
Although the {\tt FF} package has been extensively used and
checked by many authors, we have also tested its accuracy and
implementation in {\tt GRACE-loop} by comparing its results
against our own numerical approach to loop
integrals\cite{checkFF}.
\begin{figure}[htb]
\caption{\label{boxir}{\em The left panel shows the general
configuration of the infrared four-point functions of the
scattering of incoming particles with masses $m_1, m_2$ to
particles with masses $m_3, m_4$. $s, t$ are  the usual Mandelstam
variables. The right panel shows some examples that need to be
treated carefully. In the first example, the $Z$ can be close to
the resonance.}}
\begin{minipage}[c]{6.5cm}
\begin{center}
\includegraphics[width=6cm,height=6cm]{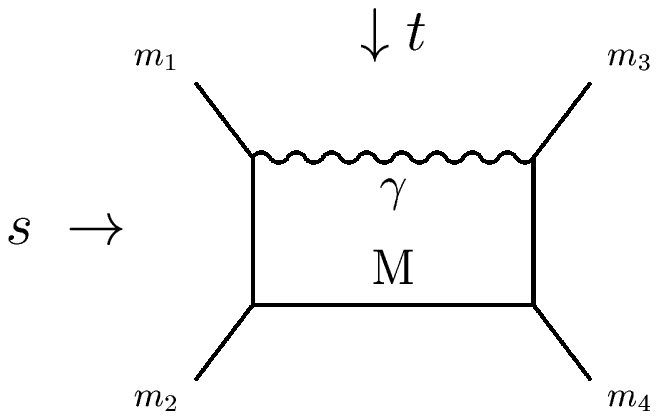}
\end{center}
\end{minipage}
\begin{minipage}[c]{2.cm}
\begin{center}
{\LARGE \bf $  \Longrightarrow$}
\end{center}
\end{minipage}
\begin{minipage}[r]{6.5cm}
\begin{center}
\includegraphics[width=4cm,height=4cm]{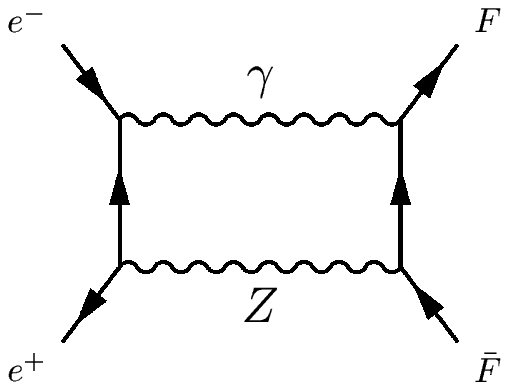}
\includegraphics[width=4cm,height=4cm]{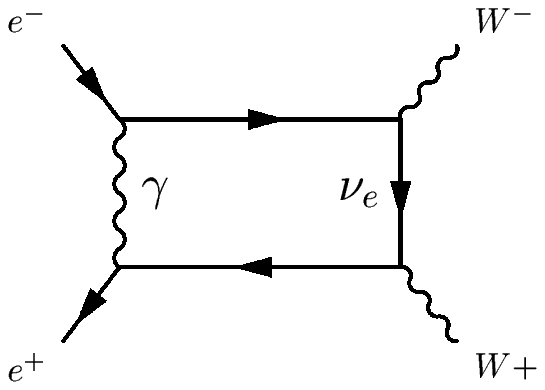}
\includegraphics[width=4cm,height=4cm]{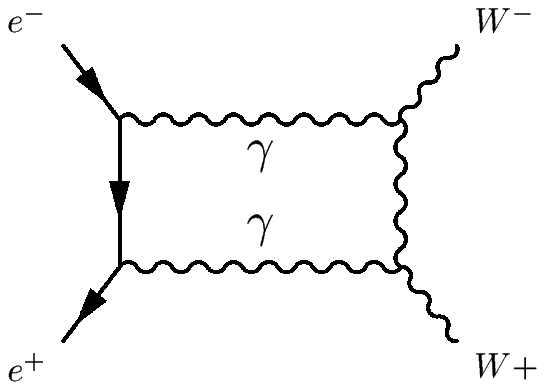}
\end{center}
\end{minipage}
\end{figure}

For the infrared four-point function, see Fig.~\ref{boxir}, we
supply our own optimised routines. A purely numerical approach
would lead to instabilities and would prevent a complete and
satisfactory cancellation of infrared divergences between these
loop functions and the infrared factors from the real soft
bremsstrahlung part. Luckily some rather simple analytical results
have been derived in this case \cite{supplement100,Dennerboxir}.
These can be further simplified when the box involves quite
separate mass scales as often occurs in $\epem$ (smallness of the
electron mass). In the example $\epem \ra F \bar F$ shown in
Fig.~\ref{boxir}, close to the $Z$ resonance, $s\simeq M_Z^2$, one
needs to take into account the width of the $Z$ circulating in the
loop. This implementation ensures that, even close to the
resonance, the infrared divergent part exactly cancels against the
soft bremsstrahlung correction. The calculation of some of these
photonic boxes is detailed in \cite{supplement100} and
\cite{Dennerboxir}.

\subsection{Reduction of the tensor integrals for $N\leq 4$}
\label{tensorintl4}
The tensor integral of rank $M$ corresponding to a
$N$-point graph is defined in Eq.~\ref{int-M-N}. The use of
Feynman's parameterisation  combines all propagators such that

\beqn
\frac{1}{D_0 D_1 \cdots D_{N-1}}&=& \Gamma(N) \; \int [dx]
\frac{1}{\left(D_1x_1 \;+\; D_2 x_2+\cdots D_0 (1-\sum_{i=1}^{N-1}
x_i)\right)^N}
\nonumber \\
\int [dx]&=& \int_0^1 dx_1 \int_0^{1-x_1} dx_2 \cdots
\int_0^{1-\sum\limits_{i=1}^{N-2} x_i} dx_{N-1}.
\eeqn
\noi Because the integrals are regulated, we first deal with the
loop momenta, before handling the integration over the parametric
variables and write

\beqn
T_{\underbrace{\mu \nu \cdots \rho}_{M}}^{(N)}&=&\Gamma(N)
\int[dx] \; {{\cal T}}_{\underbrace{\mu \nu \cdots \rho}_{M}}^{(N)},  \quad {\rm with} \nonumber \\
{{\cal T}}_{\underbrace{\mu \nu \cdots \rho}_{M}}^{(N)}&=&\int
\frac{d^n l}{(2\pi)^n} \frac{l_\mu l_\nu \cdots
l_\rho}{\left(l^2-2 l.P(x_i) -M^2(x_i)\right)^N}, \quad M \leq N
\eeqn

\noi Integration over the loop momenta $l$ is done trivially. We
may write a compact formula that applies up to the boxes.
Introducing

\beqn
\label{Li}
\Delta&=& \sum_{i,j=1}^{N-1} Q_{ij} x_i x_j + \sum_{i=1}^{N-1}L_i
x_i + \Delta_0, \quad Q_{ij}=s_i.s_j,
\quad L_i=-s_i^2+ (M_i^2-M_0^2),  \nonumber \\
 \Delta_0&=&M_0^2, \quad P=-\sum_{i=1}^{N-1} s_i x_i
\eeqn

\noi we can write

\beqn
\label{tensorI} {{\cal T}}^{(N)}&=&\widetilde{{{\cal T}}}^{(N)}
\Gamma(N-n/2)\quad {\rm with} \quad \widetilde{{{\cal T}}}^{(N)}=
\frac{(-1)^N i \pi^{n/2}}{(2\pi)^n \Gamma(N)} \Delta^{-(N-n/2)},
\nonumber
\\
{{\cal T}}_{\mu}^{(N)}&=&{{\cal T}}^{(N)} P_\mu, \nonumber
\\
{{\cal T}}_{\mu \nu }^{(N)}&=&\widetilde{{{\cal T}}}^{(N)}\;\left(
\Gamma(N-n/2)  P_\mu P_\nu-\frac{1}{2} g_{\mu \nu} \Delta
\Gamma(N-1-n/2) \right), \nonumber
\\ {{\cal T}}_{\mu \nu \rho}^{(N)}&=&\widetilde{{{\cal T}}}^{(N)}\;\left(
\Gamma(N-n/2)  P_\mu P_\nu P_\rho- \frac{\Delta}{2} (g_{\mu \nu}
P_\rho+ g_{\mu \rho} P_\nu+ g_{\nu \rho} P_\mu)
 \Gamma(N-1-n/2) \right), \nonumber
\\
{{\cal T}}_{\mu \nu \rho \sigma}^{(N)} &=&\widetilde{{{\cal
T}}}^{(N)}\;\left( \Gamma(N-n/2)  P_\mu P_\nu P_\rho P_\sigma
+\frac{\Delta^2}{4}( g_{\mu \nu} g_{\rho \sigma}+ g_{\mu \rho}
g_{\nu \sigma}+g_{\mu \sigma} g_{\nu \rho}  ) \Gamma(N-2-n/2)
\right. \nonumber
\\ &-&\left.\frac{\Delta}{2}( g_{\mu \nu} P_\rho
P_\sigma + g_{\mu \rho} P_\nu P_\sigma + g_{\mu \sigma} P_\nu
P_\rho+ g_{\nu \rho} P_\mu P_\sigma + g_{\nu \sigma} P_\mu P_\rho
+ g_{\rho \sigma} P_\mu P_\nu) \Gamma(N-1-n/2)\right) \nonumber \\
\eeqn

\noi It now rests to integrate over the Feynman parameters
contained in the momenta $P=P(\{x_i\})$. We show how this is done
for the box ($N=4$) and  triangle ($N=3$) integrals. As pointed
out earlier the case $N=2$ is straightforward and is implemented
analytically, some examples are given in
Appendix~\ref{app-onetwoptfct}. The problem now turns into finding
solutions for the parametric integrals
\beqn
\label{Inm} I_{\underbrace{i  \cdots k}_{M}}^{(N)}&=&\int  [dx]
\frac{x_i\cdots x_k }{\Delta^{(N-2)}} \quad {\rm and}  \\
J_{i;\alpha}^{(N)}&=&\int  [dx] x_i^\alpha \log\Delta \quad
\alpha=0,1.
\eeqn
solely in terms of the scalar integral $I^{(N)}=\int  [dx]
\Delta^{-(N-2)}$ (for which we use the {\tt FF} package\cite{ff}).
The appearance of the integrals $J_{i;\alpha}^{(N)}$ stems from
expanding Eq.~\ref{tensorI} around $n=4-2\epsilon$ and originates
from the $\epsilon$ independent terms in $\epsilon (1/\epsilon +
{\cal O}(\epsilon^{0,1}))$. In fact one only needs
$J^{(4)}=J_{i;0}^{(4)}$ and $J_{i;(0,1)}^{(3)}$. All these
integrals are derived recursively. The integral in Eq.~\ref{Inm}
will be referred to as the parametric integral of rank-$M$ for the
$N$-point function. It will then be expressed in terms of lower
rank tensors and lower $N$ integrals.

\subsection{Reduction of the higher rank parametric box integrals}
\label{reducboxtensor}
Let us first show how the tensor box integrals are implemented.
Note that one needs $15$ different integrals for the rank-$4$ box,
$10$ for the rank-$3$, $6$ for the rank-$2$ and $3$ for the
rank-$1$. The trick is to use the fact that
\newcommand{\deli}{\partial_i}
\newcommand{\delj}{\partial_j}
\beqn & &\int [dx]\; \deli \left(\frac{x_k^\alpha x_l^\beta
x_m^\gamma}{\Delta}\right),\nonumber \\
 {\rm with} \quad
\deli=\frac{\partial}{\partial x_i},  & &\quad 1 \leq \alpha +
\beta + \gamma=M \leq N-1
\eeqn

\noi is a surface term that can be derived from the parametric
integrals of the triangle. On the other hand expanding the partial
derivative generates parametric integrals of order $M+1$, beside
parametric integrals of rank $M-1$ and rank $M$. To wit,

\beqn
\label{master-para1} \deli \left(\frac{x_k^\alpha x_l^\beta
x_m^\gamma}{\Delta}\right)&=&-\frac{x_k^\alpha x_l^\beta
x_m^\gamma}{\Delta^2}(L_i+ 2 \sum_{j} Q_{ij}x_j)
+\frac{1}{\Delta^2} \left(\Delta_0+ \sum_j L_j x_j + \sum_{jn}
Q_{jn} x_j x_n\right) \times \nonumber \\ & & \left(\alpha
x_k^{\alpha-1} x_l^\beta x_m^\gamma \delta_{ki} + \beta
x_l^{\beta-1} x_k^\alpha x_m^\gamma \delta_{li}+ \gamma
x_m^{\gamma-1} x_k^\alpha x_l^\beta \delta_{mi} \right)
\eeqn
\noi The term on the left-hand side can be trivially integrated
and can be expressed in terms of the triangle integral of rank
$M\leq 3$. The terms proportional to $L_i$ on the right-hand side
are box integrals of rank $M$ whereas the term proportional to
$\Delta_0$ corresponds to boxes of rank $M-1$. We combine all
these terms into $C_{i;jkl}$, where the first index $i$ shows that
a derivative has been applied on the index $i$ . The terms
proportional to $Q_{ij}$ are boxes of rank $M+1$ that we want to
derive. In particular, to generate the highest rank integrals for
the box $M=4$, we apply  Eq.~\ref{master-para1} with
$\alpha=\beta=\gamma=1, (M=3)$. This amounts to solving a system
of equations for the integrals $I^{(4)}_{ijkl}$,

\beqn
C_{i;klm}&=&- 2 \sum_{j}Q_{ij}I^{(4)}_{jklm} + \sum_{jn} Q_{jn}
\left(\delta_{ki} I^{(4)}_{jnlm} +  \delta_{li} I^{(4)}_{jnkm} +
\delta_{mi}
I^{(4)}_{jnkl}\right), \nonumber \\
I^{(4)}_{ijkl}&=&\int [dx] \frac{x_i x_j x_k x_l}{\Delta^2}.
\eeqn

One can thus solve for a system of equations for the higher rank
parametric integrals of order $M+1$ in terms of (previously
derived) integrals of rank $M-1$ and $M$ for the box and rank $M$
for the triangle. One drawback of this approach is that one ends
up with a larger set of equations than needed to solve the system,
especially as the rank of the parametric integral increases. This
however also shows that one can in principle carry consistency
checks. To pick up a system of linearly independent equations we
first construct
\beqn
\sum_i C_{i;ikk}=3 \sum_{jn} Q_{jn} I^{(4)}_{jnkk},
\eeqn
in order to form the set\footnote{This could have also been
arrived at more directly had we  used
\beqn
\sum_j x_j\; \delj \Delta=2\Delta-2\Delta_0-\sum_j L_j x_j.
\nonumber
\eeqn
\noi This enables to  re-express the second terms on the
right-hand side of Eq.~\ref{master-para1}, involving the Kronecker
symbols for the case $\alpha=\beta=\gamma=1$, as lower rank terms
and triangle integrals. Namely we can write
\beqn
3 \frac{x_l x_m}{\Delta}=-2 \frac{\Delta_0}{\Delta^2}
+\sum_j\left\{ \delj\; \left(\frac{x_j x_l x_m}{\Delta}\right)-L_j
\left(\frac{x_j x_l x_m}{\Delta^2}\right)\right\}. \nonumber
\eeqn
}

\beqn
\label{decomposition} \tilde{C}_{i;kkk}=-\frac{1}{2}\left(
C_{i;kkk} - \delta_{ki} \sum_i C_{i;ikk} \right) = \sum_{j} Q_{ij}
I^{(4)}_{jkkk}.
\eeqn

\noi For the highest rank,  $M=4$, this provides $3$ independent
sets (one for each value of $k$) each consisting of three
independent integrals $I^{(4)}_{jkkk}$ (for $j=1,2,3$). Therefore
one only deals with $3$ simple $3\times 3$ matrices which help
solve $9$ out of $15$ integrals. We may refer to this set as the
diagonal integrals. The remaining integrals are provided by the
set of the $6$ independent equations $C_{i;jkk}$ where $i,j,k$ are
all different from each other,
\beqn
C_{i;jkk}=-2  \sum_n Q_{in} I^{(4)}_{njkk}, \quad i\not=j \not= k
\eeqn

\noi It is obvious that the same trick applies to solving
$I^{(4)}_{jkk}$ and provides $9$ out of the $10$ independent
integrals. The remaining integral in this case is provided by any
$C_{i;jk}$ where all indices are different. We also apply
Eq.~\ref{decomposition} to the set $I^{(4)}_{jk}$ and obtain
$\tilde{C}_{i;k}=\sum_j Q_{ij} I^{(4)}_{jk}$.\\


\noi This method shows that to solve for the $15$ independent
integrals of rank-$4$ one does not deal with a $15\times 15$
matrix. Rather the previous formulation shows that this splits
into simplified $3 \bigoplus 3 \bigoplus 3 \bigoplus 6$ systems of
equations. The $6\times 6$ matrix is also easy to deal with since
each row consists of only $2$ non-zero elements. For the rank-$3$
integrals the system of $10$ equations decomposes into $3
\bigoplus 3 \bigoplus 3 \bigoplus 1$, while for the rank-$2$, the
system of $6$ equations decomposes into $3 \bigoplus  2 \bigoplus
1$.

\subsection{$\log(\Delta)$ terms for the box and triangle}

To extract $J^{(4)}$ for the box and most of the results for the
 reduction of the higher rank integrals for the triangle, we start by giving a general
representation for the logarithm.  Take $\Delta_{N}$ where $N$ is
to just remind us that it comes from a $N$-point function. We can
write
\beqn
x_i^\alpha \log\Delta_{N}&=&\frac{1}{N+\alpha-1} \sum_{j=1}^{N-1}
\left\{\delj\left(x_i^\alpha x_j \log\Delta_N\right)-x_i^\alpha
x_j \delj\left(\log\Delta_{N}\right) \right\}
\eeqn
\noi Specific formulae needed for the boxes and triangles are
\beqn
\label{j13} x_i \log\Delta&=&\frac{1}{3} \sum_{j=1}^{2}
\left\{\delj\left(x_i x_j \log\Delta\right)-x_i+ \frac{x_i (L_j
x_j +
\Delta_0)}{\Delta}\right\}, N=3, \; \alpha=1  \\
\label{j03}\log\Delta&=&\frac{1}{2} \sum_{j=1}^{2}
\left\{\delj\left( x_j \log\Delta\right)-1+ \frac{L_j x_j +
\Delta_0}{\Delta}\right\},
N=3,\; \alpha=0 \\
\label{j4}\log\Delta&=& \frac{1}{3} \sum_{j=1}^{3}
\left\{\delj\left( x_j \log\Delta\right)-\frac{2}{3}+
\frac{\Delta(L_j x_j +2/3 \Delta_0)}{\Delta^2}\right\},  N=4, \;
\alpha=0
\eeqn

\noi Eq.~\ref{j4} shows that $J^{(4)}$ can be expressed in terms
of the ``lower" integrals $J_{i;1}^{(3)}$ and
$I^{(4)}_{M=0,1,2,3}$. In turn,  all $J_{i;(0,1)}^{(3)}$ are
expressed in terms of two-point functions and the integrals
$I^{(3)}_{M=0,1,2}$.

\subsection{Reduction of the higher rank parametric integrals for the triangle}
\noi To generate  the triangle  $I^{(3)}_{M=1,2,3}$, in analogy
with Eq.~\ref{master-para1},  we use

\beqn
\label{master-para2} \deli (x_k^\alpha x_l^\beta \log \Delta ).
\eeqn

\noi For example, for $M=3$ exploiting Eq.~\ref{j03} we get

\beqn
\deli(x_k x_l \log\Delta)&=& \frac{x_k x_l}{\Delta} \left( L_i+2
\sum_j Q_{ij}x_j \right)\nonumber \\
&+&\frac{1}{3}\left\{\delta_{ki}\sum_j \left( \delj(x_l x_j
\log\Delta)-x_l  \left( 1-\frac{L_j x_j
+\Delta_0}{\Delta}\right)\right)+ (k \leftrightarrow l) \right\}.
\nonumber
\\
\eeqn

\noi All terms with $L_i, \Delta_0$ or partial derivatives are
lower order terms (either in $N$ or $M$). Grouping all these as
$C_{i;kl}$ leads to the master equation

\beqn
C_{i;kl}=\sum_j Q_{ij} I_{jkl}^{(3)}.
\eeqn

\noi Following the same strategy as with the box, we choose the
set $C_{i;kk}$ which furnishes $2$ ``orthogonal" systems of $2$
equations each. Therefore instead of handling a $4\times 4$ matrix
we only deal with simple $2\times 2$ matrices.  Similar results
are obtained for $I_{jk}^{(3)}$. The three needed integrals are
arrived at by first solving for a reduced system of only two
independent integrals and then deriving the third from a single
equation.

Finally for $M=1$, one solves a system with a  $2\times 2$ matrix.
Note that the solution of all these equations involves the
determinant of the same $2\times 2$ matrix, namely $Q_{ij}$.

Note also that   the system of equations as described here leads
to analytic solutions in terms of the scalar integrals. In {\tt
GRACE-loop} we implement these analytical solutions.

\subsection{Reduction of 5- and 6-point integrals}
\label{fivesixint} Five point functions are calculated as linear
combinations of four point functions\cite{fivetofour,nogueira5pt}.
Variants and new techniques have also been worked out very
recently\cite{newNtofour,thomasfatpaper,DDgram1}. We will describe
the methods that have been  implemented in {\tt GRACE-loop}. The
reduction takes advantage of the fact that for $N>4$ not all the
external momenta are linearly independent. For $N=5$,  the set of
vectors $\{s_i\}$ with $i=1,\cdots 4$, see Eq.~\ref{Di}, forms an
independent basis of 4-vectors, which allows to expand any
4-momentum, particularly the loop momentum $l$ as \eqn
l^\mu=\sum_{i,j=1}^4(Q^{-1})_{ij}(l\cdot s_i)s_j^\mu,
\label{expan1} \eqne where the $4\times4$ matrix $Q_{ij}$ is
defined as in Eq.~\ref{Li} with $i,j=1,\cdots 4$.

\noi From Eq.(\ref{expan1}) we express $l^2$ as \eqn
l^2=\sum_{i,j=1}^4(Q^{-1})_{ij}(l\cdot s_i)(l\cdot s_j). \eqne
which helps define an identity between the denominators of the
propagators in the $5$-point function. Using the same notation as
in Eqs.~\ref{Di}-~\ref{Li} we  rewrite \eqn
D_0+M_0^2={1\over2}\sum_{i,j=1}^4(Q^{-1})_{ij}(D_i-D_0+ L_i)
(l\cdot s_j), \eqne to arrive at the identity \eqn
4M_0^2-\sum_{i,j=1}^4(Q^{-1})_{ij}(L_i)(D_j-D_0+ L_j)
=-4D_0+2\sum_{i,j=1}^4(Q^{-1})_{ij}(D_i-D_0)(l\cdot s_j). \eqne
This demonstrates that a 5-point function with a numerator of the
form $N(l)=l^{\mu_1}\cdots l^{\mu_k}$  is expressed as a sum of
five box integrals, \eqa \label{decomp5to4-1} &&\left(
4M_0^2-\sum_{i,j=1}^4(Q^{-1})_{ij}L_i L_j\right)
\int{\d^4l\over(2\pi)^4i}{N(l)\over\bfD_{\bf 5}}
\;=\; \non\\
&& \int{\d^4l\over(2\pi)^4i}N(l)
\biggl(-{4D_0\over\bfD_{\bf 5}}
\;+\; \sum_{i,j=1}^4(Q^{-1})_{ij} L_i {D_j-D_0\over\bfD_{\bf 5}}
+2\sum_{i,j=1}^4(Q^{-1})_{ij}
{(D_i-D_0)(l\cdot s_j)\over\bfD_{\bf 5}}\biggr),\non\\
\eqae
where
\eqn
\bfD_{\bf 5}=D_0\prod_{i=1}^4D_i.
\eqne
Putting $N(l)=1$ one gets the reduction formula of the 5-point
scalar integral to a sum of four-point integrals. This method has
been applied to \eennht\cite{eennhletter} and
\eettht\cite{eetthgrace} where the highest rank tensor of the pentagon is $M=2$.
We will refer to this technique, for short, as the {\it
scalar-derived} reduction.
\\
\noi Although this technique can be directly applied to a 5-point
integral of any rank $M\leq N=5$, we should note that the presence
of the term $l\cdot s_j$ in Eq.~\ref{decomp5to4-1} raises the rank
of the integral by one unit. This causes a superficial UV
divergence for  $M\ge3$. For $M=4$ the reduction requires the
evaluation of $M=5$ box diagrams, a case that is not covered by
our reduction formulae of the tensor boxes, see section
\ref{reducboxtensor}. Furthermore when this formula is used to get
the matrix elements in a symbolic way, the resultant {\tt FORTRAN}
code usually becomes very lengthy. We have developed another
algorithm for the  reduction of higher rank tensors which we first
applied to \eezhht\cite{eezhhgrace}. We apply the identity
Eq.(\ref{expan1}) to the numerator $N(l)$.  We have \eqa
N(l)&=&l^{\mu_1}l^{\mu_2}\cdots l^{\mu_k}
=\sum_{i,j=1}^4(Q^{-1})_{ij}
(l\cdot s_i)s_j^{\mu_1}l^{\mu_2}\cdots l^{\mu_k},\non\\
&=&{1\over2}\sum_{i,j=1}^4(Q^{-1})_{ij}(D_i-D_0+ L_i)
s_j^{\mu_1}l^{\mu_2}\cdots l^{\mu_k}.   \label{expan2} \eqae Then
\eqa \int{\d^4l\over(2\pi)^4i}{N(l)\over\bfD_{\bf 5}}
&=&{1\over2}\sum_{i,j=1}^4(Q^{-1})_{ij}s_j^{\mu_1}
\int{\d^4l\over(2\pi)^4i} {(D_i-D_0)l^{\mu_2}\cdots
l^{\mu_k}\over\bfD_{\bf 5}}
\non\\
&+&{1\over2}\sum_{i,j=1}^4(Q^{-1})_{ij} L_i s_j^{\mu_1}
\int{\d^4l\over(2\pi)^4i}{l^{\mu_2}\cdots l^{\mu_k}\over\bfD_{\bf
5}}. \eqae On the right-hand-side the rank of the numerator is
lowered by one unit, though there still remains a sum of 5-point
functions. This reduction can be repeated until one is left with a
scalar 5-point function and box integrals. We will, for short,
refer to this method as the {\it vector-derived} reduction. An
advantage of this method is that the final expression in {\tt
FORTRAN} code is about ten times shorter than that obtained by the
previous technique.

Let us also very briefly describe how the 6-point functions are
implemented in {\tt GRACE-loop}\cite{eemnudgrace}. In this case,
we first express $s_5$ in terms of the set of four linearly
independent vectors $s_i,~i=1,\cdots,4$ and construct the product
\eqn l\cdot s_5=\sum_{i,j=1}^4(Q^{-1})_{ij}(s_5\cdot s_i)(l\cdot
s_j), \eqne hence \eqn -L_5+\sum_{i,j=1}^4(Q^{-1})_{ij}(s_5\cdot
s_i) L_j =D_5-D_0-\sum_{i,j=1}^4(Q^{-1})_{ij}(s_5\cdot
s_i)(D_j-D_0). \eqne Combining with \eqn \bfD_{\bf
6}=D_0\prod_{i=1}^5D_i, \eqne we find \eqa \label{s5reduc}
&&\left(-L_5 +\sum_{i,j=1}^4(Q^{-1})_{ij}(s_5\cdot s_i) L_j
\right)
\int{\d^4l\over(2\pi)^4i}{N(l)\over\bfD_{\bf 6}}\non\\
&&\qquad\qquad\qquad=
\int{\d^4l\over(2\pi)^4i}{(D_5-D_0)N(l)\over\bfD_{\bf 6}}
-\sum_{i,j=1}^4(Q^{-1})_{ij}(s_5\cdot s_i)
\int{\d^4l\over(2\pi)^4i}{(D_j-D_0)N(l)\over\mathbf{{\cal
{D}}_6}}.
\non\\
\eqae

This is the standard reduction of a general 6-point function to a
sum of 5-point integrals. Moreover for further reduction of the
5-point integrals we use a combination of both the scalar-derived
Eq.~\ref{expan1} and vector-derived Eq.~\ref{expan2} reductions
for the loop tensor $N(l)$ of rank-$M$ to arrive at a reduction
which in a compact form writes as
\beqn
\label{reduc6to4}
\overbrace{N(l)}^{M}=\sum_{i,j=1}^4 \overbrace{R_{i,j}(l)}^{M-2}
D_i D_j + \sum_{i} \overbrace{S_i}^{M=0} D_i +
\overbrace{T_6}^{M=0}
\eeqn

The first term of the right-hand side is tensor of rank $M-2$
which corresponds to box diagrams. The next two-terms correspond
to scalar $5$-point and $6$-point function for which we use the
algorithm of the scalar-derived reduction and the one based on
Eq.~\ref{s5reduc} for the 6-point function. Once the reduction of
the 5-point and 6-point tensor function have been brought down to
box integrals with a lower tensor rank, we use the algorithm
developed for $N\leq 4$ as described in detail in
section~\ref{scalarintl4} and \ref{tensorintl4}. Infrared resonant
5-point and 6-point functions that require the introduction of a
width in the loop integrals require more care,
see\cite{eemnudgrace}.\\

In actual computations the matrix element for 5- and 6-point
diagrams is most time consuming, since as we have just seen the
reduction algorithms are quite involved and go through various
steps. For example the percentage of CPU time needed to calculate
$N$-point diagrams is summarized in table \ref{tab:cpu} for a
$2\ra 3$ and $2\ra 4$ process. One can see for instance, in the
case of $e^+e^-\to \nu_e\bar\nu_e HH$ that almost as many as
$2000$ three-point loop integrals require $8\%$ of the CPU time
whereas the $74$ 6-point integrals require $67\%$ of the CPU time.

\vspace{0.5cm}
\begin{table}[htb]
\caption{{\em Percentage of CPU time among various $N$-point
loops. The number of diagrams in each class of $N$-point diagram
is shown in parenthesis. ``Others" stands for two-point functions
and counterterms.}} \label{tab:cpu} \ct
\begin{tabular}{|c|c|c|c|c|c|}
\hline
process&6-point&5-point&4-point&3-point&others\\
\hline
$e^+e^-\to e^+e^-H$
&-&33\%&11\%&47\% & 9\%\\
&-&(20)&(44)&(348)&(98)\\
\hline
$e^+e^-\to \nu_e\bar\nu_e HH$
&67\%&13\%&10\%&8\%&2\%\\
&(74)&(218)&(734)&(1804)&(586)\\
\hline
\end{tabular}
\cte
\end{table}


As we need a lot of computer power, it is essential to develop
software engineering techniques in order to reduce the execution
time. In {\tt GRACE} we have developed a {\sl parallelised}
version that exploits  message passing libraries, such as {\tt
PVM} or {\tt MPI}\cite{parallelgrace}. Concerning the loop
calculation, it is efficient to distribute calculations of each
Feynman diagram among many CPUs because it is hard to create a
single executable file from too long source codes.

Another technique must be {\sl vectorisation} which can make CPU
much shorter. Thanks to the recent rapid development of
microprocessors and their easy availability even for the common
public one should think of building or adapting codes to run as
vectorised codes that can be quite effective. Once one has enough
memory/cache, the vectorisation of the amplitudes could lead to a
much more powerful tool. It must be said however, that
vectorisation of existing codes is not always straightforward or
even possible. For example, {\tt GRACE} relies on the package {\tt
FF}\cite{ff} for the evaluation of the one-loop scalar integrals
(boxes, triangles). Because of its structure that employs too many
{\tt if ... then ... else ... end if} statements, the {\tt FF}
package is not fully vectorised  in the current {\tt GRACE}
system.



\subsection{New techniques for the loop integrals}
\label{new-loop-int}
 The reduction formalism that we have outlined both for the
reduction of  the $N$-point functions with $N\geq 4$ to the lower
$N$-point functions and the tensorial reduction, even for $N \leq
4$,  to the scalar integrals involves implicitly  the inverse of
the determinant of the  matrix $Q_{ij}$,  the Gram determinant.
This is most apparent in our formulation of
section~\ref{fivesixint}, see for example Eq.~\ref{expan2} which
is expressed in terms of $Q^{-1}$. For kinematical configurations
where $Det Q$ is very small or vanishes this can lead to severe
numerical instabilities. It must be said that these exceptional
configurations are, in a Monte Carlo sampling, hardly met.
Moreover, the numerical instability around these singularities can
be cured if one reverts to quadruple, or higher, precision. It has
been shown\cite{fujimoto-acat05} how this solution can be
optimised with a dedicated {\tt FORTRAN} library for
multiprecision operations such that it does not require much CPU
time while keeping the benefit of exploiting the same standard
reduction formalism. \\
\noi For $N \leq 4$ and for $Det Q=0$, there also exist very
efficient algorithms\cite{robin-reduc,Sloopsgg,Aguila-Pittau} that
are amenable to an automatic computer implementation. Expansions
about vanishing Gram determinants are also
possible\cite{Sloopsgg,Zanderighi-gram,DDgram1}. Recently there
has been a lot of activity to improve this aspect of the loop
calculation by finding new, improved and efficient algorithms for
the loop integrals especially to avoid this problem. One approach
is, instead of reducing the system to the master set  of the
scalar integrals with $N\leq 4$,  to use other bases for the
master integrals which can include a tensorial integral for
example, therefore avoiding the appearance of Gram determinants
before a numerical calculation is performed. These
approaches\cite{thomasfatpaper,DDgram1,newNtofour,Aguila-Pittau,GieleGloverbasis}
combine both an algebraic reduction with an efficient numerical
implementation. Other approaches are, to a very large extent,
essentially based on a  numerical computation of all the loop
integrals\cite{Hameren-massless,Feroglia-Passarino,Grace-Doncker,Binoth-Kauer-contour,Kurihara-Kaneko-contour,Anastasiou-Daleo}.
One example is  based on the contour deformation of the
multi-dimensional parameter
integrals\cite{Kurihara-Kaneko-contour} and may be implemented in
{\tt GRACE-loop}. On the other hand, for loops with internal
massless particles as would be the case for applications to QCD,
some powerful algebraic methods of the loop integrals are being
derived\cite{Duplancic,Kurihara-masslessloops}. Let us also
mention that most of the methods extract the ultraviolet and
infrared divergences, so that the set of basis integrals is
amenable to an efficient implementation.\\
\noi It should be kept in mind that the majority of the new
techniques have not been implemented as fully working codes yet
nor has their robustness  been tested in practical calculations,
especially as concerns multi-leg processes. Apart
from\cite{newNtofour,DDgram1} which, in fact, is an extension of
the standard reduction that has been applied successfully to
$\epem \ra 4f$\cite{eeto4fdenner}, it remains to be seen how the
other new techniques perform when handling the {\em complete}
one-loop calculation of a physical process, of interest for the
LHC or the LC.

\setcounter{equation}{0}
\section{Tests on the loop calculation}
\label{sec:looptests}
 The results of the calculations are checked
by performing three kinds of tests.  This concerns the ultraviolet
and infrared finiteness as well as the gauge-parameter
independence. These tests are performed   at the level of the
differential cross section before any phase space integration is
performed for several points in phase space. These tests points
are chosen at random. Usually for these tests one keeps all
diagrams involving the couplings of the Goldstones to the light
fermions, such as $\chi_3 e^+ e^-$. For these tests to be passed
one works in quadruple precision. After these tests have been
passed one can switch off these very small couplings, involving
the scalars and the light fermions, when calculating the total
(integrated) cross section and hence speeding up the computation
time. Results of these tests on a selection of the $26$ processes
for the $2 \ra 2$ reactions displayed in
Table.~\ref{tablenlgall27} are made available at this web
location\cite{2to2nlgchecks}. This list involves both purely
vector bosons scattering, heavy as well as massless fermions
scattering into gauge bosons as well as a few processes involving
the Higgs. Therefore, as we will see, all the ingredients that
enter the calculation of radiative corrections in the \sm are
covered by this list.

\subsection{Ultraviolet and infrared finiteness checks}
\subsubsection{Ultraviolet finiteness}
\label{cuvtest}
We first check the ultraviolet finiteness of the results. This
test applies to the whole set  of the virtual one-loop diagrams.
The ultraviolet finiteness test gives a result that is stable over
$30$ digits when one varies the dimensional regularisation
parameter $C_{UV}$ defined in Eq.~\ref{cuvdef}. This parameter is
kept in the code as a free parameter. This parameter could then be
set to $0$ in further computations once the finiteness test, or
$C_{UV}$ independence test, is passed. When conducting this test
we regularise any infrared divergence by giving the photon a
fictitious mass that we fix at $\lambda=10^{-15}$GeV. The
finiteness test is carried out for a random series of the gauge
fixing parameters that include the linear gauge as a special case.

\subsubsection{Infrared finiteness and calculation  of the
soft-bremsstrahlung factor}
\label{irtest}

When discussing the calculation and implementation of the loop
integrals, some diagrams involving a photon exchange require
special treatment.  These diagrams lead to an infrared divergence
caused by the fact that the photon is massless so that its energy
could vanish. These infrared divergences in the case of the
photon, either in QED or in the electroweak theory, can be
regulated by giving the photon a small mass $\lambda$. As known
the dependence in this fictitious mass cancels against the one
contained in the soft bremsstrahlung\cite{Bloch} and do not hinder
the renormalisation procedure. For a textbook introduction see for
instance\cite{PeskinBook} or \cite{ItzyksonZuber}. In the
non-Abelian case where there is no smooth mass limit, an example
being QCD, this regularisation of the infrared divergence by
giving the gauge boson a mass would
fail\cite{smoothmasslimit-norman}. In this case one can revert to
dimensional regularisation\cite{reg-ir-qcd}. In \grcp, for the
electroweak radiative corrections we use the simple trick of the
fictitious photon mass to regulate the infrared divergences.

The second test that we perform relates to the infrared finiteness
by checking that when the virtual loop correction and
bremsstrahlung contributions are added there is no dependence on
the fictitious photon mass $\lambda$. We indeed find results that
are stable over $23$ digits, or better, when varying $\lambda$.

The soft bremsstrahlung part consists of the tree-level process
with an additional photon of very small energy, $E_\gamma < k_c$,
and requires the introduction of the photon mass regulator,
$\lambda$. The hard photon radiation with $E_\gamma
> k_c$ is regular and will be discussed in section~\ref{seckctest}.
The soft photon contribution is implemented in the system
following an analytical result based on factorisation and which
can be generalised to any process. The bremsstrahlung differential
cross section factorises as

\beqn
d\sigma_{\rm soft}(\lambda,E_\gamma<k_c)=d\sigma_0 \times
\delta_{\rm soft}(\lambda,E_\gamma<k_c) \;.
\eeqn
$k_c$ is assumed sufficiently small so that the tree-level
$d\sigma_0$ does not change rapidly when  the soft photon is
emitted. In some cases, for instance  around a resonance, special
care must be exercised, see for example\cite{supplement100}. The
factor $\delta_{\rm soft}$ is completely determined from the
classical (convection) current of a charged particle and does not
involve the spin connection. Therefore this factor is universal
and only depends on the charge $Q_i$ and momentum $p_i$ of the
particles of the tree-level process,

\beqn
\label{irpropagator} \delta_{\rm soft}=-e^2 \int_{|k|<k_c}
\frac{d^3 k}{2 E_\gamma (2\pi)^3}\sum_{ij} \varepsilon_i
\varepsilon_i Q_i Q_j \frac{p_i.p_j}{(k.p_i)( k.p_j)} =\sum_{ij}
R_{ij} \; , \quad E_\gamma=\sqrt{k^2+\lambda^2} \;.
\eeqn
\noi where $\varepsilon_i=\pm 1$ depending on whether the particle
is incoming ($+1$) or outgoing ($-1$). Very general expressions
for $R_{ij}$ have been derived most elegantly in
\cite{Veltmanscalarintegrals}. Let us here just recall a few
special cases and refer the reader to \cite{supplement100} for
more details. For instance, for the diagonal term $R_{ii}$ from a
charged particle with $|Q|=1$ of momentum $p=(E,
\overrightarrow{p}), p^2=m^2$ and $P=|\overrightarrow{p}|$, one
gets the very simple result

\beqn
\label{radiator1e}
 R_{ii}=-e^2 \int_{|k|<k_c} \frac{d^3 k}{2 E_\gamma (2\pi)^3}
\frac{m^2}{(k.p)^2}= -\frac{\alpha}{\pi} \left\{ \ln\left(
\frac{2k_c}{\lambda}\right) + \frac{E}{P}\ln \left(
\frac{m}{E+P}\right) \right\} \;.
\eeqn

Another quite useful result is the contribution, $R_{pair}$, from
a pair of particle-antiparticle of mass $m$ and charge $\pm 1$ in
their centre-of-mass system with total energy $\sqrt{s}$. The
radiator function writes, with $\beta=\sqrt{1-4m^2/s}$

\beqn
\label{radiatoree} R_{{\rm pair}}&=&\frac{2 \alpha}{\pi} \left\{
\left(\frac{s-2m^2}{s\beta}\ln
\left(\frac{1+\beta}{1-\beta}\right) -1\right) \ln
\left(\frac{2k_c}{\lambda}\right)+
\frac{1}{2\beta}\ln\left(\frac{1+\beta}{1-\beta}\right) \right. \nonumber \\
& & \left. \quad \quad - \frac{s-2m^2}{2 s\beta} \left({\rm
Li}_2\left(\frac{2\beta}{1+\beta}\right)-{\rm
Li}_2\left(\frac{-2\beta}{1-\beta}\right) \right) \right\}\;,
\eeqn
and
\beqn
{\rm Li}_2(z)=-\int_0^z dt \frac{\ln(1-t)}{t},
\eeqn
is the Spence function. This factor would represent the initial
state  bremsstrahlung part in \epemt processes and is usually
written (for $s \gg m_e^2$) as

\beqn
\label{softfactoree} R^{e^+ e^-}_{{\rm pair}}=\frac{2 \alpha}{\pi}
\left\{ \left(\ln \left(\frac{s}{m_e^2}\right) -1\right) \ln
\left(\frac{2k_c}{\lambda}\right)-\frac{1}{4} \ln^2
\left(\frac{s}{m_e^2}\right)+\frac{1}{2}\ln
\left(\frac{s}{m_e^2}\right)-\frac{\pi^2}{6} \right\} \;.
\eeqn

The same factor in Eq.~\ref{radiatoree} can be used as the
bremsstrahlung contribution for $\gamma \gamma \ra W^+ W^-$ ($m
\ra M_W$).

\subsection{Gauge-parameter independence checks}
\label{sec:nlg-tests}

For this check we set the value of the ultraviolet parameter
$\cuv$ to some fixed value. To tame the infrared divergence
contained in the virtual corrections we give the photon  a
fictitious mass $\lambda=10^{-15}$GeV. Moreover we also set all
widths to zero so that no extra gauge breaking due to the
introduction of a width is generated. We thus choose a
non-singular point in phase space, away from any resonance, for
this check on the differential cross section.

\begin{table*}[hbtp]
\caption{\label{tablenlgall27} {\em Accuracy measured by the
number of digits for the gauge-parameter checks on the $26$
processes for all five gauge parameters. The numbers that appear
in the last five columns represent the number of digits which are
stable when varying  the corresponding gauge parameter. An empty
entry means that the process does not depend on the gauge
parameter. Only one parameter is varied at a time here. We also
show the number of diagrams both at tree-level and at the one-loop
level. The number of diagrams depends on the choice of the gauge
parameter, for examples in some gauges some vertices are absent.
The number of diagrams that we list corresponds to the gauge which
leads to  the maximum number of diagrams.}}
\begin{center}
\begin{tabular}{|c||c||c|c|c|c|c|}
\hline processes &\# of graphs(Loop $\times $ Tree) &
$\tilde{\alpha}$ & $\tilde{\beta}$ & $\tilde{\delta}$ &
$\tilde{\epsilon}$ & $\tilde{\kappa}$
\\
\hline

$\nu_e \bar{\nu}_e \rightarrow \nu_e \bar{\nu}_e$ &46$\;  \times
\; $2 &-- &30 &-- &-- &--
\\
$e^+e^- \rightarrow \nu_e \bar{\nu_e} $ &112$\;  \times \; $3 &31
& 31 &31 &-- &32
\\
$e^+e^- \rightarrow t \bar{t} $ &150$\;  \times \; $4 &32 & 31 &
31 &31 & 31
\\
$e^+e^- \rightarrow e^+e^-$ &288$\;  \times \; $4 & 32 & 30 & 30
&31 & 31
\\
$e^+e^- \rightarrow W^+W^-$ &334$\;  \times \; $4 & 27 & 27 & 30
&31 &--
\\
$e^+e^- \rightarrow Z^0 Z^0$ &336$\;  \times \; $3 & 33 & 29 & 31
&31 &--
\\
$e^+e^- \rightarrow H^0 Z^0$ &341$\;  \times \; $3 &30 & 30 & 31
&31 & 30
\\
$\mu \bar{\nu_{\mu}} \rightarrow W^- \gamma$ &162$\;  \times \; $3
& 27 & 27 &28 &-- &28
\\
$\mu \bar{\nu_{\mu}} \rightarrow W^- Z^0$ &213$\;  \times \; $4
&31 & 29 &30 &-- &30
\\

$\mu \bar{\nu_{\mu}} \rightarrow W^- H^0$ &196$\;  \times \; $3
&30 & 29 & 31 & 31 &31
\\
$t \bar{b} \rightarrow W^+ \gamma$ &239$\;  \times \; $4 &22 &25
&29 &-- &29
\\
$t \bar{b} \rightarrow W^+ Z^0$ &284$\;  \times \; $4 & 31 &22 &31
&-- & 32
\\
$t \bar{b} \rightarrow W^+ H^0$ &285$\;  \times \; $4 &29 &28 &21
&26 &30
\\
$\gamma \gamma \rightarrow t \bar{t}$ &267$\;  \times \; $2 &24 &
34 &30 &-- &--
\\
$Z^0 Z^0\rightarrow t \bar{t}$ &338$\;  \times \; $3 &30 &29 &31
&31 &--
\\
$W^+ W^-\rightarrow t \bar{t}$ &354$\;  \times \; $4 &30 &26 &31
&31 &--
\\

$Z^0 H^0\rightarrow t \bar{t}$ &355$\;  \times \; $4 &30 &28 &29
&29 &31
\\

$\gamma \gamma \rightarrow W^+W^- $ &619$\;  \times \; $5 &22 &24
&32 &-- &31
\\

$Z^0 Z^0\rightarrow Z^0 Z^0 $ &657$\;  \times \; $3 &-- &24 &31
&31 &--
\\
$Z^0 \gamma \rightarrow W^+ W^- $ &680$\;  \times \; $5 & 28 & 28
&31 &-- &31
\\
$Z^0 W^-\rightarrow Z^0W^- $ &840$\;  \times \; $6 &26 &24 &29 &30
&29
\\
$W^+ W^-\rightarrow W^+W^- $ &925$\;  \times \; $7 &27 &26 &30 &31
&--
\\
$Z^0 H^0 \rightarrow W^+ W^- $ &823$\;  \times \; $5 & 29 &25 &29
&26 &31
\\

$Z^0 H^0\rightarrow Z^0 H^0 $ &830$\;  \times \; $6 &-- &23 &24
&20 &31
\\
$H^0 W^-\rightarrow H^0W^- $ &827$\;  \times \; $6 & 29 &23 & 22
&23 &30
\\
$H^0 H^0\rightarrow H^0 H^0 $ &805$\;  \times \; $4 &-- &-- &29
&27 &--
\\
\hline
\end{tabular}
\end{center}
\end{table*}

For each process we verify that it does not  depend on any of the
five non-linear gauge parameter of the set
$\zeta=(\tilde{\alpha},\tilde{\beta},\tilde{\delta},\tilde{\kappa},\tilde{\epsilon})$.
Let us remind the reader that we always work with
$\xi_W=\xi_Z=\xi_A=1$. The use of five parameters is not redundant
as often these different  parameters check complementary sets of
diagrams. For example the parameter $\tilde{\beta}$ is involved in
all diagrams containing the gauge $WWZ$ coupling and their
Goldstone counterpart, whereas $\tilde{\alpha}$ checks $WW\gamma$
and $\tilde{\delta}$ is implicitly present in $WWH$. For each
parameter of  the set, the first check is made  while freezing all
other four parameters to $0$. \\
\noi In a second check we give, in turn, each of the remaining $4$
parameters a non-zero value (we usually take the values
$(2,3,4,5)$ for this set) so that we also check vertices and
diagrams that involve cross terms (like $\tilde{\alpha} \times
\tilde{\delta}$). In principle checking for $2$ or $3$ values of
the gauge parameter should be convincing enough. We in fact go one
step further and perform a comprehensive gauge-parameter
independence test. To achieve this we generate for each non-linear
gauge parameter $\zeta_i$ of the set $\zeta$, the values of the
loop correction to the total differential cross section as well as
the individual contribution of each one-loop diagram $g$, ${\rm
d}\sigma_g$ for a sequence of values for $\zeta_i$, while freezing
the other parameters to a fixed value, not necessarily zero. The
one-loop diagram contribution from each loop graph $g$ to the
fully differential cross section, is defined as

\beqn
{\rm d}\sigma_g\equiv{\rm d}\sigma_g(\zeta)=\Re e
\left(T^{loop}_g\cdot {{\cal T}}^{tree\ \dagger}\right) \; .
\eeqn

${{\cal T}}^{tree}$ is the tree-level amplitude summed over all
tree-diagrams. Therefore the tree-level  amplitude does not depend
on any gauge parameter. Note that in many processes, some
individual tree diagrams do depend on a gauge parameter, however
after summing over all tree-level diagrams, the gauge-parameter
independence at tree-level for any process is exact within machine
precision. $T^{loop}_g$ is the one-loop amplitude contribution of
a one-loop diagram $g$. It is not difficult to see, from the
structure of the Feynman rules of the non-linear gauge, that for
each $2\ra 2$ process the differential cross section is a
polynomial of (at most) fourth degree in the gauge parameter.
Therefore the contribution ${\rm d}\sigma_g$ of diagram $g$ to the
one-loop differential cross section  may be written as

\beqn
\label{decomposition-zeta}
{\rm d}\sigma_g(\zeta)={\rm d}\sigma_g^{(0)}+\zeta {\rm
d}\sigma_g^{(1)} +\zeta^2 {\rm d}\sigma_g^{(2)} +\zeta^3 {\rm
d}\sigma_g^{(3)}+\zeta^4 {\rm d}\sigma_g^{(4)}\; .
\eeqn


\begin{table}[hbtp]
\vspace*{-1cm}  \caption{\label{tablecheckswwnlg}{\em Non-linear
gauge parameter checks on $\tilde{\alpha}$ (all other parameters
set to zero), for the differential cross section $W^+ W^- \ra W^+
W^-$. For details see text. }} \footnotesize
\begin{tabular}{|c|c|c|c|c|c|}
\hline Graph number &      $ {\rm d}\sigma_g^{(4)}$ & ${\rm
d}\sigma_g^{(3)}$ &
${\rm d}\sigma_g^{(2)}$& ${\rm d}\sigma_g^{(1)}$&$ {\rm d}\sigma_g^{(0)}$\\ \hline & & & & & \\
2   &&                               &   .2335514E+03
&.7789374E+03 &
 .5615925E+03\\
     5    &&                              &  -.1721616E+01 &  -.1171640E+01 &
     .2893256E+01\\
    10       &&                           &  -.4324751E+01 &   .8649502E+01 &
    -.4324751E+01\\
    13     &&                             &  -.1721616E+01 &  -.1171640E+01 &
    .2893256E+01\\
     33     &&&                                             &   .1048909E+01 &
     -.1048909E+01\\
     35        &&&                                            &   .1048909E+01 &
     -.1048909E+01\\
\multicolumn{6}{|c|}{$\wr \wr$}\\
      \multicolumn{6}{|c|}{$\wr \wr$}\\
     321 &  -.3606596E+02 &   .1243056E+03&   .6056929E+03 &-.1534141E+04   &
     -.4615316E+04\\
    322 & &                                &  -.7411780E-02 &   .2758337E+00 &
    -.2984030E+01\\
    323 & &                                &  -.2864131E+01 &  -.2648726E+02 &
    .2935139E+02\\
    324 &   .2999432E+02 &  -.1197862E+03 &  -.1886059E+03 &   .6165932E+03 &
    -.3381954E+03\\
    325 & &                                &   .7411780E-02 &  -.1453286E+00 &
    .1379168E+00\\
    326 & &                                &   .7411780E-02 &  -.1453286E+00 &
    .1379168E+00\\
    327 & &                                &  -.2864131E+01 &  -.2648726E+02 &
    .2935139E+02\\
    \multicolumn{6}{|c|}{$\wr \wr$}\\
     \multicolumn{6}{|c|}{$\wr \wr$}\\
    493 &                  &  -.1798684E+03 &   .4305188E+03 &-.2277636E+03   &
    -.8935366E+03\\
    494 &                  &   .8331849E+02 &  -.2608640E+03 &   .2717725E+03 &
    -.9422699E+02\\
    495 &                  &   .8331849E+02 &  -.2608640E+03 &   .2717725E+03 &
    -.9422699E+02\\
    496 & &                                &   .1666370E+03 &  -.3332740E+03 &
    .1666370E+03\\
    498       & & &                                                &  -.2274438E-01 &
    .2274438E-01\\
    499        & & &                                               &  -.2274438E-01 &
    .2274438E-01\\
 \multicolumn{6}{|c|}{$\wr \wr$}\\
 \multicolumn{6}{|c|}{$\wr \wr$}\\
    741 &   .3286920E-31 &  -.6573841E-31 &  -.2380925E+01 &.2380925E+01    &
    .0000000E+00\\
    743      & & &                                                 &   .3853975E+01 &
    -.8927045E+01\\
    749 &                  &   .6445007E+00 &   .4734479E+00 &   .1060865E+01 &
    -.2457305E+01\\
    755   & & &                                                    &   .2853713E+00 &
    -.2853713E+00\\
    758        & & &                                               &  -.4247529E+01 &
    .4261065E+02\\
    764 &            &   .6615728E+01 &  -.2526752E+02 &  -.7116714E+01 &
    .7139393E+02\\
    \multicolumn{6}{|c|}{$\wr \wr$}  \\
     \multicolumn{6}{|c|}{$\wr \wr$}\\
    923     & & &                                                  &  -.1479291E+01 &
    -.1127685E+02\\
    924     & & &                                                  &  -.8424135E+00
    &
    .4331788E+03\\
    \hline
    &&&&& \\
 $Max(| {\rm
d}\sigma_g^{(i)}|)$       &   36.066    &       179.87 &          605.69   &        1534.1  &         4615.3 \\
${\rm sum}_i$     & .63168E-28   &    .60757E-29 & .44209E-28&
.69380E-28   &    .20116 \\
$ \sum_g {\rm d}\sigma_g^{(i)}/ \sum_g {\rm d}\sigma_g$ &
.24538E-29  &     .11771E-29     &  .28841E-28    &   .11464E-27&
1.0000
\\
\hline \hline \multicolumn{6}{|c|}{} \\
\hline \hline \multicolumn{6}{|c|}{Results for $\sum_g {\rm d}\sigma_g$} \\
$\tilde{\alpha}= 0$ &  \multicolumn{5}{|c|}
 {928.43820021286338928513117418831577$\;\;\;\;\;$(input)}\\
$\tilde{\alpha}= 1$   & \multicolumn{5}{|c|}
  {928.43820021286338928513117432490231$\;\;\;\;\;$(input)}\\
$\tilde{\alpha}= -1$   &  \multicolumn{5}{|c|}
  {928.43820021286338928513117410983989$\;\;\;\;\;$(input)}\\
$\tilde{\alpha}= 2$  &  \multicolumn{5}{|c|}
  {928.43820021286338928513117455347002$\;\;\;\;\;$(input)}\\
$\tilde{\alpha}= -2$  &  \multicolumn{5}{|c|}
  {928.43820021286338928513117411023117$\;\;\;\;\;$(input)}\\
$\tilde{\alpha}= 5$  &  \multicolumn{5}{|c|}
  {928.43820021286338928513117695043335$\;\;\;\;\;$(derived)}\\
\hline
\end{tabular}
\end{table}

\normalsize

We have therefore chosen the sequence of the five values
$\zeta=0,\pm 1, \pm 2$. For each contribution ${\rm d}\sigma_g$,
it is a straightforward matter, given the values of ${\rm
d}\sigma_g$ for the five input $\zeta=0,\pm 1, \pm 2$, to
reconstruct $d \sigma_g^{(0,1,2,3,4)}$. For each set of parameters
we automatically pick up all those diagrams that involve a
dependence on the gauge parameter. The number of diagrams in this
set depends on the parameter chosen. Different parameters involve
different (often) complementary sets. In some cases a very large
number of diagrams is involved. An example is $Z W^+ \ra Z W^+$
with $\tilde{\beta}\neq 0,\tilde{\kappa}=1$ where the set involves
$601$ one-loop diagrams out of a total of $840$. We then
numerically verify that the (physical) differential cross section
is independent of $\zeta$
\beqn
{\rm d}\sigma=\sum_g {\rm d}\sigma_g=\sum_g {\rm
d}\sigma_g^{(0)}\; ,
\eeqn

\noi and therefore that

\beqn
{\rm sum}_i=\frac{\sum_g {\rm d}\sigma_g^{(i)}}{{\rm Max}(|{\rm
d}\sigma_g^{(i)}|)}=0\;\; ,\;\; i=1,2,3,4 \; .
\eeqn

As summarised in Table~\ref{tablenlgall27}, we find a precision of
at least $21$ digits on all ${\rm sum}_i$ for all the checks we
have done. We usually get a much better precision when the number
of diagrams involved in the check is smaller.  \\
\noi To appreciate how this level of accuracy is arrived at after
summing on all diagrams,  we show here, see
Table~\ref{tablecheckswwnlg}, in some detail the result for $W^+
W^- \ra W^+ W^-$ for the check on the $\tilde{\alpha}$
gauge-parameter independence (all other parameters set to zero).
This is extracted from the web-page where we have made these
checks public\cite{2to2nlgchecks}. This process involves  some
$925$ one-loop diagrams (and $7$ at tree-level). Even for this
particular example it is not possible to list all the entries of
the table (that is the numerical contributions for all the
diagrams) since they would not fit into a single page (the check
on $\tilde{\alpha}$ involves some $336$ diagrams), thus the skip
($\wr \wr$) on some of the data. For each graph, labelled by its
graph number in Table~\ref{tablecheckswwnlg}, we give all ${\rm
d}\sigma_g^{(i)}$. We see that although individual contributions
can be of the order of $10^2$, when summed up they give a total of
the order of $10^{-27}$. We also show, at the bottom of the table,
$\sum_g {\rm d}\sigma_g$ for the input values
$\tilde{\alpha}=0,\pm 1,\pm2$ and compare these results to the
result obtained by setting $\tilde{\alpha}=5$ in
Eq.~\ref{decomposition-zeta} after $d \sigma_g^{(0,1,2,3,4)}$ have
been reconstructed. In this example concerning $\alphat$ we have
set the values of the other gauge parameters, $\betat, \deltat,
\kappat, \epsilont$ to zero. We have also made a similar test on
$\alphat$ allowing all other parameters non-zero. The same tests
done on $\alphat$ are in turn made for all other parameters. These
tests are made on $26$ processes. More information on the check
concerning $W^+ W^- \ra W^+ W^-$ and all those listed in
Table~\ref{tablenlgall27} are to be found at \cite{2to2nlgchecks}.

One more note concerning the checks on the non-linear gauge
parameter compared to a check one would do through the Feynman
gauge parameter $\xi_{W,A,Z}$ in the usual linear gauge. Having
more parameters that clearly affect different sectors differently
helps in detecting any possible bug. Within the linear gauge, the
usual gauge-parameter dependence is not a polynomial, it also
involves $\log \xi$ and other functions of $\xi$. It is therefore
almost impossible to fit the exact $\xi$ dependence of each graph.
Moreover as pointed out earlier one needs to prepare new libraries
for handling (very) high rank tensor integrals that are not
necessary in the Feynman gauges.

\subsection{Inclusion of hard bremsstrahlung, $k_c$ stability}
\label{seckctest} A complete ${\cal O}(\alpha)$ correction
necessitates the inclusion of the contribution from hard photon
bremsstrahlung. Although this is a tree-level process, in most
cases the total cross section can not be derived analytically.
There is no difficulty in computing the matrix elements. In \grc
this is done automatically keeping all particle masses. The
integration over phase space can get tricky in many cases. In fact
in most cases of interest like for \epemt processes some care must
be exercised. The reason is that though there is no infrared
problem one still needs to very carefully control the $k_c$
dependence and also the collinear mass singularity. This $k_c$
dependence when combined with the one in the soft bremsstrahlung
part (based on a analytical implementation), see section
\ref{irtest}, should cancel leaving no dependence on the cut-off
$k_c$. This would constitute another test on the ${\cal
O}(\alpha)$ calculation of \grcp. The collinear mass singularity
is most acute when the mass of the charged particle is very small
compared to the typical energy scale of the problem, as in \epemt
at high energies. All these problems are due to the integration
over the propagators encountered in Eq.~\ref{irpropagator}. For
example, take the emission from the positron with momentum, $p$.
This propagator is defined from

\beqn
k.p=E_\gamma (E-P \cos\theta_\gamma)=E_\gamma P \left[
\frac{m_e^2}{P(E+P)}+(1-\cos\theta_\gamma) \right] \;
\eeqn
This  becomes extremely peaked in the forward direction,
$\cos\theta_\gamma=1$. For instance,  while the term in square
bracket is of order one for $\cos\theta_\gamma=-1$ it is of order
$\sim 10^{-13}$ for linear collider energies of $500$GeV. In \grc,
integration is done with {\tt BASES}\cite{bases} which is an
adaptive Monte-Carlo program.  For these particular cases one
adapts the integration variables so that one fully picks up the
singularities brought about by the hard photon collinear mass
singularities. This step is therefore not as automatic as the
previous ones in the calculation of the radiative corrections
since one needs to judiciously choose the integration variables.
For more details see\cite{supplement100}.

\noi  Stability of the result as concerns the cut-off $k_c$ is
tested by varying the value of the cut-off $k_c$. We take $k_c \gg
\lambda$ but, usually, much smaller than the maximum energy that
the photon can have, for example a few percent of the
centre-of-mass energy in \epemt processes. This is, typically, of
the order of the photon energy that can be observed by a standard
detector. In some multi-leg processes and for high energy it may
be necessary to go down to values as low as $0.1$GeV or even
$10^{-3}$GeV. One finds agreement within the precision of the
Monte-Carlo which is at least better than 4 digits. From the view
point of the computation a numerical cancellation occurs among the
contribution of  on the one hand, the virtual loop diagrams and
soft photon emission, and on the other hand the hard photon
emission. These individual contributions can be  $1\sim 10$ times
larger than the tree cross section while the full $O(\alpha)$
correction, including virtual, bremmstrahlung and hard photon
emission, is of order $1\%$. The individual contributions should
therefore be calculated extremely accurately. A loss of accuracy
can also be caused if there is some severe cancellation among loop
diagrams, as due to unitarity for example. In such cases one
reverts to quadruple precision.

In \epemt processes where corrections from initial state radiation
can be large it is possible to sum up the effect of multiple
emission of photons, either through a structure function approach
(see for instance \cite{supplement100}) or more sophisticated
approaches that even takes into account the $p_T$ of the photon
like that of the {\tt QEDPS} approach\cite{qedps}.

\setcounter{equation}{0}
\section{Checks on selected cross sections}
\label{testxs}

The previous sections have shown that the system passes highly non
trivial checks for the calculation of the one-loop radiative
corrections to \sm processes. All those tests are {\em internal}
tests within the system. To further establish the reliability of
the system we have also performed comparisons with a number of
one-loop electroweak  calculations that have appeared in the
literature. For all the comparisons we tune our input parameters
to those given by the authors. Therefore one should remember that
some of the results in the following tables are outdated due to
the use of by now obsolete input parameters. All results refer to
integrated cross sections with, in some cases, cuts on the
scattering angle so as to avoid singularities in the forward
direction. Apart from $\epem \ra t \bar t$ where a complete fully
tuned comparison was conducted with high precision and includes
the effect of hard photon radiation at ${\cal O}(\alpha)$, we
compare the results of the virtual electroweak and soft photon
bremsstrahlung ($V+S$), taking the same cut-off, $k_c$, on the
soft photon as specified in those references. We note in passing
that since the \grc system is adapted to multi-particle
production, we can, contrary to some calculations, treat both the
loop corrections and the bremsstrahlung correction within the same
system. Let us also note that for all the processes we will
consider below, we have taken the widths of all particles to zero
since we never hit a pole.

\subsection{$e^+ e^- \ra t \bar t$}
\begin{table*}[hbtp]
\caption{\label{eettcompar} {\em Comparison of the total cross
section $ e^+ e^- \ra t \bar t$ between \grcl and \cite{topfit}.
The corrections refer to the  full one-loop electroweak
corrections  including hard photon radiation.}}
\begin{center}
\begin{tabular}{|c|c|c|}
 \hline
$e^+ e^- \ra t \bar t$ &                     \grcl & \cite{topfit}
\\ \hline
&& \\

$\sqrt{s}=500$GeV & &\\
  tree-level(in pb) & 0.5122751  & 0.5122744
  \\
  ${{\cal O}}(\alpha)$ (in pb)  &0.526371
  & 0.526337\\
   $\delta$ (in $\%$)& 2.75163 &2.74513
\\
   && \\
$\sqrt{s}=1$TeV & &\\
  tree-level(in pb)  &  0.1559187 &0.1559185
  \\
  ${{\cal O}}(\alpha)$ (in pb)  &0.171931
  & 0.171916\\
  $\delta$ (in $\%$)  &10.2696
  &10.2602
  \\
  \hline
\end{tabular}
\end{center}
\end{table*}
This process is an extension of the $2$-fermion production program
that has been successfully carried at {\tt LEP/SLC}. The radiative
corrections to this process first appeared in\cite{eettrcfujimoto}
and then in\cite{eettrceurope}. A new computation has appeared
very recently \cite{topfit}. A dedicated tuned comparison between
\grcl and the program {\tt topfit}\cite{topfit} has recently been
conducted at some depth including hard photon radiation and with
the active participation of the authors of\cite{topfit}. Details
of the comparison are to be found in \cite{grcvstopfit}. Here we
only show the excellent quality of the agreement for the total
cross section including hard photons and we refer the reader to
\cite{grcvstopfit} for other comparisons concerning differential
cross sections and forward-backward asymmetries. Let us point out
however that the comparisons at the level of the differential
cross sections agree within $8$ digits before inclusion of the
hard photon correction and to $7$ digits when the latter are
included. For the totally integrated cross section including hard
photons this quality of agreement  is somehow degraded but stays
nonetheless excellent even at high energies. As Table
\ref{eettcompar} shows, the agreement is still  better than
$0.1$permil.

The authors of \cite{topfit} have also conducted a tuned
comparison with another independent calculation based
on\cite{eettrceurope}. Practically similar conclusions to the ones
presented here are reached, see\cite{eettrccomp1}.

\subsection{$e^+ e^- \ra W^+ W^-$}
\begin{table*}[hbtp]
\caption{\label{eewwcompar} {\em Comparison of the total cross
section $ e^+ e^- \ra W^+ W^-$ between \grcl and
\cite{eewwrcreview}. The calculation  includes full one-loop
electroweak corrections, but no hard photon radiation.}}
\begin{center}
\begin{tabular}{|c|c|c|}
 \hline
$e^+ e^- \ra W^+ \bar W^-$ &                     \grcl &
\cite{eewwrcreview} \\ \hline
&& \\
$\sqrt{s}=190$GeV & &\\
  tree-level(in pb) & 17.8623 & 17.863  \\
  $\delta$ (in $\%$)& $-$9.4923
   & $-$9.489
  \\
&& \\
$\sqrt{s}=500$GeV & &\\
  tree-level(in pb) & 6.5989 &6.599
  \\
   $\delta$ (in $\%$)& $-$12.743  &
   $-$12.74
   \\
   && \\
$\sqrt{s}=1$TeV & &\\
  tree-level(in pb)  & 2.4649 & 2.465
  \\
  $\delta$ (in $\%$)  & $-$15.379
  &$-$15.375
  \\
  \hline
\end{tabular}
\end{center}
\end{table*}

This process is the most important electroweak process at LEP2 and
constitutes one of the most important reactions for the linear
collider. A few independent calculations\cite{alleewwrc} exist and
the most recent ones agree better than the permil. To check the
results given by \grcp, we have set our parameters to those
appearing in Table~2 of the review \cite{eewwrcreview}. The
results refer to the total cross section but without the inclusion
of the hard photon bremsstrahlung. As we see the agreement for
energies ranging from LEP2 to $1$TeV are about at least
$0.1$permil.

\subsection{$e^+ e^- \ra Z H$}
\begin{table*}[hbtp]
\caption{\label{eezhcompar} {\em Comparison of percentage
correction to the total cross section $ e^+ e^- \ra Z H$ between
\grcl and \cite{eezhsmrcDenner}.}}
\begin{center}
\begin{tabular}{|c|c|c|}
 \hline
$e^+ e^- \ra Z H$ &                     \grcl &
\cite{eezhsmrcDenner}
\\ \hline
&& \\
$\sqrt{s}=500$GeV $\;\;\;M_H=100$GeV &  4.15239 &
   4.1524\\
 && \\
$\sqrt{s}=500$GeV $\;\;\;M_H=300$GeV & 6.90166
   &6.9017\\
 && \\
$\sqrt{s}=1000$GeV $\;\;\;M_H=100$GeV & $-$2.16561
   &$-$2.1656 \\
&& \\
$\sqrt{s}=1000$GeV $\;\;\;M_H=300$GeV & $-$2.49949
&$-$2.4995 \\
&& \\
$\sqrt{s}=1000$GeV $\;\;\;M_H=800$GeV & 26.10942
   &26.1094 \\
&& \\
$\sqrt{s}=2000$GeV $\;\;\;M_H=100$GeV &   $-$11.54131
   &$-$11.5414\\
&& \\
$\sqrt{s}=2000$GeV $\;\;\;M_H=300$GeV &  $-$12.82256
   &$-$12.8226
\\
   && \\
$\sqrt{s}=2000$GeV $\;\;\;M_H=800$GeV & 11.24680
   &11.2468 \\
&& \\
  \hline
\end{tabular}
\end{center}
\end{table*}
This process is an important discovery channel for an intermediate
mass Higgs at a moderate energy linear collider and could permit
to study the properties of the Higgs. Three independent one-loop
calculations
exist\cite{eezhsmrcDenner,eezhsmrcKniehl,eezhsmrcJeger} which all
agree beyond the precision of any future linear collider. A
comparison was conducted against the  calculation
in\cite{eezhsmrcDenner} where one of the authors has provided us
with more precise numbers than those appearing in Table~1 of
\cite{eezhsmrcDenner}\footnote{We thank A.~Denner for providing us
with the correct $M_W$ masses used in this table. Beside the input
given in \cite{eezhsmrcDenner}, $M_W$ is crucial for a precise
comparison. The following $M_W$ masses have been used:
$M_W=80.231815$GeV($M_H=100$GeV), $M_W=80.159313
$GeV($M_H=300$GeV), $M_W=80.081409$GeV($M_H=800$GeV).}. Table
\ref{eezhcompar} shows that the results given by our system \grcl
and those of \cite{eezhsmrcDenner} agree on all digits. This means
that the radiatively corrected cross sections at different
energies for a Higgs mass ranging from the light to the heavy
agree within at least $6$ digits. The corrections refer to the
full one-loop electroweak corrections but without hard photon
radiation.


\subsection{$\gamma \gamma \ra t \bar t$}
\begin{table*}[bthp]
\caption{\label{ggttcompar} {\em Comparison of the total cross
section $\gamma \gamma \ra t \bar t$ between \grcl and
\cite{ggttrc} include full one-loop electroweak corrections at
one-loop, but no hard (final) photon radiation.}}
\begin{center}
\begin{tabular}{|c|c|c|}
 \hline
$\gamma \gamma \ra t \bar t$ &                     \grcl &
\cite{ggttrc} \\ \hline
&& \\
 $\sqrt{s}=350$GeV & & \\
  tree-level (in pb) & 0.332477    &     0.33248 \\
  $\delta$ (in $\%$)  &  $-$6.889      &        $-$6.88
  \\
  && \\
$\sqrt{s}=500$GeV && \\
  tree-level(in pb) &  0.904371 &        0.90439 \\
  $\delta$ (in $\%$) & $-$4.824    &           $-$4.82
  \\
  && \\
$\sqrt{s}$=1TeV && \\
  tree-level  (in pb) &  0.434459 &         0.43447\\
 $\delta$ (in $\%$)  & $-$5.633  &             $-$5.63\\
&&\\ \hline
\end{tabular}
\end{center}
\end{table*}

The comparison has been made with Table~1 of \cite{ggttrc} without
any convolution over any photon spectra.  $M_H=150$GeV so as to
avoid the Higgs resonance. As we see the agreement is very good,
it is just limited by the precision of the numbers provided in
\cite{ggttrc}.

\subsection{$\gamma \gamma \ra W^+  W^-$}

\begin{table*}[hbtp]
\caption{\label{ggwwcompar} {\em Comparison for $\gamma \gamma \ra
W^+ W^-$ between \grcl and \cite{ggwwrc1} and \grcl and
\cite{Jikiaggwwrc}. No hard (final) photon radiation is included.
When not stated the cross sections and corrections refer to the
total cross section with no angular cut.}}
\begin{center}
\begin{tabular}{|c|c|c|}
 \hline
&& \\
$\gamma \gamma \ra W^+  W^-$, $M_W=80.36$GeV $M_H=300$GeV & \grcl
& \cite{Jikiaggwwrc}
\\ \hline
&& \\
$\sqrt{s}=500$GeV & &\\
  tree-level(in pb) & 77.497    &   77.50
  \\
$\delta$ (in $\%$)&  $-$10.06 & $-$10.1%
  \\
&& \\
$\sqrt{s}=1$TeV & &\\
  tree-level(in pb) & 79.995&79.99
  \\
$\delta$ (in $\%$)& $-$18.73 &$-$18.7
   \\
&& \\
$\sqrt{s}=2$TeV &  &\\
tree-level(in pb)  & 80.531    &80.53
 \\
$\delta$ (in $\%$)  &  $-$27.25 &$-$27.2
\\
&& \\
$\sqrt{s}=2$TeV $60^\circ< \theta < 120^\circ$&  &\\
tree-level(in pb)  &  0.39356  &0.3936
 \\
$\delta$ (in $\%$)  &  $-$75.6827 & $-$75.6 \\
  \hline
\hline
&& \\
 $M_W=80.333$GeV $M_H=250$GeV &\grcl & \cite{ggwwrc1}
\\ \hline
&& \\
$\sqrt{s}=500$GeV & &\\
  tree-level(in pb) &   77.552  &   77.55
  \\
  $\delta$ (in $\%$)& $-$3.376   & $-$3.38
  \\
&& \\
$\sqrt{s}=1$TeV & &\\
  tree-level(in pb) &80.049  &   80.05
  \\
   $\delta$ (in $\%$)& $-$7.087  & $-$7.08
   \\
  \hline
\end{tabular}
\end{center}
\end{table*}

The first complete calculation of the electroweak radiative
corrections  to $\gamma \gamma \ra W^+ W^-$ has been performed in
\cite{ggwwrc1}. Jikia has performed a full ${\cal O}(\alpha)$
calculation, including hard photon radiation \cite{Jikiaggwwrc}.
Comparison has been made on the one hand with Table~1 of
\cite{ggwwrc1} without convolution over any photon spectra as well
as with Table~2 of \cite{Jikiaggwwrc}. In both comparisons we
considered the total integrated cross section but with no
inclusion of the hard photon radiation which in any case is not
treated in \cite{ggwwrc1}. Because the fermionic contribution is
extremely small compared to the bosonic contribution in the
radiative correction to the total cross section, we also looked at
the correction with an angular cut on the outgoing $W$ as check on
the fermionic correction. As can be seen from
Table~\ref{ggwwcompar} the agreement is  just limited by the
precision of the numbers provided in \cite{ggwwrc1} and
\cite{Jikiaggwwrc}. Note that in \cite{Jikiaggwwrc}, the
correction is split between the bosonic corrections and the
fermionic corrections. When considering the total cross section
the latter are much too small and are below the precision with
which the bosonic corrections are displayed in \cite{Jikiaggwwrc}.
Therefore given the precision of the data in \cite{Jikiaggwwrc}
the corrections are essentially given by the bosonic part for the
total cross section. For the entry with the angular cut, the
fermionic corrections are not negligible.

\subsection{$e \gamma \ra W \nu_e$}
\begin{table*}[hbtp]
\caption{\label{egnwcompar} {\em Comparison of the total cross
section $ e \gamma \ra W \nu_e $ between \grcl and
\cite{egnwrc}~.}}
\begin{center}
\begin{tabular}{|c|c|c|}
 \hline
$e \gamma \ra W \nu_e $ &                     \grcl & \cite{egnwrc} \\
\hline
&& \\
$\sqrt{s}=500$GeV & &\\
  tree-level(in pb) &   36.5873  &36.587
  \\
  $\delta$ (in $\%$)& $-$12.2803  &$-$12.281
  \\
&& \\
$\sqrt{s}=2$TeV & &\\
  tree-level(in pb) & 43.9368 &43.937
  \\
   $\delta$ (in $\%$)& $-$19.0917  &$-$19.092
\\
  \hline
\end{tabular}
\end{center}
\end{table*}

The comparison shown in Table.~\ref{egnwcompar} is made on the
total cross section ($0^{\circ}\leq \theta \leq 180^{\circ}$)
based on Table~5.1 of \cite{egnwrc}. No convolution on the photon
spectra is applied nor is the hard photon bremsstrahlung included.
The agreement is rather excellent.

\subsection{$e \gamma \ra e Z$}
\begin{table*}[hbtp]
\caption{\label{egezcompar} {\em Comparison of the total cross
section $ e \gamma \ra e Z$ between \grcl and \cite{egezrc}.}}
\begin{center}
\begin{tabular}{|c|c|c|}
 \hline
$e \gamma \ra e Z  $ &                     \grcl & \cite{egezrc} \\
\hline
&& \\
$\sqrt{s}=500$GeV $\;\;,\;$   & &\\
  tree-level(in pb) &  0.70515
  &0.7051\\
  $\delta$ (in $\%$)& $-$25.689 & $-$25.69
  \\
$\sqrt{s}=500$GeV $\;\;,\;$ $1^{\circ} \leq \theta \leq 179^{\circ}$  & &\\
tree-level(in pb) & 1.7696&1.770\\
$\delta$ (in $\%$)&  $-$22.313 &$-$22.31
\\
&& \\
$\sqrt{s}=2$TeV $\;\;,\;$ $20^{\circ} \leq \theta \leq 160^{\circ}$  & &\\
  tree-level(in pb) &  0.046201  &   0.04620
 \\
  $\delta$ (in $\%$)& $-$39.529    &   $-$39.53
  \\
$\sqrt{s}=2$TeV $\;\;,\;$ $1^{\circ} \leq \theta \leq 179^{\circ}$  && \\
tree-level(in pb) & 0.1170     &    0.117 \\
$\delta$ (in $\%$)&  $-$30.845 &$-$30.84
\\
&& \\
\hline
\end{tabular}
\end{center}
\end{table*}

The comparison shown in Table.~\ref{egezcompar} is made with
Table~5.3 of \cite{egezrc}. No convolution on the photon spectra
is applied nor is the hard photon bremsstrahlung included. The
agreement is excellent.

\subsection{$W^+ W^- \ra W^- W^+$}
\begin{table*}[hbtp]
\caption{\label{wwwwcompar} {\em Comparison of the total
(unpolarised) cross section $W^+ W^- \ra W^- W^+ $ between \grcl
and \cite{wwwwrc}. $M_H=100$GeV. For the cuts see the text.}}
\begin{center}
\begin{tabular}{|c|c|c|}
 \hline
$W^+ W^- \ra W^- W^+$ &                     \grcl & \cite{wwwwrc}
\\ \hline
 \fbox{$k_c=.05\sqrt{s}$}&& \\
$\sqrt{s}=2$TeV & &\\
  tree-level(in pb) &77.17067   &77.17067
  \\
   $\delta$ (in $\%$)&$-$21.0135 &$-$21.0135
\\
   && \\
$\sqrt{s}=5$TeV & &\\
  tree-level(in pb)  & 14.2443  &14.2443
  \\
  $\delta$ (in $\%$)  &$-$57.1567
  &$-$57.1556\\
     && \\
$\sqrt{s}=10$TeV & &\\
  tree-level(in pb)  &  3.644573 &3.644573
  \\
  $\delta$ (in $\%$)  &$-$93.9942
  &$-$94.0272 \\
  && \\
\fbox{$k_c=.5\sqrt{s}$}&& \\
$\sqrt{s}=2$TeV & &\\
  tree-level(in pb) & 77.17067
  &77.17067
  \\
   $\delta$ (in $\%$)& $-$17.23988&$-$17.23989
\\
   && \\
$\sqrt{s}=5$TeV & &\\
  tree-level(in pb)  & 14.24434  &14.24434
  \\
  $\delta$ (in $\%$)  &$-$49.9736
  &$-$49.9724\\
     && \\
$\sqrt{s}=10$TeV & &\\
  tree-level(in pb)  & 3.644574 &3.644573
  \\
  $\delta$ (in $\%$)  &$-$83.9247
  & $-$83.9577\\
\hline
\end{tabular}
\end{center}
\end{table*}

  This is one of the most difficult $2 \ra 2$ processes in the \sm
ever to be calculated. As discussed previously the number of
diagrams at one-loop is of the order $1000$. Moreover very subtle
gauge cancellations take place especially as the energy of the
participating $W$'s increases. The most complete calculation has
been performed in \cite{wwwwrc} and the code is freely available
at {\tt www.hep-processes.de}. However, hard photon radiation is
not included. Following \cite{wwwwrc} we have compared our results
with those of the code by requiring a cut on the forward-backward
direction such that the integration over the scattering angle is
over $10^\circ \leq \theta \leq 170^\circ$. Moreover we have
considered two cuts on the photon energy (for the bremsstrahlung
part), $k_c=.05\sqrt{s}$ and $k_c=.5\sqrt{s}$\footnote{ These
cut-off photon energies are much higher than those recommended in
section~\ref{seckctest}. However we stick to these values to
comply with those chosen in Ref.~\cite{wwwwrc}.}. The Higgs in
this comparison is light, see \cite{wwwwrc} for a justification on
this issue. Having at our disposal the code, a tuned comparison
could be performed. We can see from Table~\ref{wwwwcompar} that at
centre-of-mass energy of the $W$ pair of $2$TeV one reaches
agreement over $6$ digits, $4$-$5$ digits for $\sqrt{s}=5$TeV but
``only" 3 digits agreement for $\sqrt{s}=10$TeV. Note that even in
this case this means that the radiative corrections are known to
about $0.1$permil. This very high energy for the $WW$ would
probably never be reached. Moreover as the authors of\cite{wwwwrc}
warn, for this kind of energy an integration in quadruple
precision is probably already mandatory. Therefore it is fair to
conclude that one has for this reaction an excellent agreement.
Note that \grcl automatic calculation is the first confirmation of
the result of \cite{wwwwrc}.

\newpage
\setcounter{equation}{0}
\section{Conclusions}
\label{sec:conclude}

Precision measurements in high-energy physics need to be matched
by very accurate theoretical predictions. This requires that one
performs calculations that go beyond the tree-level approximation.
Moreover with the increase in energy, as will be available at the
upcoming colliders, some important multi-particle final states
beyond the usual $2 \ra 2$ processes become  important physical
observables. Higgs production at the linear collider is such an
example. Calculation of multi-loop and multi-leg processes
involves the computations of thousands of diagrams, with the
property that the topologies that appear as the order of
perturbation theory increases become extremely arduous. An example
is the treatment of the $N$-point functions beyond the box or
two-loop diagrams.  A calculation by hand is obviously no longer
possible. Although a few of the steps involved in these
computations are now done with the help of computers, it has
become almost a necessity to perform the whole chain in the
calculation of these complex cross sections by a computer with a
minimum of human intervention so as to avoid any risk of error. We
have in this review taken \grcl as a prototype of such a fully
automated system and described in detail its workings and
performance in evaluating one-loop processes in the electroweak
theory. The general strategy of constructing a code like this can
of course be applied to other systems and we have discussed some
of them. We have also reviewed in some detail the most important
modules and components that an automated system must have. Since
there are a  few fully automated systems at tree-level, the
emphasis in this report has been on the extension to and the
implementation of the one-loop corrections. Central to this
implementation is the library for the reduction of the tensorial
loop integrals and also the reduction of the higher $N$-point
functions ($N=5,6$) to lower $N$-point scalar integrals. We have
presented an algorithm which is now fully functional in \grcl.\\
\noi Another crucial aspect of a fully automated calculation is
the possibility of checking the results of its output. In an
automated tree-level system this is almost trivial and is usually
provided by switching between the Feynman gauge and the unitary
gauge. As we argued  the unitary gauge or the usual general
$R_\xi$ {\em linear} gauge are not suitable at all for an
implementation in a multi-purpose one-loop automatic code. Other
important checks are the ultraviolet and infrared finiteness
tests, but the gauge-parameter independence check is most
powerful. This is the reason a part of this review has been
devoted to renormalisation in the \sm within a non-linear
gauge.The non-linear gauge that we exploit introduces $5$ gauge
parameters. This gauge fixing modifies a large number of vertices
in the bosonic sector but can be chosen so as to leave the
propagators of all gauge bosons as simple as in the standard
linear 't~Hooft-Feynman gauge. Technically this means that the
structure of any one-loop $N$-point function is not more involved
than what it is with fermionic loops and therefore that many
libraries for these functions need not be extended. To show that
one-loop automated systems have now become completely functional
and trustworthy as concerns the treatment of any $2 \ra 2$
processes we have presented conclusive tests on the finiteness,
both infrared and ultraviolet, and the gauge-parameter
independence of the results pertaining to some $26$ processes.
These checks are verified with a precision that attains at least
20 digits.  We have also used the new system to carry further
comparison on radiative corrections to a few processes that have
appeared in the literature. This selection includes heavy fermion
production, vector boson and Higgs boson production in both
$\epem, \gamma \gamma$ and $e\gamma$ machines as well as the very
challenging $W W$ scattering process. For the latter we provide
the first check to the complete calculation that has appeared in
the literature. In all cases we find excellent agreement.

This shows that the automatic system of calculating radiative
corrections numerically has now all the ingredients to tackle
$2\ra 3$ processes confidently. In the last two years, major
progress has been made in this area thanks to the automated system
\grcl and the package {\tt FeynArts-FeynCalc-FormCalc}. Most
important processes for Higgs production at the linear collider,
\eennht\cite{eennhradcor2002, eennhletter,Dennereennh1},
\eeeeht\cite{eeeehgrace}, \eezhht\cite{eezhhgrace,eezhhchinese},
\eettht\cite{eetthgrace,eetthdenner,eetthchinese}, $\gamma \gamma
\ra t \bar t H$\cite{ggtthchinese} as well as
\eenngt\cite{eennggrace} have been computed and for most of them
checked against each other, thanks to the automatic systems. \grcl
is now in a position to compute one-loop $2 \ra 4$ processes. We
have in fact already presented some  results pertaining to
\eennhht\cite{eennhhgrace} and some preliminary ones for the
four-fermion final state $\epem \ra \mu^- \bar\nu_\mu u \bar d$ at
one-loop \cite{eemnudgrace}. Based on the {\tt
FeynArts-FeynCalc-FormCalc} package the one-loop correction for a
certain class of $\epem \ra 4f$ has also been
achieved\cite{eeto4fdenner} recently.

 To improve the efficiency of the system for applications to
one-loop corrections for processes with more than $2$ particles in
the final state, one should seek a derivation based on helicity
amplitudes. At tree-level this has been nicely implemented in \grc
and applied to processes up to $6$ particles in the final state.
Moreover a derivation based on helicity amplitudes allows the
implementation of full spin-correlation, for processes when the
final particle is unstable. At the one-loop level, another
advantage is that it would allow the calculation by the system of
processes that are not generated at tree-level, such as $\gamma
\gamma \ra \gamma \gamma, \gamma \gamma \ra ZZ, Z\ra 3 \gamma, H
\ra \gamma \gamma$. The working version of \grcl can not handle
such processes since the one-loop correction are calculated as
products of tree-level and one-loop matrix elements. A version of
\grcl which is being developed is based on helicity amplitudes.
Preliminary results are encouraging. This new version will also be
used as an additional test on the results given by the traditional
method of squaring matrix elements.

\vspace*{1cm} \noi {\bf \large Acknowledgment}

This work is part of a collaboration between the {\tt GRACE}
project in the Minami-Tateya group and LAPTH. D.~Perret-Gallix and
Y.~Kurihara deserve special thanks for their contribution. We also
thank M.~Kuroda and J.A.M.~Vermaseren for a critical reading of
the manuscript. This work was supported in part by the Japan
Society for Promotion of Science (JSPS) under the Grant-in-Aid for
scientific Research B(N$^{{\rm o}}$.~14340081), PICS 397 and The
GDRi-ACPP of the French National Centre for Scientific Research
(CNRS).

\cleardoublepage
\addcontentsline{toc}{section}{Appendices}

\renewcommand{\thesection}{\Alph{section}}
\setcounter{section}{0}

\renewcommand{\theequation}{\thesection.\arabic{equation}}
\setcounter{equation}{0}

\noi {\Large {\bf Appendices}}

\def\db{\delta_{\rm BRS}}
\section{Specific form of the BRST transformations}
\label{app-brstrans}

The action of the BRST transformations is derived as a
generalisation of the usual gauge transformations. The ghost
fields corresponding to the four gauge bosons write in terms of
the ghosts of the $SU(2)\times U(1)$ fields as:

\beqn
c^\pm&=&\frac{1}{\sqrt{2}} (c^1 \mp
ic^2)\nonumber\\
c^A&=&\sw c^3+ \cw c^B \nonumber \\
c^Z&=&\cw c^3-\sw c^B \; .
\eeqn

One obtains
\beqn
\db W_\mu^\pm&=&\partial_\mu c^\pm \mp\;ie \left[ \left( (A_\mu
+\frac{\cw}{\sw} Z_\mu\right) c^\pm - \left(c^A +\frac{\cw}{\sw}
c^Z\right) W^\pm_\mu\right]
\nonumber \\
\db Z_\mu&=&\partial_\mu c^Z -ig \cw \left(
W_\mu^+c^- - W_\mu^-c^+ \right)\nonumber \\
 \db A_\mu&=&\partial_\mu c^A -i e \left(
 W_\mu^+c^- - W_\mu^-c^+ \right)\; .
\eeqn

Likewise by considering the gauge transformation on the Higgs
doublet one gets

\beqn
\db H &=&-\frac{ g}{2} \left(c^- \chi^+ +c^+ \chi^-
\right) - \frac{e}{2 \sw \cw} c^Z \chi_3 \nonumber \; ,\\
\db \chi_3 &=&-\frac{i g}{2} \left(-c^+ \chi^- +c^- \chi^+
\right) + \frac{e}{2 \sw \cw} c^Z (v+H) \nonumber \; ,\\
\db \chi^\pm &=&  \frac{ g}{2} \left(v+H \mp i\chi_3\right) c^\pm
\pm ie \chi^\pm \left(c^A + \frac{\cw^2-\sw^2}{2\sw \cw}
c^Z\right) \; .
\eeqn

To find the transformation for the ghost fields, notice that the
BRST transformation is nilpotent. For instance from $(\db)^2
W^i_\mu=0$ one gets $\db c^i$. Indeed more generally one has, for
any group,
\beqn
\db A_\mu^i= D_\mu c^i= \partial_\mu c^i +g [A_\mu, c]^i \ra \db
c^i=-g \frac{1}{2} [c,c]^i
\eeqn
Care should be taken that $\db$ being a fermion operator the
graded Leibnitz rule applies: $\db (XY)=(\db X)Y\pm X(\db Y)$
where the minus sign applies if X has an odd number of ghosts or
antighosts, note also that $(c^i)^2=0$.

In our case this implies
\beqn
\db c^B=0 \;\;\;\;\;\;\;\; \db c^i=- \frac{1}{2} g
\;\epsilon_{ijk} \; c^j c^k
\eeqn

and thus

\beqn
\db c^\pm&=&\mp i\;g\; c^\pm ( \sw c^A + \cw
c^Z) \nonumber \\
\db c^A&=&+i \;e\; c^+ c^- \nonumber \\
\db c^Z&=&+i \;g\; \cw\; c^+ c^-
\eeqn

The transformation for the anti-ghost field is defined through the
auxiliary $B$ field of the gauge functions,

\beqn \label{antigtransfapp} \db \bar c^i= B^i \, . \eeqn

\setcounter{equation}{0}
\section{Feynman Rules}
\label{sec:frule}

The basic Feynman rules follow the so-called Kyoto
convention\cite{kyotorc}. A particle at the endpoint
\textit{enters} the vertex. For instance, if a line is denoted as
$W^+$, then the line shows either the incoming $W^+$ or the
outgoing $W^-$. The momentum assigned to a particle is defined as
\textit{inward} except for the case of a ghost particle for which
the momentum is defined \textit{along the flow of its ghost
number}, as will be shown in the figures.

\subsection{Propagators}

\begin{tabular}{cll}
\hline $ W^{\pm} $ \rule[-5mm]{0mm}{12mm}
 & $\displaystyle{ \frac{1}{k^2-M_W^2}
 \left( g_{\mu\nu}-(1-\xiw)\frac{k_{\mu}k_{\nu}}{k^2-\xiw M_W^2}\right) }$ \\
$ Z $ \rule[-5mm]{0mm}{12mm}
 & $\displaystyle{ \frac{1}{k^2-M_Z^2}
 \left( g_{\mu\nu}-(1-\xiz)\frac{k_{\mu}k_{\nu}}{k^2-\xiz M_Z^2}\right) }$ \\
$ A $ \rule[-5mm]{0mm}{12mm}
 & $\displaystyle{ \frac{1}{k^2}
 \left( g_{\mu\nu}-(1-\xi_A)\frac{k_{\mu}k_{\nu}}{k^2}\right) }$ \\
\hline $ f $ \rule[-5mm]{0mm}{12mm}
& $\displaystyle{\frac{-1}{\ksl-m_f}}$ \\
\hline $ H $ \rule[-5mm]{0mm}{12mm}
 & $\displaystyle{\frac{-1}{k^2-M_H^2}}$ \\
\hline $ \chi^{\pm} $ \rule[-5mm]{0mm}{12mm}
 & $\displaystyle{\frac{-1}{k^2-\xiw M_W^2}}$ \\
\hline $ \chi_3 $ \rule[-5mm]{0mm}{12mm}
 & $\displaystyle{\frac{-1}{k^2-\xiz M_Z^2}}$ \\
\hline $ c^{\pm} $ \rule[-5mm]{0mm}{12mm}
 & $\displaystyle{\frac{-1}{k^2-\xiw M_W^2}}$ \\
\hline $ c^{Z} $ \rule[-5mm]{0mm}{12mm}
 & $\displaystyle{\frac{-1}{k^2-\xiz M_Z^2}}$ \\
\hline $ c^{A} $ \rule[-5mm]{0mm}{12mm}
 & $\displaystyle{\frac{-1}{k^2}}$ \\
\hline
\end{tabular}

\subsection{Vector-Vector-Vector}

\begin{minipage}[c]{5cm}
\begin{center}
\includegraphics[width=5cm,height=5cm]{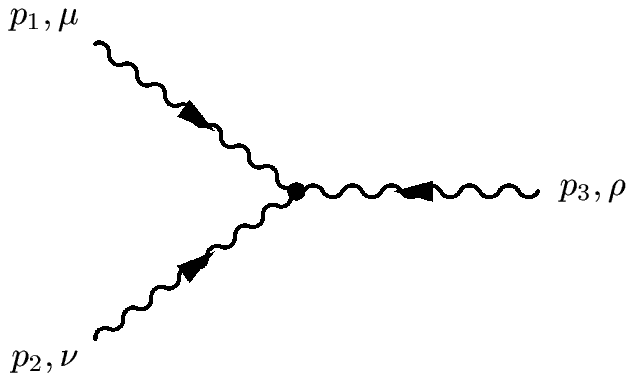}
\end{center}
\end{minipage}
\hspace*{3mm}
\begin{minipage}[c]{10cm}
\begin{tabular}{cccl}
\hline $p_1 \ (\mu)$ & $p_2 \ (\nu)$ &
$p_3 \ (\rho)$ & \\
\hline
& & & \\
$W^-$ & $W^+$ & $A$ & $\displaystyle{e\Bigl[
g^{\mu\nu}(p_1-p_2)^{\rho} }$\\
& & & $\displaystyle{
+(1+\anlg/\xiw)(p_3^{\nu}g^{\mu\rho}-p_3^{\mu}g^{\nu\rho}) }$ \\
& & & $\displaystyle{
+(1-\anlg/\xiw)(p_2^{\mu}g^{\nu\rho}-p_1^{\nu}g^{\mu\rho})
\Bigr]}$ \\
& & & \\
\hline
& & & \\
$W^-$ & $W^+$ & $Z$ & $\displaystyle{e\frac{c_W}{s_W}\Bigl[
g^{\mu\nu}(p_1-p_2)^{\rho} }$\\
& & & $\displaystyle{
+(1+\bnlg/\xiw)(p_3^{\nu}g^{\mu\rho}-p_3^{\mu}g^{\nu\rho}) }$ \\
& & & $\displaystyle{
+(1-\bnlg/\xiw)(p_2^{\mu}g^{\nu\rho}-p_1^{\nu}g^{\mu\rho})
\Bigr]}$ \\
& & & \\
\hline
\end{tabular}
\end{minipage}

\subsection{Vector-Vector-Scalar}

\begin{minipage}[c]{5cm}
\begin{center}
\includegraphics[width=5cm,height=5cm]{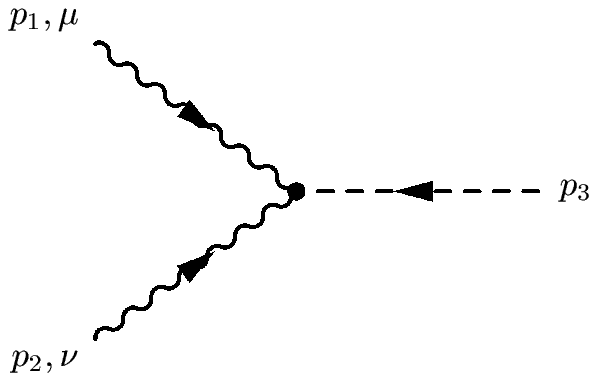}
\end{center}
\end{minipage}
\hspace*{3mm}
\begin{minipage}[c]{10cm}
\begin{tabular}{cccl}
\hline $p_1 \ (\mu)$ & $p_2 \ (\nu)$ &
$p_3$ & \\
\hline
& & & \\
$W^{\pm}$ & $A$ & $\chi^{\mp}$ &
$\displaystyle{\mp i e M_W(1-\anlg)g^{\mu\nu}}$ \\
& & & \\
\hline
& & & \\
$W^{\pm}$ & $Z$ & $\chi^{\mp}$ & $\displaystyle{\pm i
e\frac{1}{s_Wc_W} M_W
\left(1-c_W^2(1-\bnlg)\right)g^{\mu\nu}}$ \\
& & & \\
\hline
& & & \\
$W^-$ & $W^+$ & $H$ &
$\displaystyle{e\frac{1}{s_W}M_W g^{\mu\nu}}$ \\
& & & \\
\hline
& & & \\
$Z$ & $Z$ & $H$ &
$\displaystyle{e\frac{1}{s_Wc_W^2}M_W g^{\mu\nu}}$ \\
& & & \\
\hline
\end{tabular}
\end{minipage}

\subsection{Scalar-Scalar-Vector}

\begin{minipage}[c]{5cm}
\begin{center}
\includegraphics[width=5cm,height=5cm]{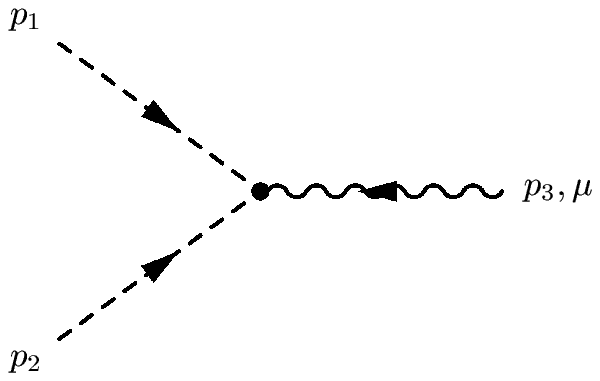}
\end{center}
\end{minipage}
\hspace*{3mm}
\begin{minipage}[c]{10cm}
\begin{tabular}{cccl}
\hline $p_1$ & $p_2$ &
$p_3 \ (\mu)$ & \\
\hline
& & & \\
$H$ & $\chi^{\mp}$ & $W^{\pm}$ & $\displaystyle{i
e\frac{1}{2s_W}\left[ (1-\dnlg)p_2^{\mu}-(1+\dnlg)p_{1}^{\mu}
\right]}$ \\
& & & \\
\hline
& & & \\
$\chi_3$ & $\chi^{\mp}$ & $W^{\pm}$ & $\displaystyle{\pm e
\frac{1}{2s_W}\left[ (1-\knlg)p_2^{\mu}-(1+\knlg)p_{1}^{\mu}
\right]}$ \\
& & & \\
\hline
& & & \\
$\chi^-$ & $\chi^+$ & $A$ &
$\displaystyle{e(p_2-p_1)^{\mu}}$ \\
& & & \\
\hline
& & & \\
$\chi^-$ & $\chi^+$ & $Z$ &
$\displaystyle{e\frac{c_W^2-s_W^2}{2s_Wc_W}(p_2-p_1)^{\mu}}$ \\
& & & \\
\hline
& & & \\
$H$ & $\chi_3$ & $Z$ & $\displaystyle{i e \frac{1}{2s_Wc_W}\left[
(1-\enlg)p_2^{\mu}-(1+\enlg)p_{1}^{\mu}
\right]}$ \\
& & & \\
\hline
\end{tabular}
\end{minipage}

\subsection{Scalar-Scalar-Scalar}

\begin{minipage}[c]{5cm}
\begin{center}
\includegraphics[width=5cm,height=5cm]{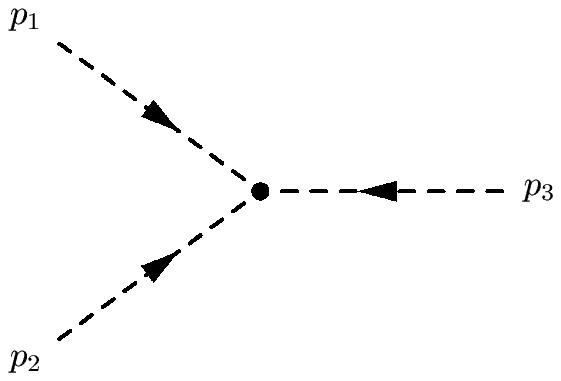}
\end{center}
\end{minipage}
\hspace*{3mm}
\begin{minipage}[c]{10cm}
\begin{tabular}{cccl}
\hline $p_1$ & $p_2$ &
$p_3$ & \\
\hline
& & & \\
$H$ & $H$ & $H$ &
$\displaystyle{-e\frac{3}{2s_W M_W}M_H^2}$ \\
& & & \\
\hline
& & & \\
$H$ & $\chi^-$ & $\chi^+$ &
$\displaystyle{-e\frac{1}{2s_W M_W}(M_H^2+2\dnlg M_W^2\cdot\xiw)}$ \\
& & & \\
\hline
& & & \\
$H$ & $\chi_3$ & $\chi_3$ &
$\displaystyle{-e\frac{1}{2s_W M_W}(M_H^2+2\enlg M_Z^2\cdot\xiz)}$ \\
& & & \\
\hline
\end{tabular}
\end{minipage}

\subsection{Vector-Vector-Vector-Vector}

\begin{minipage}[c]{4.5cm}
\begin{center}
\includegraphics[width=4cm,height=4cm]{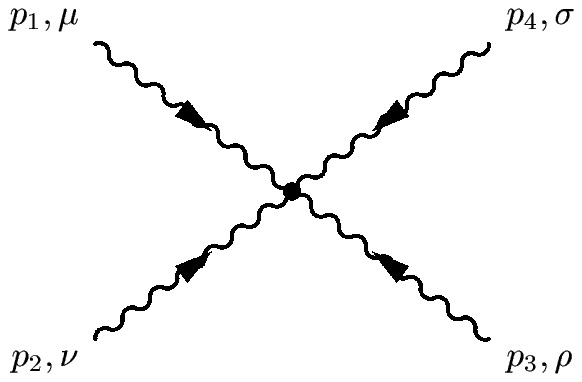}
\end{center}
\end{minipage}
\begin{minipage}[c]{10.8cm}
\begin{tabular}{ccccl}
\hline $p_1 \ (\mu)$ & $p_2 \ (\nu)$ & $p_3 \ (\rho)$ &
$p_4 \ (\sigma)$ \\
\hline
$W^+$ & $W^-$ & $A$ & $A$ \\
& & & \\
\multicolumn{5}{r}{ $\displaystyle{e^2\left[
-2g^{\mu\nu}g^{\rho\sigma} +(1-\anlg^2/\xiw)
(g^{\mu\rho}g^{\nu\sigma}+g^{\mu\sigma}g^{\nu\rho})
\right]}$} \\
& & &  \\
\hline
$W^+$ & $W^-$ & $A$ & $Z$ \\
& & & \\
\multicolumn{5}{r}{ $\displaystyle{e^2\frac{c_W}{s_W}\left[
-2g^{\mu\nu}g^{\rho\sigma} +(1-\anlg\bnlg/\xiw)
(g^{\mu\rho}g^{\nu\sigma}+g^{\mu\sigma}g^{\nu\rho})
\right]}$} \\
& & & \\
\hline
$W^+$ & $W^-$ & $Z$ & $Z$ \\
& & & \\
\multicolumn{5}{r}{ $\displaystyle{e^2\frac{c_W^2}{s_W^2}\left[
-2g^{\mu\nu}g^{\rho\sigma} +(1-\bnlg^2/\xiw)
(g^{\mu\rho}g^{\nu\sigma}+g^{\mu\sigma}g^{\nu\rho})
\right]}$} \\
& & &  \\
\hline
$W^+$ & $W^-$ & $W^-$ & $W^+$ \\
& & &  \\
\multicolumn{5}{r}{ $\displaystyle{-e^2\frac{1}{s_W^2}\left[
-2g^{\mu\sigma}g^{\nu\rho}
+(g^{\mu\rho}g^{\nu\sigma}+g^{\mu\nu}g^{\rho\sigma})
\right]}$ }\\
& & & \\
\hline
\end{tabular}
\end{minipage}

\subsection{Vector-Vector-Scalar-Scalar}

\begin{minipage}[c]{4.5cm}
\begin{center}
\includegraphics[width=4cm,height=4cm]{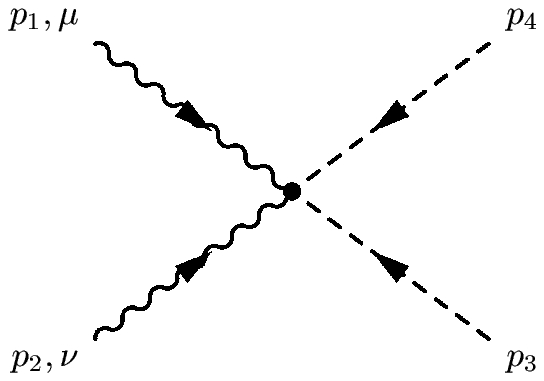}
\end{center}
\end{minipage}
\begin{minipage}[c]{10.8cm}
\begin{tabular}{ccccl}
\hline $p_1 \ (\mu)$ & $p_2 \ (\nu)$ & $p_3 $ &
$p_4 $  \\
\hline $A$ & $W^{\pm}$ & $H$ & $\chi^{\mp}$ &
\rule[-5mm]{0mm}{12mm}
$\displaystyle{\mp i e^2\frac{1}{2s_W}(1-\anlg\dnlg)g^{\mu\nu}}$ \\
\hline \rule[-5mm]{0mm}{12mm} $A$ & $W^{\pm}$ & $\chi_3$ &
$\chi^{\mp}$ &
$\displaystyle{- e^2\frac{1}{2s_W}(1-\anlg\knlg)g^{\mu\nu}}$ \\
\hline \rule[-5mm]{0mm}{12mm} $Z$ & $W^{\pm}$ & $H$ & $\chi^{\mp}$
& $\displaystyle{\pm i e^2\frac{1}{2s_W^2c_W}
\left(1-c_W^2(1-\bnlg\dnlg)\right)g^{\mu\nu}}$ \\
\hline \rule[-5mm]{0mm}{12mm} $Z$ & $W^{\pm}$ & $\chi_3$ &
$\chi^{\mp}$ & $\displaystyle{ e^2\frac{1}{2s_W^2c_W}
\left(1-c_W^2(1-\bnlg\knlg)\right)g^{\mu\nu}}$ \\
\hline \rule[-5mm]{0mm}{12mm} $A$ & $A$ & $\chi^+$ & $\chi^-$ &
$\displaystyle{ 2 e^2 g^{\mu\nu}}$ \\
\hline \rule[-5mm]{0mm}{12mm} $Z$ & $A$ & $\chi^+$ & $\chi^-$ &
$\displaystyle{ 2e^2\frac{c_W^2-s_W^2}{2s_Wc_W}g^{\mu\nu}}$ \\
\hline \rule[-5mm]{0mm}{12mm} $Z$ & $Z$ & $\chi^+$ & $\chi^-$ &
$\displaystyle{ 2e^2\left(\frac{c_W^2-s_W^2}{2s_Wc_W}\right)^2g^{\mu\nu}}$ \\
\hline \rule[-5mm]{0mm}{12mm} $W^+$ & $W^-$ & $H$ & $H$ &
$\displaystyle{ e^2\frac{1}{2s_W^2}g^{\mu\nu}}$ \\
\hline \rule[-5mm]{0mm}{12mm} $W^+$ & $W^-$ & $\chi_3$ & $\chi_3$
&
$\displaystyle{ e^2\frac{1}{2s_W^2}g^{\mu\nu}}$ \\
\hline \rule[-5mm]{0mm}{12mm} $W^+$ & $W^-$ & $\chi^-$ & $\chi^+$
&
$\displaystyle{ e^2\frac{1}{2s_W^2}g^{\mu\nu}}$ \\
\hline \rule[-5mm]{0mm}{12mm} $Z$ & $Z$ & $H$ & $H$ &
$\displaystyle{ e^2\frac{1}{2s_W^2c_W^2}g^{\mu\nu}}$ \\
\hline \rule[-5mm]{0mm}{12mm} $Z$ & $Z$ & $\chi_3$ & $\chi_3$ &
$\displaystyle{ e^2\frac{1}{2s_W^2c_W^2}g^{\mu\nu}}$ \\
\hline
\end{tabular}
\end{minipage}

\subsection{Scalar-Scalar-Scalar-Scalar}

\begin{minipage}[c]{5cm}
\begin{center}
\includegraphics[width=5cm,height=5cm]{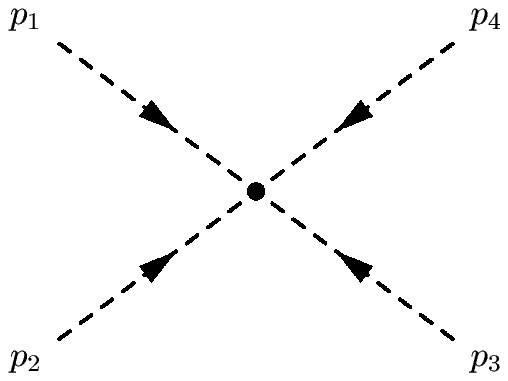}
\end{center}
\end{minipage}
\hspace*{3mm}
\begin{minipage}[c]{10cm}
\begin{tabular}{ccccl}
\hline $p_1 $ & $p_2 $ & $p_3 $ &
$p_4 $ \\
\hline \rule[-5mm]{0mm}{12mm} $H$ & $H$ & $H$ & $H$ &
$\displaystyle{-e^2\frac{3M_H^2}{4s_W^2 M_W^2} }$ \\
\hline \rule[-5mm]{0mm}{12mm} $\chi_3$ & $\chi_3$ & $\chi_3$ &
$\chi_3$ &
$\displaystyle{-e^2\frac{3M_H^2}{4s_W^2 M_W^2} }$ \\
\hline \rule[-5mm]{0mm}{12mm} $\chi^{\pm}$ & $\chi^{\mp}$ &
$\chi^{\mp}$ & $\chi^{\pm}$ &
$\displaystyle{-e^2\frac{M_H^2}{2s_W^2 M_W^2} }$ \\
\hline \rule[-5mm]{0mm}{12mm} $H$ & $H$ & $\chi_3$ & $\chi_3$ &
$\displaystyle{-e^2\frac{M_H^2+2\enlg^2M_Z^2\cdot\xiz}{4s_W^2M_W^2} }$ \\
\hline \rule[-5mm]{0mm}{12mm} $H$ & $H$ & $\chi^+$ & $\chi^-$ &
$\displaystyle{-e^2\frac{M_H^2+2\dnlg^2M_W^2\cdot\xiw}{4s_W^2M_W^2} }$ \\
\hline \rule[-5mm]{0mm}{12mm} $\chi^+$ & $\chi^-$ & $\chi_3$ &
$\chi_3$ &
$\displaystyle{-e^2\frac{M_H^2+2\knlg^2M_W^2\cdot\xiw}{4s_W^2M_W^2} }$ \\
\hline
\end{tabular}
\end{minipage}

\subsection{Fermion-Fermion-Vector}

Fermion mixing is not shown here. Colour for quarks is also not
explicit and should be taken into account when appropriate.

\vspace{3mm}

\begin{center}
\begin{tabular}{|c|c|cc|c|cc|}
\hline \rule[-2mm]{0mm}{5mm}
  $f$    &           & $I_3$ & $Q_f$ &
                      & $I_3$ & $Q_f$ \\
\hline
 $U$ & $u, c, t$ & \rule[-5mm]{0mm}{12mm}
      $\frac{1}{2}$ & $\frac{2}{3}$
    & $\nu_e, \nu_{\mu}, \nu_{\tau}$ & $\frac{1}{2}$ & $0$ \\
\hline
 $D$ & $d, s, b$ & \rule[-5mm]{0mm}{12mm}
      $-\frac{1}{2}$ & $-\frac{1}{3}$
    & $e, \mu, \tau$ & $-\frac{1}{2}$ & $-1$ \\
\hline
\end{tabular}
\end{center}

\vspace{3mm}

\begin{minipage}[c]{5cm}
\begin{center}
\includegraphics[width=5cm,height=5cm]{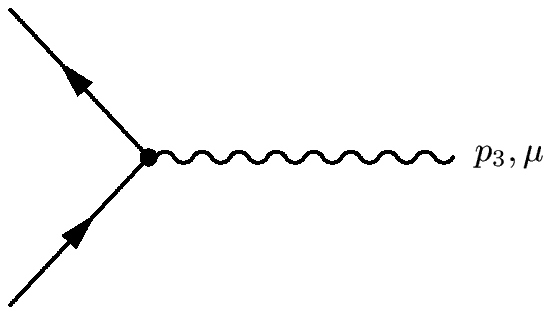}
\end{center}
\end{minipage}
\hspace*{3mm}
\begin{minipage}[c]{10cm}
\begin{tabular}{cccl}
\hline $p_1 $ & $p_2 $ &
$p_3 \ (\mu)$ & \\
\hline \rule[-5mm]{0mm}{12mm} $\bar{f}$ & $f$ & $A$ &
$\displaystyle{e Q_f \gamma^{\mu}}$ \\
\hline \rule[-1mm]{0mm}{5mm}
$\bar{f}$ & $f$ & $Z$ & \\
\multicolumn{4}{r}{ $\displaystyle{e \frac{1}{2s_Wc_W}
\gamma^{\mu}\left(I_3(1-\gamma_5)-2s_W^2 Q_f\right) }$ }\\
\hline \rule[-5mm]{0mm}{12mm} $\bar{U}/\bar{D}$ & $D/U$ &
$W^{+}/W^{-}$ &
$\displaystyle{e \frac{1}{2\sqrt{2}s_W} \gamma^{\mu}(1-\gamma_5)}$ \\
\hline
\end{tabular}
\end{minipage}

\subsection{Fermion-Fermion-Scalar}

\begin{minipage}[c]{5cm}
\begin{center}
\includegraphics[width=5cm,height=5cm]{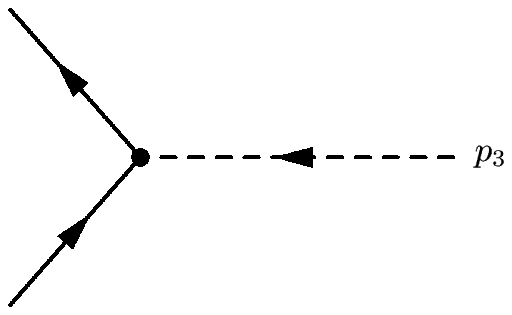}
\end{center}
\end{minipage}
\hspace*{3mm}
\begin{minipage}[c]{10cm}
\begin{tabular}{cccl}
\hline $p_1 $ & $p_2 $ &
$p_3 $ & \\
\hline \rule[-5mm]{0mm}{12mm} $\bar{f}$ & $f$ & $H$ &
$\displaystyle{-e\frac{1}{2s_W}\frac{m_f}{M_W} }$ \\
\hline \rule[-5mm]{0mm}{12mm} $\bar{U}/\bar{D}$ & $U/D$ & $\chi_3$
&
$\displaystyle{(-/+) i e\frac{1}{2s_W}\frac{m_f}{M_W}\ \gamma_5}$ \\
\hline \rule[-1mm]{0mm}{5mm}
$\bar{U}$ & $D$ & $\chi^+$ & \\
\multicolumn{4}{r}{ \rule[-5mm]{0mm}{12mm} $\displaystyle{-i
e\frac{1}{2\sqrt{2}s_W}\frac{1}{M_W}
\left[ (m_D-m_U)+(m_D+m_U)\gamma_5\right] }$} \\
\hline \rule[-1mm]{0mm}{5mm}
$\bar{D}$ & $U$ & $\chi^-$ & \\
\multicolumn{4}{r}{ \rule[-5mm]{0mm}{12mm} $\displaystyle{-i
e\frac{1}{2\sqrt{2}s_W}\frac{1}{M_W}
\left[ (m_U-m_D)+(m_U+m_D)\gamma_5\right] }$ } \\
\hline
\end{tabular}
\end{minipage}

\subsection{Ghost-Ghost-Vector}

\begin{minipage}[c]{5cm}
\begin{center}
\includegraphics[width=5cm,height=5cm]{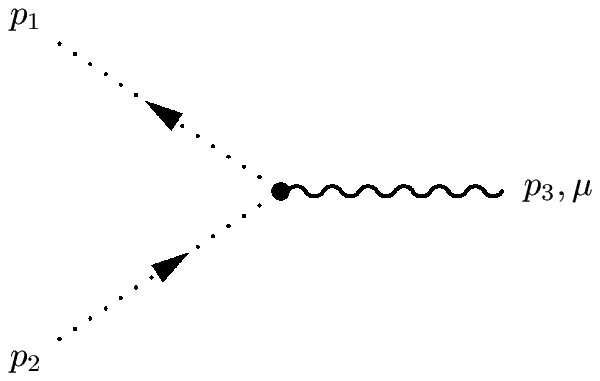}
\end{center}
\end{minipage}
\hspace*{3mm}
\begin{minipage}[c]{10cm}
\begin{tabular}{cccl}
\hline $p_1 $ & $p_2 $ &
$p_3 \ (\mu)$ & \\
\hline \rule[-5mm]{0mm}{12mm} $\bar{c}^{A}$ & $c^{\mp}$ &
$W^{\pm}$ &
$\displaystyle{\pm e p_1^{\mu}}$ \\
\hline \rule[-5mm]{0mm}{12mm} $\bar{c}^{Z}$ & $c^{\mp}$ &
$W^{\pm}$ &
$\displaystyle{\pm e \frac{c_W}{s_W}p_1^{\mu}}$ \\
\hline \rule[-5mm]{0mm}{12mm} $\bar{c}^{\mp}$ & $c^{A}$ &
$W^{\pm}$ &
$\displaystyle{\mp e( p_1^{\mu}-\anlg p_2^{\mu})}$ \\
\hline \rule[-5mm]{0mm}{12mm} $\bar{c}^{\mp}$ & $c^{Z}$ &
$W^{\pm}$ &
$\displaystyle{\mp e \frac{c_W}{s_W}( p_1^{\mu}-\bnlg p_2^{\mu})}$ \\
\hline \rule[-5mm]{0mm}{12mm} $\bar{c}^{\mp}$ & $c^{\pm}$ & $A$ &
$\displaystyle{\pm e( p_1^{\mu}+\anlg p_2^{\mu})}$ \\
\hline \rule[-5mm]{0mm}{12mm} $\bar{c}^{\mp}$ & $c^{\pm}$ & $Z$ &
$\displaystyle{\pm e \frac{c_W}{s_W}( p_1^{\mu}+\bnlg p_2^{\mu})}$ \\
\hline
\end{tabular}
\end{minipage}

\subsection{Ghost-Ghost-Scalar}

\begin{minipage}[c]{5cm}
\begin{center}
\includegraphics[width=5cm,height=5cm]{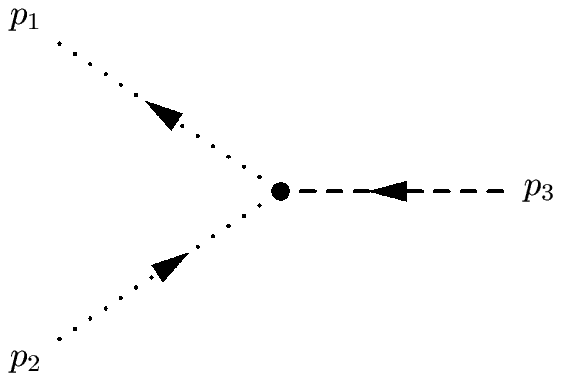}
\end{center}
\end{minipage}
\hspace*{3mm}
\begin{minipage}[c]{10cm}
\begin{tabular}{cccl}
\hline $p_1 $ & $p_2 $ &
$p_3 $ & \\
\hline \rule[-5mm]{0mm}{12mm} $\bar{c}^{Z}$ & $c^{Z}$ & $H$ &
$\displaystyle{- e \frac{1}{2s_Wc_W^2}(1+\enlg)M_W\cdot\xiz}$ \\
\hline \rule[-5mm]{0mm}{12mm} $\bar{c}^{Z}$ & $c^{\mp}$ &
$\chi^{\pm}$ &
$\displaystyle{\pm i e \frac{1}{2s_Wc_W}M_W\cdot\xiz}$ \\
\hline \rule[-5mm]{0mm}{12mm} $\bar{c}^{\mp}$ & $c^{A}$ &
$\chi^{\pm}$ &
$\displaystyle{\mp i e M_W\cdot\xiw}$ \\
\hline \rule[-5mm]{0mm}{12mm} $\bar{c}^{\mp}$ & $c^{Z}$ &
$\chi^{\pm}$ &
$\displaystyle{\mp i e \frac{1}{2s_Wc_W}(c_W^2-s_W^2+\knlg)M_W\cdot\xiw}$ \\
\hline \rule[-5mm]{0mm}{12mm} $\bar{c}^{\mp}$ & $c^{\pm}$ & $H$ &
$\displaystyle{- e \frac{1}{2s_W}(1+\dnlg)M_W\cdot\xiw}$ \\
\hline \rule[-5mm]{0mm}{12mm} $\bar{c}^{\mp}$ & $c^{\pm}$ &
$\chi_3$ &
$\displaystyle{\pm i e \frac{1}{2s_W}(1-\knlg)M_W\cdot\xiw}$ \\
\hline
\end{tabular}
\end{minipage}

\subsection{Ghost-Ghost-Vector-Vector}

\begin{minipage}[c]{5cm}
\begin{center}
\includegraphics[width=5cm,height=5cm]{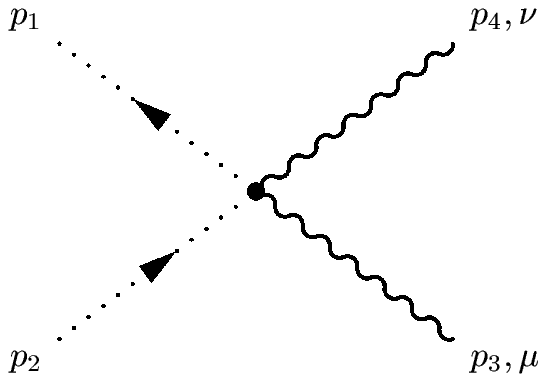}
\end{center}
\end{minipage}
\hspace*{3mm}
\begin{minipage}[c]{10cm}
\begin{tabular}{ccccl}
\hline $p_1 $ & $p_2 $ & $p_3 \ (\mu)$ &
$p_4 \ (\nu)$ & \\
\hline \rule[-5mm]{0mm}{12mm} $\bar{c}^{\mp}$ & $c^{A}$ & $A$ &
$W^{\pm}$ &
$\displaystyle{-e^2\anlg g^{\mu\nu}}$ \\
\hline \rule[-5mm]{0mm}{12mm} $\bar{c}^{\mp}$ & $c^{A}$ & $Z$ &
$W^{\pm}$ &
$\displaystyle{-e^2 \frac{c_W}{s_W} \bnlg g^{\mu\nu}}$ \\
\hline \rule[-5mm]{0mm}{12mm} $\bar{c}^{\mp}$ & $c^{Z}$ & $A$ &
$W^{\pm}$ &
$\displaystyle{-e^2\frac{c_W}{s_W} \anlg g^{\mu\nu}}$ \\
\hline \rule[-5mm]{0mm}{12mm} $\bar{c}^{\mp}$ & $c^{Z}$ & $Z$ &
$W^{\pm}$ &
$\displaystyle{-e^2 \frac{c_W^2}{s_W^2} \bnlg g^{\mu\nu}}$ \\
\hline \rule[-5mm]{0mm}{12mm} $\bar{c}^{\mp}$ & $c^{\pm}$ &
$W^{\mp}$ & $W^{\pm}$ & $\displaystyle{-e^2\left(\anlg
+\frac{c_W^2}{s_W^2}\bnlg\right)
 g^{\mu\nu}}$ \\
\hline \rule[-5mm]{0mm}{12mm} $\bar{c}^{\mp}$ & $c^{\mp}$ &
$W^{\pm}$ & $W^{\pm}$ & $\displaystyle{2e^2\left(\anlg
+\frac{c_W^2}{s_W^2}\bnlg\right)
 g^{\mu\nu}}$ \\
\hline \rule[-5mm]{0mm}{12mm} $\bar{c}^{\mp}$ & $c^{\pm}$ & $A$ &
$A$ &
$\displaystyle{2e^2\anlg g^{\mu\nu}}$ \\
\hline \rule[-5mm]{0mm}{12mm} $\bar{c}^{\mp}$ & $c^{\pm}$ & $Z$ &
$A$ &
$\displaystyle{e^2 \frac{c_W}{s_W}(\anlg +\bnlg) g^{\mu\nu}}$ \\
\hline \rule[-5mm]{0mm}{12mm} $\bar{c}^{\mp}$ & $c^{\pm}$ & $Z$ &
$Z$ &
$\displaystyle{2e^2 \frac{c_W^2}{s_W^2} \bnlg g^{\mu\nu}}$ \\
\hline
\end{tabular}
\end{minipage}

\subsection{Ghost-Ghost-Scalar-Scalar}

\begin{minipage}[c]{5cm}
\begin{center}
\includegraphics[width=5cm,height=5cm]{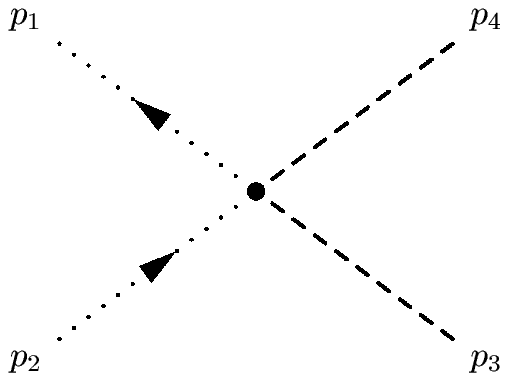}
\end{center}
\end{minipage}
\hspace*{3mm}
\begin{minipage}[c]{10cm}
\begin{tabular}{ccccl}
\hline $p_1 $ & $p_2 $ & $p_3 \ (\mu)$ &
$p_4 \ (\nu)$ & \\
\hline \rule[-5mm]{0mm}{12mm} $\bar{c}^{Z}$ & $c^{Z}$ & $H$ & $H$
&
$\displaystyle{-e^2 \frac{1}{2s_W^2c_W^2} \enlg \cdot\xiz}$ \\
\hline \rule[-5mm]{0mm}{12mm} $\bar{c}^{Z}$ & $c^{Z}$ & $\chi_3$ &
$\chi_3$ &
$\displaystyle{ e^2 \frac{1}{2s_W^2c_W^2} \enlg \cdot\xiz}$ \\
\hline \rule[-5mm]{0mm}{12mm} $\bar{c}^{Z}$ & $c^{\pm}$ &
$\chi^{\mp}$ & $H$ &
$\displaystyle{\mp i e^2 \frac{1}{4s_W^2c_W} \enlg \cdot\xiz}$ \\
\hline \rule[-5mm]{0mm}{12mm} $\bar{c}^{Z}$ & $c^{\pm}$ &
$\chi^{\mp}$ & $\chi_3$ &
$\displaystyle{ e^2 \frac{1}{4s_W^2c_W} \enlg \cdot\xiz}$ \\
\hline \rule[-5mm]{0mm}{12mm} $\bar{c}^{\mp}$ & $c^{A}$ &
$\chi^{\pm}$ & $H$ &
$\displaystyle{\mp i e^2 \frac{1}{2s_W} \dnlg \cdot\xiw}$ \\
\hline \rule[-5mm]{0mm}{12mm} $\bar{c}^{\mp}$ & $c^{A}$ &
$\chi^{\pm}$ & $\chi_3$ &
$\displaystyle{ e^2 \frac{1}{2s_W} \knlg \cdot\xiw}$ \\
\hline \rule[-5mm]{0mm}{12mm}
$\bar{c}^{\mp}$ & $c^{Z}$ & $\chi^{\pm}$ & $H$ & \\
\multicolumn{5}{r}{ \rule[-5mm]{0mm}{12mm} $\displaystyle{\mp i
e^2 \frac{1}{4s_W^2c_W}
\left(\knlg+ \dnlg(c_W^2-s_W^2)\right) \cdot\xiw}$ }\\
\hline \rule[-5mm]{0mm}{12mm}
$\bar{c}^{\mp}$ & $c^{Z}$ & $\chi^{\pm}$ & $\chi_3$ & \\
\multicolumn{5}{r}{ \rule[-5mm]{0mm}{12mm} $\displaystyle{ e^2
\frac{1}{4s_W^2c_W}
\left(\dnlg+ \knlg(c_W^2-s_W^2)\right)\cdot\xiw }$ }\\
\hline \rule[-5mm]{0mm}{12mm} $\bar{c}^{\mp}$ & $c^{\pm}$ & $H$ &
$H$ &
$\displaystyle{ -e^2 \frac{1}{2s_W^2} \dnlg \cdot\xiw }$ \\
\hline \rule[-5mm]{0mm}{12mm} $\bar{c}^{\mp}$ & $c^{\pm}$ &
$\chi_3$ & $\chi_3$ &
$\displaystyle{ -e^2 \frac{1}{2s_W^2} \knlg \cdot\xiw}$ \\
\hline \rule[-5mm]{0mm}{12mm} $\bar{c}^{\mp}$ & $c^{\pm}$ &
$\chi_3$ & $H$ &
$\displaystyle{ \mp i e^2 \frac{1}{4s_W^2} (\knlg-\dnlg)\cdot\xiw }$ \\
\hline \rule[-5mm]{0mm}{12mm} $\bar{c}^{\mp}$ & $c^{\pm}$ &
$\chi^-$ & $\chi^+$ &
$\displaystyle{ e^2 \frac{1}{4s_W^2} (\dnlg+\knlg)\cdot\xiw }$ \\
\hline \rule[-5mm]{0mm}{12mm} $\bar{c}^{\mp}$ & $c^{\mp}$ &
$\chi^{\pm}$ & $\chi^{\pm}$ &
$\displaystyle{ -e^2 \frac{1}{2s_W^2} (\knlg-\dnlg) \cdot\xiw}$ \\
\hline
\end{tabular}
\end{minipage}

\setcounter{equation}{0}
\section{Counterterms in the ghost sector}
\label{app:rcountergh}

Since we deal specifically with processes at one-loop, there is no
need to dwell on the renormalisation of the ghost sector. We only
briefly sketch the procedure without giving explicit formulae for
the various counterterms and the renormalisation constants.

There is some freedom for the introduction of renormalisation
constants for the ghost fields.  We use the following convention.

\beqn
\BARE{c}^{\pm} &=&\ZFT{3} {c}^{\pm}  \nonumber \\
 \left( \begin{array}{c} \BARE{c}^{Z} \\
\BARE{c}^{A} \end{array} \right) &=&
\left( \begin{array}{cc} \ZFT{ZZ} & \ZFT{ZA} \\
                        \ZFT{AZ} & \ZFT{AA} \end{array} \right)
\left( \begin{array}{c} {c}^{Z} \\ {c}_{A} \end{array} \right)  \nonumber \\
\BARE{\bar{c}}^{\pm}& =& {\bar{c}}^{\pm}  \\
\BARE{\bar{c}}^{Z} &= &{\bar{c}}^{Z} \nonumber \\
\BARE{\bar{c}}^{A} &= &{\bar{c}}^{A}\nonumber
\eeqn

To derive the full counterterm Lagrangian for the ghost we appeal
to the auxiliary $B$-field formulation of the gauge-fixing
Lagrangian ${\cal L}_{GF}$, see Eq.~\ref{lgfB} in
section~\ref{quantisation}. As stressed previously ${\cal L}_{GF}$
is written in terms of renormalised fields and as such does not
induce any counterterm. However the BRST transformation are
defined for bare fields. Therefore in order to generate the ghost
Lagrangian including counterterms one needs to re-express ${\cal
L}_{GF}$ in terms of bare fields to first generate ${\cal L}_{Gh}$
with bare fields. From there one can then derive the counterterm
ghost Lagrangian. One exploits the freedom in the renormalisation
of the $B$-fields so that the combination of the $B$ fields and
gauge fields shows no explicit dependence on wave function
renormalisation. We therefore define

\begin{equation}
\BARE{B}^{\pm}={\ZF{W}}^{-1} B^{\pm},\quad \left( \begin{array}{c}
\BARE{B}^Z \\ \BARE{B}^A \end{array} \right) =
\left( \begin{array}{cc} \ZF{ZZ} & \ZF{ZA} \\
                         \ZF{AZ} & \ZF{AA} \end{array} \right)^{-1}
\left( \begin{array}{c} {B}^Z \\ {B}^A \end{array} \right)
\end{equation}
The relation is just the inverse of that for gauge fields.

Then Eq.~\ref{lgfB} is
\begin{equation}
\begin{array}{ll}
{\cal L}_{GF}= &
\BARE{B}^+\partial^{\mu}\BARE{W}_{\mu}^- + B^+ \xi_W M_W \chi^- + (h.c.) \\
 &+ \BARE{B}^Z \partial^{\mu} \BARE{Z}_{\mu} + B^Z \xi_Z M_Z \chi_3
  + \;\BARE{B}^A \partial \BARE{A}_{\mu} \\
& + \; {\rm non}\mbox{-}{\rm linear\ gauge\ terms} (\alphat,\betat,\deltat,\epsilont,\kappat)\\
& + \; \BARE{B}\mbox{-}{\rm linear\ terms}.
\end{array}
\label{eq:gfixtm2}
\end{equation}

One then makes the identifications
\begin{equation}
B^+ \xi_W M_W \chi^- = \BARE{B}^+ \BARE{\xi}_W \BARE{M}_W
\BARE{\chi}^-,
\end{equation}
\begin{equation}
B^Z \xi_Z M_Z \chi_3 = \BARE{B}^Z \BARE{\xi}_Z \BARE{M}_Z
\BARE{\chi}_3
 + \BARE{B}^A \BARE{\xi}_{ZA} \BARE{M}_Z \BARE{\chi}_3
\end{equation}
where we defined the renormalisation of the gauge parameters as
\begin{equation}
\label{barexighosts}
\begin{array}{l}
\BARE{\xi}_W=\xi_W(M_W/\BARE{M}_W) \ZF{W}\ZF{\chi}^{-1} \\
\BARE{\xi}_Z = \xi_Z (M_Z/\BARE{M}_Z) \ZF{ZZ} \ZF{\chi_3}^{-1} \\
\BARE{\xi}_{ZA} = \xi_Z (M_Z/\BARE{M}_Z) \ZF{ZA} \ZF{\chi_3}^{-1}
\end{array}
\end{equation}
$\BARE{\xi}_{ZA}$ is not an independent parameter but the
short-hand notation given by Eq.~\ref{barexighosts}. Non-linear
gauge terms can be transformed in a similar way by the
renormalisation of ($\alphat,\betat,\deltat,\epsilont,\kappat$).
Note that the terms bilinear in the $B$ fields would get extra
factors. However this does not affect the renormalisation program
since $\db B=0$. This helps define the bare $G$ functions.

We obtain bare $G$ terms by the above equations.
\begin{equation}
\begin{array}{l}
\BARE{G}^{\mp} = \partial^{\mu}\BARE{W}_{\mu}^{\mp}
  +\BARE{\xi}_W \BARE{M}_W \BARE{\chi}^{\mp}
  + \mathrm{non-linear\ gauge\ terms} \\
\BARE{G}^Z = \partial^{\mu}\BARE{Z}_{\mu}
  +\BARE{\xi}_Z \BARE{M}_Z \BARE{\chi}_3
  + \mathrm{non-linear\ gauge\ terms} \\
\BARE{G}^A = \partial^{\mu}\BARE{A}_{\mu}
  +\BARE{\xi}_{ZA} \BARE{M}_Z \BARE{\chi}_3
  + \mathrm{non-linear\ gauge\ terms}
\end{array}
\end{equation}
Except for $G^A$, these are the same as those in
\ref{fullnonlineargauge} assuming that quantities are bare ones.
With these, one defines the bare ghost Lagrangian, that contains
extra terms than those obtained at tree-level due to the induced
mixing $\xi_{ZA}$. One then readily obtains the renormalised ghost
Lagrangian.

\section{Auxiliary Fields and Generalised Ward-Takahashi
Identities in the unphysical scalar sector}
\label{ward-id-goldstones}

Although the renormalisation of this sector is not essential if
one wants to arrive at finite S-matrix elements, the various
two-point functions of the Goldstones and the longitudinal gauge
bosons as well as their mixing are related.

To easily derive these generalised Ward-Takahashi identities that
constrain the different propagators  it is very useful to
introduce ${\cal L}_{GF}$ via the auxiliary fields as done in
Eq.~\ref{lgfB}. The Ward-Takahashi identities are particularly
easy to derive if one works with the $B$-fields and considers the
BRST transformations on some specific Green's functions (vacuum
expectation values of time ordered products). For the two-point
function of any two fields $\phi_A$ and $\phi_B$, we use the
short-hand notation:

\beqn
\langle  \phi_A\; \phi_B \rangle=\langle  0|\left( T \phi_A(x)
\phi_B(y) \right)|0\rangle
\eeqn

For example, take the generic Green's function $\langle  \bar
c^i\; B^j \rangle$ which in fact is zero (it has a non vanishing
ghost number). Subjecting it to a BRST transformation one gets:

\beqn \db \langle  \bar{c}^i\; B^j \rangle&=& \langle
(\db\bar{c}^i)\; B^j\rangle - \langle  \bar{c}^i\; (\db
B^j)\rangle =i \langle  B^i\; B^j\rangle =0
 \nonumber \\
&{\rm or}& \langle  G^i \; G^j\rangle =0
\eeqn
where in the last part we have used the equation of motion for the
$B^i$'s. The above relation leads directly to a constraint on the
two-point functions of the gauge vector boson, the gauge-Goldstone
mixing and the Goldstone two-point functions. One novelty compared
to the usual linear gauge is that these identities also involve
correlation functions with composite operators. Indeed if we
specialise to the $ZZ$ functions, one has with $\xi_Z=1$

\begin{eqnarray}
\label{wtidzchi3} & & \langle G_Z(x) G_Z(y)\rangle = \nonumber \\
&\langle & \left((\partial. Z(x)+M_Z \chi_3(x)+\frac{g}{2
c_W}\tilde{\epsilon} H(x) \chi_3(x)\right) \left((\partial.
Z(y)+M_Z \chi_3(y)+\frac{g}{2
c_W}\tilde{\epsilon} H(y) \chi_3(y)\right)\rangle \nonumber \\
&=&\partial^\mu_x
\partial^\nu_y \langle  Z_\mu(x) Z_\nu(y)\rangle +2 M_Z
\partial^\mu_x \langle Z_\mu(x) \chi_3(y)\rangle + M_Z^2 \langle \chi_3(x) \chi_3(y)\rangle
\nonumber \\
&+&\tilde{\epsilon} \frac{g}{c_W} \left[ \partial^\mu_x \langle
Z_\mu(x) (H(y) \chi_3(y))\rangle+M_Z\langle \chi_3(x)(H(y)
\chi_3(y))\rangle \right] \nonumber
\\
&+&(\frac{g \tilde{\epsilon}}{2c_W})^2 \langle H(x) \chi_3(x) H(y)
\chi_3(y)\rangle=0 \; .
\end{eqnarray}

It is important to realise that these are the full Green's
function and therefore the external legs are not {\it amputated}.
Therefore it is crucial to note that the last two terms (in the
last two lines) do not have the double pole structure.

This translates into the following identity, in momentum space,
\begin{eqnarray}
\label{ward-id-nogf}
 \biggl( q^2 (\Pi_L^{ZZ}-2
M_Z \Pi^{Z\chi_3} ) +M_Z^2 \Pi^{\chi_3 \chi_3} \biggr) \equiv
A_{ZZ}\; .
\end{eqnarray}

Note that in the approach where the gauge-fixing Lagrangian is
expressed in terms of renormalised quantities,
Eq.~\ref{ward-id-nogf} also holds for the renormalised two-point
function.
 In the linear gauge $A_{ZZ}=0$ for any $q^2$ (tadpole
contributions must be included here). An explicit calculation
gives
\begin{eqnarray}
A_{ZZ}&=&
\frac{\alpha \tilde{\epsilon}}{16\pi s_W^2 c_W^2} (q^2-M_Z^2)
\Biggl\{ \tilde{\epsilon} (q^2-3 M_Z^2) (C_{UV}-F_0(ZH)) \nonumber \\
& &  +2\left[q^2 (F_0(ZH)-2F_1(ZH))-M_H^2(C_{UV}-F_0(ZH)) \right]
\biggl\}\; .
\end{eqnarray}

The functions $F_{0,1}$ are defined in Eq.~\ref{Fnfctdef}. We see
clearly that $A_{ZZ}$ does not vanish in the non-linear gauge, but
at the $Z$ pole. In fact the contribution $A_{ZZ}$ can be derived
directly from the last two terms of Eq.~\ref{wtidzchi3}. In a
diagrammatic form, at one-loop, Eq.~\ref{wtidzchi3} can be
described as

\setbox1=\hbox to
3cm{{\resizebox*{3cm}{!}{\includegraphics{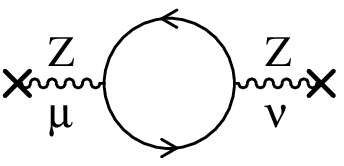}}}}
\setbox2=\hbox to
3cm{{\resizebox*{3cm}{!}{\includegraphics{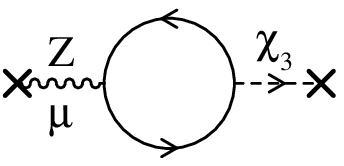}}}}
\setbox3=\hbox to
3cm{{\resizebox*{3cm}{!}{\includegraphics{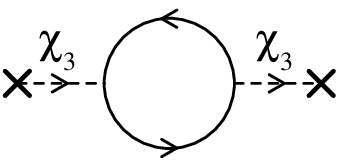}}}}
\setbox4=\hbox to
2.2cm{{\resizebox*{2.2cm}{!}{\includegraphics{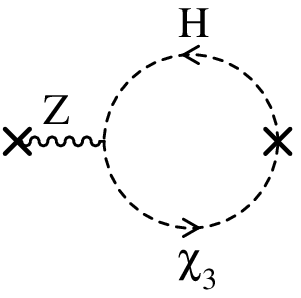}}}}
\setbox5=\hbox to
2.2cm{{\resizebox*{2.2cm}{!}{\includegraphics{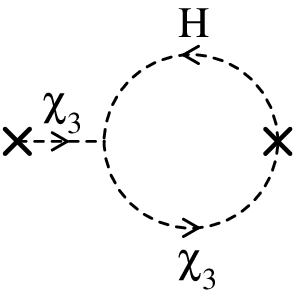}}}}
\setbox6=\hbox to
1.5cm{{\resizebox*{1.5cm}{!}{\includegraphics{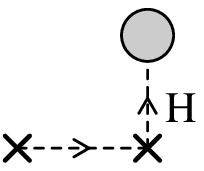}}}}
\setbox7=\hbox to
1.5cm{{\resizebox*{1.5cm}{!}{\includegraphics{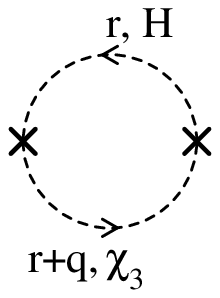}}}}

\begin{eqnarray}
  &&\left({1\over{q^2-M_Z^2}}\right)^2\times \nonumber\\
&&  \left[q_\mu q_\nu\;\raise -7mm\box1\;
  +2iM_Z q_\mu\;\raise -7mm\box2
+M^2_Z\;\raise -7mm\box3\;\right]_{\rm amp.} \nonumber\\
&&=
  \left({{g\tilde{\epsilon}}\over{2c_W}}\right)^2
  \left[\;\raise -9mm\box7\;\right]
+\left({{g\tilde{\epsilon}}\over{c_W}}\right)
  \left[iq_\mu\;\raise -10mm\box4\;+M_Z\;\raise -10mm\box5\;
\right]_{\rm amp.} \times{1\over{q^2-M_Z^2}}
\nonumber \\
\end{eqnarray}

Calculating the new graphs explicitly confirms the identity and is
a check on our calculation of $\Pi_L^{ZZ}, \Pi^{Z\chi_3}$ and
$\Pi^{\chi_3\chi_3}$. Note that for the genuine two-point
functions (contributing to $A_{ZZ}$) the loops include all
possible particles including matter fields. For the latter one
gets the same contribution as in the linear gauge.

For the charged sector the identities go along the same line.

\noi For the photon, the identities give
\beqn
\label{wardpiaa} \Pi_L^{AA}=0 \quad {\rm at \;any} \; q^2.
\eeqn

For the $AZ$ transition we get (from $\langle  G_A G_Z\rangle=0$)
that
\beqn
\label{wardpiaz} \Pi_L^{AZ}-M_Z \Pi^{A\chi_3} =0\; .
\eeqn
This identity holds at any $q^2$, but only at one-loop thanks to
the fact that the vertex $A H \chi_3$ does not exist at
tree-level. This can be easily checked by looking up
the explicit formulae in sections~\ref{explicitaa} and~\ref{explicitaz} and section~\ref{explicitachi3}. \\

\setcounter{equation}{0}
\section{Renormalising the gauge-fixing functions}
\label{app-ren-gf}

\newcommand{\xw}{\xi_W}
\newcommand{\xiwp}{\xi_W^\prime}
\newcommand{\xiwt}{\tilde{\xi}_W}
\newcommand{\xz}{\xi_Z}
\newcommand{\xizp}{\xi_Z^\prime}
\newcommand{\xizt}{\tilde{\xi}_Z}
\def\db{\delta_{\rm BRS}}

 So far we have chosen to take the gauge-fixing term as being
written in terms of renormalised quantities. Although this is
quite practical and avoids the introduction of more counterterms
for the (unphysical) parameters entering the gauge-fixing
Lagrangian, it does not lead to finite Green's functions, in the
general case of the non-linear gauge, even when all Feynman
parameters are set to one, $\xi_{A,Z,W}=1$. However all S-matrix
elements are finite and gauge-parameter independent. Therefore as
we argued, this approach of taking the gauge-fixing Lagrangian as
renormalised from the outset is easy to implement and at the same
time acts as a good test on our system since many divergences in a
few Green's functions cancel at the level of  the S-matrix. If one
wants to have finite Green's functions one also needs to consider
counterterms to the gauge-fixing Lagrangian,
Eq.~\ref{fullnonlineargauge}. The purpose of this Appendix is to
show how all two-point functions can be made finite if one also
introduces counterterms for the gauge parameters, beside the
renormalisation of the physical parameters (masses, electric
charge) and the tadpole as well as the wave-function
renormalisation for all fields as defined in the main text, see
section  \ref{sec:rconst}. For the two-point functions,  the
difference between this approach (taking a bare gauge fixing
Lagrangian) and that of the paper (taking the gauge fixing as
renormalised), only concerns the unphysical scalars (and
longitudinal part of the vector bosons). Instead of
Eq.~\ref{fullnonlineargauge} we write for the charged sector the
gauge fixing terms in bare quantities,

\beqn
\label{fullnonlineargaugexw} {{\cal
L}}_{GF,W}&=&-\frac{1}{\BARE{\xi_W}} |(\partial_\mu\;-\;i \BARE{e}
\; \BARE{\tilde{\alpha}} \;\BARE{A}_\mu\;-\;i\BARE{g} \;\BARE{c_W}
\; \BARE{\tilde{\beta}} \; \BARE{Z}_\mu) \BARE{W}^{\mu+} +
\BARE{\xi_W^\prime} \frac{\BARE{g}}{2}\left( (\BARE{M_W} +
\frac{\BARE{g}}{2}(\BARE{\tilde{\delta}}\; \BARE{H} +i
\BARE{\tilde{\kappa}} \BARE{\chi}_3)\right)\BARE{\chi}^{+}|^{2} \;
.\nonumber
\\
\eeqn

\newcommand{\dxaz}{\delta\eta_{AZ}}
\newcommand{\dxazt}{\delta\tilde{\eta}_{AZ}}
\noi For the neutral sector one has to  allow for $A\mbox{-}Z$ and
$A\mbox{-}\chi_3$ mixing. We take \footnote{${{\cal
L}}_{GF,(Z,A)}$ be derived through the auxiliary B-field
formulation and the gauge functions $G^A$ and $G^Z$. One writes
${{\cal
L}}_{GF,(Z,A)}=(\BARE{\xi_Z}/2)(\BARE{B}^Z)^2+(\BARE{\xi_{A}}/2)(\BARE{B}^A)^2+\dxazt
\BARE{B}^A \BARE{B}^Z+ \BARE{B}^Z \BARE{G}^Z + \BARE{B}^A
\BARE{G}^A$ with $\BARE{G}^Z=\partial.\BARE{Z} +
\BARE{\xi_Z^\prime}
 (\BARE{M_Z}+(\BARE{g}/2 \BARE{c_W})
 \BARE{\tilde\varepsilon}\;
 \BARE{H})
\BARE{\chi}_3$ and $\BARE{G}^A=\partial.\BARE{A} +\delta
\xi_{A\chi}^\prime \BARE{M_Z} \BARE{\chi_3}$. For the purpose of
generating the counterterms for the two-point functions at the
one-loop order,  one makes the identification,
$\dxaz=\dxazt/(\xi_A \xi_Z)$ and $\delta\xi_A^\prime=\delta
\xi_{A\chi}-(\xi_Z^\prime/\xi_Z)\dxazt$.}

\beqn \label{fullnonlineargaugeneut} {{\cal
L}}_{GF,(Z,A)}&=&-\frac{1}{2 \BARE{\xi_Z}} (\partial.\BARE{Z} +
\BARE{\xi_Z^\prime}
 (\BARE{M_Z}+\frac{\BARE{g}}{ 2 \BARE{c_W}} \BARE{\tilde\varepsilon}\;  \BARE{H})
\BARE{\chi}_3)^2 \;-\frac{1}{2 \BARE{\xi_A}} (\partial.\BARE{A}
+\delta \xi_A^\prime \BARE{M_Z} \BARE{\chi_3} )^2 \nonumber \\
& & \quad + \; \dxaz \; \partial.\BARE{A}\;
\partial.\BARE{Z}\;.
\eeqn

\noi It will also be useful to introduce $\BARE{\xiwt}$ and
$\BARE{\xizt}$ such that $\BARE{\xiwp}=\sqrt{\BARE{\xw}\;
\BARE{\xiwt}}, \BARE{\xizp}=\sqrt{\BARE{\xz} \;\BARE{\xizt}}$. At
tree-level our implementation requires that
$\BARE{\xi}_{A,Z,W}=\BARE{\xi}_{Z,W}^\prime=1, \delta
\xi_A^\prime=\dxaz=0$. $\delta \xi_A^\prime$ and $\dxaz$ should be
considered of order ${{\cal O}}(\alpha)$ and are introduced to
avoid that ${{\cal L}}_{GF,(Z,A)}$ does not induce any non
diagonal photon transition. Apart from the $\delta \xi_A^\prime$
and $\dxaz$ terms, these bare gauge fixing conditions are, of
course, formally the same as those we introduced in
Eq.~\ref{fullnonlineargauge}. In fact since the gauge fixing for
the photon is still linear and that we are working with $\xi_A=1$,
we could still take the gauge-fixing for the photon to be
renormalised. As we will see later, the two approaches lead to the
same form of the two-point functions for $\hat{\Pi}_L^{AA,AZ}$ and
$\hat{\Pi}^{A\chi_3}$. Since we only seek to show how finite
two-point functions can be arrived at, it is sufficient to only
consider the addition of the counterterms to the Feynman gauge
parameters $\xi$ and $\xi^\prime$.  The bare and renormalised
Feynman gauge parameters are related as

\beqn
\BARE{\xi_i}= \xi_i + \delta \xi_i\;;\; \quad
\BARE{\xi_i^{\prime}}= \xi_i^{\prime}  + \delta
\xi_i^{\prime}\;;\; \quad i=W,Z,A .
\eeqn

Since we will only work in the 't~Hooft-Feynman gauge, it is
sufficient to  only consider the case with all
$\xi_i=\xi_i^{\prime}=1$ apart from $\xi_A^{\prime}=\dxaz=0$.

The generated counterterms for the two-point functions, allowing
for the renormalisation of the gauge fixing Lagrangian, and for
the approach we take in \grcl of considering the gauge fixing
Lagrangian in Eq.~\ref{fullnonlineargauge} renormalised, are shown
below.

\begin{enumerate}
\item Vector-Vector

\begin{center}
\begin{tabular}{l||l|l}
&
${{\cal L}}_{GF}$ with renormalised quantities& ${{\cal L}}_{GF}$ with bare quantities \\
\hline
$WW$ & $\hat{\Pi}_T^W = \DM{W} + 2(M_W^2-q^2)\ZH{W} $ & unchanged \\
     & $\hat{\Pi}_L^W = \DM{W} + 2M_W^2\ZH{W} $ &$\hat{\Pi}_L^W = 2(M_W^2-q^2)\ZH{W}+ \DM{W} + q^2 \delta \xw $ \\
\hline
$ZZ$ & $\hat{\Pi}_T^{ZZ} = \DM{Z} + 2(M_Z^2-q^2)\ZH{ZZ} $ &unchanged\\
     & $\hat{\Pi}_L^{ZZ} = \DM{Z} + 2M_Z^2\ZH{ZZ} $ & $\hat{\Pi}_L^{ZZ}  = 2(M_Z^2-q^2)\ZH{ZZ}+ \DM{Z} + q^2 \delta \xi_Z $\\
\hline
$ZA$ & $\hat{\Pi}_T^{ZA} = (M_Z^2-q^2)\ZH{ZA}-q^2\ZH{AZ} $& unchanged\\
     & $\hat{\Pi}_L^{ZA} = M_Z^2\ZH{ZA} $ & $M_Z^2 \ZH{ZA}-q^2 (\ZH{ZA}+\ZH{AZ}-\dxaz)$\\
\hline
$AA$ & $\hat{\Pi}_T^{AA} = -2q^2\ZH{AA} $ &unchanged\\
     & $\hat{\Pi}_L^{AA} = 0 $ &$\hat{\Pi}_L^{AA}= -2q^2\ZH{AA} + q^2\delta \xi_A$ \\
\hline
\end{tabular}
\end{center}

\item Scalar-Scalar

\begin{center}
\begin{tabular}{l||l|l}
&
${{\cal L}}_{GF}$ with renormalised quantities& ${{\cal L}}_{GF}$ with bare quantities \\
\hline \rule[-2mm]{0mm}{8mm} $HH$ & $\hat{\Pi}^{H}
 = 2(q^2-M_H^2)\ZH{H} - \DM{H} + \frac{3\delta T}{v}$ &unchanged\\
\hline \rule[-2mm]{0mm}{8mm} $\chi_3\chi_3$ & $\hat{\Pi}^{\chi_3}
 = 2q^2\ZH{\chi_3} + \frac{\delta T}{v}$ &$\hat{\Pi}^{\chi_3}
 = 2(q^2-M_Z^2)\ZH{\chi_3} -M_Z^2 \delta \tilde{\xi_Z} $\\
 \rule[-2mm]{0mm}{8mm}  &  & $\;\;\;\;\;\;\;\;\;\; - \DM{Z}+\frac{\delta T}{v}$\\
\hline \rule[-2mm]{0mm}{8mm} $\chi\chi$ & $\hat{\Pi}^{\chi}
 = 2q^2\ZH{\chi} + \frac{\delta T}{v}$ & $\hat{\Pi}^{\chi}
 = 2(q^2-M_W^2)\ZH{\chi} -M_W^2 \delta \xiwt $\\
\rule[-2mm]{0mm}{8mm}  &  & $\;\;\;\;\;\;\;\;\;\; - \DM{W}+\frac{\delta T}{v}$\\
\hline
\end{tabular}
\end{center}

\item Vector-Scalar

\begin{center}
\begin{tabular}{l||l|l}
&
${{\cal L}}_{GF}$ with renormalised quantities& ${{\cal L}}_{GF}$ with bare quantities \\
\hline $W\chi$ & $\hat{\Pi}^{W\chi}
 = M_W (\delta G_W + \ZH{W}+\ZH{\chi}) $ &$\hat{\Pi}^{W\chi}=\frac{M_W}{2}(\delta \xw-\delta \xiwt)$\\
\hline $Z\chi_3$ & $\hat{\Pi}^{Z\chi_3}
 = M_Z (\delta G_Z + \ZH{ZZ}+\ZH{\chi_3}) $ &$\hat{\Pi}^{Z\chi_3}=\frac{M_Z}{2}(\delta \xi_Z-\delta \tilde{\xi}_Z)$\\
\hline $A\chi_3$ & $\hat{\Pi}^{A\chi_3}
 = M_Z \ZH{ZA} $ &$\hat{\Pi}^{A\chi_3}
 = - M_Z \delta \xi^\prime_A$\\
\hline
\end{tabular}
\end{center}

\item Fermion-Fermion

This remains the same in both approaches and is given by
Eq.~\ref{zctfermions}.

\end{enumerate}

Note that in both approaches one has
$\hat{\Pi}_L^V(0)=\hat{\Pi}_T^V(0)$ for all vectors as should be.
Note also that for the 2-point functions involving photons,
$\hat{\Pi}_L^{AA,ZA}$  and $\hat{\Pi}^{A\chi_3}$, $\delta \xi_A$,
$\delta \xi_A^\prime$ and $\dxaz$ can be chosen so that the
counterterms in both approaches are the same. In particular the
Ward identities $\Pi_L^{AA}=0$ and $\Pi_L^{AZ}-M_Z \Pi^{A\chi_3}
=0$, see Eqs.~\ref{wardpiaa}-\ref{wardpiaz}, are maintained after
renormalisation. Therefore we can take

\beqn
\delta\xi_A&=&2 \ZH{AA}, \quad \delta\xi_A^\prime=- \ZH{ZA}\; ,
\nonumber \\
\dxaz&=&\ZH{ZA}+\ZH{AZ}\;.
\eeqn

\noi Let us turn to the charged sector (the $Z$ transitions go
along the same line). Defining

\beqn
A_{WW}^{CT}=q^2 (\hat{\Pi}_L^W-2M_W \hat{\Pi}^{W\chi})+M_W^2
\hat{\Pi}^{\chi \chi}\; ,
\eeqn

\noi we find that \beqn A_{WW}^{CT}&=& 0 \;\;\; {\rm in \; our \;
approach \;( as \; expected)}, \nonumber
\\
&=& (q^2-M_W^2) \left( q^2 \delta \xw +\mww \delta \xiwt
+\DM{W}+2\mww \ZH{\chi}-2 q^2 \ZH{W} \right)\; \nonumber \\ & &
{\rm with \;} {{\cal L}}_{GF} {\rm \;in\; terms\; of\; bare\;
fields.} \eeqn

This again means that there are constraints on $\xw$ and $\xiwt$,
{\it i.e} they are not independent once the other wave functions
have been set. Exactly the same applies for the $ZZ$ transition.
Also this  means that on-shell renormalisation for the unphysical
sector is possible. That is, that the scalars have poles at the
same location as the physical vector bosons. For example for the
$W$, this condition
($\tilde{\Pi}_L(\mww)=\tilde{\Pi}^\chi(\mww)=0)$ gives that

\beqn
\label{dxiw} \mww \delta \xw +\DM{W}&=&-\Pi_L^W(\mww) \;
,\nonumber
\\
\mww \delta \xiwt +\DM{W}&=&+\Pi^{\chi}(\mww) + \frac{\delta
T}{v}\; .
\eeqn

This shows that the renormalised $W\mbox{-}\chi$ transition
becomes finite in the non-linear gauge:

\beqn \tilde{\Pi}^{W\chi}=\Pi^{W\chi}+\frac{M_W}{2}(\delta
\xw-\delta \xiwt)=\Pi^{W\chi}-\frac{1}{2 M_W}
\left(\Pi_L^W(\mww)+\Pi^{ \chi}(\mww) + \frac{\delta
T}{v}\right)={\rm finite} \; .\nonumber \\
\eeqn

\noi This can be explicitly shown by using the full expressions
for the two-point functions at one-loop given in
Eqs.~\ref{pilww},\ref{pilwchi} and \ref{pichichi}.

\noi Using Eq.~\ref{dxiw}, the renormalised $\chi \chi$ writes

\beqn
\label{piccr} \tilde{\Pi}^{\chi} = 2 (q^2-\mww)\left( \ZH{\chi}+
\frac{\alpha}{16 \pi s_W^2} (\kappat +\deltat -(2+1/c_W^2))\cuv
\;+\;{\rm finite} \right) \; .
\eeqn

In our approach we define $\ZH{\chi}$ so that all $C_{UV}$ terms
proportional to $q^2$ vanish:

\beqn
\label{explicitzchi} \ZH{\chi}\equiv-
\frac{\Pi^{\chi}_{{\cuv}-{\rm part}} }{2q^2}=\frac{\alpha}{16\pi
\sww}\left((2+1/c_W^2)-\kappat -\deltat \right)
 \cuv \; .
\eeqn

 This is {\em exactly} the
same result had we required $\tilde{\Pi}^{\chi}$ in
Eq.~\ref{piccr} to be finite. This result would have been arrived
at directly had we required that the residue at the pole of the
$\chi \chi$ propagator be 1. The $C_{UV}$  part of $\ZH{\chi}$
would be the same, differences would appear in finite terms that
have no incidence on $S$-matrix.

\noi Having constrained $\delta \xiwt$ and $\ZH{\chi}$, $\delta
\xw$ is fixed. Taking for example only the $C_{UV}$ part of
$\Pi_L^W$, from Eq.~\ref{dxiw}, one has that
\beqn
\delta \xw =-\frac{\delta \mww}{\mww}-\frac{\Pi_L^W(\mww)}{\mww}=2
\ZH{W}-\frac{\alpha}{4\pi s_W^2} \left(5 \alphat^2 s_W^2+5
\betat^2 c_W^2 +\frac{\deltat^2}{4}
+\frac{\kappat^2}{4}\right)\cuv \; .\nonumber \\
\eeqn

We indeed find, by explicit calculations, this to be verified.
Similar results hold for the other combinations of two-point
functions.

\setcounter{equation}{0}
\section{A library of counterterms for the vertices}
\label{sec:vtxcnt}

 Here, we list the full counterterms to the vertices after applying
 the field
redefinitions. Those for the ghost vertices are not shown since
they are not required at one-loop. Those for the two-point
functions (propagators) and the tadpole have been discussed
separately for a proper definition of the renormalisation
conditions. \\
\noi $\tree{\cdots}$ will refer to the  tree-level expression of
the vertex defined in \setu{frule} but with
$\anlg=\bnlg=\dnlg=\enlg=\knlg=0$. As a result of $Z-\gamma$
mixing new vertices, denoted as {\itshape{(new)}}, appear.

To help write our results in a compact form, we introduce, as is
done, in \cite{kyotorc}  the following ``correction" factors
\beqn
\delta G_{mj}&=& \frac{\delta m_j}{m_j} \nonumber \\
\delta G_H &=& \frac{\delta M_H^2}{M_H^2} \nonumber \\
\delta G_W &=& \frac{\delta M_W^2}{2M_W^2}\nonumber \\
\delta G_Z &=& \frac{\delta M_Z^2}{2M_Z^2} \nonumber \\
\delta H &=& \frac{\delta M_Z^2- \delta M_W^2}{2(M_Z^2-M_W^2)} \nonumber \\
\delta G_1 &=& \delta G_W - \delta H \nonumber \\
\delta G_2 &=& \delta G_Z - \delta H \nonumber \\
\delta G_3 &=& \delta G_Z - \delta G_W \nonumber \\
\delta G_4 &=& \frac{2 \delta M_W^2- \delta M_Z^2}{2M_W^2-M_Z^2}
           - \delta G_W - \delta H
\eeqn

\subsection{Vector-Vector-Vector}

\begin{tabular}{cccl}
\hline $p_1 \ (\mu)$ & $p_2 \ (\nu)$ &
$p_3 \ (\rho)$ & \\
\hline $W^-$ & $W^+$ & $A$ & $(\delta Y + 2\ZH{W} +
\ZH{AA})\tree{WWA}
+ \ZH{ZA}\tree{WWZ} $ \\
\hline $W^-$ & $W^+$ & $Z$ & $(\delta Y + \delta G_1+ 2\ZH{W} +
\ZH{ZZ})\tree{WWZ}
+ \ZH{AZ}\tree{WWA} $ \\
\hline
\end{tabular}

\subsection{Vector-Vector-Scalar}

\begin{tabular}{cccl}
\hline $p_1 \ (\mu)$ & $p_2 \ (\nu)$ &
$p_3$ & \\
\hline $W^{\pm}$ & $A$ & $\chi^{\mp}$ & $(\delta Y + \delta G_W+
\ZH{W} + \ZH{\chi} + \ZH{AA})\tree{WA\chi}
+\ZH{ZA}\tree{WZ\chi}$ \\
\hline $W^{\pm}$ & $Z$ & $\chi^{\mp}$ & $(\delta Y + \delta H+
\ZH{W} + \ZH{\chi} + \ZH{ZZ})\tree{WZ\chi}
+\ZH{AZ}\tree{WA\chi}$ \\
\hline $W^-$ & $W^+$ & $H$ &
$(\delta Y + \delta G_2+ \delta G_W + 2\ZH{W}  + \ZH{H})\tree{WWH}$ \\
\hline $Z$ & $Z$ & $H$ &
$(\delta Y + \delta G_2+\delta G_3+ \delta G_Z +2\ZH{ZZ} +\ZH{H})\tree{ZZH}$ \\
\hline $Z$ & $A$ & $H$ &
$\ZH{ZA}\tree{ZZH}$  \ \ \itshape{(new)}\\
\hline
\end{tabular}

\subsection{Scalar-Scalar-Vector}

\begin{tabular}{cccl}
\hline $p_1$ & $p_2$ &
$p_3 \ (\mu)$ & \\
\hline $H$ & $\chi^{\mp}$ & $W^{\pm}$ & $(\delta Y + \delta G_2
 + \ZH{H} + \ZH{\chi} +\ZH{W})\tree{H\chi W}$ \\
\hline $\chi_3$ & $\chi^{\mp}$ & $W^{\pm}$ & $(\delta Y + \delta
G_2
 + \ZH{\chi_3} + \ZH{\chi} +\ZH{W})\tree{\chi_3\chi W}$ \\
\hline $\chi^-$ & $\chi^+$ & $A$ & $(\delta Y
 + 2\ZH{\chi} + \ZH{AA})\tree{\chi\chi A} + \ZH{ZA}\tree{\chi\chi Z}$ \\
\hline $\chi^-$ & $\chi^+$ & $Z$ & $(\delta Y + \delta G_4
 + 2\ZH{\chi} + \ZH{ZZ})\tree{\chi\chi Z} + \ZH{AZ}\tree{\chi\chi A}$ \\
\hline $H$ & $\chi_3$ & $Z$ & $(\delta Y + \delta G_2+\delta G_3
 + \ZH{H} + \ZH{\chi_3} +\ZH{ZZ})\tree{H\chi_3 Z}$ \\
\hline $H$ & $\chi_3$ & $A$ &
$ \ZH{ZA}\tree{H\chi_3 Z}$\ \ \itshape{(new)} \\
\hline
\end{tabular}

\subsection{Scalar-Scalar-Scalar}


\noindent
\begin{tabular}{cccl}
\hline $p_1$ & $p_2$ &
$p_3$ & \\
\hline \rule[-5mm]{0mm}{12mm} $H$ & $H$ & $H$ &
$\displaystyle{\left[(\delta Y + \delta G_2-\delta G_W+ \delta G_H
  +3\ZH{H} ) -\delta T \frac{e}{s_W M_W M_H^2} \right]\tree{HHH}}$ \\
\hline \rule[-5mm]{0mm}{12mm} $H$ & $\chi^-$ & $\chi^+$ &
$\displaystyle{\left[(\delta Y + \delta G_2-\delta G_W+ \delta G_H
  +\ZH{H} +2\ZH{\chi})-\delta T \frac{e}{s_W M_W M_H^2} \right]
  \tree{H\chi\chi}}$ \\
\hline \rule[-5mm]{0mm}{12mm} $H$ & $\chi_3$ & $\chi_3$ &
$\displaystyle{\left[(\delta Y + \delta G_2-\delta G_W+ \delta G_H
  +\ZH{H} +2\ZH{\chi_3} )-\delta T \frac{e}{s_W M_W M_H^2} \right]
  \tree{H\chi_3\chi_3}}$ \\
\hline
\end{tabular}

\subsection{Vector-Vector-Vector-Vector}

\begin{tabular}{ccccl}
\hline $p_1 \ (\mu)$ & $p_2 \ (\nu)$ & $p_3 \ (\rho)$ &
$p_4 \ (\sigma)$ \\
\hline $W^+$ & $W^-$ & $A$ & $A$ & $(2\delta Y
  +2\ZH{W} +2\ZH{AA} )\tree{WWAA} $ \\
  & & & & $+ 2 \ZH{ZA}\tree{WWAZ} $ \\
\hline $W^+$ & $W^-$ & $A$ & $Z$  & $(2\delta Y + \delta G_1
  +2\ZH{W} +\ZH{AA}+\ZH{ZZ} )\tree{WWAZ} $ \\
  & & & & $+ \ZH{AZ}\tree{WWAA} + \ZH{ZA}\tree{WWZZ} $ \\
\hline $W^+$ & $W^-$ & $Z$ & $Z$  & $(2\delta Y + 2\delta G_1
  +2\ZH{W} +2\ZH{ZZ} )\tree{WWZZ} $ \\
  & & & & $ + 2 \ZH{AZ}\tree{WWAZ} $ \\
\hline $W^+$ & $W^-$ & $W^-$ & $W^+$  & $(2\delta Y + 2\delta G_2
  +4\ZH{W}  )\tree{WWWW}$ \\
\hline
\end{tabular}

\subsection{Vector-Vector-Scalar-Scalar}

\begin{tabular}{ccccl}
\hline $p_1 \ (\mu)$ & $p_2 \ (\nu)$ & $p_3 $ &
$p_4 $  \\
\hline $A$ & $W^{\pm}$ & $H$ & $\chi^{\mp}$ & $(2\delta Y + \delta
G_2
  +\ZH{AA}+\ZH{W}+\ZH{H} +\ZH{\chi} )\tree{AWH\chi}
  $ \\
  & & & & $ +\ZH{ZA}\tree{ZWH\chi}$\\
\hline $A$ & $W^{\pm}$ & $\chi_3$ & $\chi^{\mp}$ & $(2\delta Y +
\delta G_2
  +\ZH{AA}+\ZH{W}+\ZH{\chi_3} +\ZH{\chi} )\tree{AW\chi_3\chi}
  $ \\
  & & & & $ +\ZH{ZA}\tree{ZW\chi_3\chi}$ \\
\hline $Z$ & $W^{\pm}$ & $H$ & $\chi^{\mp}$ & $(2\delta Y  +\delta
G_3
  +\ZH{ZZ}+\ZH{W}+\ZH{H} +\ZH{\chi} )\tree{ZWH\chi}
  $ \\
  & & & & $ +\ZH{AZ}\tree{AWH\chi}$ \\
\hline $Z$ & $W^{\pm}$ & $\chi_3$ & $\chi^{\mp}$ & $(2\delta Y
+\delta G_3
  +\ZH{ZZ}+\ZH{W}+\ZH{\chi_3} +\ZH{\chi} )\tree{ZW\chi_3\chi}
  $ \\
  & & & & $ +\ZH{AZ}\tree{AW\chi_3\chi}$ \\
\hline $A$ & $A$ & $\chi^+$ & $\chi^-$ & $(2\delta Y
  +2\ZH{AA}+2\ZH{\chi} )\tree{AA\chi\chi}
  +2 \ZH{ZA}\tree{ZA\chi\chi}$ \\
\hline $Z$ & $A$ & $\chi^+$ & $\chi^-$ & $(2\delta Y + \delta G_4
  +\ZH{ZZ}+\ZH{AA}+2\ZH{\chi} )\tree{ZA\chi\chi}
 $ \\
 & & & & $  +\ZH{ZA}\tree{ZZ\chi\chi}+\ZH{AZ}\tree{AA\chi\chi}$ \\
\hline $Z$ & $Z$ & $\chi^+$ & $\chi^-$ & $(2\delta Y + 2\delta G_4
  +2\ZH{ZZ}+2\ZH{\chi} )\tree{ZZ\chi\chi}
$ \\
& & & & $   + 2 \ZH{AZ}\tree{ZA\chi\chi}$ \\
\hline $W^+$ & $W^-$ & $H$ & $H$ & $(2\delta Y + 2\delta G_2
  +2\ZH{W}+2\ZH{H} )\tree{WWHH}$ \\
\hline $W^+$ & $W^-$ & $\chi_3$ & $\chi_3$ & $(2\delta Y + 2\delta
G_2
  +2\ZH{W}+2\ZH{\chi_3} )\tree{WW\chi_3\chi_3}$ \\
\hline $W^+$ & $W^-$ & $\chi^-$ & $\chi^+$ & $(2\delta Y + 2\delta
G_2
  +2\ZH{W}+2\ZH{\chi} )\tree{WW\chi\chi}$ \\
\hline $Z$ & $Z$ & $H$ & $H$ & $(2\delta Y +2\delta G_2 +2\delta
G_3
  +2\ZH{ZZ}+2\ZH{H} )\tree{ZZHH}$ \\
\hline $Z$ & $Z$ & $\chi_3$ & $\chi_3$ & $(2\delta Y +2\delta G_2
+2\delta G_3
  +2\ZH{ZZ}+2\ZH{\chi_3} )\tree{ZZ\chi_3\chi_3}$ \\
\hline $Z$ & $A$ & $H$ & $H$ &
$\ZH{ZA}\tree{ZZHH}$ \ \ \itshape{(new)}\\
\hline $Z$ & $A$ & $\chi_3$ & $\chi_3$ &
$\ZH{ZA} \tree{ZZ\chi_3\chi_3}$ \ \ \itshape{(new)}\\
\hline
\end{tabular}

\subsection{Scalar-Scalar-Scalar-Scalar}
\noindent
\begin{tabular}{ccccl}
\hline $p_1 $ & $p_2 $ & $p_3 $ &
$p_4 $ \\
\hline $H$ & $H$ & $H$ & $H$ & $\Big[ (2\delta Y +2\delta G_2
-2\delta G_W +\delta G_H
  +4\ZH{H} ) $ \\
& & & & $
  -\delta T \frac{e}{s_W M_W M_H^2} \Big]\tree{HHHH}$ \\
\hline $\chi_3$ & $\chi_3$ & $\chi_3$ & $\chi_3$ & $\Big[ (2\delta
Y +2\delta G_2 -2\delta G_W +\delta G_H
  +4\ZH{\chi_3} ) $ \\
& & & & $
  -\delta T \frac{e}{s_W M_W M_H^2} \Big]\tree{\chi_3\chi_3\chi_3\chi_3}$ \\
\hline $\chi^{\pm}$ & $\chi^{\mp}$ & $\chi^{\mp}$ & $\chi^{\pm}$ &
$\Big[ (2\delta Y +2\delta G_2 -2\delta G_W +\delta G_H
  +4\ZH{\chi} ) $ \\
& & & & $
  -\delta T \frac{e}{s_W M_W M_H^2} \Big]\tree{\chi\chi\chi\chi}$ \\
\hline $H$ & $H$ & $\chi_3$ & $\chi_3$ & $\Big[ (2\delta Y
+2\delta G_2 -2\delta G_W +\delta G_H
  +2\ZH{H}+2\ZH{\chi_3} ) $ \\
& & & & $
  -\delta T \frac{e}{s_W M_W M_H^2} \Big]\tree{HH\chi_3\chi_3}$ \\
\hline $H$ & $H$ & $\chi^+$ & $\chi^-$ & $\Big[ (2\delta Y
+2\delta G_2 -2\delta G_W +\delta G_H
  +2\ZH{H}+2\ZH{\chi} ) $ \\
& & & & $
  -\delta T \frac{e}{s_W M_W M_H^2} \Big]\tree{HH\chi\chi}$ \\
\hline $\chi^+$ & $\chi^-$ & $\chi_3$ & $\chi_3$ & $\Big[ (2\delta
Y +2\delta G_2 -2\delta G_W +\delta G_H
  +2\ZH{\chi}+2\ZH{\chi_3} ) $ \\
& & & & $
  -\delta T \frac{e}{s_W M_W M_H^2} \Big]\tree{\chi\chi\chi_3\chi_3}$ \\
\hline
\end{tabular}

\subsection{Fermion-Fermion-Vector}
\label{sec:ctffv} We  define $L, R = (1 \mp \gamma_5)/2 $.\\

\noindent
\begin{tabular}{cccl}
\hline $p_1 $ & $p_2 $ &
$p_3 \ (\mu)$ & \\
\hline $\bar{f}$ & $f$ & $A$ & $ \displaystyle{ (\delta Y +
\ZH{AA} + 2 \ZH{fL} )
e Q_f \gamma^{\mu}L} $ \\
 & & &
$ \displaystyle{ +(\delta Y + \ZH{AA} + 2 \ZH{fR} )
e Q_f \gamma^{\mu}R} $ \\
\rule[-5mm]{0mm}{12mm} & & &
 $+ \ZH{ZA}\displaystyle{\frac{e}{2s_Wc_W}
\left(2I_3\gamma^{\mu}L-2s_W^2 Q_f\gamma^{\mu}(L+R) \right) }$
 \\
\hline $\bar{f}$ & $f$ & $Z$ & \rule[-5mm]{0mm}{12mm}
$\displaystyle{(\delta Y + \delta G_2 + \delta G_3 +\ZH{ZZ}
 + 2\ZH{fL} ) \frac{e}{2s_Wc_W}2I_3\gamma^{\mu}L }$ \\
\rule[-5mm]{0mm}{12mm} & & \multicolumn{2}{l}{
$+\displaystyle{(\delta Y - \delta G_2 + \delta G_3 +\ZH{ZZ}
 + 2 \ZH{fL} ) \frac{e}{2s_Wc_W}(-2s_W^2 Q_f\gamma^{\mu}L) }$ }\\
\rule[-5mm]{0mm}{12mm} & & \multicolumn{2}{l}{
$+\displaystyle{(\delta Y - \delta G_2 + \delta G_3 +\ZH{ZZ}
 + 2 \ZH{fR} ) \frac{e}{2s_Wc_W}(-2s_W^2 Q_f\gamma^{\mu}R) }$ }\\
& & &
 $+ \ZH{AZ}\displaystyle{e Q_f \gamma^{\mu}(L+R)}$ \\
\hline \rule[-5mm]{0mm}{12mm} $\bar{U}/\bar{D}$ & $D/U$ &
$W^{+}/W^{-}$ & $\displaystyle{(\delta Y + \delta G_2
 + \ZH{(U/D)L}+ \ZH{(D/U)L} +\ZH{W})
 \frac{e}{\sqrt{2}s_W} \gamma^{\mu}L}$ \\
\hline
\end{tabular}

\subsection{Fermion-Fermion-Scalar}

$L, R = (1 \mp \gamma_5)/2 $

\noindent
\begin{tabular}{cccl}
\hline $p_1 $ & $p_2 $ &
$p_3 $ & \\
\hline \rule[-1mm]{0mm}{6mm}
$\bar{f}$ & $f$ & $H$ & \\
\multicolumn{4}{c}{ \rule[-5mm]{0mm}{12mm} $\displaystyle{(\delta
Y + \delta G_2 + \delta G_{mf} - \delta G_W
 + {\ZH{fR}} + \ZH{fL} +\ZH{H})
\left(-\frac{e}{2s_W}\frac{m_f}{M_W}\right)L }$} \\
\multicolumn{4}{c}{ \rule[-5mm]{0mm}{12mm} $\displaystyle{+(\delta
Y + \delta G_2 + \delta G_{mf} - \delta G_W
 + {\ZH{fL}} + \ZH{fR} +\ZH{H})
\left(-\frac{e}{2s_W}\frac{m_f}{M_W}\right)R }$ } \\
\hline \rule[-1mm]{0mm}{6mm}
$\bar{U}/\bar{D}$ & $U/D$ & $\chi_3$ & \\
\multicolumn{4}{c}{ \rule[-5mm]{0mm}{12mm} $(\displaystyle{\delta
Y + \delta G_2 + \delta G_{mf} - \delta G_W
 + {\ZH{(U/D)R}}+ \ZH{(U/D)L} +\ZH{\chi_3})
\left((-/+) \frac{i e}{2s_W}\frac{m_f}{M_W} \right)(-L)  }$ } \\
\multicolumn{4}{c}{ \rule[-5mm]{0mm}{12mm} $\displaystyle{+(\delta
Y + \delta G_2 + \delta G_{mf} - \delta G_W
 + {\ZH{(U/D)L}} + \ZH{(U/D)R} +\ZH{\chi_3})
\left((-/+) \frac{i e}{2s_W}\frac{m_f}{M_W} \right)R  }$ } \\
\hline $\bar{U}$ & $D$ & $\chi^+$ & \rule[-5mm]{0mm}{12mm}
$(\delta Y + \delta G_2 + \delta G_{mU} - \delta G_W
 + {\ZH{UR}} + \ZH{DL} +\ZH{\chi})
 \displaystyle{\frac{-i e}{\sqrt{2}s_W}\frac{m_U}{M_W}(-L)}$ \\
 & & &
\rule[-5mm]{0mm}{12mm} $+(\delta Y + \delta G_2 + \delta G_{mD} -
\delta G_W
 + {\ZH{UL}} + \ZH{DR} +\ZH{\chi})
 \displaystyle{\frac{-i e}{\sqrt{2}s_W}\frac{m_D}{M_W}R}$ \\
\hline $\bar{D}$ & $U$ & $\chi^-$ & \rule[-5mm]{0mm}{12mm}
$(\delta Y + \delta G_2 + \delta G_{mD} - \delta G_W
 + {\ZH{DR}} + \ZH{UL} +\ZH{\chi})
 \displaystyle{\frac{-i e}{\sqrt{2}s_W}\frac{m_D}{M_W}(-L)}$ \\
 & & &
\rule[-5mm]{0mm}{12mm} $+(\delta Y + \delta G_2 + \delta G_{mU} -
\delta G_W
 + {\ZH{DL}} + \ZH{UR} +\ZH{\chi})
 \displaystyle{\frac{-i e}{\sqrt{2}s_W}\frac{m_U}{M_W}R}$ \\
\hline
\end{tabular}


\setcounter{equation}{0}
\section{Properties of two-point functions}
\label{app-onetwoptfct} As mentioned earlier, loop integrals are
calculated using dimensional regularisation. In the following $l$
will be the loop momentum. Since one-point and two-point functions
(tadpoles and self-energies) are essential in the derivation of
the counterterms, we list here the properties of these functions.

For the one-point function, which corresponds for example to the
diagram shown in \zu{one-two-pntx}-(a), we have the well known
result:

\begin{figure}[htbp]
\caption{{\em Diagrams for the one-point (a) and two-point
functions (b). }} \label{fig:one-two-pntx}
\begin{center}
\hspace*{-5cm} \unitlength 0.1in
\begin{picture}(38.20,14.80)(3.80,-21.60)

\special{pn 8}%
\special{pa 2800 1800}%
\special{pa 4200 1800}%
\special{fp}%
%
\special{pn 8}%
\special{ar 3480 1520 283 283  0.0000000 6.2831853}%
%
\special{pn 20}%
\special{sh 1}%
\special{ar 3470 1800 10 10 0  6.28318530717959E+0000}%
\special{sh 1}%
\special{ar 3470 1800 10 10 0  6.28318530717959E+0000}%
%
\special{pn 8}%
\special{pa 2830 1740}%
\special{pa 3060 1740}%
\special{fp}%
\special{sh 1}%
\special{pa 3060 1740}%
\special{pa 2993 1720}%
\special{pa 3007 1740}%
\special{pa 2993 1760}%
\special{pa 3060 1740}%
\special{fp}%
\put(28.0000,-16.7000){\makebox(0,0)[lb]{$q$}}%
\put(33.4000,-11.6000){\makebox(0,0)[lb]{$\ell$}}%
%
\special{pn 8}%
\special{pa 3670 1190}%
\special{pa 3270 1190}%
\special{fp}%
\special{sh 1}%
\special{pa 3270 1190}%
\special{pa 3337 1210}%
\special{pa 3323 1190}%
\special{pa 3337 1170}%
\special{pa 3270 1190}%
\special{fp}%
\put(37.9000,-15.5000){\makebox(0,0)[lb]{$M_A$}}%
\put(33.4000,-22.7){\makebox(0,0)[lb]{$(a)$}}%
\end{picture}%
\hspace*{1cm}
\unitlength 0.1in
\begin{picture}(38.20,14.80)(3.80,-21.60)
%
\special{pn 8}%
\special{pa 400 1400}%
\special{pa 800 1400}%
\special{fp}%
\special{pa 2000 1400}%
\special{pa 2400 1400}%
\special{fp}%
%
\special{pn 8}%
\special{ar 1400 1400 600 400  0.0000000 6.2831853}%
%
\special{pn 20}%
\special{sh 1}%
\special{ar 800 1400 10 10 0  6.28318530717959E+0000}%
\special{sh 1}%
\special{ar 2000 1400 10 10 0  6.28318530717959E+0000}%
%
\special{pn 8}%
\special{pa 410 1310}%
\special{pa 640 1310}%
\special{fp}%
\special{sh 1}%
\special{pa 640 1310}%
\special{pa 573 1290}%
\special{pa 587 1310}%
\special{pa 573 1330}%
\special{pa 640 1310}%
\special{fp}%
%
\special{pn 8}%
\special{pa 2170 1310}%
\special{pa 2400 1310}%
\special{fp}%
\special{sh 1}%
\special{pa 2400 1310}%
\special{pa 2333 1290}%
\special{pa 2347 1310}%
\special{pa 2333 1330}%
\special{pa 2400 1310}%
\special{fp}%
\put(21.8000,-12.4000){\makebox(0,0)[lb]{$q$}}%
\put(3.8000,-12.4000){\makebox(0,0)[lb]{$q$}}%
%
\special{pn 8}%
\special{pa 1200 1870}%
\special{pa 1600 1870}%
\special{fp}%
\special{sh 1}%
\special{pa 1600 1870}%
\special{pa 1533 1850}%
\special{pa 1547 1870}%
\special{pa 1533 1890}%
\special{pa 1600 1870}%
\special{fp}%
%
\special{pn 8}%
\special{pa 1590 930}%
\special{pa 1190 930}%
\special{fp}%
\special{sh 1}%
\special{pa 1190 930}%
\special{pa 1257 950}%
\special{pa 1243 930}%
\special{pa 1257 910}%
\special{pa 1190 930}%
\special{fp}%
\put(12.7000,-20.8000){\makebox(0,0)[lb]{$\ell$}}%
\put(12.2000,-8.5000){\makebox(0,0)[lb]{$\ell-q$}}%
\put(10.7000,-17.0000){\makebox(0,0)[lb]{$1-x, M_A$}}%
\put(11.9000,-12.5000){\makebox(0,0)[lb]{$x, M_B$}}%
\put(12.7000,-22.7){\makebox(0,0)[lb]{$(b)$}}%
%
\end{picture}%

\end{center}
\end{figure}
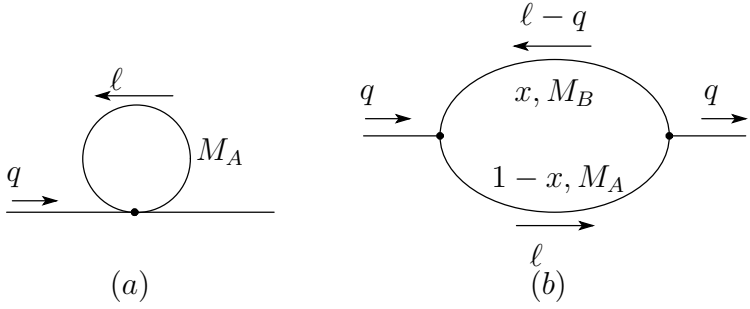

\begin{equation}
\int \frac{d^n \ell}{i(2\pi)^n}
  \frac{1}{\ell^2-m_A^2}= \frac{1}{16\pi^2}
  m_A^2 \left( \Cuv - \log m_A^2 + 1
  \right)
\label{eq:oneint}
\end{equation}
\noi where $\cuv$ is defined in Eq.~\ref{cuvdef}.

A typical two-point function refers to a diagram as shown in
\zu{one-two-pntx}.

This leads to the calculation of
\begin{equation}
\begin{array}{l}
\displaystyle{ \int \frac{d^n \ell}{i(2\pi)^n}
\frac{N(l)}{(\ell^2-M_A^2)((\ell-q)^2-M_B^2)}} \\
 \\
\displaystyle{ = \int \frac{d^n \ell}{i(2\pi)^n} \int_0^1 dx\;
\frac{N}{[(1-x)(\ell^2-M_A^2)+x((\ell-q)^2-M_B^2)]^2} } .
\end{array}
\end{equation}

\noi where $N(l)$ depends, in general, on the momenta $l,q$ and
the masses. Defining $D_2$ as

\begin{equation}
D_2=(1-x)M_A^2+x M_B^2 - x(1-x) s, \quad (s=q^2)
\end{equation}
\noi we usually need to compute
\begin{equation}
\int \frac{d^n \ell}{i(2\pi)^n}
  \frac{1}{(\ell^2-D_2)^2}= \frac{1}{16\pi^2}
  \left(
  \Cuv - \log D_2
  \right)
\end{equation}

\begin{equation}
\int \frac{d^n \ell}{i(2\pi)^n}
  \frac{\ell^2}{(\ell^2-D_2)^2}= \frac{1}{16\pi^2}
  2 D_2 \left(
  \Cuv + \frac{1}{2} - \log D_2
  \right)
\end{equation}

\begin{equation}
\int \frac{d^n \ell}{i(2\pi)^n}
  \frac{\ell^{\mu}\ell^{\nu}}{(\ell^2-D_2)^2}= \frac{1}{16\pi^2}
  \frac{D_2}{2}  \left(
  \Cuv + 1 - \log D_2
  \right) g^{\mu\nu}
\end{equation}

\vspace{5mm}

Then the integral over the parameter $x$ gives

\begin{equation}
\int_0^1 dx\; D_2  = \frac{1}{2}(M_A^2+M_B^2)-\frac{1}{6}s
\end{equation}

\begin{equation}
\label{Fnfctdef} F_n(A,B)=\int_0^1 dx\; x^n \log D_2  =\int_0^1
dx\; x^n \log \left[(1-x)M_A^2+x M_B^2 - x(1-x) s\right]
\end{equation}

We do not show the explicit form of $F_n$ in terms of elementary
functions (the result is well known). We only encounter $n=0,1,2$.

The notation $F(A,B)=F_1(A,B)-F_2(A,B)=F(B,A)$ is sometimes used.

\begin{equation}
\begin{array}{ll}
\tilde{F}(A,B)=  & \displaystyle{\int_0^1 dx\; D_2 \log D_2 } \\
 & \\
& \displaystyle{
=M_A^2\left(F_0(A,B)-F_1(A,B)\right)+M_B^2F_1(A,B) -sF(A,B) }
\end{array}
\end{equation}

\vspace{5mm} We have several relations for the $F_n$ functions as
shown below. All $F_n$ can be reduced into $F_0$.

\noindent \underline{Exchange of A and B}

\begin{equation}
\begin{array}{ll}
F_0(B,A)= & F_0(A,B) \\
F_1(B,A)= & F_0(A,B)-F_1(A,B) \\
F_2(B,A)= & F_0(A,B)-2 F_1(A,B)+F_2(A,B) \\
\end{array}
\end{equation}

\vspace{5mm}

\noindent \underline{Reduction into $F_0$, $A\ne B$ }

\begin{equation}
F_1(A,B)=\frac{1}{2}\left(1+\frac{M_A^2-M_B^2}{s}\right)F_0(A,B)
+\frac{1}{2s}\left( M_B^2\log M_B^2 - M_A^2\log M_A^2 - M_B^2 +
M_A^2 \right)
\end{equation}

\begin{equation}
\begin{array}{ll}
F_2(A,B)= &
\displaystyle{\frac{2}{3}\left(1+\frac{M_A^2-M_B^2}{s}\right)F_1(A,B)
-\frac{M_A^2}{3s}F_0(A,B)} \\
{ } & { } \\
& \displaystyle{ +\frac{1}{3s}\left(M_B^2\log
M_B^2+\frac{1}{2}(M_A^2-M_B^2)\right) -\frac{1}{18}}
\end{array}
\end{equation}

\vspace{5mm}

\noindent \underline{Reduction into $F_0$, $A=B$ }

\begin{equation}
F_1(A,A)=\frac{1}{2}F_0(A,A)
\end{equation}

\begin{equation}
F_2(A,A)=\frac{1}{3}\left(1-\frac{M_A^2}{s} \right)F_0(A,A)
+\frac{M_A^2}{3s}\log M_A^2 -\frac{1}{18}
\end{equation}

\vspace{5mm}

\noindent \underline{$G$ functions (Derivative of $F$)}

\begin{equation}
G_n(A,B)=\frac{d}{ds}F_n(A,B)= \int_0^1  dx\; \frac{-x^n\cdot
x(1-x)}{D_2}
\end{equation}

\vspace{5mm}

\noindent \underline{$F$ and $G$ at special energy}

\begin{equation}
\begin{array}{l}
\displaystyle{ \left.  F_n(A,B;C)=F_n(A,B)\right|_{s=M_C^2} }\\
\displaystyle{ \left.  F_n(A,B;0)=F_n(A,B)\right|_{s=0} }\\
\displaystyle{ \left.  G_n(A,B;C)=G_n(A,B)\right|_{s=M_C^2} }\\
\displaystyle{ \left.  G_n(A,B;0)=G_n(A,B)\right|_{s=0} }
\end{array}
\end{equation}

\vspace{5mm}

\noindent \underline{$F_0$ for $s=0$}
\begin{equation}
F_0(A,B,0) = \left\{
\begin{array}{ll}
\log M_A^2 & (A=B) \\
& \\
\displaystyle{
 \frac{M_B^2\log M_B^2 -M_A^2\log M_A^2 }{M_B^2 - M_A^2} - 1 } &
(A\ne B)
\end{array}
\right. \label{eq:fnzero}
\end{equation}


\setcounter{equation}{0}
\section{Results for the one-loop corrections to the propagators}
\label{sec:prop-corr}

 We here give the details on the calculation
of the loop corrections to the various propagators and mixings.
For the vector bosons we present both the transverse and
longitudinal part as defined in Section~\ref{sec:ren-cdts}. We
also show the various contributions by classifying them according
to the diagrams of Fig.~\ref{fig:one-two-pntx}. Therefore for each
propagator we show a table containing the two types of diagrams.
The fermion contributions are summed over all fermion species in
the case of the neutral sector and over all doublets in the
charged sector. In both cases summing over colour for quarks is
implied. Before presenting the results for the two-point function,
we first start by presenting the one-loop contribution to the
tadpole. Although we will require this contribution to vanish
against the tadpole counterterm we give its full expression for
completeness. Also, the latter is needed for the Ward identities.

\subsection{The tadpole}

The  tadpole contribution $T^{loop}$ only receives contributions
from diagrams of the type  shown in
Fig.~\ref{fig:one-two-pntx}-(a) where $A= W, Z, \chi, \chi_3, H,
c, c^{Z}, f$. The result is as follows:

\begin{equation}
\begin{array}{ll}
T^{loop} = & \displaystyle{\frac{e}{16\pi^2 s_W M_W}
 \Bigl[
 \MWt \left( (\Cuv - \logw + 1) (3\MWt+\frac{1}{2}\MHt) - 2 \MWt \right) } \\
 & \\
 &  \displaystyle{
+ \MZt \left( (\Cuv - \logz + 1) (\frac{3}{2}\MZt+\frac{1}{4}\MHt) - \MZt \right)  } \\
 & \\
 & \displaystyle{
 + \frac{3}{4} M_H^4  (\Cuv - \log \MHt + 1)  }\\
 & \\
 &  \displaystyle{
- \sum_{f}  2m_f^4 (\Cuv-\log m_f^2 +1) \Bigr] }\; .
\end{array}
\end{equation}

It is important to note that all dependence on the non-linear
gauge parameters (namely $\enlg$ and $\dnlg$) vanishes among all
diagrams and is therefore the same as in the usual linear gauge.
This can be considered as a check on the calculation, since the
tadpole $T$ can be considered as a basic parameter of the theory.

From this expression we immediately derive the tadpole
counterterm:
\beqn
\delta T=-T^{loop}\; .
\eeqn

\subsection{$A-A$}
\label{explicitaa}
For all the two-point functions we will list, as done in the table
below, the contributing diagrams where (a) corresponds to the type
shown in Fig.~\ref{fig:one-two-pntx}-(a) and (b) to
Fig.~\ref{fig:one-two-pntx}-(b)
\begin{center}
\begin{tabular}{l|l}
\hline (b) & $(A,B)=(W,W),  (W,\chi), (\chi,W), (\chi,\chi),
 (c^+,c^+), (c^-,c^-), (f,f)$
\\
\hline (a) & $A=W, \chi, c^+, c^-$
\\
\hline
\end{tabular}
\end{center}

\vspace{2mm}

\beqn
\Pi^{AA}_T(q^2)&=&\frac{\alpha}{4\pi} q^2 \left[7C_{UV} -5
F_0(W,W) -12 F(W,W)+\frac{2}{3}-4(1-\alphat)(\cuv-F_0(W,W))
\right. \nonumber \\
 & -& \left.
8 \sum_{f} Q_f^2 \left(\frac{1}{6}\Cuv - F(f,f) \right) \right]\;
.
\eeqn

Note that independently of the gauge parameter $\Pi^{AA}_T(0)=0$.
This is just a remnant of the QED gauge invariance which is
explicit at one-loop. This also gives  $\Pi^{AA}_L(q^2)=0$ which
we explicitly verify. More generally we will also check explicitly
that in both the linear and non-linear gauges  $\Pi_T(0)=\Pi_L(0)$
for all vector-vector transitions. This is another check on the
calculation and encodes the property that there is no spurious
pole in the propagators essential for the Goldstone mechanism.

\subsection{$Z-A$}
\label{explicitaz}
\begin{center}
\begin{tabular}{l|l}
\hline (b) & $(A,B)=(W,W),  (W,\chi), (\chi,W), (\chi,\chi),
 (c^+,c^+), (c^-,c^-), (f,f)$
\\
\hline (a) & $A=W, \chi, c^+, c^-$
\\
\hline
\end{tabular}
\end{center}

\vspace{2mm}

\beqn
\Pi_T^{ZA}=\frac{\alpha}{4\pi} \frac{c_W}{s_W}& \Biggl\{ & q^2
\Biggl[C_{UV} \left(7+\frac{1}{6c_W^2} \right)- 4
\left(3-\frac{1}{2c_W^2}\right)F(W,W)
 +\frac{2}{3}
 -\left(5+\frac{1}{2c_W^2}\right)F_0(W,W)
\nonumber\\
&&-2(1-\betat) \left( \cuv-F_0(W,W) \right)
\nonumber\\
&&-\frac{2}{c_W^2}\sum_{f} |Q_f| \left(1 -4|Q_f| s_W^2\right)
\left(\frac{1}{6}\Cuv - F(f,f)
\right) \Biggr] \nonumber\\
&&-2(1-\alphat) (q^2-\mzz)\left( \cuv-F_0(W,W) \right) \Biggr\}\; ,\nonumber\\
\eeqn
\beqn
\label{pilza} \Pi_L^{ZA}=\frac{\alpha}{2\pi} \frac{c_W}{s_W}
 (1-\alphat)\mzz
\left( C_{UV}-F_0(W,W)\right)\; .
\eeqn

\noi Note that we do get as a check that
$\Pi_T^{ZA}(0)=\Pi_L^{ZA}(0)$. Moreover for $\alphat=1$ this
condition is even stronger since we get
$\Pi_T^{ZA}(0)=\Pi_L^{ZA}=0$. This is due to the fact that for
this particular choice of the parameter, the gauge-fixing in the
charged sector which contributes here (note that fermions do not
contribute to $\Pi_L^{ZA}$), there is an additional $U(1)_{\rm
QED}$ gauge invariance. This choice is therefore very useful. As
we will see this is also responsible for the vanishing of the
induced $A-\chi_3$ transition, see section~\ref{explicitachi3}. It
is also important to remark that at the $Z$-pole the $\anlg$
dependence vanishes. This is also responsible for the fact that
the counterterms needed for the mass definitions do not depend on
the gauge fixing.

\subsection{$Z-Z$}

\begin{center}
\begin{tabular}{l|l}
\hline (b) & $(A,B)=(W,W),  (W,\chi), (\chi,W), (\chi,\chi),
(H,\chi), (H,Z), (c^+,c^+), (c^-,c^-), (f,f)$
\\
\hline (a) & $A=W, H, \chi_3, \chi, c^+, c^-$
\\
\hline
\end{tabular}
\end{center}

\vspace{2mm}

\beqn
\Pi_T^{ZZ}=\frac{\alpha}{4\pi s_W^2 c_W^2}\left( T^{ZZ}_b +
T^{ZZ}_f + (1-\betat)\left(q^2-\mzz\right)\Delta T^{ZZ}\right)\; ,
\eeqn

\beqn
T^{ZZ}_b&=& C_{UV}\Biggl[q^2\left(7c_W^4-\frac{1-2c_W^2}{6}\right)
-2\mww-\mzz\Biggr]
+\frac{2}{3}q^2 c_W^4-\frac{q^2}{12}\nonumber\\
&&-8 q^2 c_W^4 F_0(W,W) +q^2(F_0(W,W)-4F(W,W))
\Biggl(3c_W^4+\frac{1-4c_W^2}{4}\Biggr)\nonumber\\
&+&2\mww F_0(W,W)+\frac{q^2}{2}F(H,Z)-
\frac{M_H^2}{2} F_0(H,Z)-\frac{\mzz-M_H^2}{2} F_1(H,Z)\nonumber\\
&+&   \mzz F_0(H,Z) + \frac{1}{4}(M_H^2 \log M_H^2+ \mzz \logz), \nonumber\\
T^{ZZ}_f&=& -\frac{1}{2} \sum_{f} \Bigl[\Biggl( (1 -4|Q_f|
s_W^2)^2+1\Biggr) \left(\frac{1}{6}\Cuv - F(f,f) \right)q^2
 -m_f^2 \left(  \Cuv - F_0(f,f)
\right) \Bigr], \; \nonumber\\
\eeqn
\beqn
\Delta T^{ZZ}= -4c_W^4 (C_{UV}- F_0(W,W)).
\eeqn

\noi We have, $$\Pi^T_{{\rm NonLinear}}(\mzz)=\Pi^T_{{\rm
Linear}}(\mzz).$$

\beqn
\Pi_L^{ZZ}&=&\frac{\alpha}{16\pi s_W^2 c_W^2} \Biggl\{ q^2
\Biggl[C_{UV}\tilde\epsilon^2 -\frac{1}{3}+2  F(H,Z)
-(1-\tilde\epsilon)^2 F_0(H,Z)-4 F_2(H,Z)\nonumber \\
& & \;\;\;\;\;\;\;\;\;\;\;\;+4(1-\tilde\epsilon)F_1(H,Z)\Biggr]
\nonumber\\
&-& 4\mzz (C_{UV}-F_0(H,Z)) -2 M_H^2 F_0(H,Z) -2(\mzz-M_H^2)
F_1(H,Z)
\nonumber\\
&+&(M_H^2 \log M_H^2+ \mzz \log \mzz)\;-\;8\mww
\left(C_{UV}-F_0(W,W)\right)
\left(1-2c_W^2(1-\betat)\right)  \nonumber \\
&+& 2 \sum_{f}
 m_f^2 \left(  \Cuv - F_0(f,f) \right)
\Biggr\}.
\eeqn

It is easy to see that, for any choice of the gauge parameters,
$\Pi_L^{ZZ}(0)=\Pi_T^{ZZ}(0)$ which is a check on the calculation.
Also note that  the $\epsilont$ dependence is proportional to
$q^2$.

\subsection{$W-W$}
\begin{center}
\begin{tabular}{l|l}
\hline (b) & $(A,B)=(Z,W),  (Z,\chi), (A,W), (A,\chi),
 (H,\chi), (H,W), (\chi_3,\chi), $ \\
 & $(c^Z,c^+), (c^Z,c^-), (c^A,c^+), (c^A,c^-), (f,f^\prime)$
\\
\hline (a) & $A=A, Z, W, H, \chi_3, \chi, c^+, c^-$
\\
\hline
\end{tabular}
\end{center}

\vspace{2mm}

\beqn
\Pi_T^{WW}=\frac{\alpha}{4\pi s_W^2}\left( T^{WW}_b + T^{WW}_f +
\left(q^2-\mww\right)\Delta T^{WW}_{\alphat,\betat}\right)\; ,
\eeqn

\renewcommand{\sw}{s_W^2}
\renewcommand{\cw}{c_W^2}
\beqn
T^{WW}_b&=&
\cuv\biggl(\frac{19}{6}q^2+2\mww-\mzz\biggr)-\frac{q^2}{6}
\nonumber\\ &&+4\sw\biggl[q^2(F(A,W)-F_0(A,W))-\mww
F_1(A,W)\biggr] \nonumber\\
&&+\cw\biggl[4q^2(F(Z,W)-F_0(Z,W))-4(\mww-\mzz)F_1(Z,W) \nonumber\\
&&+\biggl(-7\mzz+\frac{\mzz}{\cw}\biggr) F_0(Z,W)\biggr]+
\frac{q^2}{2}\left(F(H,W)+F(Z,W) \right)+\mww F_0(H,W)\nonumber\\
&&-\frac{1}{2}\left[\mhh F_0(H,W)+\mzz
F_0(Z,W)+(\mww-\mhh)F_1(H,W)\right. \nonumber\\
&& \left. + (\mww-\mzz)F_1(Z,W)\right] \nonumber\\
&&+\frac{5\mww}{2}\logw+\frac{\mzz}{4}\logz+\frac{\mhh}{4}\logh
+2\mww\logz \; ,\\
T^{WW}_f&=&- \frac{1}{2} \sum_{doublet} \left\{
4\left(\frac{1}{6}\Cuv - F(f,f') \right)q^2
-(m_f^2+m_f'^2)\Cuv+2 m_f^2 F_1(f',f) \right.\nonumber \\
& &\left. \;\;\;\;\;\;+ 2 m_f'^2 F_1(f,f') \right\}\; ,\nonumber \\
\\
\Delta T^{WW}_{\alphat,\betat}&=&-2
\left(\sw\alphat(\cuv-F_0(A,W)) +\cw\betat(\cuv-F_0(Z,W))\right)\;
.
\eeqn

\noi Here also we check that , $$\Pi^T_{{\rm
NonLinear}}(\mww)=\Pi^T_{{\rm Linear}}(\mww)\; .$$

\beqn
\label{pilww} \Pi_L^{WW}&=&\frac{\alpha}{4\pi \sw} \Biggl\{
\cuv\biggl(2\mww-\mzz\biggr)-\frac{5}{6}q^2 \nonumber\\
&+&2 \sw \left[ q^2 (6 F(A,W) - F_0(A,W))-2\mww F_1(A,W) \right]
\nonumber
\\ &+& \cw \left[ 2 q^2 (6 F(Z,W) - F_0(Z,W))-4 (\mww-\mzz) F_1(Z,W) \right.\nonumber\\
& & \left. + (\frac{\mzz}{\cw}-7\mzz) F_0(Z,W) \right] \nonumber \\
&+& \frac{1}{2} \left[q^2 (3F(H,W)-\frac{F_0(H,W)}{2})+2\mww
F_0(H,W) -\mhh F_0(H,W) \right. \nonumber \\
& & \left. \;\;\;\;-(\mww-\mhh)F_1(H,W) \right] \nonumber \\
&+& \frac{1}{2} \left[q^2 (3F(Z,W)-\frac{F_0(Z,W)}{2})-\mzz
F_0(Z,W)
-(\mww-\mzz)F_1(Z,W) \right] \nonumber \\
&+&\frac{5\mww}{2}\logw+\frac{\mzz}{4}\logz+\frac{\mhh}{4}\logh
+2\mww\logz \nonumber\\ &+&2\sw\alphat \left[3 q^2
(F_0(A,W)-2F_1(A,W)) + \mww (\cuv-F_0(A,W))\right] \nonumber \\
&+&2\cw\betat \left[3 q^2 (F_0(Z,W)-2F_1(Z,W)) + \mww
(\cuv-F_0(Z,W))\right] \nonumber \\ &+&\sw \alphat^2 q^2
(5\cuv-6F_0(A,W)+2F_1(A,W)-2)\nonumber\\ &+&\cw \betat^2
q^2(5\cuv-6F_0(Z,W)+2F_1(Z,W)-2)\nonumber\\ &+& \frac{\deltat}{2}
q^2 (F_0(H,W)-2F_1(H,W))\;+\;\frac{\kappat}{2} q^2
(F_0(Z,W)-2F_1(Z,W)) \nonumber \\ &+& \frac{\deltat^2}{4} q^2
(\cuv-F_0(H,W))+\frac{\kappat^2}{4} q^2 (\cuv-F_0(Z,W))\nonumber\\
&+& \frac{1}{2} \sum_{doublet} \left\{ (m_f^2+m_f'^2)\Cuv-2m_f^2
F_1(f',f)-2m_f'^2 F_1(f,f') \right\} \Biggr\}\; .
\eeqn

We again have that, for any choice of the gauge parameters,
$\Pi_L^{WW}(0)=\Pi_T^{WW}(0)$ which is a check on the calculation.
Also note that the $\deltat,\kappat$ dependence is proportional to
$q^2$, as is any dependence quadratic in $\alphat, \betat$. All
these dependencies will be present in the propagators/mixings of
the Goldstones.

Note that for both the $WW$ and $ZZ$ transition the tadpole
contribution is not included as it will be canceled against that
of  the tadpole counterterms. Moreover note that such contribution
do not depend on the gauge parameter.  On the other hand the
inclusion of the tadpole one-loop correction is needed for the
Ward identities relating the bosonic two-point functions.

\subsection{$H-H$}

\begin{center}
\begin{tabular}{l|l}
\hline (b) & $(A,B)=(W,W),  (W,\chi), (\chi,W), (\chi,\chi),
(Z,\chi_3), (\chi_3,\chi_3), (Z,Z), (H,H),$ \\
& $ (c^+,c^+), (c^-,c^-), (c^Z,c^Z), (f,f)$
\\
\hline (a) & $A=W, Z, H, \chi, \chi_3, c^+, c^-, c^Z$
\\
\hline
\end{tabular}
\end{center}

\vspace{2mm}
Here we explicitly add the tadpole contribution.

\beqn
\Pi^{H}(q^2)+\frac{3\delta
T}{v}=\frac{\alpha}{4\pi\sw}\left(\Pi^{H}_b+\Pi^{H}_f+(q^2-\mhh)\Pi^{H}_{\deltat,\epsilont}\right)\;
,
\eeqn

with

\beqn
\Pi^H_b&=&
\cuv\Biggl[-\left(q^2+\frac{\mhh}{2}\right)\left(1+\frac{1}{2\cw}\right)
+\frac{3M_H^4}{4\mww}
\Biggr] -\frac{9M_H^4}{8\mww}F_0(H,H)\nonumber\\
&&-F_0(W,W)\left(-q^2+3\mww+\frac{M_H^4}{4\mww}\right)
-\frac{F_0(Z,Z)}{2\cw}\left(-q^2+3\mzz+\frac{M_H^4}{4\mzz}\right)\nonumber\\
&& -\left(\frac{\mhh}{2}+3\mww\right)(1-\logw)-\frac{1}{2
\cw}\left(\frac{\mhh}{2}+3\mzz\right)(1-\logz) \nonumber
\\
& & \;\;\;\;\; -\frac{3M_H^4}{4\mww} (1-\logh)\; ,
\\
\Pi^H_f&=&\sum_f \frac{m_f^2}{\mww}  \Biggl\{\frac{q^2}{2}
(\cuv-F_0(f,f))+ 2m_f^2 (1- \log m_f^2+F_0(f,f)) \Biggr\}\; , \\
\Pi^{H}_{\deltat,\epsilont}&=&\left(-\cuv+F_0(W,W)\right)\deltat
+\left(-\cuv+F_0(Z,Z)\right)\frac{\epsilont} {2\cw}\; .
\eeqn

\noi Again at $q^2=M_H^2$ the self-energy is independent of the
gauge parameter, which means that the shift in the Higgs mass will
also not depend on the gauge parameters.

\subsection{$f-f$}

At one-loop the result is the same as in the linear gauge, but we
give here the full result that includes mass effects as well as
the contribution of the Goldstones. We have neglected all fermion
mixing.  The $K_j^f$ have been introduced in
section~\ref{sec:ct2pt} and correspond to the different Lorentz
structures of the fermion propagator. Since we are neglecting
mixing and assuming \cpviol-invariance $K_5^f=0$ holds. We have
also found it convenient to express each of these Lorentz
coefficients in a basis that corresponds to the various
contributions to the self energy (photon exchange, W-exchange,
etc...).

\begin{center}
\begin{tabular}{l|l}
\hline (b) & $(A,B)= (f,A), (f,Z), (f',W), (f,H), (f,\chi_3),
(f',\chi)$
\\
\hline (a) & None
\\
\hline
\end{tabular}
\end{center}

\vspace{2mm}

\begin{equation}
\begin{array}{l}
\displaystyle{ K_j^f (s) = \frac{\alpha}{4\pi} \left[ Q_f^2 K_j^A
+ \frac{1}{c_W^2}Q_f^2 s_W^2  K_j^{Z(1)} -
  \frac{1}{2c_W^2} |Q_f| K_j^{Z(2)}
 +   \frac{1}{8s_W^2c_W^2}   K_j^{Z(3)}\right.  } \\
{ } \\
\displaystyle{\left. +\frac{1}{4s_W^2}K_j^W +
\frac{1}{4s_W^2c_W^2} \frac{m_f^2}{M_Z^2} K_j^S\right]} \qquad
(j=1,\gamma, 5\gamma),
\end{array}
\end{equation}

\vspace{4mm}

\begin{equation}
\begin{array}{l}
K_1^A=m_f[-4\Cuv+2 + 4 F_0(f,A)] ,\\
{ } \\
K_{\gamma}^A=\Cuv-1-2 F_1(f,A), \\
{ } \\
K_{5\gamma}^A=0,
\end{array}
\end{equation}

\begin{equation}
\begin{array}{l}
K_1^{Z(1)}=m_f[-4\Cuv+2+4F_0(f,Z)] ,
\quad K_1^{Z(2)}=K_1^{Z(1)}, \quad K_1^{Z(3)}=0, \\
{ } \\
K_{\gamma}^{Z(1)}=\Cuv-1-2 F_1(f,Z) , \quad
K_{\gamma}^{Z(2)}=K_{\gamma}^{Z(1)},
\quad K_{\gamma}^{Z(3)}=K_{\gamma}^{Z(1)}, \\
{ } \\
K_{5\gamma}^{Z(1)}=0, \quad K_{5\gamma}^{Z(2)}=-K_{\gamma}^{Z(1)},
\quad K_{5\gamma}^{Z(3)}=-K_{\gamma}^{Z(1)},
\end{array}
\end{equation}

\begin{equation}
\begin{array}{l}
K_1^W=0, \\
{ } \\
K_{\gamma}^W=\Cuv-1-2 F_1(f',W), \\
{ } \\
K_{5\gamma}^W=-K_{\gamma}^W,
\end{array}
\end{equation}

\begin{equation}
\begin{array}{l}
\displaystyle{ K_1^S=m_f\left[
  -F_0(f,H)+F_0(f,Z)-2\frac{m_f'^2}{m_f^2}(\Cuv-F_0(f',W))
\right] } ,\\
{ } \\
\displaystyle{ K_{\gamma}^S=\Cuv-F_1(f,H)-F_1(f,Z)
+\frac{1}{2}\left(1+\frac{m_f'^2}{m_f^2}\right)(\Cuv-2F_1(f',W))}, \\
{ } \\
\displaystyle{ K_{5\gamma}^S=
+\frac{1}{2}\left(1-\frac{m_f'^2}{m_f^2}\right)(\Cuv-2F_1(f',W))}.
\end{array}
\end{equation}

\subsection{The Goldstone sector}

We do not need to be explicit about the renormalisation of this
sector in order to arrive at finite S-matrix elements. Nonetheless
we list all the vector-Goldstone mixings and Goldstone
propagators.

\subsubsection{$A-\chi_3$} \label{explicitachi3}
\begin{center}
\begin{tabular}{l|l}
\hline (b) & $(A,B)=  (W,\chi), (\chi,W),
 (c^+,c^+), (c^-,c^-), (f,f)$
\\
\hline (a) & None
\\
\hline
\end{tabular}
\end{center}

\vspace{2mm}

There is no fermionic contribution.

\begin{equation}
\Pi^{A\chi_3}(q^2)= \frac{\alpha M_W}{2\pi s_W} (1-\anlg)\left(
\Cuv - F_0(W,W) \right)\; .
\end{equation}

\noi As expected this does vanish for $\anlg=1$ and is a remnant
of the $U(1)_{\rm QED}$ gauge invariance for this value of the
parameter.

\subsubsection{$Z-\chi_3$}
\begin{center}
\begin{tabular}{l|l}
\hline (b) & $(A,B)=  (W,\chi), (\chi,W), (H,\chi_3), (H,Z),
 (c^+,c^+), (c^-,c^-), (f,f)$
\\
\hline (a) & None
\\
\hline
\end{tabular}
\end{center}

\vspace{2mm}

\beqn
\Pi_{Z\chi_3}&=& \frac{\alpha \mz}{8\pi\sww\cww} \biggl\{\cww
(-3+4\cww(1-\betat)-\kappat)\left(\cuv-F_0(W,W)\right)\nonumber\\
&+&\;\frac{\mhh}{2\mzz}\left[F_0(H,Z)-2F_1(H,Z)\right]-
\left[\frac{3\cuv}{2}-F_1(H,Z)-F_0(H,Z)\right]\nonumber\\
&+&\;\epsilont\left[\frac{\cuv}{2}-F_1(H,Z)+\frac{M_H^2}{2M_Z^2}(\cuv-F_0(H,Z)\right]
\nonumber\\ &+&\epsilont^2\left(\cuv-F_0(H,Z)\right)\ \nonumber \\
&+& \sum_f\frac{m_f^2}{M_Z^2} \left(\Cuv-F_0(f,f)\right)
\biggr\}\; .
\eeqn

\subsubsection{$W-\chi$}
\begin{center}
\begin{tabular}{l|l}
\hline (b) & $(A,B)= (H,W), (H,\chi),  (Z,\chi), (Z,W), (A,\chi),
(A,W),$ \\
    & $(c^A,c), (c^Z,c^+), (c^Z,c^-), (f,f^\prime)$
\\
\hline (a) & None
\\
\hline
\end{tabular}
\end{center}

\vspace{2mm}
\beqn
\label{pilwchi} \Pi_{W\chi_+}&=& \frac{\alpha\mw}{16\pi\sww}
\Biggl\{\cuv\left(2-\frac{3}{\cww}\right)\nonumber\\
&+&\;\cuv\left(\frac{\mhh}{\mww}\deltat+\kappat+\deltat+2\deltat^2+
\sww(18\alphat^2-12\alphat)+4\betat(4-3\cww)+18\cww
\betat^2\right)\nonumber\\ &+&\;4\sww
\left(4F_1(A,W)-F_0(A,W)\right)\nonumber\\
&+&\;4\sww\alphat \;\left(-6F_1(A,W)+6F_0(A,W)-5\alphat
F_0(A,W)+\alphat F_1(A,W)\right)\nonumber\\
&+&\;F_1(Z,W)\left(
\sww(-16-\frac{2}{\cww})-2\kappat-24\cww\betat+4\cww\betat^2\right)
\nonumber\\ &+&\;F_0(Z,W)
\left(2-4\cww+4\frac{\sww}{\cww}-16\betat+24\cww\betat-20\cww\betat^2\right)
\nonumber\\
&+&\;\frac{\mhh}{\mww}\left(2F_1(W,H)-F_0(W,H)(1+\deltat)\right)+2\left(2F_0(W,H)-F_1(W,H)\right)
\nonumber\\
&+&\;2\deltat\left(F_1(W,H)-F_0(W,H)(1+\deltat)\right)\nonumber\\
&+&\;8\alphat\sww(1-\alphat)-8\betat\left(1-\cww(1-\betat)\right)
\nonumber \\ &+& 4 \sum_{doublet}  \frac{1}{M_W^2}
\left[\frac{m_f^2+m_f'^2}{2}\Cuv
-\left(m_f'^2F_0(f',f)+m_f'^2F_1(f,f')\right) \right] \Biggr\}\; .
\eeqn

\subsubsection{$\chi_3-\chi_3$}
\begin{center}
\begin{tabular}{l|l}
\hline (b) & $(A,B)= (W,\chi), (\chi,W), (H,Z), (H,\chi_3),
 (c^+,c^+), (c^-,c^-), (f,f)$
\\
\hline (a) & $A=W, Z, \chi_3, \chi, H, c^+, c^-, c^Z$
\\
\hline
\end{tabular}
\end{center}

\vspace{2mm}

The second line of $\Pi^{\chi_3}$ shows the fermionic
contributions. Since the tadpole contribution appears with
$\Pi^{\chi_3}$, we present the formula for the sum. We note that
$C^{\chi_3}$, the coefficient for the divergent part, is
proportional to $s$ in the linear gauge.

\begin{equation}
\begin{array}{ll}
\displaystyle { \Pi^{\chi_3}(q^2)+ \frac{\delta T}{v} = } &
\displaystyle{ \frac{\alpha}{16\pi \swt} \left[
\rule[-1mm]{0mm}{6mm} C^{\chi_3} \Cuv + d_{WW}^{\chi_3} F_0(W,W) +
d_{HZ}^{\chi_3} F_0(H,Z) + d_0^{\chi_3} \right. } \\
 & \\
 & \displaystyle{ \left.
+2 q^2\sum_f \frac{m_f^2}{M_W^2} \left( \Cuv- F_0(f,f) \right)
\rule[-1mm]{0mm}{6mm} \right] }\; ,
\end{array}
\end{equation}

\begin{equation}
C^{\chi_3}= -\left( \frac{2}{\cwt}+4 \right) q^2 +\enlg
\frac{2}{\cwt}(\MHt+q^2) + \enlg^2 \frac{3}{\cwt}\MZt - 4 \knlg
q^2,
\end{equation}

\begin{equation}
d_{WW}^{\chi_3}= 4q^2 (1 + \knlg),
\end{equation}

\begin{equation}
d_{HZ}^{\chi_3}= \frac{1}{\cwt}\left( 2 \MHt - \MZt + 2q^2 -
\frac{(\MHt)^2}{\MZt} -2\enlg(\MHt+q^2) - 3 \enlg^2\MZt \right),
\end{equation}

\begin{equation}
d_{0}^{\chi_3}= \frac{1}{\cwt} \left[ \left( \frac{(\MHt)^2}{\MZt}
-\MHt \right) \logh  +  \left( \MZt -\MHt   \right)\logz -
\frac{(\MHt)^2}{\MZt} + 2\MHt - \MZt \right]\; .
\end{equation}

\subsubsection{$\chi-\chi$}
\begin{center}
\begin{tabular}{l|l}
\hline (b) & $(A,B)= (H,W), (H,\chi), (\chi_3,W), (Z,\chi), (Z,W),
(A,\chi), (A,W), (c^Z,c^+), (c^Z,c^-), (f,f^\prime)$
\\
\hline (a) & $A=A, Z, W, H, \chi_3, \chi, c^+, c^-$
\\
\hline
\end{tabular}
\end{center}

\vspace{2mm}

The second line of $\Pi^{\chi}$ shows the fermionic contribution.
Since the tadpole contribution appears with $\Pi^{\chi}$, we
present the formula for the sum. We note that $C^{\chi}$, the
coefficient for the divergent part, is proportional to $s$ in the
linear gauge.

\beqn
\label{pichichi}
\begin{array}{ll}
\displaystyle { \Pi^{\chi}(q^2) + \frac{\delta T}{v} = } &
\displaystyle{ \frac{\alpha}{16\pi\swt}
\left[\rule[-1mm]{0mm}{6mm} C^{\chi} \Cuv + d_{ZW}^{\chi} F_0(Z,W)
+ d_{HW}^{\chi} F_0(H,W) + d_{AW}^{\chi} F_0(A,W) + d_0^{\chi}
\right. } \\
 & \\
 & \displaystyle{ \left.
+ 2 q^2\sum_f \frac{m_f^2}{M_W^2} \left( \Cuv -2 F_1(f',f) \right)
 \rule[-1mm]{0mm}{6mm}\right] },
\end{array}
\eeqn

\begin{equation}
\begin{array}{ll}
C^{\chi}= & \displaystyle{ -\left( \frac{2}{\cwt}+4 \right) q^2
-32\anlg\swt\MWt +32\bnlg\swt\MWt
+16\anlg^2\swt\MWt+16\bnlg^2\cwt\MWt }\\ &
+2\dnlg(q^2+\MHt)+3\dnlg^2\MWt +2\knlg q^2 -\knlg^2\MWt,
\end{array}
\end{equation}

\begin{equation}
\begin{array}{ll}
d_{ZW}^{\chi}= & \displaystyle{
\left(2-8\swt+\frac{2}{\cwt}\right)q^2 -16\cwt\MWt
-\frac{6}{\cwt}\MWt -\frac{1}{\cwt}\MZt -8\swt \MWt +23\MWt }
\\ & -32\bnlg\swt\MWt -16\bnlg^2\cwt\MWt -2\knlg q^2
+\knlg^2\MWt,
\end{array}
\end{equation}

\begin{equation}
d_{HW}^{\chi}= 2 q^2 - \MWt + 2\MHt - \frac{(\MHt)^2}{\MWt} - 2
\dnlg(s+\MHt) - 3\dnlg^2\MWt,
\end{equation}

\begin{equation}
d_{AW}^{\chi}= 8\swt(s-\MWt) +32\anlg\swt\MWt -16\anlg^2\swt\MWt,
\end{equation}

\begin{equation}
\begin{array}{ll}
d_{0}^{\chi}=  & \displaystyle{ \logw \left( -\MHt +2\MWt -\MZt
\right)
+ \logh \left(  \frac{(\MHt)^2}{\MWt} -\MHt \right) } \\
 & \\
& \displaystyle{ + \logz \left( \frac{\MZt}{\cwt} + 8\swt\MZt-\MZt
\right)
 - \frac{(\MHt)^2}{\MWt} +2\MHt
 +6\MWt
- \frac{\MZt}{\cwt} - 6\MZt
} \\
 & \\
& \displaystyle{ + 16\anlg\swt\MWt - 16\bnlg\swt\MWt -
8\anlg^2\swt\MWt - 8\bnlg^2\cwt\MWt }\; .
\end{array}
\end{equation}

\setcounter{equation}{0}
\section{Direct determination of the charge counterterm}
\label{sec:deltaY}
\begin{figure}[htb]
\caption{Electron self energy and $e^+ e^- A$ vertex.}
\label{fig:chren}
\begin{center}
\includegraphics[width=10cm,height=7cm]{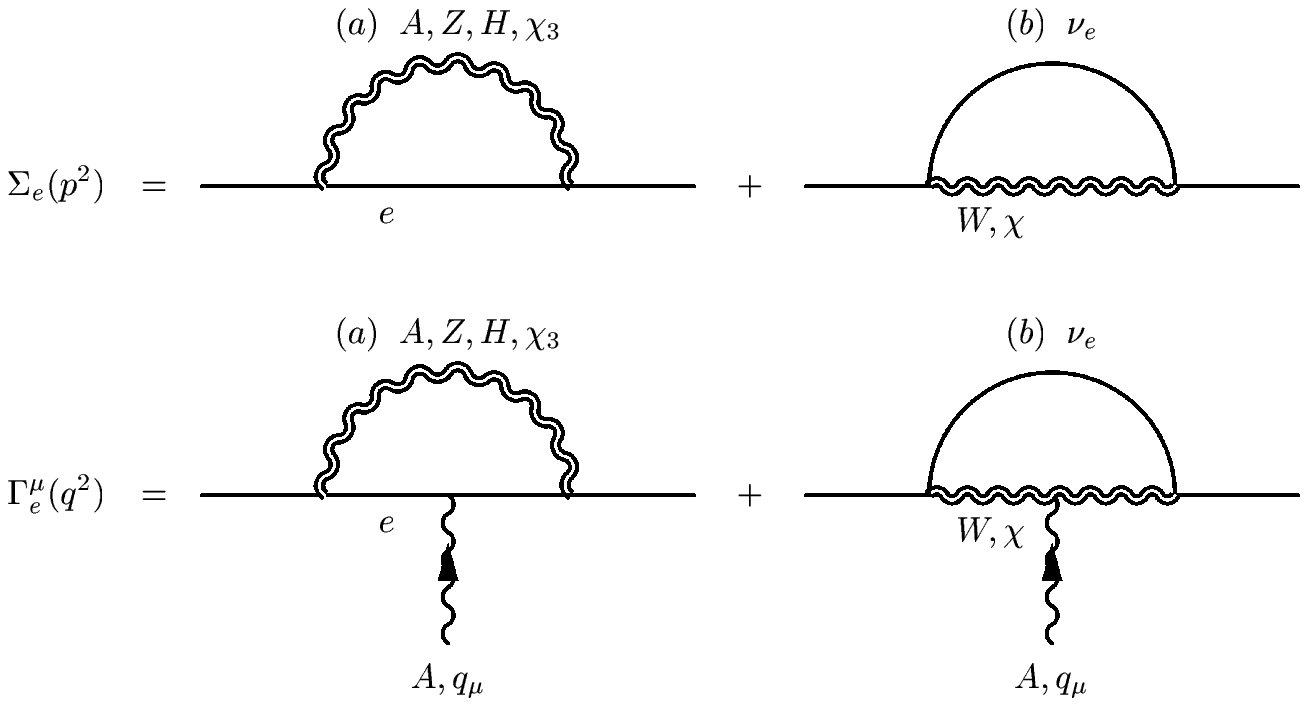}
\end{center}
\end{figure}
A direct derivation of the charge counterterm $\delta Y$
necessitates the calculation of the vertex $e^+ e^- A$ with
on-shell electrons and in the Thomson limit where the photon
momentum $q \ra 0$, $\Gamma_{e}^{\mu}(0)$. In this limit one can
relate the vertex to the electron self-energy as in depicted
\zu{chren} ($p$ is the electron momentum and $l$ is the
integration momentum). Due to the identities
\begin{equation}
\frac{\partial}{\partial p_{\mu}} \left(
\frac{-1}{\VECsl{p}+\VECsl{\ell}-m}\right) =
\frac{-1}{\VECsl{p}+\VECsl{\ell}-m} \gamma^{\mu}
\frac{-1}{\VECsl{p}+\VECsl{\ell}-m}\; ,
\end{equation}
and
\begin{equation}
\frac{\partial}{\partial p_{\mu}} \left(\frac{1}{(p+\ell)^2-M^2}
\right)= \frac{-2(p+\ell)^{\mu}}{((p+\ell)^2-M^2)^2},
\end{equation}
The major part of  $\Gamma_{e}^{\mu}(0)$ can be calculated by the
corresponding fermion self energy. Using the notation in
\zu{chren},
\begin{equation}
\mathrm{(a)}\;\; A, Z, H, \chi_3 \qquad \Gamma_{e}^{\mu}(0)=
(-e)\left.\frac{\partial}{\partial
p_{\mu}}\Sigma(p^2)\right|_{\VECsl{p}=m},
\end{equation}
\begin{equation}
\label{Wardnaive} \mathrm{(b)}\;\; W, \chi \qquad
\Gamma_{e}^{\mu}(0)= (-e)\left.\frac{\partial}{\partial
p_{\mu}}\Sigma(p^2)\right|_{\VECsl{p}=m} + G_e^{\mu}.
\end{equation}
It is only $G_e^{\mu}$ which is gauge-parameter dependent, in fact
it only depends on $\alphat$ and vanishes for $\alphat=1$.

\begin{equation}
G_e^{\mu}=
-\frac{\alpha}{4\pi}\frac{e}{s_W^2}(1-\anlg)(\Cuv-\logw)
\gamma^{\mu} L \; . \label{eq:geform}
\end{equation}

We note that for this particular value of the gauge parameter
there is a residual $U(1)_{{\rm QED}}$ symmetry and therefore it
is  no wonder that the naive Ward identity is verified in this
case in Eq.~\ref{Wardnaive}.

The gauge-parameter independent part of $\Gamma_\mu(0)$ is derived
from the (on-shell) electron self-energy. We obtain

\begin{equation}
\begin{array}{ll}
\displaystyle{
 \left.\frac{\partial}{\partial p_{\mu}}\Sigma(p^2)\right|_{\VECsl{p}=m} }&
 \displaystyle{=
(2m_e K_1'(m_e^2)+2m_e^2  K_{\gamma}'(m_e^2) + K_{\gamma}(m_e^2))
\gamma^{\mu}
+ K_{5\gamma}(m_e^2) \gamma^{\mu}\gamma_5 } \\
{ } & { } \\
& \displaystyle{= - 2\ZH{eL} \gamma^{\mu} L -  2\ZH{eR}
\gamma^{\mu} R}\; .
\end{array}
\end{equation}
Here \siki{crnrmf} is used.

The counterterm for the $e^+ e^- A$ vertex, $\Gamma_{e}^{\mu}(0)$,
is defined in \setu{ctffv}. Adding this to the loop calculation we
get
\begin{equation}
\begin{array}{ll}
\displaystyle{ \tilde{\Gamma}_{e}^{\mu}(0) } & \displaystyle{ =
\Gamma_{e}^{\mu}(0) + \hat{\Gamma}_{e}^{\mu}(0)}
\\
{ } & { } \\
& \displaystyle{ = (-e\gamma^{\mu})\left(\delta Y + \ZH{AA}-
\frac{s_W}{c_W}\ZH{ZA}\right) +  \left[ - \frac{e}{2s_Wc_W}\ZH{ZA}
\gamma^{\mu}L  + G_e^{\mu} \right]}\; .
\end{array}
\end{equation}
The second term (within square brackets) vanishes identically.
Imposing the renormalisation condition (\siki{chargerc})
$\tilde{\Gamma}_{e}^{\mu}(0)=0$,
\begin{equation}
\delta Y = - \ZH{AA} + \frac{s_W}{c_W}\ZH{ZA}\; .
\end{equation}
and we find the linear gauge result
\begin{equation}
\delta Y= \frac{\alpha}{4\pi} \left\{
-\frac{7}{2}(\Cuv-\logw)-\frac{1}{3} +\frac{2}{3}\sum_f Q_f^2
(\Cuv-\log m_f^2) \right\}.
\end{equation}

\setcounter{equation}{0}
\section{Graph theory and optimization in the generation of Feynman diagrams}
\label{sec:graph-gen-app}

We shall first introduce some technical terms. In the following
let us call a vertex or an external particle a \textsl{node}.
Similarly, let an \textsl{edge} be a connection between two nodes,
which may be a propagator or a connection between a vertex and an
external particle. Thus an edge is expressed by a pair of two
nodes (which are connected by the edge). The graph generation
process is to construct edges in all possible ways. Although
vertices of the same kind are not distinguished from each other,
they will be distinguished in a program, usually through a
sequence of numbered labels. Since a Feynman graph is a
topological object, it is independent of the way one assigns the
sequence of numbers to nodes. Therefore two graphs are
topologically the same when there is a permutation in the
sequential numbers in a graph which produces another graph. Some
permutations acting on a graph will keep the graph unchanged.
These permutations form a symmetry group of the graph whose number
of elements is needed to calculate the symmetry factor required
for the calculation of the Feynman amplitudes.

Let us consider three-loop vacuum graphs in the \(\phi^3\) model
for example.  There are two one-particle irreducible Feynman
graphs as shown in Fig.\ref{fig:e0l3}. A permutation of any two
nodes in the graph of Fig.\ref{fig:e0l3}a produces the same graph.
However in graph Fig.\ref{fig:e0l3}b, the exchange of nodes $1$
and $2$ gives different numbering of the topologically equivalent
graph, while the simultaneous exchange of nodes \(1
\leftrightarrow 2\) and
\(3 \leftrightarrow 4\) results in the same graph with the same numbering.
When a permutation of nodes in a graph produces a different
representation of the graph, one must remove either the original
graph or the new one obtained by the permutation. When the
permutation keeps a graph invariant, it is found to be an element
of the symmetry group of the graph.
\begin{figure}[htb]
\caption{{\em Three-loop vacuum graphs in \(\phi^3\) model}}
    \label{fig:e0l3}
    \begin{center}
    \includegraphics[width=8cm,clip=true]{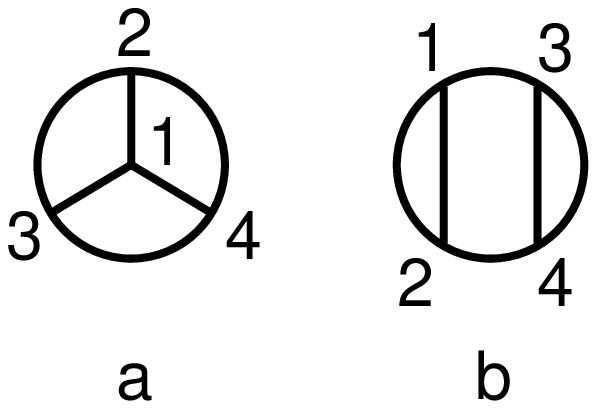}
    \end{center}
\end{figure}
\noi This symmetry structure of a graph originates from  the fact that
a vertex is symmetric under the exchange of the three legs. When
the  vertex does not have such a symmetry as in the case of QED,
it is possible to avoid the duplication of graphs at all orders of
the coupling constant\cite{grand}. Generation of tree-level graphs
is also easy since a node is distinguished from others by the set
of momenta entering the node. The most difficult cases are vacuum
graphs of the \(\phi^3\) or \(\phi^4\) model at the multi-loop
level, since all nodes in a graph are equivalent. Fortunately, the
calculation of decay or scattering processes involves external
particles that make the distinction simpler. This also makes the
code for the generation of graphs  simpler and faster.

\noi The first idea to avoid duplication is to keep already
generated graphs in the main memory or the hard disks in order to
compare them with the newly generated ones \cite{feynart}.
However as the number of graphs grows as the factorial of the
number of nodes, it is impossible to keep all the graphs in the
main memory. One is then forced to move all the generated graphs
onto the hard disk. This very much slows down the code. A more
efficient  method has been developed by graph theorists to avoid
generating duplicated graphs
\cite{orderly}. It checks a generated graph with the one obtained through
permutation and therefore requires memory space just for two
graphs. The outline of this method is  the following:
\begin{enumerate}
\item Let \(P\) be the set of all permutations among nodes of
      a graph \(G\) and \(\sigma(G)\) be a graph obtained
      by applying a permutation
      \(\sigma \in P\) to a graph \(G\).
      The set of all topologically equivalent graphs of \(G\) is
      \(\{\sigma(G) \;|\; \sigma \in P\}\).

\item We introduce the connection matrix \(M\) of a graph \(G\),
      whose matrix element \(M_{ij}\) is 1 when nodes \(i\) and
      \(j\) are connected by an edge and is 0 otherwise.
      We apply the permutation \(\sigma\) to this matrix.
      When the permuted matrix of a graph is identical to
      the matrix of another graph, these two graphs are
      topologically the same.

\item By lexicographical comparison of matrix elements among two
      connection matrices we can introduce an ordering relation,
      \(\succ\),
      among graphs. Among all topologically identical graphs, we keep only the ``larger" graph in
      this comparison .
      This means that we discard graph \(G\) when there exists  a
      permutation \(\sigma\) such that \(\sigma(G) \succ G\).

\item The resulting algorithm is as  follows:

  \begin{enumerate}
  \item When a graph \(G\) is generated, generate all possible
        permutations \(\sigma\) of the nodes.

  \item The connection matrix of \(G\) is compared with  one of
        \(\sigma(G)\).

  \item If \(\sigma(G) \succ G\), the graph \(G\) is discarded.
        Otherwise another permutation is tried in the same way.

  \item When \(G\) is found to satisfy \(G \succeq \sigma(G)\) for
        all possible permutations, \(G\) is kept.

  \item At the same time, the number \(N_s\) of such permutations \(\sigma\)
        is counted that satisfy \(G = \sigma(G)\).
  \end{enumerate}

\item The symmetry factor of \(G\) is given by  \(1/(N_s \times N_e)\),
      where \(N_e\) is the number of  permutations among edges
      that keep \(G\) invariant.

\end{enumerate}

This method was first applied to the generation of Feynman graphs
in the code
\texttt{QGRAF}\cite{qgraf}.
Even with this method, it is necessary to compare two graphs.
Unfortunately, no efficient algorithm which avoids the factorial
growth from all permutations and  that terminates within a number
of steps which is a polynomial function of the size of the graph
is known\cite{isograph}\footnote{ Such efficient algorithms in
graph theory are known as polynomial-time algorithms.}. The size
is of course determined by the number of nodes. In order to speed
up the code, one has to decrease the number of generated graphs
and the number of permutations to be tested in a consistent
way\footnote{\texttt{QGRAPH} seems to include some acceleration
method but this is not described in
\cite{qgraf}.}. A method for acceleration by a systematic classification of
graphs proposed in \cite{grc} is used in the
\texttt{GRACE} system.

Programs can be checked by two counting methods \cite{qgraf}.
First the number of generated graphs for the \(\phi^3\) or
\(\phi^4\) model is compared with what is  predicted by a graph
theoretical method. Second the sum of the number of graphs
weighted by the symmetry factor is compared with the value
calculated in zero-dimensional field theory.

\cleardoublepage
\addcontentsline{toc}{section}{Bibliography}

\end{document}